\documentclass[11pt]{article}

\usepackage[english]{babel}

\usepackage[letterpaper,top=2cm,bottom=2cm,left=3cm,right=3cm,marginparwidth=1.75cm]{geometry}

\usepackage{amsmath}
\usepackage{graphicx}
\usepackage[colorlinks=true, allcolors=blue]{hyperref}
\usepackage{setspace}
\usepackage{natbib}
\usepackage{booktabs}
\usepackage{multirow}
\usepackage{tabularx}
\usepackage{bm}
\usepackage{array}
\usepackage{amsthm}
\usepackage{amssymb}
\usepackage{hyperref}
\usepackage[title]{appendix}
\usepackage{caption} 
\usepackage[perpage]{footmisc}
\setcitestyle{authoryear,round}
\newcolumntype{L}[1]{>{\raggedright\arraybackslash}p{#1}}
\newcolumntype{C}[1]{>{\centering\arraybackslash}p{#1}}
\newtheorem{Theorem}{Theorem}  
  
\newtheorem{Hypothesis}{Hypothesis}

\title{\textbf{Trade Dynamics with Heterogeneous Fluctuations}}
\author{Yongheng Hu\footnote{School of International Business, Zhejiang International Studies University, Liuhe Road, Hangzhou 310023, China. Correspondence to: Yongheng Hu (22030101043@st.zisu.edu.cn). This working paper is the \textit{Essays on International Trade and Macroeconomics}, it is originally completed in September 2023. We have two chapters, Chapter \MakeUppercase{\expandafter{\romannumeral1}} is from Section 2 to Section 5, Chapter \MakeUppercase{\expandafter{\romannumeral2}} is from Section 6 to Section 9. I am deeply grateful to Professor.Longzheng Du, Dr.Mingyi Yang of Washington State University and Professor.Shuli Song of Zhejiang International Studies University for their many helpful suggestions and discussions. It is incomplete and still being improved, all comments and opinions about the article are welcome. Of course all remaining omissions and errors in statement or technique are mine.}}

\begin{document}
\maketitle

\begin{abstract}
In this paper, we design two chapters to discuss trade dynamics with heterogeneous fluctuations, contributing new insights to macroeconomic issues related to international trade. In the first chapter, we model general exchange rate fluctuations through stochastic processes and analyze the impact of heterogeneous price shocks on export competitiveness. We find that monetary policy and innovation both show positive effects on export trade, while monetary policy stabilizes exchange rate fluctuations to comprehensively boost provincial export competitiveness, innovation reduces its reliance on exchange rate mechanisms. The optimal policy according to exchange rate fluctuations aims to solve the wealth distribution of exporters, and it suggests that optimal policy should promote dynamic transitions in trade patterns rather than maintain existing comparative advantages in heterogeneous trade structures. In the second chapter, we model labor market fluctuations and the ability to utilize production factors through stochastic processes, and we analyze the impact of heterogeneous aggregate production shocks on general international trade. We find that labor market fluctuations only benefit international trade under the cooperation policy. Moreover, for both sanction and cooperation policy scenarios, positive shocks (i.e., shocks where average wage growth in the labor market exceeds unemployment) strengthen their impact on import trade while weakening their impact on export trade, and vice versa. Regarding the theories proposed in these two chapters, we prove them through empirical analyses using the provincial data of China.


\end{abstract}

\newpage

\section{Introduction}

The CNY exchange rate (it refers to bilateral exchange rate of the USD-CNY in this paper) serves as a crucial indicator of China's participation in the global economic cycle. It profoundly influences not only domestic trade output but also China's competitiveness in international export trade. Over the past two decades, China's export volume has shown an overall upward trend\footnote{Figure 1 illustrates the annual manufacturing industry export value changes from 2002 to 2021 for the entire nation (China) and its 31 provincial-level administrative units, measured in Billions of RMB.}. Although manufacturing industry export patterns vary across provinces, they generally adhere to the overarching theme of “short-term fluctuations and long-term growth” as illustrated in Figure 1.

\begin{figure}[htbp]
\centering
\includegraphics[width=15cm]{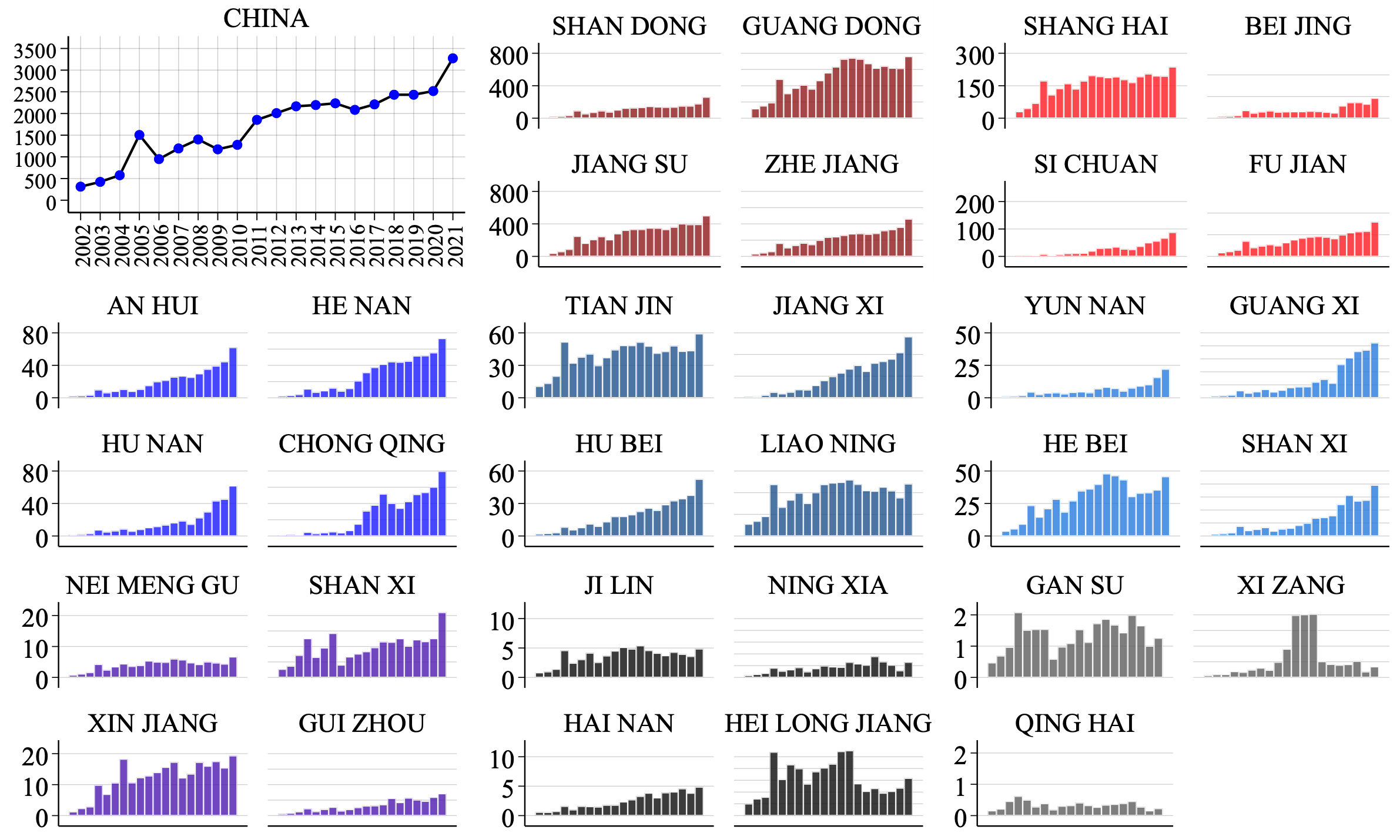}
\caption{\label{fig:F1}Manufacturing Industry Export Situation from 2002 to 2021 in China}
\end{figure}

The impact of exchange rate fluctuations on macroeconomics is remarkable (\citealp{fukui2025macroeconomic}). Existing research has found that exchange rate appreciation generally helps strengthen market competition, promotes improvements in resource allocation efficiency, and ultimately enhances manufacturing firms' productivity and product quality (\citealp{fung2008large}; \citealp{jeanneney2011does}; \citealp{ekholm2012manufacturing}; \citealp{mouradian2013real}; \citealp{tomlin2014exchange}; \citealp{hu2021exchange}). In contrast, several theoretical and empirical studies have similarly found that during currency depreciation, firms can more readily secure cash flows and profits. This environment is more conducive to firms increasing their investment in $R\&D$ and skilled labor, thereby promoting industrial innovation activities and enhancing productivity and product quality (\citealp{verhoogen2008trade}; \citealp{alvarez2009skill}; \citealp{cimoli2013technological}; \citealp{missio2016real}; \citealp{blaum2017importing}; \citealp{alfaro2018real}). However, recent research challenges this scenario. Currency depreciation reduces the real cost of borrowing from abroad, meaning that while the current exchange rate depreciates, future expected interest rates rise. Capital inflows stimulate domestic demand, triggering an economic prosperity characterized by significant growth in domestic production, consumption, and investment. And this also leads to a decline in net exports (\citealp{fukui2025macroeconomic}). Beyond the linear trends of exchange rate discussed above, academic consensus holds that freely fluctuating exchange rates are crucial (\citealp{meese1983empirical}; \citealp{engel2005exchange}; \citealp{atkeson2008trade}). Fixed exchange rate regimes are one of the primary factors contributing to currency crises (\citealp{krugman1979model}; \citealp{obstfeld1996models}). Exchange rate formation mechanisms characterized by high institutional control and limited short-term flexibility imply a divergence between the market-determined exchange rate and its theoretical equilibrium level. This divergence creates opportunities for arbitrage, currency speculation, and capital flight, leading to distortions in resource allocation and posing threats to financial stability. Increasing the flexibility of exchange rate fluctuations and allowing exchange rates to adjust automatically under market mechanisms can better respond to external economic shocks while safeguarding the independence of domestic monetary policy and avoiding imported inflation or deflation. 

In this paper, monetary policy specifically refers to exchange rate policy, which is a macroeconomic policy that adjusts the direction and magnitude of exchange rate fluctuations. It does not refer to the traditional study of interest rate policy (\citealp{kaplan2018monetary}; \citealp{Acharya}). From the “7.21 Exchange Rate Reform\footnote{Announcement of the People's Bank of China on Improving the Reform of the CNY Exchange Rate Formation Mechanism (2005). See \href{http://www.pbc.gov.cn/zhengcehuobisi/125207/125217/125922/125945/2816977/index.html}{Official Document Website}.}” on July 21, 2005, to the “8.11 Exchange Rate Reform\footnote{Statement by the People's Bank of China on Improving the Quotation of the USD-CNY Central Parity Rate (2015). See \href{http://www.pbc.gov.cn/goutongjiaoliu/113456/113469/2927054/index.html}{Official Document Website}.}” on August 11, 2015, China has consistently advanced market-oriented reforms of its exchange rate system. The “7.21 Exchange Rate Reform” marked the turning point in China's transition from a fixed exchange rate system to a managed floating exchange rate system. Since the implementation of the July 21 exchange rate reform system, technical analysts in the foreign exchange market, holding increasingly strong expectations of appreciation, have gradually replaced the intervention of the People's Bank of China, thereby gaining the dominant influence over fluctuations in the CNY exchange rate. After the “8.11 Exchange Rate Reform,” due to the unpredictability of the USD index and other currencies significantly increased uncertainty in CNY exchange rate movements, this weakened unilateral speculative forces and made the CNY's two-way floating characteristics to the USD more pronounced, i.e., USD-CNY bilateral exchange rate's volatility rises significantly. Therefore, the “8.11 Exchange Rate Reform” reversed the CNY exchange rate's single linear trend, making it in a state of long-term fluctuation.

Previous studies have examined the impact of exchange rate reform policies on the economy (\citealp{Auer}). However, research on the impact of monetary policy shocks on the competitiveness of provincial manufacturing exports in China remains scarce. Most studies examining the effects of exchange rate fluctuations focus on micro-level enterprises (\citealp{foster2008reallocation}), with few concentrating on the macro-level provincial regional dimension. Furthermore, most of the research employs variables such as “export volume,” “export quality,” and “export technological sophistication” as dependent variables for exports (\citealp{feenstra2014international}; \citealp{martin2014low}), with few employing “export competitiveness” as the dependent variable to examine CNY exchange rate impacts on exports. Regarding CNY exchange rate stability, most studies analyze the effects of CNY exchange rate fluctuations on exports through mathematical models, lacking empirical evidence. Moreover, existing studies generally assume exchange rate stability as a prerequisite when examining the economic effects of CNY exchange rate fluctuations, overlooking the control effect of monetary policy on exchange rate stability. Specifically, when the CNY experiences abnormal appreciation or depreciation, the government implements policies to restore exchange rate movements to normal trajectories. What is the mathematical logic behind such monetary policies (e.g., the 8.11 exchange rate reform) in stabilizing exchange rate fluctuations? How does this affect export competitiveness? Do policy expectations align with reality? Existing literature and academic analyses can not answer these questions. Our paper aims to fill this gap by combining theoretical modeling with empirical analysis.

We design two chapters to discuss trade dynamics with heterogeneous fluctuations. The first chapter comprises \hyperref[Section 2]{Section 2}, \hyperref[Section 3]{Section 3}, \hyperref[Section 4]{Section 4} and \hyperref[Section 5]{Section 5}. Referring to \citet{itskhoki2021exchange}, we employ stochastic processes to model exchange rate fluctuations and study the impact of heterogeneous price shocks on export competitiveness. The second chapter comprises \hyperref[Section 6]{Section 6}, \hyperref[Section 7]{Section 7}, \hyperref[Section 8]{Section 8} and \hyperref[Section 9]{Section 9}. We analyze general trade dynamics and labor market fluctuations in a small open economy through stochastic processes. And we study the impact of heterogeneous production shocks on international trade.

\section{Model Economy: Dynamics of Export Competitiveness}\label{Section 2}
\subsection{Monetary Policy and Exchange Rate Fluctuations}
Consider a simple bilateral trade scenario where the partner country's demand for domestic goods $i\in(0,1)$ is denoted as $X_t(i)$, and the substitution elasticity between different goods is $\varepsilon$. Domestic firms' production of goods $X_t$ is modeled using the \textit{Dixit-Stiglitz} aggregate equation:
\[X_t=(\int_0^1X_t(i)^{\frac{\varepsilon-1}{\varepsilon}}di)^{\frac{\varepsilon}{\varepsilon-1}}\]

Assuming that the price $P_t(i)$ of export commodity $i$ set by the domestic exporter is only influenced by the exchange rate $S_t$, and that the overall price level $P_t$ in the foreign country remains constant, the profit maximization problem for the foreign country is as follows:
\[\pi_a=\max_{X_t(i)}\left[P_tX_t-\int_0^1P_t(i)X_t(i)di\right]\]

\textit{FOC}:
\[P_t\frac{\varepsilon}{\varepsilon-1}\left(\int_0^1X_t\left(i\right)^{\frac{\varepsilon-1}{\varepsilon}}di\right)^{\left(\frac{\varepsilon}{\varepsilon-1}-1\right)}\frac{\varepsilon-1}{\varepsilon}X_t\left(i\right)^{\left(\frac{\varepsilon-1}{\varepsilon}-1\right)}=P_t\left(i\right)\]

And:
\[\left(\int_0^1X_t\left(i\right)^{\frac{\varepsilon-1}{\varepsilon}}di\right)^{\left(\frac{\varepsilon}{\varepsilon-1}-1\right)}=X_t^{\frac{1}{\varepsilon}}\]

Hence:
\[P_tX_t^{\frac{1}{\varepsilon}}X_t(i)^{-\frac{1}{\varepsilon}}=P_t(i)\]

Therefore, we can get the optimal demand $X_t(i)$ for the domestic commodity $i$ in a foreign country:
\[X_t(i)=\left(\frac{P_t(i)}{P_t}\right)^{-\varepsilon}X_t\]

Now consider the exchange rate $S_t$ for domestic exporters:
\[P_t(i)=\frac{P_t^d(i)}{S_t}\]

$P_t^d(i)$ represents the domestic price of commodity $i$. Assuming that the marginal cost of product $i$ is $MC_t$, the profit maximization problem for exporters is as follows:
\[\pi_b=\max_{P_t(i)}[(P_t(i)S_t-MC_t)X_t(i)]\]

\textit{FOC}:
\[P_t(i)=\frac{\varepsilon}{\varepsilon-1}\times\frac{MC_t}{S_t}\]

Hence:
\[X_t(i)=\left(\frac{\varepsilon}{\varepsilon-1}\times\frac{MC_t}{S_tP_t}\right)^{-\varepsilon}X_t=\left(\frac{\varepsilon}{\varepsilon-1}\right)^{-\varepsilon}\left(\frac{MC_t}{S_tP_t}\right)^{-\varepsilon}X_t\]

That is:
\[X_i(t)=K\left(\frac{MC_t}{S_tP_t}\right)^{-\varepsilon}X_t\]

Where $K$ is a constant. To simplify the problem, the RCA (Revealed Comparative Advantage Index) of domestic commodity $i$'s exports $X_t(i)$ relative to global exports $X_t^w(i)$ is defined as follows:
\[RCA_t=\frac{X_t(i)/\sum_iX_t(i)}{X_t^w(i)/\sum_iX_t^w(i)}=\frac{X_t(i)}{\sum_iX_t(i)}\times\frac{\sum_iX_t^w(i)}{X_t^w(i)}=\frac{X_t(i)}{X_t}\times\frac{Y_t}{Y_t(i)}=\left(\frac{MC_t}{S_tP_t}\right)^{-\varepsilon}\frac{KY_t}{Y_t(i)}\]

Taking the logarithm of $RCA_t$, we get $rca_t=\ln(RCA_t)$:
\[\begin{gathered}rca_t=\ln\left[\left(\frac{MC_t}{S_tP_t}\right)^{-\varepsilon}\frac{KY_t}{Y_t(i)}\right]=k+(y_t-y_t(i))-\varepsilon(mc_t-s_t-p_t)\\rca_t=\varepsilon s_t+\varepsilon p_t+k+(y_t-y_t(i))-\varepsilon mc_t\end{gathered}\]

It can be seen that if currency depreciation increases $s_t$, it enhances export competitiveness $rca_t$. Assuming the exchange rate dynamics $s_t$ is as follows:
\[s_t=\rho_ss_{t-1}+(1-\rho_s)\bar{s}_i+\theta_{st}+\zeta_t\]

Where $\rho_s\in(0,1)$, $\bar{s}_i$ represents the average (stable) exchange rate, with $i\in\{L, M, H\}$ indicating whether the long-term stable state of the exchange rate is depreciation or appreciation. An increase in $\bar{s}_i$ indicates that the currency's long-term state is depreciation. $\theta_{st}\sim \mathcal{N}(0,\sigma_{st}^2)$, where $\zeta_t$ represents monetary policy, primarily designed to reduce exchange rate volatility and diminish linear trends. Consequently, $\sigma_{st}$ is influenced by policy $\zeta_t$. Specifically, assuming policy implementation occurs at time $t^*$, then $\zeta_{t<t^*}=0$ and $\zeta_{t>t^*}=1$. Setting the policy intensity as a constant $\gamma>0$, we have:
\[\sigma_{st}=\sigma_{s0}e^{(-\gamma\zeta_t)}\]

Therefore, $RCA_t$ could be written as:
\[RCA_t=e^{\left(k+(y_t-y_t(i))-\varepsilon(mc_t-(\rho_ss_{t-1}+(1-\rho_s)\bar{s}_i+\theta_{st}+\zeta_t)-p_t)\right)}\]

\subsection{Numerical Simulation}
Now, we simplify the problem for numerical simulation: Assuming global total exports of all products remain stable without fluctuations, $y_t=1$. The global total export $y_t(i)$ of product $i$ follows a distribution: $y_t(i)=a\mathcal{Z}$, where $\mathcal{Z}\sim \mathcal{N}(0,1)$. The marginal cost $MC_t$ of the product follows a distribution: $mc_t=b+c\mathcal{Z}$, where $\mathcal{Z}\sim \mathcal{N}(0,1)$. Assuming the foreign price level remains stable with no inflation or deflation, the foreign price $P_t=1$. The remaining parameter settings are as shown in Table 1:

\begin{table}[h]
\centering
\caption{Parameter Table}
\label{tab:table1}
\small
\setlength{\tabcolsep}{15pt}
\renewcommand{\arraystretch}{1.5}
\begin{tabular}{@{}l c c@{}}
\toprule
\textbf{Parameter} & \textbf{Notation} & \textbf{Value} \\
\midrule
Total Time & \( T \) & 1000 \\
Policy Shock Time & \( t^* \) & 300 \\
Substitution Elasticity & \( \varepsilon \) & 2 \\
Persistence of Exchange Rate & \( \rho_s \) & 0.89 \\
Mean of Exchange Rate & \( \bar{s}_L \) & \( \ln(1) = 0 \) \\
& \( \bar{s}_M \) & \( \ln(e^{0.3}) = 0.3 \) \\
& \( \bar{s}_H \) & \( \ln(e^{0.6}) = 0.6 \) \\
Initial Fluctuation & \( \sigma_{s0} \) & 0.05 \\
Policy Effect & \( \gamma \) & 2 \\
World Total Export Index & \( a \) & 0.02 \\
Marginal Cost Index & \( b \) & 0.8 \\
& \( c \) & 0.05 \\
\bottomrule
\end{tabular}
\end{table}

We can get three groups of figures through numerical simulation. Figure 2, Figure 3 and Figure 4 correspond to the cases where $\bar{s}_L=0$, $\bar{s}_M=0.3$ and $\bar{s}_H=0.6$, respectively. In each group of figures:

(\textbf{A}) The top left figure shows the log of the bilateral exchange rate over time, with the red dashed line indicating the time of the monetary policy shock.

(\textbf{B}) The bottom left figure shows the change in export competitiveness $RCA_t$ over time, with the red dashed line indicating the time of the monetary policy shock.

(\textbf{C}) The top right figure shows changes in $RCA_t$ within the 95$\%$ confidence interval before and after monetary policy shock. The red dashed line indicates the policy implementation period, while the red horizontal line represents the mean value of $RCA_t$ prior to the policy shock.

(\textbf{D}) The bottom right figure shows the kernel density distributions of $RCA_t$ before and after the policy shock.

Then, three groups of figures are shown as follows:

\begin{figure}[htbp]
\centering
\includegraphics[width=15cm]{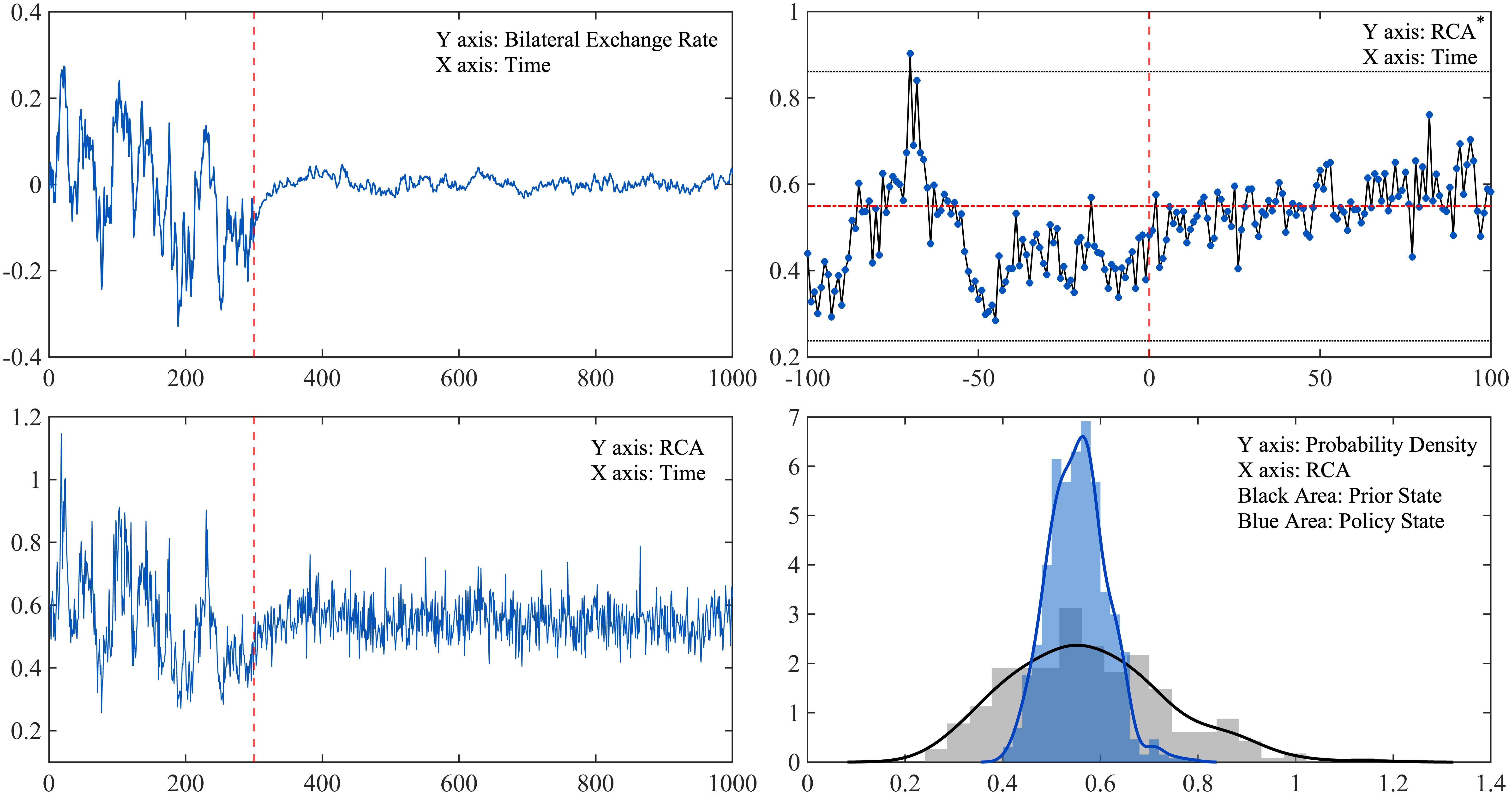}
\caption{\label{fig:F2}$RCA_t$ and Exchange Rate Fluctuations with $\bar{s}_L$}
\end{figure}

\begin{figure}[htbp]
\centering
\includegraphics[width=15cm]{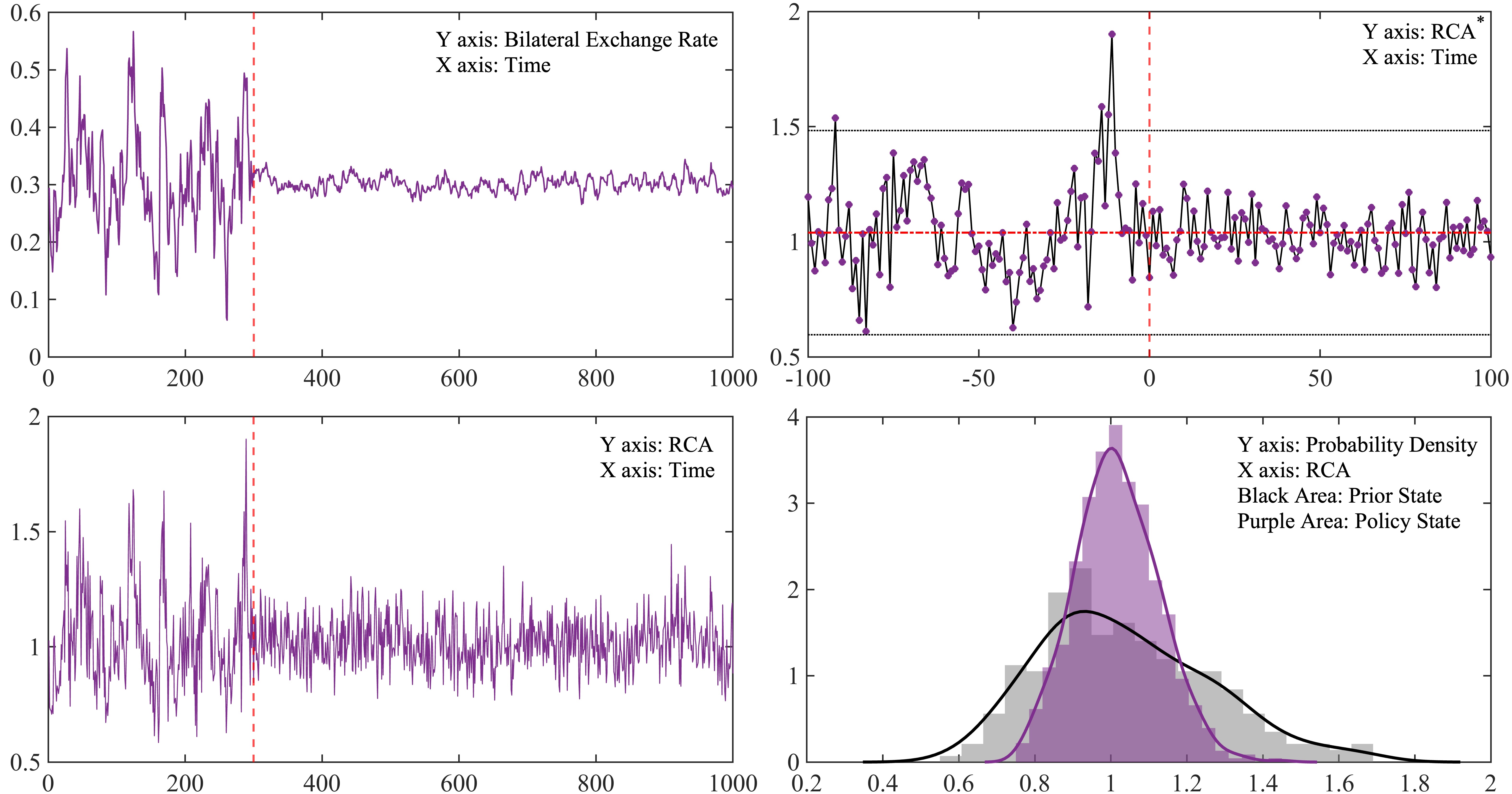}
\caption{\label{fig:F3}$RCA_t$ and Exchange Rate Fluctuations with $\bar{s}_M$}
\end{figure}

\begin{figure}[htbp]
\centering
\includegraphics[width=15cm]{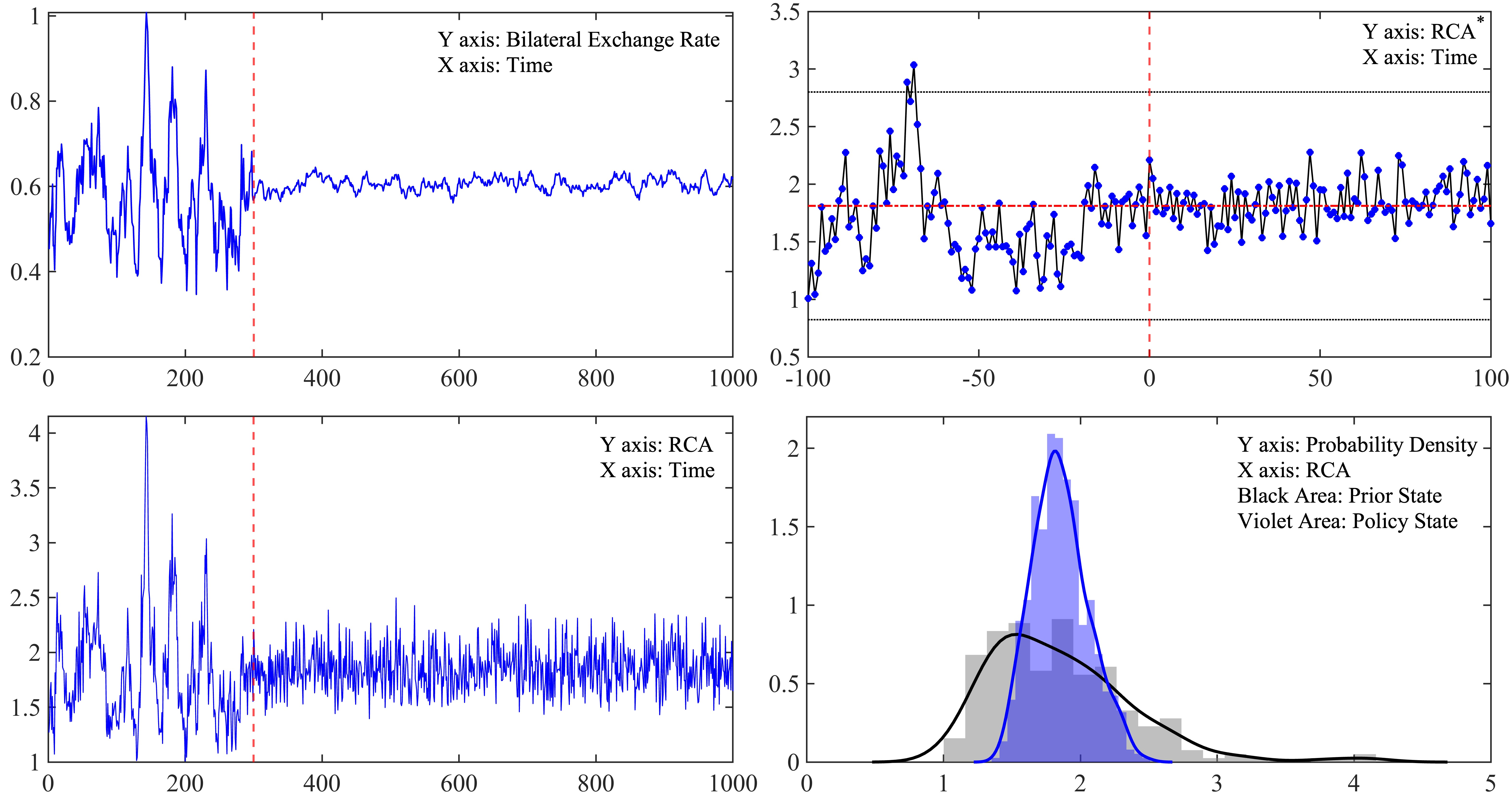}
\caption{\label{fig:F4}$RCA_t$ and Exchange Rate Fluctuations with $\bar{s}_H$}
\end{figure}

As shown in Figure 2, Figure 3 and Figure 4: After the implementation of monetary policy $\zeta_t$, the volatility of the bilateral exchange rate decreases, and the fluctuation in export competitiveness $RCA_t$ also diminishes accordingly. $RCA_{t>t^*}$ gradually converges toward the mean $\mathbb{E}[RCA_{t<t^*}]$ prior to the policy shock. Moreover, the variance of the kernel density distribution graph for $RCA_t$ decreases after the policy shock, indicating a more concentrated distribution. Additionally, when the currency's long-term state depreciates, that is, as $\bar{s}_i$ increases from $\bar{s}_L=0$ to $\bar{s}_H=0.6$, the mean of $RCA_t$ gradually increases, and the kernel density distribution graph shifts to the right overall, signifying enhanced export competitiveness.

\section{Theoretical Analysis and Research Design}\label{Section 3}
\subsection{Research Hypothesis}

\textbf{Exchange Rate Fluctuations and Export Competitiveness}: More generally, under the conditions of international perfect competition, the demand for export processing is a function of the external real exchange rate, real wages, real interest rates, and foreign real income. A depreciation of the external real exchange rate benefits exports, enabling enterprises to secure more global orders, thus promoting investment and increasing employment. This is because a depreciation of the exchange rate signifies a decline in the value of the domestic currency relative to foreign currencies. This makes the prices of domestic export commodities relatively lower in the international market. As a result of enhanced price competitiveness, the purchasing desire of foreign importers is stimulated, enabling exporters to exchange more domestic currency when receiving foreign currencies, thus increasing export profits. Moreover, the relative price reduction of export commodities in international markets enables exporters to secure more orders and market share. Thus, they can expand production scale and enhance production efficiency, further promoting the development of international trade. However, excessive currency depreciation will have negative effects on export trade. While domestic currency depreciation may temporarily boost export competitiveness by making exports priced in foreign currencies relatively cheaper, excessive and prolonged devaluation can significantly increase costs for imported raw materials. which may drive up domestic prices and trigger the inflation from cost increasing. Moreover, if a country's currency remains excessively depreciated for a long period, it may give international markets the impression that the country's economy is unstable and carries higher risks. International buyers may become concerned about whether exporters' ability to maintain consistent supply and quality control could be compromised by domestic economic conditions affected by currency depreciation. Consequently, they may be more cautious in selecting suppliers, preferring exporters from countries with relatively stable currencies. This leads to a crisis of confidence for domestic exporters in international competition, resulting in the loss of potential customers and market share. Therefore, combined with the theoretical model presented in the paper and the analysis above, we propose the first hypothesis:

\begin{Hypothesis}
Within a certain range, depreciation of the USD-CNY bilateral exchange rate can enhance provincial export competitiveness. However, continued depreciation beyond this range will be detrimental to further improving provincial export competitiveness.
\end{Hypothesis}

\noindent\textbf{Monetary Policy and Export Competitiveness}: This paper selects the “8.11 Exchange Rate Reform” as the policy shock\footnote{The “8.11 Exchange Rate Reform” significantly impacted the foreign exchange market, triggering sharp fluctuations in the short term. On December 11, 2015, the People’s Bank of China authorized the China Foreign Exchange Trade System (CFETS) to begin publishing the “CFETS CNY Exchange Rate Index.” This moves effectively reduced market influence in the exchange rate formation mechanism. The “8.11 Exchange Rate Reform” and the “12.11 Exchange Rate Reform” collectively refined the formation mechanism for the USD-CNY central parity rate. For these reasons, and since this paper utilizes annual CNY exchange rate data, the year of monetary policy impact is set as 2016.}. Here, we obtained Figures 5 and Figure 6 through empirical data analysis: 

(\textbf{A}) Figure 5 illustrates the trend of the USD-CNY bilateral exchange rate from January 2008 to June 2024. We selected the period from January 2008 to December 2021 as the research sample to avoid the influence of more additional exogenous shocks on the bilateral exchange rate. The dashed line on the y-axis represents the average CNY exchange rate of 6.58. This value was calculated by taking the annual average of monthly closing rates from January 2008 to December 2021, then computing the overall arithmetic mean of these annual averages. The dashed line on the x-axis marks the timing of exchange rate reform policy shocks.

(\textbf{B}) Left graph in Figure 6: The size of each scatter point represents the weight of provincial administrative units. A larger scatter point indicates more provinces approaching that RCA level in the given year, while a smaller point indicates fewer provinces. This graph reflects annual changes in overall provincial RCA levels. The solid line on the y-axis denotes the range of relatively higher RCA values $(0.8-1.3)$, while the dashed line on the x-axis marks the timing of the exchange rate reform policy shock. Right graph in Figure 6: The kernel density plot represents the distribution of provincial RCA levels. The black area indicates RCA levels prior to the policy shock $(t<2016)$, while the blue area shows RCA levels after the policy shock $(t>2016)$. This reveals that RCA levels became more concentrated after the policy shock, which matches all results from the numerical simulation in section 2.

\begin{figure}[htbp]
\centering
\includegraphics[width=11cm]{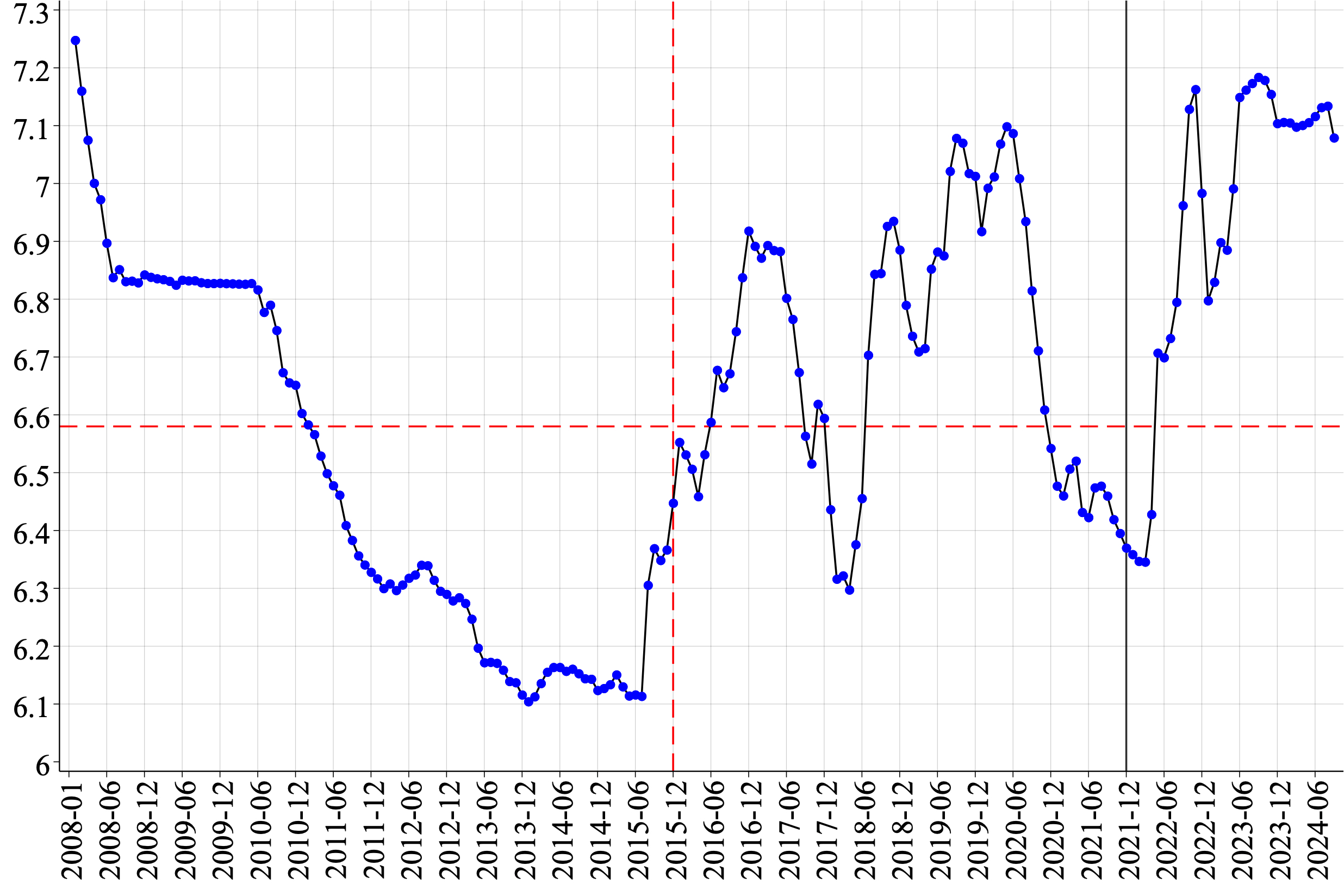}
\caption{\label{fig:F5}RMB-USD Bilateral Exchange Rate Fluctuation with Time}
\end{figure}

\begin{figure}[htbp]
\centering
\includegraphics[width=15cm]{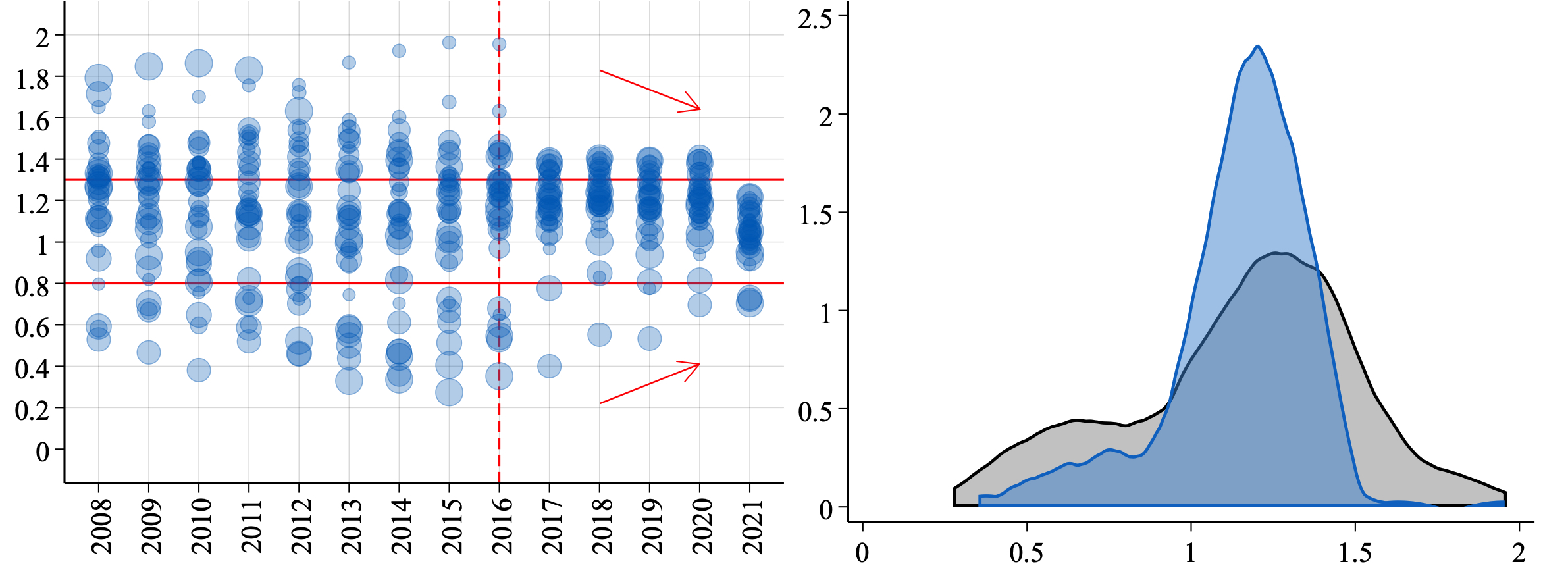}
\caption{\label{fig:F6}The Situation of Provincial RCA Changes with Time}
\end{figure}

By analyzing and studying the monthly trends in the CNY exchange rate from 2008 to 2021, as shown in Figure 5, and the annual changes in the provincial export competitiveness index (RCA), as shown in Figure 6, we found that between 2008 and 2016, the CNY exchange rate experienced an overall abnormal appreciation with a strong and wide-ranging linear trend. During this period, the RCA index exhibited significant variation, with the export competitiveness of most provinces remaining at relatively low levels. After 2016, the linear trend of the CNY exchange rate weakened, fluctuating around the average value of “6.58,” with most exchange rates remaining above this level. Furthermore, between 2016 and 2021, the range of provincial RCA indices gradually narrowed, with provincial export competitiveness converging toward a more homogeneous and relatively higher level. This is consistent with our analysis in Figures 2, Figure 3 and Figure 4 of the numerical simulation in section 2, which examined the impact of monetary policy on export competitiveness, i.e., RCA. It is obvious that the “8.11 Exchange Rate Reform” policy not only moderated extreme fluctuations in the CNY exchange rate, stabilizing its fluctuations, but also may played a role in enhancing overall export competitiveness. Based on this, we propose the second hypothesis:

\begin{Hypothesis}
Monetary policy, i.e., the “8.11 Exchange Rate Reform”, effectively stabilized the USD-CNY bilateral exchange rate and enhanced the overall competitiveness of exports.
\end{Hypothesis}

\noindent\textbf{Mechanism of Innovation}: Random fluctuations in exchange rates introduce uncertainty risks. Influenced by factors such as balance of payments conditions, shifts in national macroeconomic policies, market expectations, government interventions and geopolitical conflicts, exchange rates are constantly in a fluctuation state. For export enterprises, it is difficult to accurately predict the future trend of their domestic currency against the currencies of trading partner countries over a given period. Therefore, in practice, the impact of exogenous factors on exchange rates often offsets the effects of deliberate interventions. Relying solely on currency depreciation to significantly enhance export competitiveness seems difficult.

Hence, beyond identifying exchange rate mechanisms that enhance export competitiveness, this paper focuses more on identifying variables that can substitute for exchange rates in positively boosting export competitiveness. Such variables can exhibit desirable trends aligned with deliberate development plans or objectives, and can not only significantly enhance export competitiveness but also mitigate the negative impact of random exchange rate fluctuations on export competitiveness. Accordingly, we select the “\textit{Innovation}” factor as the moderating variable for the experimental section of this paper:

Innovative products feature convenience and intelligence, enabling rapid information acquisition, application and storage, thereby significantly enhancing operational efficiency. Furthermore, these products offer personalized customization, delivering more precise and tailored services to better satisfy user needs and elevate satisfaction levels.

Innovative technologies serve as the production methods for innovative products. The widespread application of innovative technologies such as artificial intelligence and blockchain makes the product manufacturing process more intelligent and automated, enabling comprehensive monitoring of equipment status, production efficiency and the supply chain. AI technology, powered by data analysis, builds an intelligent decision support system. By exploring vast data elements, it delivers deep insights, helping enterprises understand production processes and the entire supply chain operation. Simultaneously, AI technology autonomously adjusts production schedules, optimizes resource allocation, and even predicts uncertainties in production, driving the intelligent transformation of manufacturing.

Innovative patterns represent innovative approaches to product sales. Through internet platforms and mobile communication technologies, innovative products can enter global markets as data elements, enabling transnational sales and services. This attracts more users, enhances brand awareness and increases market share. Furthermore, by monitoring product sales and trade in real time via platforms, a user feedback loop is established to gather data from both the supply and demand sides. Through data analysis and forecasting, more scientific, rational and consistent production decisions aligned with dynamic market changes can be made, leading to a better guidance of trade strategies.

Theoretically, innovation endows export products with high value-added attributes and enhances product quality. Simultaneously, innovation-driven intelligent production, sales and feedback patterns can significantly boost economic efficiency and reduce costs. Therefore, from these two perspectives, innovation exerts a positive influence on export competitiveness through both “expanding demand” and “reducing costs”.

Based on the model and parameter assumptions in Section 2.2, we will analyze how the exchange rate mechanism for export competitiveness is affected by innovation shocks. We assume that innovation $\Gamma_t>0$ is stable and follows the following stochastic process:
\[\ln(\Gamma_{it})=\rho_\Gamma\ln(\Gamma_{t-1})+(1-\rho_\Gamma)\ln(\bar{\Gamma}_i)+\omega_{\Gamma t}\]

Where $\omega_{\Gamma_t}\sim\mathcal{N}(0,\sigma_{\Gamma}^2)$, $\bar{\Gamma}_i$ represents the average level of innovation, which is associated with regional development capabilities characterized by heterogeneity.

Innovation can enhance production efficiency and reduce marginal costs, then:
\[MC_t^{new}=\frac{MC_t}{\Gamma_{it}^{\psi_1}}\]

Where $\psi_1\in(0,1)$ measures the elasticity of innovation to marginal cost reduction.

Innovation enhances product quality and expands export demand, then:
\[X_t^{new}(i)=X_t(i)\Gamma_{it}^{\psi_2}\]

Where $\psi_2\in(0,1)$ measures the direct contribution of innovation driven product quality improvements to export growth.

Therefore, the dynamic of $RCA_t^{new}$ under innovation shocks can be expressed as:
\[RCA_t^{new}=\Gamma_{it}^{\psi_2}\left(\frac{MC_t}{\Gamma_{it}^{\psi_1}S_tP_t}\right)^{-\varepsilon}\frac{KY_t}{Y_t(i)}\]

Taking the partial derivative of $RCA_t^{new}$ regarding $S_t$, we get:
\[\frac{\partial\log(RCA_t^{new})}{\partial\log(S_t)}=\varepsilon+(\psi_2+\varepsilon\psi_1)\times\frac{\partial\log(\Gamma_{it})}{\partial\log(S_t)}\]

We assume that the innovation shock is exogenous, meaning innovation is independent of the exchange rate, i.e., $(\partial \log(\Gamma_{it}))/(\partial\log(S_t))=0$. Then the dependence of $RCA_t^{new}$ on the exchange rate can be expressed as $(\partial \log(RCA_t^{new}))/(\partial\log(S_t))=\varepsilon$. 

However, following the innovation shock, we must consider the moderating effect of innovation on the exchange rate sensitivity of $RCA_t^{new}$. Based on the theoretical analysis in this paper, we argue that innovation will help reduce the dependence of export competitiveness on the exchange rate. Therefore, the greater the degree of innovation, the weaker the exchange rate sensitivity of $RCA_t^{new}$. Accordingly, we set:
\[\frac{\partial\log(RCA_t^{new})}{\partial\log(S_t)}=\varepsilon e^{-\phi_\Gamma\Gamma_{it}}\]

Where $\phi_{\Gamma}>0$ measures the moderating intensity of innovation on the exchange rate sensitivity of $RCA_t^{new}$. Thus, we obtain the complete dynamic:
\[RCA_t^{new}=\Gamma_{it}^{\psi_2+\varepsilon\psi_1}\left(\frac{MC_t}{S_tP_t}\right)^{-\varepsilon e^{-\phi_\Gamma\Gamma_{it}}}\frac{KY_t}{Y_t(i)}\]

We assume that export competitiveness prior to the innovation shock is stable, i.e., $RCA_t=\overline{RCA}$. Then:
\[\frac{KY_t}{Y_t(i)}=\overline{RCA}\left(\frac{MC_t}{S_tP_t}\right)^\varepsilon\]

Hence:
\[RCA_t^{new}=\overline{RCA}\times\Gamma_{it}^{\psi_2+\varepsilon\psi_1}\times\left(\frac{MC_t}{S_tP_t}\right)^{-\varepsilon e^{-\phi_{\Gamma}{\Gamma}it}+\varepsilon}=\overline{RCA}\times\Gamma_{it}^{\psi_2+\varepsilon\psi_1}\times\left(\frac{MC_t}{S_tP_t}\right)^{-\varepsilon\left(e^{-\phi_{\Gamma}\Gamma_{it}}-1\right)}\]

We set $rca_t^{new}=\ln(RCA_t^{new})$, then:
\[rca_t^{new}=\overline{rca}+(\psi_2+\varepsilon\psi_1)\gamma_{it}-\varepsilon(e^{-\phi_\Gamma\Gamma_{it}}-1)(mc_t-s_t-p_t)\]

Let $\Delta^i=e^{-\phi_\Gamma\Gamma_{it}}-1$, where $\Delta^i\in(0,1)$. The stronger the innovation capability, the greater the reduction in exchange rate sensitivity. Therefore:
\[rca_t^{new}=\overline{rca}+(\psi_2+\varepsilon\psi_1)\gamma_{it}+\varepsilon\Delta^is_t+\varepsilon\Delta^i(p_t-mc_t)\]

Accordingly, for regions with high innovation capabilities, we get:
\[\begin{gathered}
\ln(\Gamma_t)=\rho_\Gamma\ln(\Gamma_{t-1})+(1-\rho_\Gamma)\ln(\bar{\Gamma}_h)+\omega_{\Gamma t}\\rca_t^{new}=\overline{rca}+(\psi_2+\varepsilon\psi_1)\gamma_t^h+\varepsilon\Delta^hs_t+\varepsilon\Delta^h(p_t-mc_t)
\end{gathered}\]

Similarly, for regions with medium innovation capabilities, we get:
\[\begin{gathered}
\ln(\Gamma_t)=\rho_\Gamma\ln(\Gamma_{t-1})+(1-\rho_\Gamma)\ln(\bar{\Gamma}_m)+\omega_{\Gamma t}\\rca_t^{new}=\overline{rca}+(\psi_2+\varepsilon\psi_1)\gamma_t^m+\varepsilon\Delta^ms_t+\varepsilon\Delta^m(p_t-mc_t)
\end{gathered}\]

Ultimately, for regions with low innovation capacity, we get:
\[\begin{gathered}
\ln(\Gamma_t)=\rho_\Gamma\ln(\Gamma_{t-1})+(1-\rho_\Gamma)\ln(\bar{\Gamma}_l)+\omega_{\Gamma t}\\rca_t^{new}=\overline{rca}+(\psi_2+\varepsilon\psi_1)\gamma_t^l+\varepsilon\Delta^ls_t+\varepsilon\Delta^l(p_t-mc_t)
\end{gathered}\]

Now, we assume that export competitiveness is exogenous and stable prior to the innovation shock, i.e., $\overline{rca}=1$. The marginal effect of innovation on the economy is exogenous, i.e., $\psi_2+\varepsilon\psi_1=1.5$. For heterogeneous innovation capabilities, we always have $\bar{\Gamma}_h>\bar{\Gamma}_m>\bar{\Gamma}_l$. Therefore, we set $(\bar{\Gamma}_h,\bar{\Gamma}_m,\bar{\Gamma}_l)=(e^{1.5},e^{0.9},e^{0.4})$. Regarding innovation's moderating effect, we assume that regions with higher innovation capacity exhibit greater reduction in export competitiveness's exchange rate sensitivity. Therefore, we have $1>\Delta^l >\Delta^m>\Delta^h>0$. Assuming an initial exchange rate sensitivity $\varepsilon=0.3$, we set $(\varepsilon\Delta^h, \varepsilon\Delta ^m, \varepsilon\Delta ^l) = (0.01, 0.05, 0.1)$.

Referring to Section 2.1, we assume both economic costs and prices are exogenous. Here, $(p_t-mc_t)$ follows a standard normal random variable function. For exchange rates, unlike the random process fluctuating around the mean in Section 2.1, we are more concerned with the sensitivity of export competitiveness to exchange rates. Therefore, we model exchange rate dynamics as a stochastic process with a stronger linear trend\footnote{This can also be interpreted as coming from the component of the aggregate exchange rate dynamics in Section 2.1 with a stronger linear trend, and using this part of the exchange rate dynamics more intuitively shows the impact of the exchange rate on export competitiveness.}, i.e., $s_t=s_{t-1}+\theta_{st}$, where $\theta_{st}\sim\mathcal{N}(0,0.05)$. For innovation dynamics, we set $\omega_{\Gamma_t}\sim\mathcal{N}(0,0.01)$. We assume that the innovation process, as a strategic decision in economic development, is stable, hence $\rho_{\Gamma}=0.1$.

\begin{figure}[htbp]
\centering
\includegraphics[width=15.5cm]{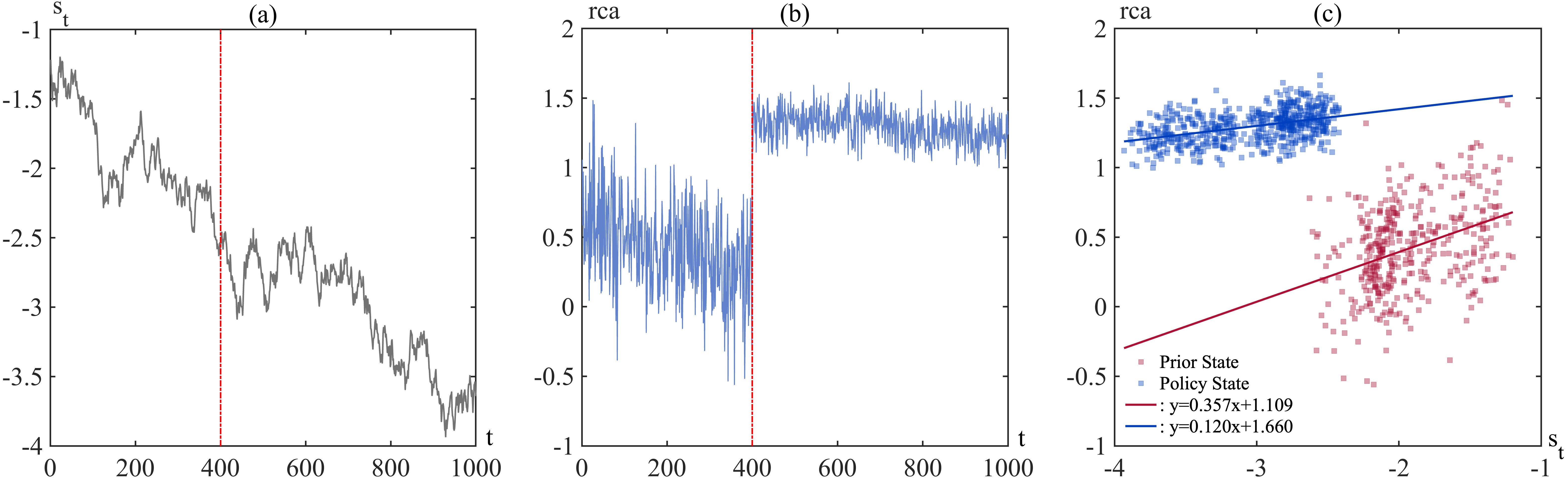}
\caption{\label{fig:F7}Moderator with Low Level of Innovation}
\end{figure}
\vspace{-0.3cm}
\begin{figure}[htbp]
\centering
\includegraphics[width=15.5cm]{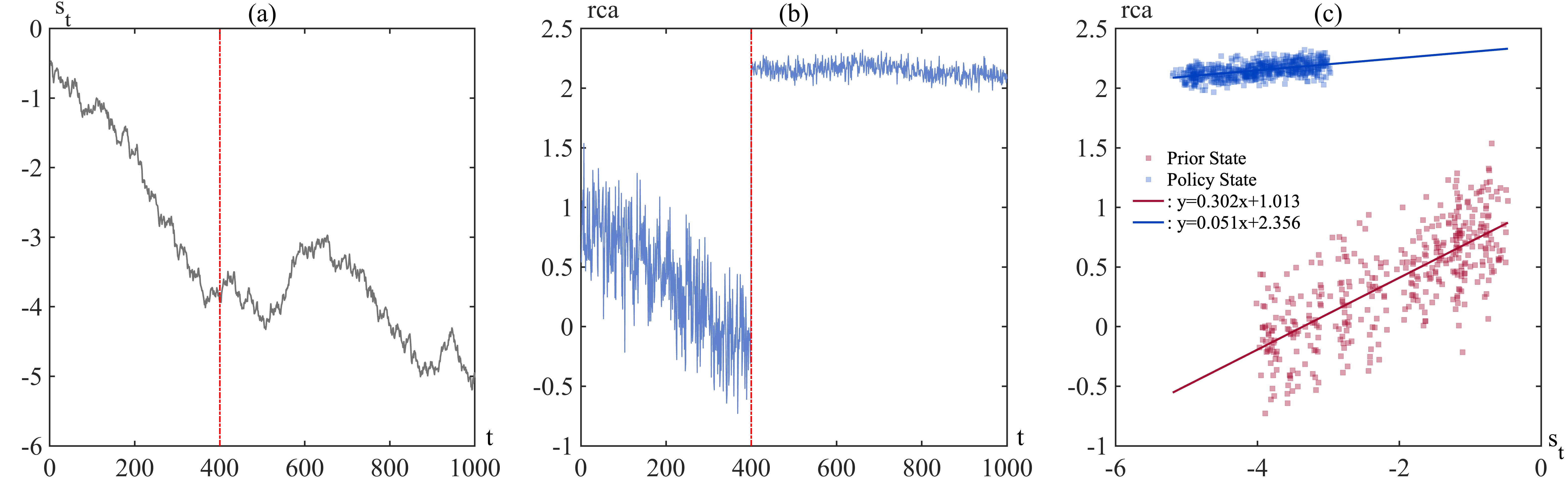}
\caption{\label{fig:F8}Moderator with Medium Level of Innovation}
\end{figure}
\vspace{-0.3cm}
\begin{figure}[htbp]
\centering
\includegraphics[width=15.5cm]{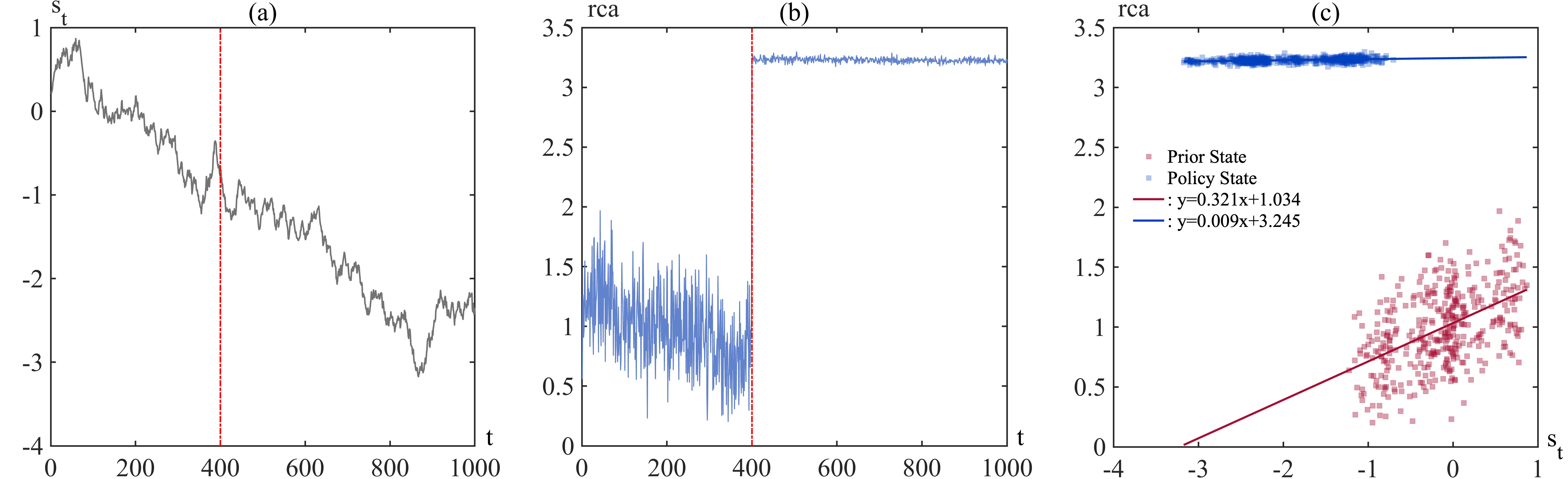}
\caption{\label{fig:F9}Moderator with High Level of Innovation}
\end{figure}

The numerical simulation results are shown in Figure 7, Figure 8 and Figure 9. Here, a dynamic process with a total duration of 1000 periods is observed, when an innovation shock occurs at period 400.

Figure 7, Figure 8 and Figure 9 illustrate low-level innovation, medium-level innovation and high-level innovation, respectively. In each figure: 

Figure (a) shows the dynamics of the logarithmic exchange rate, which maintains a relatively declining trend regardless of whether it is affected by innovation shocks, indicating a bilateral exchange rate appreciation. 

Figure (b) shows the dynamics of logarithmic export competitiveness. Prior to the innovation shock, its trend aligns with the dynamics of the logarithmic exchange rate, indicating that bilateral exchange rate appreciation would diminish domestic export competitiveness. Following the innovation shock, the level of logarithmic export competitiveness increases, demonstrating innovation's positive impact on export competitiveness. Furthermore, as innovation levels rise, the dynamics of logarithmic export competitiveness become increasingly less influenced by exchange rate trends and gradually stabilize.

Figure (c) measures the moderating effect of innovation using a random sample. We find that innovation indeed reduces the dependence of export competitiveness on exchange rates. Following an innovation shock, export competitiveness remains at a high level even when bilateral exchange rates appreciate. Furthermore, by analyzing the slope of the linear fit, we observe that the correlation between export competitiveness and exchange rates progressively diminishes as innovation levels increase. Notably, under high innovation levels, the linear relationship between the two variables is only 0.009. This preliminary finding suggests that within regions possessing strong innovation capabilities, exchange rates cease to be the primary determinant of export competitiveness.

Therefore, based on the analysis above, this paper argues that innovation can enhance export competitiveness and reduce its dependence on exchange rates. Accordingly, Hypothesis 3 is proposed:

\begin{Hypothesis}
Innovation development can enhance export competitiveness. And exchange rate fluctuations cease to be the primary factor affecting export competitiveness in the regions with higher level of innovation.
\end{Hypothesis}

Therefore, the following main task of this paper is to design econometric experiments to prove Hypothesis 1, Hypothesis 2 and Hypothesis 3.

\subsection{Research Variable}
In empirical section, the primary research variables are as follows: the independent variable is the dynamics of the USD-CNY bilateral exchange rate, the dependent variable is the RCA index measuring provincial manufacturing export competitiveness, the policy variable is “8.11” exchange rate reform policy of The People's Bank of China (PBC) and the moderating variable is innovation.

\noindent\textbf{RCA}: This paper adopts the Revealed Comparative Advantage (RCA) index as the dependent variable for analysis, which serves as the most persuasive indicator for measuring the international competitiveness of a country or region's commodities and industries. The RCA index represents the ratio of a country's or region's export value in a specific sector during a given period to its total export value during that period, compared to the share of that sector's global export value during the same period. We refer to the WIOD2016 classification\footnote{The 18 manufacturing industries in WIOD2016 are: C5 Food, beverages, and tobacco products, C6 Textiles, apparel, leather, and related products, C7 Wood, wood products, and cork products (excluding furniture), straw and woven goods, C8 Paper and paper products, C9 Printing and reproduction of recorded media, C10 Coke and refined petroleum products, C11 Chemicals and chemical products, C12 Basic pharmaceutical products and pharmaceutical preparations, C13 Rubber and plastic products, C14 Other non-metallic mineral products, C15 Basic metals, C16 Metal products, except machinery and equipment, C17 Computers, electronic, and optical products, C18 Electrical equipment, C19 Machinery and equipment not elsewhere classified, C20 Motor vehicles, trailers, and semi-trailers, C21 Other transport equipment, C22 Furniture (Other Manufacturing).} for manufacturing to calculate provincial manufacturing export competitiveness indices:

\[RCA_{ijt}=\frac{X_{ijt}/\sum_i^nX_{ijt}}{\sum_j^GX_{ijt}/\sum_j^G\sum_i^n(X_{ijt})}\]

$RCA_{ijt}$ denotes the Revealed Comparative Advantage Index for industry $i$ in province $j$ at time $t$. $X_{ijt}$ represents the export value of industry $i$ in province $j$ to foreign markets at time $t$. $\sum_i^nX_{ijt}$ signifies the total export value of all industries in province $j$ to foreign markets at time $t$. $\sum_j^GX_{ijt}$ represents the total export value of industry $i$ from all countries in the world market at time $t$. $\sum_j^G\sum_i^n(X_{ijt})$ represents the total export value of all industries from all countries in the world market at time $t$. In general, $RCA\rightarrow1$ indicates a neutral relative comparative advantage, with no discernible relative strength or weakness. $RCA>1$ signifies that the export share of this commodity exceeds its global export share, indicating that the commodity possesses a comparative advantage in the international market. $0<RCA<1$ indicates a lack of comparative advantage in the international market\footnote{Specifically: $RCA>2.5$ indicates extremely strong export competitiveness. $1.25<RCA<2.5$ indicates strong export competitiveness. $0.8<RCA<1.25$ indicates moderate export competitiveness. $0<RCA<0.8$ indicates weak export competitiveness.}.

\noindent\textbf{EXRATE}: This paper selects the bilateral exchange rate of the USD-CNY as the core independent variable. By analyzing monthly closing CNY exchange rate data (monthly closing CNY equivalent per USD) from 2008 to 2021, the arithmetic mean is calculated to derive annual CNY exchange rate data (annual average CNY equivalent per USD).

\noindent\textbf{INNOVATION}: This study utilizes data from the 2008–2021 China Regional Innovation Capability Evaluation Report. It employs the logarithm of provincial innovation capability scores as the variable measuring innovation across provinces (\textbf{lnInnovation}). Furthermore, it selects the logarithm of patent application approvals at the end of each year from 2008 to 2021 for each province, also serving as a variable of provincial innovation levels (\textbf{lnPatent}). 

Hence, two groups of variables exist in the regression analysis: the first group (\textbf{Group-a}) is classified based on innovation capability scores, with provinces whose average innovation capability scores exceed the overall average categorized as having high innovation capability, and those below the overall average categorized as provinces with low innovation capability. The second group (\textbf{Group-b}) is classified based on the volume of innovation-related patent authorizations, with provinces whose average patent authorization volumes exceed the overall average categorized as having high innovation capability, and vice versa.

The weighting for China's regional innovation capability assessment is as follows: Knowledge Creation 0.15, Knowledge Acquisition 0.15, Enterprise Innovation 0.25, Innovation Environment 0.25, Innovation Performance 0.2. Regional innovation capability is evaluated through innovation capability scoring. From 2001 to 2021, China's average regional innovation capability score was 28.8. Between 2008 and 2021, provinces with average scores exceeding 28.8 included: \textit{Beijing}, \textit{Tianjin}, \textit{Liaoning}, \textit{Shanghai}, \textit{Jiangsu}, \textit{Zhejiang}, \textit{Anhui}, \textit{Fujian}, \textit{Shandong}, \textit{Hubei}, \textit{Guangdong}, \textit{Chongqing}, \textit{Sichuan}, and \textit{Sh$\check{a}$nxi}. Provinces with scores below this average include: \textit{Hebei}, \textit{Hainan}, \textit{Jilin}, \textit{Heilongjiang}, \textit{Henan}, \textit{Hunan}, \textit{Sh$\bar{a}$nxi}, \textit{Jiangxi}, \textit{Inner Mongolia}, \textit{Guangxi}, \textit{Guizhou}, \textit{Yunnan}, \textit{Gansu}, \textit{Qinghai}, \textit{Ningxia}, and \textit{Xinjiang}. From 2008 to 2021, the average annual year-end patent application volume across Chinese provinces was approximately 50,000. Therefore, provinces with an average annual year-end patent application volume exceeding 50,000 during this period were classified as high innovation provinces, while those below this threshold were categorized as low innovation provinces\footnote{Due to space constraints, specific provinces will not be listed for the second group. If information is needed, please contact the author for details.}.

We further incorporate provincial macro control variables to mitigate endogeneity issues from omitted variables: Urban registered unemployment rate ($\%$), \textbf{Unemployment}. Logarithm of permanent resident population (10000 persons), \textbf{lnPopulation}. Logarithm of total retail sales of consumer goods (100 million RMB), \textbf{lnRetail}. Logarithm of total agricultural machinery power (10000 kWh), \textbf{lnPower}. GDP growth rate ($\%$), \textbf{Vgdp}. Development index of market intermediary organizations and legal environment, \textbf{Law}. Local government general budget expenditure (100 million RMB), \textbf{lnGovernment}. And total output value of primary industries, \textbf{lnFirst}.

Considering data availability and completeness, this study utilizes macroeconomic data from China's provincial regions for the period 2008–2021 (excluding Tibet, encompassing 30 provincial administrative units). Variable construction data primarily originates from: The National Bureau of Statistics official website, China Statistical Yearbook (2008–2021), Bank of China website, State Administration of Foreign Exchange website, China Money Network, CSMAR database, Wind database and official website of the local government.

\subsection{Econometrics Model}
Our empirical analysis employs a one-way fixed model\footnote{We do not employ time-fixed effects model for the following reasons: First, the core independent variable “bilateral exchange rate of the USD-CNY” exhibits significant time varying effects, constituting a time series fluctuation (stochastic process). Therefore, the time effect itself is an inherent characteristic of the exchange rate variable and should not be fixed again (For details, see the analytical proof in \hyperref[Appendix A]{Appendix A}). Second, time-fixed effects may absorb the risk of exchange rate changes caused by external shocks, presenting only the average effect of exchange rate changes over time in the empirical results. This obscures the specific exchange rate changes at a certain policy point in time, which is the focus of this paper: the exchange rate reform policy regulates extreme trends in the USD-CNY exchange rate, stabilizing its fluctuations. Therefore, we do not introduce time-fixed effects in the benchmark regression model for a more direct observation and capture of the USD-CNY exchange rate trend changes before and after the policy shock.}, fixing $Province_i$. To control the potential upward or downward trend inherent in the dependent variable itself, a time trend term $Year_t$ is incorporated into the regression equation, i.e., in practice, the export competitiveness variable may also be a stochastic process exhibiting a trend of increasing or decreasing over time. We need to incorporate a time trend term to mitigate such issues.

\noindent\textbf{Basic Regression Model}: We denote the control variables group as $\mathbb{C}_{it}$ ($Controls_{it}$) and the province variables group as $\mathbb{P}_{i}$ ($Provinces_{it}$). To examine the impact of CNY exchange rate fluctuations on provincial export competitiveness, we establish the following benchmark model:
\[\begin{gathered}RCA_{it}=\alpha_0+\alpha_1EXRATE_t+\sum_i\alpha_i\mathbb{C}_{it}+\sum_i\mathbb{P}_i+Year_t+\varepsilon_{it}\\RCA_{it}=\beta_0+\beta_1D.[EXRATE_t]+\sum_i\beta_i\mathbb{C}_{it}+\sum_i\mathbb{P}_i+Year_t+\varepsilon_{it}\end{gathered}\]

Here, subscripts $i$ and $t$ denote provincial administrative units and years, respectively. $\varepsilon_{it}$ represents the residual term following a normal distribution. $RCA_{it}$ represents the provincial annual export competitiveness index. $EXRATE_t$ denotes the CNY exchange rate. Since we aim to empirically examine the impact of CNY exchange rate fluctuations on provincial export competitiveness, both appreciation and depreciation are considered manifestations of volatility. Through differential operation, we try to use the absolute value of the first order difference of the CNY exchange rate as the independent variable reflecting the magnitude of exchange rate fluctuations affecting $RCA_{it}$, that is:
\[D.[EXRATE_t]=|(EXRATE_{t+1})-(EXRATE_t)|\]

\noindent\textbf{Difference in Differences Model}: This paper selects the “8.11 Exchange Rate Reform” as the monetary policy shock. We constructed a DID model and a dynamic effect testing model, as shown below:
\[\begin{gathered}\begin{gathered}RCA_{it}=\gamma_0+\gamma_1[Treat_t\times Post_{it}]+\sum_i\gamma_i\mathbb{C}_{it}+\sum_i\mathbb{P}_i+Year_t+\varepsilon_{it}\\RCA_{it}=\delta_{0}+\delta_{1}[Treat_{t}\times Post(n)_{it}]+\sum_{i}\delta_{i}\mathbb{C}_{it}+\sum_{i}\mathbb{P}_{i}+Year_{t}+\varepsilon_{it}\end{gathered}\end{gathered}\]

The exchange rate effect of the reform policy is defined as the CNY exchange rate exceeding the window period average of 6.58. Here, Treat denotes the dummy variable for the treatment group, taking $Treat=1$ when the CNY exchange rate in the statistical year exceeds 6.58, and $Treat=0$ otherwise. Post represents the policy shock effect of the exchange rate reform. We select 2012–2020 as the window period\footnote{The selection of 2012-2020 as the window period aims to consider the impact of the 2008 global financial crisis and to exclude the effects of various exogenous shocks occurring after 2020 as far as possible.}, with 2016 as the policy shock point, covering the four periods before and after the policy. The period before 2016 constitutes the control group, where $Post=0$, while the period after 2016 forms the treatment group, where $Post=1$. $Post(n)$ denotes the n period before or after the implementation of the exchange rate reform policy, with $Post(0)$ representing the period of policy implementation.

\noindent\textbf{Moderating Effect Model}: For the moderating effect analysis, we will incorporate variables measuring innovation and interaction terms $X_i$ into the master regression equation. We have two groups of variables, for Group-a, the econometric model is as follows:
\[\begin{gathered}
X_a=EXRATE_t\times \ln(Innovation_{it})\\RCA_{it}=\theta_1^aEXRATE_t+\theta_2^aX_a+\theta_3^a\ln(Innovation_{it})+\sum_i\theta_i^a\mathbb{C}_{it}+\sum_i\mathbb{P}_i+Year_t+\varepsilon_{it}
\end{gathered}\]

Similarly, for Group-b:
\[\begin{gathered}
X_b=EXRATE_t\times \ln(Patent_{it})\\RCA_{it}=\theta_1^bEXRATE_t+\theta_2^bX_b+\theta_3^b\ln(Patent_{it})+\sum_i\theta_i^b\mathbb{C}_{it}+\sum_i\mathbb{P}_i+Year_t+\varepsilon_{it}
\end{gathered}\]

Now, we need data to measure coefficients $\theta_2^j$ and $\theta_3^j$, where $j\in\{a,b\}$. If $\theta_2^j<0$, it indicates that innovation can reduce the dependence of export competitiveness on the exchange rate. If $\theta_2^j<0$ and $\theta_3^j>0$, it suggests that the innovation mechanism enhancing export competitiveness will exert a significant substitution effect on the exchange rate mechanism.

\section{Empirical Evidence and Analysis}\label{Section 4}
\subsection{Benchmark Regression Results}
Figure 10 illustrates the linear and nonlinear effects of $EXRATE$ and $D.[EXRATE]$ on $RCA$. As shown in the left graph of Figure 10: A depreciation of the CNY within the range of 6 to 7 effectively enhances provincial export competitiveness (the linear and nonlinear trend lines are nearly identical, indicating a strong positive correlation between the variables). However, the right graph reveals that the impact of CNY exchange rate fluctuations on provincial export competitiveness follows an inverted U-shaped trend overall, with the inflection point occurring around $D.[EXRATE]=0.2$. Further analysis of exchange rate fluctuations' impact on export competitiveness before and after this inflection point reveals: When $D.[EXRATE]<0.2$, CNY exchange rate fluctuations are relatively small, and $D.[EXRATE]$ exhibits a positive correlation with $RCA$. When $D.[EXRATE]>0.2$, CNY exchange rate fluctuations are relatively large, and $D.[EXRATE]$ exhibits a negative correlation with $RCA$.

\begin{figure}[htbp]
\centering
\includegraphics[width=15cm]{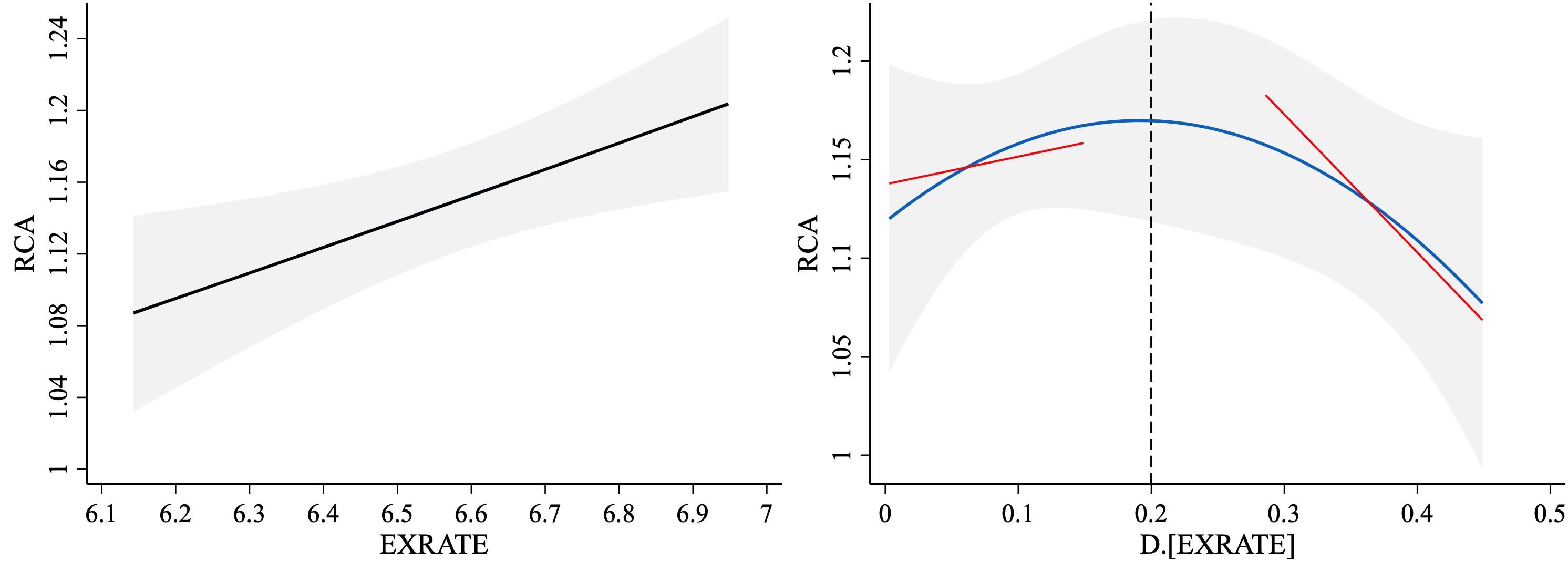}
\caption{\label{fig:F10}Descriptive Statistics and Analysis}
\end{figure}

Based on the results in Figure 10, Table 2 reports the benchmark regression results\footnote{ Note: ***, ** and * indicate significance at the 1$\%$, 5$\%$, and 10$\%$ levels, respectively. Standard errors are shown in parentheses. The same applies to the table below.} for the influence of $EXRATE$ and $D.[EXRATE]$ on $RCA$. 

\begin{table}[ht]
\centering
\caption{Benchmark Regression Results}
\label{tab:rca_results}
\small
\setlength{\tabcolsep}{3pt}
\renewcommand{\arraystretch}{1}
\begin{tabular}{l*{6}{c}}
\toprule
 & \multicolumn{6}{c}{RCA} \\
\cmidrule(lr){2-7}
 & (1) & (2) & (3) & (4) & (5) & (6) \\
 & ols & ols-fe & l.(exrate) & tobit-fe & d.(exrate)$<$0.2 & d.(exrate)$>$0.2 \\
\midrule
EXRATE & $0.132^{**}$ & $0.144^{***}$ &  & $0.111^{**}$ &  &  \\
 & (0.048) & (0.039) &  & (0.035) &  &  \\[0.4em]
L.[EXRATE] &  &  & $0.155^{***}$ &  &  &  \\
 &  &  & (0.043) &  &  &  \\[0.4em]
D.[EXRATE] &  &  &  &  & $0.760^{**}$ & $-0.677^{*}$ \\
 &  &  &  &  & (0.283) & (0.274) \\[0.4em]
lnPopulation & $-0.011$ & $-0.448$ & $-0.913^{**}$ & $-0.002$ & $-0.652$ & $-0.832$ \\
 & (0.053) & (0.296) & (0.331) & (0.076) & (0.404) & (0.682) \\[0.4em]
lnRetail & $0.250^{***}$ & $0.179^{**}$ & $0.250^{***}$ & $0.221^{***}$ & $0.266^{**}$ & $0.219$ \\
 & (0.048) & (0.066) & (0.067) & (0.056) & (0.084) & (0.150) \\[0.4em]
Vgdp & $-0.006$ & $-0.013^{**}$ & $-0.019^{***}$ & $-0.012^{**}$ & $-0.014^{*}$ & $-0.010$ \\
 & (0.005) & (0.004) & (0.005) & (0.004) & (0.006) & (0.010) \\[0.4em]
lnGovernment & $-0.147^{**}$ & $-0.151$ & $-0.030$ & $-0.250^{***}$ & $-0.441^{***}$ & $-0.220$ \\
 & (0.054) & (0.092) & (0.114) & (0.062) & (0.092) & (0.176) \\[0.4em]
Law & $0.004$ & $0.004$ & $0.0003$ & $0.003$ & $0.026^{**}$ & $-0.010$ \\
 & (0.006) & (0.006) & (0.006) & (0.005) & (0.009) & (0.012) \\[0.4em]
lnFirst & $-0.236^{***}$ & $0.036$ & $0.019$ & $-0.119^{*}$ & $0.050$ & $0.055$ \\
 & (0.029) & (0.078) & (0.079) & (0.057) & (0.105) & (0.158) \\[0.4em]
lnPower & $0.215^{***}$ & $0.118$ & $0.105$ & $0.162^{***}$ & $0.093$ & $0.096$ \\
 & (0.029) & (0.063) & (0.064) & (0.048) & (0.079) & (0.128) \\[0.4em]
Unemployment & $0.114^{***}$ & $0.049$ & $0.049$ & $0.055^{*}$ & $0.013$ & $0.073$ \\
 & (0.020) & (0.027) & (0.026) & (0.025) & (0.041) & (0.044) \\[0.4em]
Constants & $-0.924^{*}$ & $36.264$ & $72.745^{**}$ & $0.073$ & $6.554^{*}$ & $6.913$ \\
 & (0.442) & (20.885) & (24.231) & (0.547) & (3.236) & (5.241) \\[0.4em]
Province & $\times$ & $\checkmark$ & $\checkmark$ & $\checkmark$ & $\checkmark$ & $\checkmark$ \\[0.4em]
Time Trend & $\times$ & $\checkmark$ & $\checkmark$ & $\times$ & $\times$ & $\times$ \\[0.4em]
\midrule
$R^{2}$ & 0.342 & 0.139 & 0.139 & $-$ & 0.153 & 0.192 \\[0.4em]
Number & 420 & 420 & 390 & 420 & 270 & 120 \\
\bottomrule
\end{tabular}

\end{table}

Columns (1) to (4) in Table 2 present the regression analysis results for the impact of the CNY exchange rate $EXRATE$ on the provincial export competitiveness index $RCA$. Column (3) presents the regression analysis for the first order lagged CNY exchange rate variable $L.[EXRATE]$, while column (4) shows the regression result for the xttobit model with $0<RCA<2$. Columns (5) and (6) present the regression analysis results for the impact of CNY exchange rate volatility $D.[EXRATE]$ on the provincial export competitiveness index $RCA$, with $D.[EXRATE]=0.2$ serving as the inflection point for the benchmark regression.

As shown in columns (1) to (2) of Table 2, a depreciation of the CNY within the range of 6 to 7 significantly enhances provincial export competitiveness. After incorporating provincial fixed effect, the marginal effect of $EXRATE$ depreciation on $RCA$ is 0.144, with the estimated coefficient being statistically significant at the 1$\%$ level. Results from columns (5) and (6) indicate that when $D.[EXRATE]<0.2$, the impact of CNY exchange rate fluctuations on $RCA$ is 0.760. Conversely, when $D.[EXRATE]>0.2$, the impact becomes $-$0.677. Both coefficients are statistically significant. This indicates that moderate and stable fluctuations in the CNY exchange rate enhance provincial export competitiveness. Conversely, significant exchange rate volatility, particularly abnormal appreciation or depreciation, suppresses provincial export competitiveness, hence, Hypothesis 1 is proved.

For robustness, subsequent analysis is mainly based on the empirical results in column (2) of Table 2 to illustrate the role of CNY depreciation in enhancing export competitiveness.

\subsection{Robustness Test}
To ensure the robustness of our results, we conducted a series of stability tests\footnote{We have also tested the robustness of the benchmark regression results by changing the control variables. After changing the control variables, the benchmark regression results remained significantly positive. Due to space constraints, the detailed results are available upon request.}. As shown in columns (3) and (4) of Table 2, after incorporating provincial fixed effects, introducing lagged independent variable and replacing the baseline OLS model with an Tobit model, the coefficient of the primary independent variable $EXRATE$ remained positively significant. This indicates that the conclusion which shows a depreciation of the CNY exchange rate within the 6–7 range significantly enhances provincial export competitiveness is robust.

Furthermore, we employ a nonparametric permutation method to conduct a placebo test on the regression results between the CNY exchange rate and provincial export competitiveness. Figure 11 presents the placebo test results. 

\begin{figure}[htbp]
\centering
\includegraphics[width=15cm]{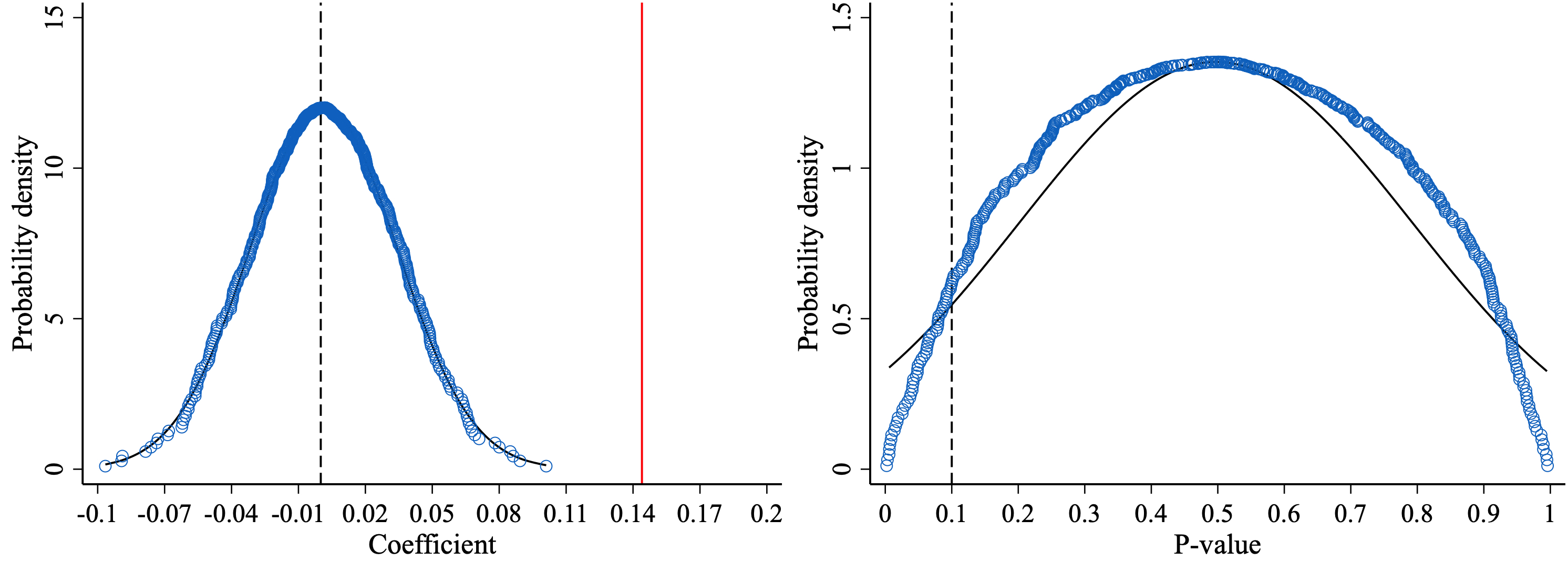}
\caption{\label{fig:F11}Placebo Test Results and Analysis}
\end{figure}

As shown in the left graph of Figure 11: The mean of the estimated coefficients from 500 random samples is close to zero, while the benchmark regression coefficient in Table 2 is 0.144, indicating a significant difference from the estimated coefficient obtained through the nonparametric test. The right graph of Figure 11 indicates that since the threshold for a significant P-value is 0.1, most random samples in the right graph have P-values exceeding 0.1, passing the P-value test. Thus, the placebo test results exclude interference from other events on the benchmark regression, further confirming the robustness of the benchmark regression results.

To mitigate endogeneity problem caused by omitted variables, this paper has already included a series of control variables in the benchmark regression analysis. However, endogeneity challenges may still exist. Therefore, we will next employ the 2SLS (Two Stage Least Square) method to attempt to mitigate potential endogeneity risk.

\subsection{Endogeneity Risk and 2SLS Method}
A country's fiscal and monetary policies influence currency exchange rates. By adjusting interest rates, intervening in markets, and altering the money supply, governments can affect the supply and demand dynamics of their currency, thus impacting exchange rates. Additionally, geopolitical risks and market sentiment also influence currency exchange rates: Unstable geopolitical situations or market panic can drive capital flows toward risk free assets, thus affecting currency supply, demand and exchange rates. 

The Geopolitical Risk Index quantifies geopolitical risks by comprehensively considering political stability, international relations, military conflicts, economic interdependence, and other relevant factors. The Economic Policy Uncertainty Index quantifies policy uncertainty by analyzing the number of articles and word frequency related to economic and monetary policy uncertainty in major newspapers\footnote{The calculation of the U.S. monetary policy uncertainty index and geopolitical risk index in this article primarily references data from ten newspapers: USA Today, The Miami Herald, The Chicago Tribune, The Washington Post, The Los Angeles Times, The Boston Globe, The San Francisco Chronicle, The Dallas Morning News, The Houston Chronicle and The Wall Street Journal.} or social media.

Therefore, given that the main independent variable in this paper is bilateral exchange rate of the USD-CNY, we introduce Economic Policy Uncertainty ($EPU_t$) index\footnote{The Economic Policy Uncertainty Index is calculated and compiled by Scott R. Baker, Nicholas Bloom, and Steven J. Davis from Stanford University and the University of Chicago. It primarily reflects economic policy uncertainty among major global economies. The monetary policy uncertainty index referenced in this paper is derived from the categorized economic policy uncertainty index.} and Geopolitical Risk ($GPR_t$) Index\footnote{The Geopolitical Risk Index is compiled by American economists Dario Caldara and Matteo Iacoviello, calculated by measuring the proportion of negative geopolitical events discussed in internationally renowned newspapers and magazines.} of the USA from 2008 to 2021. Then, we employ the logarithm of the product $\ln(EPU_t\times GPR_t)$ as the instrumental variable of the 2SLS method.

On one hand, U.S. monetary policy and geopolitical risks are correlated with our bilateral exchange rate, aligning with the correlation between the independent variable (USD-CNY exchange rate) and the instrumental variable. On the other hand, U.S. monetary policy and geopolitical risks collectively represent issues within and surrounding the United States, exerting no direct influence on provincial export trade in China. Thus, it satisfies the requirement of exogeneity. 

In addition, we believe that the magnitude of influence exerted by geopolitical risks and economic policy uncertainties on the USD-CNY exchange rate is contingent upon the degree of marketization within domestic regional markets. Specifically, only under the assumption of free markets or highly marketized markets, where currency exchange rates are solely determined by market mechanisms, the USD-CNY exchange rate will be significantly correlated with the U.S. monetary policy uncertainty and geopolitical risks. Moreover, since we employ provincial data of China while the product of $EPU_t$ and $GPR_t$ constitutes annual cross-sectional data, we multiply the variable $\ln(EPU_t\times GPR_t)$ by each province's annual marketization index $Mrket_{i,t}$ to construct the final instrumental variable $Tool_{i,t}$:
\[Tool_{i,t}=Market_{i,t}\times \ln(EPU_{t}\times GPR_t)\]

Then, it is generally believed that current exchange rate changes are significantly influenced by the previous period's geopolitical risk, economic policy uncertainty index and marketization index. Therefore, the instrumental variable should be lagged by one period in the main regression of 2SLS method.

Table 3 columns (1) and (2) report the results of the 2SLS regression with $Tool_{i,t-1}$ as the instrumental variable. Column (1) presents the first stage regression result, with a positive and significant coefficient. Column (2) shows the second stage regression result, also featuring a positive and significant coefficient. The results from both columns indicate: The \textit{Kleibergen-Paap rk LM} statistic is significant at the 1$\%$ level, rejecting the null hypothesis of insufficient instrument identification. The \textit{Cragg-Donald Wald F} statistic also rejects the null hypothesis of weak instruments. Thus, the selected instruments in this study are reasonable. After considering and mitigating potential endogeneity risks, the depreciation of the CNY exchange rate still has a significantly positive impact on provincial export competitiveness, hence, the main content of Hypothesis 1 remains valid.

\subsection{Monetary Policy Shock and DID Model}
\textbf{Reality and Policy Motivation Analysis}: As shown in Figure 5, from 2008 to June 2015, the CNY exchange rate exhibited an overall appreciation trend, primarily driven by China's external economic imbalance manifested in the form of a substantial current account surplus during this period. The driving force stemmed from the economic rebalancing process centered on “expanding domestic demand, adjusting economic structure, reducing the trade surplus and promoting balance.” On August 11, 2015, China announced reforms to refine the USD-CNY central parity rate quotation mechanism, aiming to enhance its market nature and benchmark status.

During the initial phase of the “8.11 exchange rate reform”, China's foreign exchange market experienced increasing volatility, facing the impact of cross-border capital flows characterized by “capital outflows, reserve depletion and exchange rate depreciation”. By the end of 2016, the CNY exchange rate was approaching 7 and foreign exchange reserves were about to fall below 3 trillion dollars. However, the introduction of counter cyclical factors in May 2017 enabled the CNY exchange rate to not only defend the 7 threshold but also appreciate nearly 7$\%$ throughout the year. This restored credibility in exchange rate policy and achieved a reversal of the 8.11 reform's initial challenges. 

Entering 2018, due to international trade friction between China and America, which put renewed pressure on the CNY starting in April 2018. By early August 2019, as the China-US trade negotiations reached another impasse, the CNY exchange rate broke the 7 threshold despite the Federal Reserve initiating a new round of interest rate cuts. Subsequently, with the signing of the Phase One trade agreement between China and the US, the CNY exchange rate returned to below 7 by the end of 2019. The outbreak of the pandemic in early 2020 caused the CNY exchange rate to break through the 7 threshold again in February. Later that year in May, geopolitical factors pushed the CNY exchange rate to depreciate further to around 7.2. After that, the widening interest rate differential between China and the United States and the softening of the U.S. dollar index triggered a brief period of extreme appreciation in the CNY exchange rate starting in early June 2020.

Figure 12 utilizes the annual average exchange rate of the USD-CNY. The blue scatter points connected by lines represent the trend in exchange rate fluctuations, while the thin gray solid lines indicate the trend in export competitiveness (RCA) of each province. The x-axis spans the time from 2008 to 2021. The left y-axis displays the provincial export competitiveness index, ranging from 0 to 2. The right y-axis shows the USD-CNY exchange rate, ranging from 6 to 7.

According to Figure 12, the changes in the CNY exchange rate and provincial export competitiveness are divided by the year that the “8.11 Exchange Rate Reform” was implemented (end of 2015): Prior to the implementation of the exchange rate reform policy (2008–2015), the CNY exchange rate showed an overall appreciation trend\footnote{Following the outbreak of the U.S. subprime mortgage crisis in 2007 and the global financial crisis in 2008, China's 10 year government bond yield remained above 3$\%$ for most of the period, while the U.S. 10 year government bond yield stayed below 3$\%$ for the majority of the time. Consequently, a positive interest rate differential between China and the U.S. persisted throughout most of this period. From July 2008 to June 2010, China proactively narrowed the fluctuation range of the CNY exchange rate, maintaining the central parity rate within a tight band of 6.8 to 6.84. On June 19, 2010, China resumed exchange rate reform to enhance CNY flexibility. By the end of 2013, the CNY had gradually appreciated to around 6.1, representing a cumulative increase of 27.6$\%$ compared to the end of 2006. Foreign exchange reserves reached 3.8213 trillion dollars, increased by 2.58 times compared to the end of 2006. Except for the impact of the European sovereign debt crisis in 2012, all other years saw a “double surplus” in the balance of payments, with foreign exchange reserve assets continuing to increase substantially.}, with weak short term fluctuations and a strong long term growth trend. At this time, the variance in export competitiveness among provinces was significant, with a relatively dispersed distribution and the export competitiveness of most provinces remained at a low level.

Following the policy implementation (2016–2021), the CNY exchange rate exhibited overall volatility compared to the previous period (see Figure 5), with pronounced short term fluctuations and a weak long term depreciation trend. During this phase, provincial export competitiveness showed a tendency to converge toward a higher level, with the distribution curve gradually concentrating at higher competitiveness levels (see Figure 6), indicating an overall improvement in export competitiveness.

Therefore, this paper argues that linear appreciation or depreciation trends in the CNY exchange rate are not conducive to enhancing provincial export competitiveness. Only when the CNY exchange rate exhibits small range fluctuations overall does it facilitate improvements in provincial export competitiveness\footnote{Referring to the benchmark regression Table 2, when the short term fluctuation of the CNY exchange rate is less than 0.2 RMB, it helps enhance export competitiveness.}. The exchange rate reform policy (the 8.11 reform) achieves this by regulating the flexibility of the CNY exchange rate, expanding the market space for exchange rate fluctuations, avoiding long term linear trends, smoothing out extreme volatility, and enabling a “soft landing” under external shocks. This keeps the exchange rate in a state of stable long term fluctuation and thus indirectly enhances provincial export competitiveness.

\begin{figure}[htbp]
\centering
\includegraphics[width=10cm]{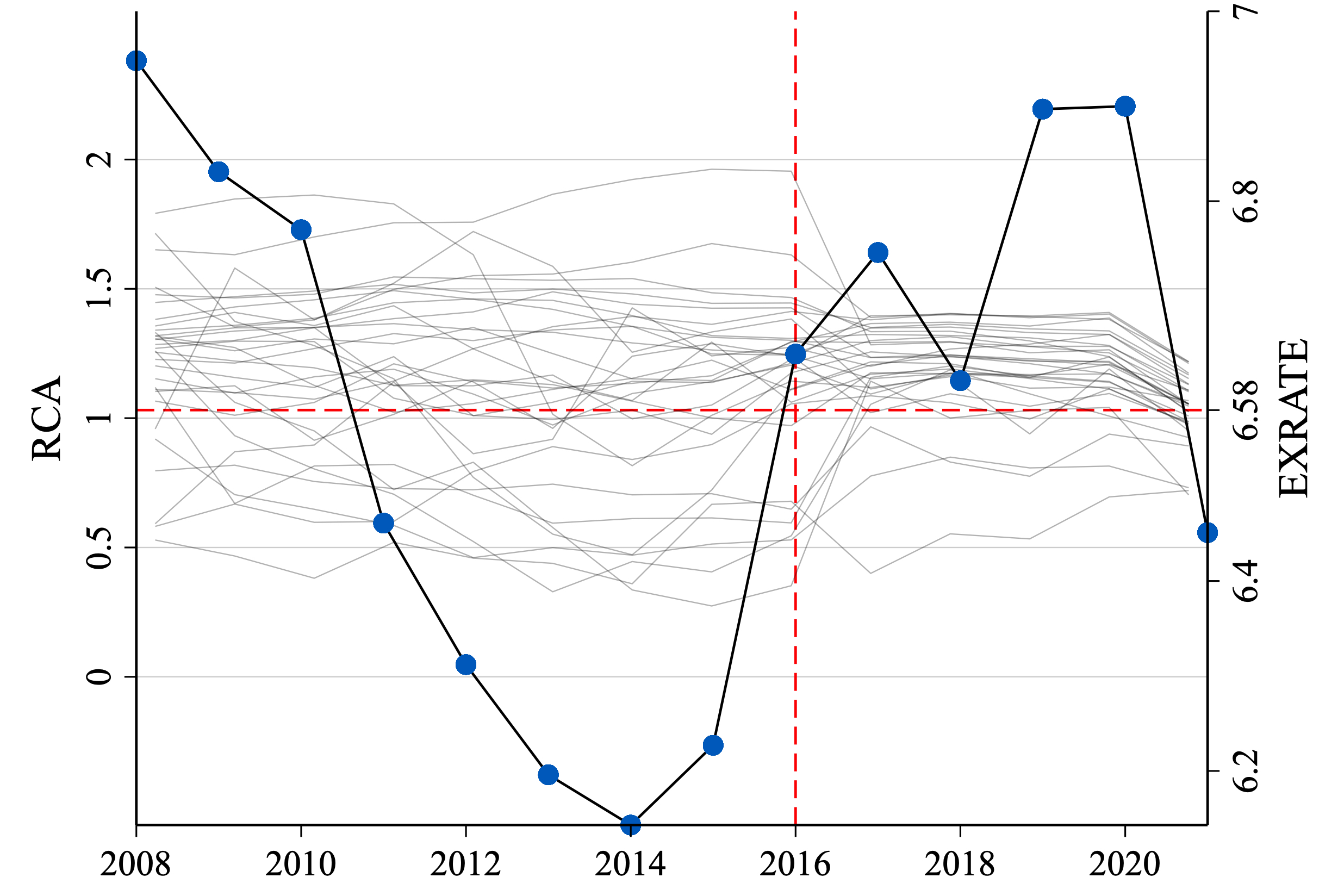}
\caption{\label{fig:F12}The Joint Evolution of Exchange Rate and RCA with Time}
\end{figure}

\noindent\textbf{DID Experiment}: The results of the difference-in-differences experiment are shown in column (3) of Table 3: According to column (3) of Table 3, the regression coefficient for the interaction term $Treat\times Post$ is significantly positive at the 1$\%$ level. This indicates that the provincial export competitiveness has been significantly enhanced following the impact of the exchange rate reform policy.

We employ event analysis to examine the parallel trend and the policy dynamic effects, as illustrated in Figure 13.

The test results reveal that during 2012–2016 (i.e., $Treat\times Post(-4)$ to $Treat\times Post(-1)$), the regression coefficients are not significant. However, after $Treat\times Post(0)$, the regression coefficients for the period 2017–2020 are all significantly positive. This indicates that the DID model used in this paper satisfies the parallel trend assumption. This verifies that the “8.11 Exchange Rate Reform” policy significantly enhanced provincial export competitiveness by regulating the fluctuation trend of the CNY exchange rate.

The results demonstrate that after using the DID model to identify the dynamic impact of the CNY exchange rate on provincial export competitiveness, the main conclusion of this paper remains valid. Specifically, the exchange rate reform policy reversed the appreciation trend of the CNY exchange rate between 2008 and 2015, leading to a depreciation of the CNY exchange rate in 2016–2020 compared to the previous period, followed by an overall fluctuating state stably, which enhanced provincial export competitiveness.

Therefore, the DID experiment results demonstrate the effectiveness of the “8.11” exchange rate reform policy and the main contents of Hypothesis 2 are verified.

\begin{figure}[htbp]
\centering
\includegraphics[width=9cm]{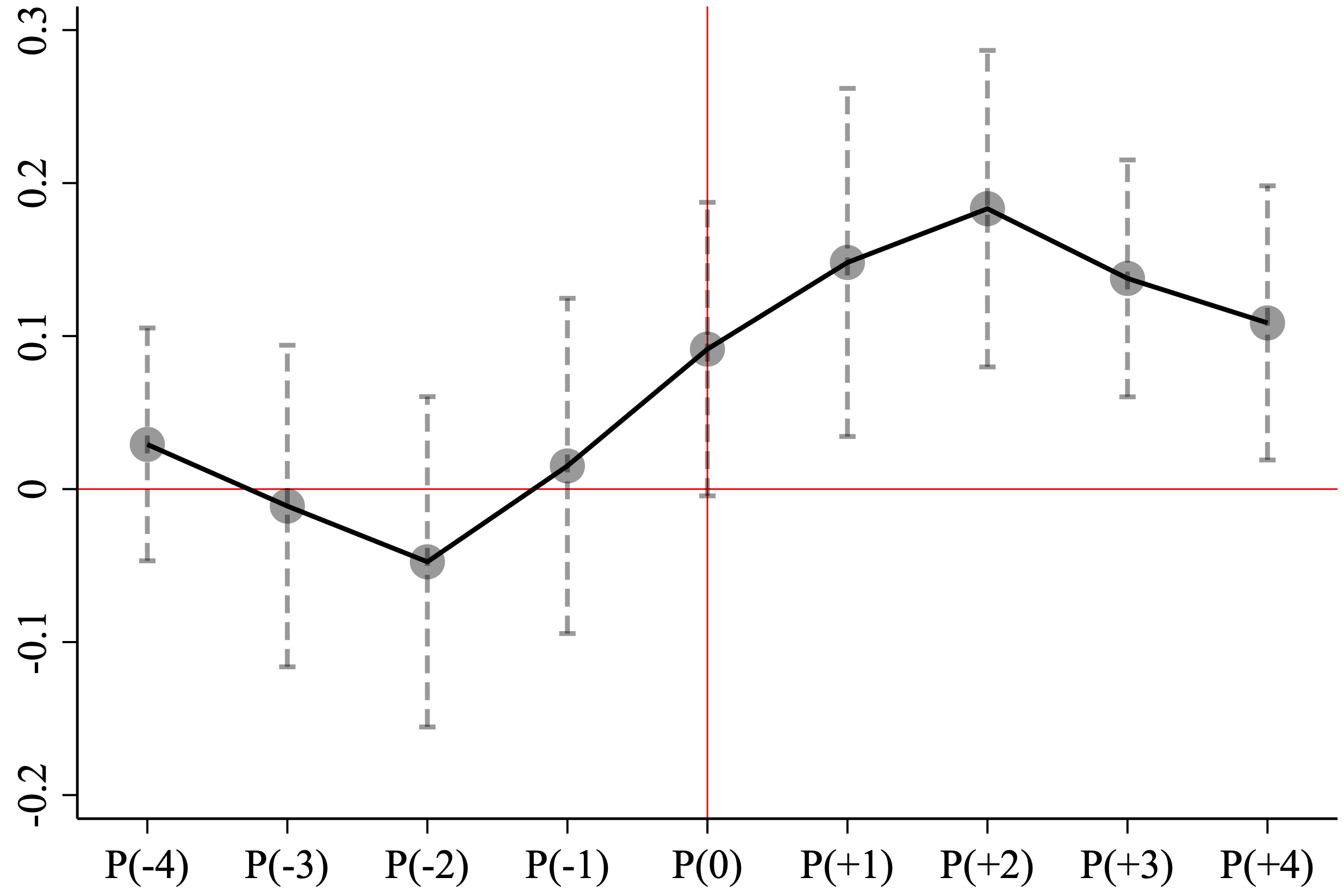}
\caption{\label{fig:F13}Dynamic Effects and Parallel Trend Tests}
\end{figure}

\noindent\textbf{Heterogeneity}: Before conducting moderating effect analysis, we first identify heterogeneity in policy impacts to preliminarily estimate the effects of innovation. Based on the results of Group-a and Group-b in Section 3.2, we separately test the dynamic effects of the policy, with results shown in Figures 14 and 15:

\begin{figure}[htbp]
\centering
\includegraphics[width=15cm]{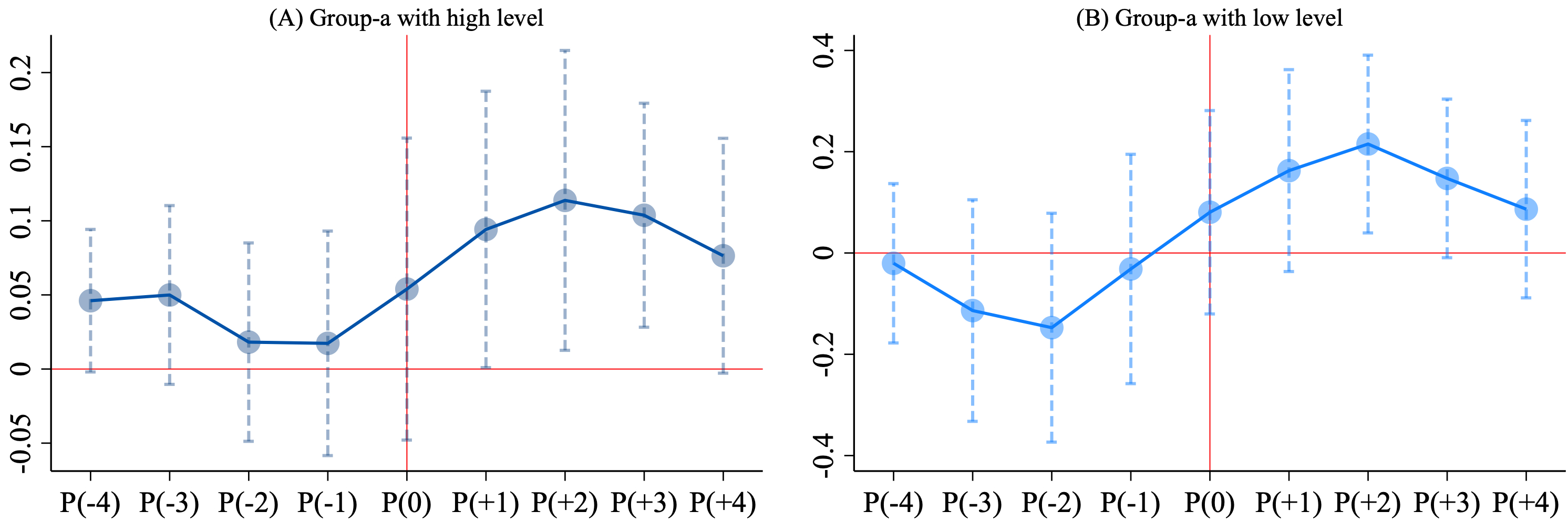}
\caption{\label{fig:F14}Dynamic Effects in Group-a}
\end{figure}
\vspace{-0.2cm}
\begin{figure}[!ht]
\centering
\includegraphics[width=15cm]{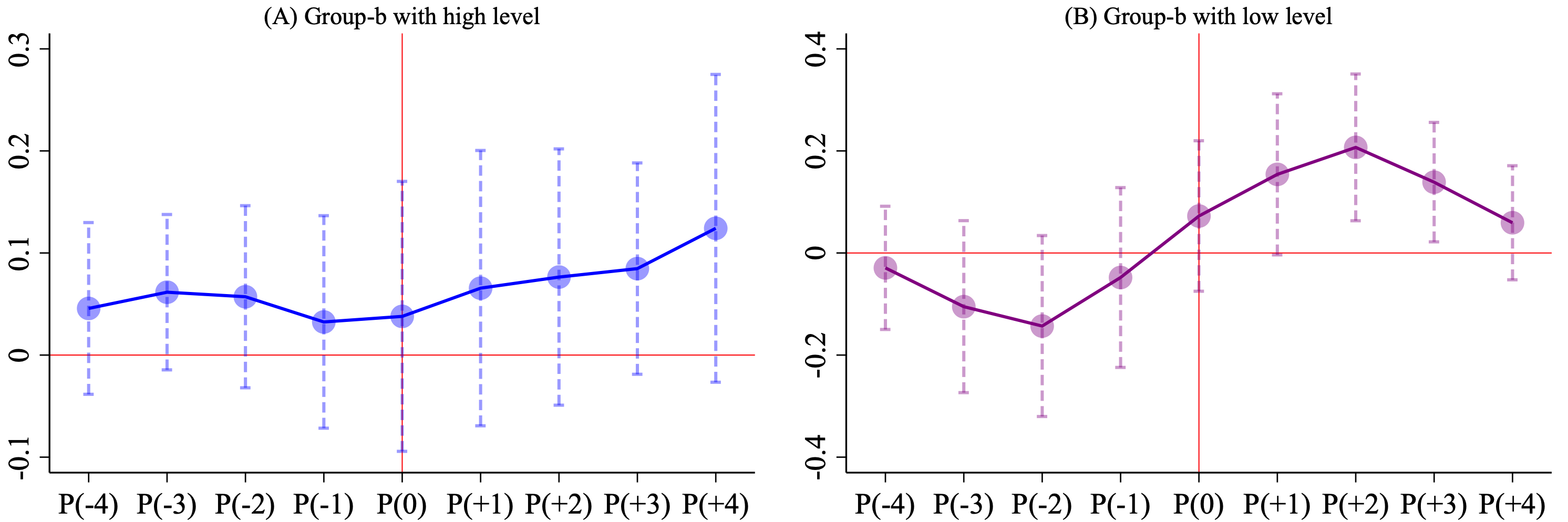}
\caption{\label{fig:F15}Dynamic Effects in Group-b}
\end{figure}

\begin{table}[!ht]
\centering
\caption{2SLS Method and DID Model Results}
\label{tab:iv_did_results}
\small
\setlength{\tabcolsep}{6pt}
\renewcommand{\arraystretch}{1}
\begin{tabular}{lccc}
\toprule
 & (1) & (2) & (3) \\
 & the first stage & the second stage & did \\
 \midrule
 & EXRATE & RCA & RCA \\
$\mathrm{Tool}_{i,t-1}$ & $0.235^{***}$ &  &  \\
 & (0.023) &  &  \\[0.4em]
EXRATE &  & $0.097^{*}$ &  \\
 &  & (0.046) &  \\[0.4em]
Treat$\times$Post &  &  & $0.138^{***}$ \\
 &  &  & (0.026) \\[0.4em]
lnPopulation & $0.499$ & $-0.813$ & $-0.316$ \\
 & (0.394) & (0.565) & (0.291) \\[0.4em]
lnRetail & $0.064$ & $0.232^{*}$ & $0.147^{*}$ \\
 & (0.081) & (0.095) & (0.065) \\[0.4em]
Vgdp & $-0.012^{*}$ & $-0.014^{**}$ & $-0.006$ \\
 & (0.005) & (0.004) & (0.005) \\[0.4em]
lnGovernment & $-0.835^{***}$ & $-0.172$ & $-0.304^{***}$ \\
 & (0.112) & (0.142) & (0.081) \\[0.4em]
Law & $-0.017^{*}$ & $0.003$ & $0.013^{*}$ \\
 & (0.008) & (0.010) & (0.006) \\[0.4em]
lnFirst & $-0.331^{***}$ & $0.047$ & $0.064$ \\
 & (0.095) & (0.100) & (0.077) \\[0.4em]
lnPower & $-0.181^{*}$ & $0.109$ & $0.151^{*}$ \\
 & (0.076) & (0.094) & (0.063) \\[0.4em]
Unemployment & $-0.016$ & $0.048$ & $0.048$ \\
 & (0.032) & (0.045) & (0.026) \\[0.4em]
Constants & $-215.584^{***}$ & $42.772$ & $25.491$ \\
 & (23.928) & (30.194) & (19.404) \\[0.4em]
\midrule
Anderson canon. corr. LM & \multicolumn{2}{c}{$64.29^{***}$} &  \\[0.4em]
Kleibergen-Paap rk LM & \multicolumn{2}{c}{$66.02^{***}$} &  \\[0.4em]
Cragg-Donald Wald F & \multicolumn{2}{c}{$76.31^{***}$} &  \\[0.4em]
Kleibergen-Paap Wald rk F & \multicolumn{2}{c}{$107.65^{***}$} &  \\[0.4em]
\midrule
Province & $\checkmark$ & $\checkmark$ & $\checkmark$ \\[0.4em]
Time Trend & $\checkmark$ & $\checkmark$ & $\checkmark$ \\[0.4em]
\midrule
$R^{2}$ & 0.479 & 0.150 & 0.169 \\[0.4em]
Number & 390 & 390 & 420 \\
\bottomrule
\end{tabular}
\end{table}

Figure 14 presents the results for Group-a. Figure 14-A shows the dynamic effects of the policy in regions with high innovation levels, while Figure 14-B shows the dynamic effects in regions with low innovation levels. It is clear that incorporating heterogeneous innovation levels seriously threatens the significance of the policy effects. Furthermore, by comparing the magnitude of the coefficients, we find that the policy effects are stronger in regions with low innovation levels. This indicates that the exchange rate policy plays a more pronounced role in enhancing export competitiveness within these regions.

Figure 15 presents the results for Group-b. Figure 15-A illustrates the dynamic effects of the policy in regions with high innovation levels, while Figure 15-B shows the dynamic effects in regions with low innovation levels. It is apparent that after incorporating heterogeneous innovation levels, the policy no longer exhibits a significantly positive effect on export competitiveness in high-innovation regions. Conversely, in the regions with low innovation levels, the exchange rate policy could continue to significantly enhance export competitiveness.

The above results preliminarily indicate that innovation may disrupt the effectiveness of exchange rate policy. As a further analysis according to Figure 6, we examined and discussed the impact of exchange rate policy on the unbalanced development of provincial export competitiveness. For details, see \hyperref[Appendix B]{Appendix B}.

\subsection{Moderating Effect of Innovation}
Based on the model specification derived from the moderating effect analysis in Section 3.3, Tables 4 and Table 5 report the impact of innovation development on the exchange rate mechanism for export competitiveness.

From the column (1) and column (2) of Table 4 and Table 5: The interaction terms $X_a$ ($X_b$) between the logarithm of provincial innovation capability scores $\ln(Innovation_{it})$ (the logarithm of year-end patent application authorizations $\ln(Patent_{it})$) and the USD-CNY bilateral exchange rate $EXRATE_t$ are both significantly negative at the $1\%$ level. This indicates that innovation weakens the positive effect of USD-CNY depreciation on enhancing the export competitiveness of the provinces in China.

Column (3) and column (4) in Table 4 and Table 5 respectively represent the high level innovation and low level innovation groups classified based on provincial innovation capability scores (Group-a) and whether the number of patent applications authorized at the end of the year exceeds the overall average level (Group-b). Results indicate that in low level innovation provinces, exchange rate depreciation significantly and positively affects export competitiveness at the $1\%$ level. Conversely, in high level innovation provinces, the marginal coefficient of exchange rate depreciation on export competitiveness is statistically insignificant. This suggests that in highly innovative provinces, exchange rates cease to be a primary determinant of export competitiveness.

Furthermore, as shown in column (2) of Table 4 and Table 5, the interaction terms $X_a$ and $X_b$ between innovation variables and exchange rates are both significantly negative at the $1\%$ level. Moreover, the marginal effects of both $\ln(Innovation_{it})$ and $\ln(Patent_{it})$ on export competitiveness are significantly positive at the $1\%$ level. This indicates that innovation development plays a substitutive role in enhancing export competitiveness for exchange rate depreciation. High levels of innovation often promote the improvement of export competitiveness. This aligns with the theoretical analysis involved in Hypothesis 3.

Therefore, encouraging innovative development and technological innovation enables enterprises to deeply improve and upgrade their products, transforming them from homogeneous products with low added value into heterogeneous products with high added value. High value added products offer relatively larger profit margins and can better withstand price pressures from exchange rate fluctuations, unlike low value added products that are highly sensitive to price changes. Furthermore, through innovative business models, enterprises can establish their own online sales platforms or join cross border e-commerce platforms to directly connect with overseas consumers, reducing intermediary links and lowering sales costs. At the same time, using digital marketing tools such as social media marketing and search engine optimization to precisely target customer groups, expand brand awareness and product influence. In this way, even if exchange rate fluctuations lead to a decrease in offline channel orders, sales can be increased through online channels to maintain the stable development of export businesses and mitigate the adverse impact of exchange rate volatility on overall export competitiveness.

\begin{table}[!ht]
\centering
\caption{Moderating Effective in Group-a}
\label{Moderating-a}
\small
\setlength{\tabcolsep}{15pt}
\renewcommand{\arraystretch}{1.2}
\begin{tabular}{lcccc}
\toprule
 & \multicolumn{4}{c}{RCA} \\
\cmidrule(lr){2-5}
 & $(1)$ & $(2)$ & $(3)$ & $(4)$ \\
 & Group-a & Group-a & High & Low \\
\midrule
EXRATE & $0.144^{***}$ & $1.816^{***}$ & $0.026$ & $0.277^{***}$ \\
 & $(0.039)$ & $(0.327)$ & $(0.030)$ & $(0.070)$ \\
$\mathrm{X_{a}}$ &  & $-0.500^{***}$ &  &  \\
 &  & $(0.097)$ &  &  \\
lnInnovation &  & $3.171^{***}$ &  &  \\
 &  & $(0.655)$ &  &  \\
Fisher's test &  &  & \multicolumn{2}{c}{$0.252^{**}$}  \\
lnPopulation & $-0.448$ & $-0.612^{*}$ & $-0.401$ & $-0.980$ \\
 & $(0.296)$ & $(0.296)$ & $(0.285)$ & $(0.521)$ \\
lnRetail & $0.179^{**}$ & $0.216^{***}$ & $0.373^{***}$ & $0.065$ \\
 & $(0.066)$ & $(0.065)$ & $(0.060)$ & $(0.124)$ \\
Vgdp & $-0.013^{**}$ & $-0.012^{**}$ & $-0.015^{***}$ & $-0.012$ \\
 & $(0.004)$ & $(0.004)$ & $(0.003)$ & $(0.008)$ \\
lnGovernment & $-0.151$ & $-0.125$ & $-0.176^{*}$ & $-0.132$ \\
 & $(0.092)$ & $(0.090)$ & $(0.077)$ & $(0.159)$ \\
Law & $0.004$ & $0.001$ & $-0.006$ & $0.005$ \\
 & $(0.006)$ & $(0.006)$ & $(0.006)$ & $(0.010)$ \\
lnFirst & $0.036$ & $0.035$ & $-0.055$ & $0.364^{*}$ \\
 & $(0.078)$ & $(0.076)$ & $(0.065)$ & $(0.151)$ \\
lnPower & $0.118$ & $0.121^{*}$ & $0.005$ & $0.136$ \\
 & $(0.063)$ & $(0.061)$ & $(0.071)$ & $(0.095)$ \\
Unemployment & $0.049$ & $0.060^{*}$ & $0.060^{**}$ & $0.046$ \\
 & $(0.027)$ & $(0.026)$ & $(0.022)$ & $(0.044)$ \\
Constants & $36.264$ & $33.508$ & $44.775^{**}$ & $75.593$ \\
 & $(20.885)$ & $(20.358)$ & $(16.924)$ & $(40.291)$ \\
Province & $\checkmark$ & $\checkmark$ & $\checkmark$ & $\checkmark$ \\
Time Trend & $\checkmark$ & $\checkmark$ & $\checkmark$ & $\checkmark$ \\
$R^{2}$ & $0.139$ & $0.200$ & $0.309$ & $0.198$ \\
Number & $420$ & $420$ & $196$ & $224$ \\
\bottomrule
\end{tabular}
\end{table}

\newpage

\begin{table}[!ht]
\centering
\caption{Moderating Effective in Group-b}
\label{Moderating-b}
\small
\setlength{\tabcolsep}{15pt}
\renewcommand{\arraystretch}{1.2}
\begin{tabular}{lcccc}
\toprule
 & \multicolumn{4}{c}{RCA} \\
\cmidrule(l){2-5}
 & $(1)$ & $(2)$ & $(3)$ & $(4)$ \\
 & Group-a & Group-a & High & Low \\
\midrule
EXRATE & $0.144^{***}$ & $1.419^{***}$ & $-0.012$ & $0.250^{***}$ \\
 & $(0.039)$ & $(0.231)$ & $(0.037)$ & $(0.056)$ \\
$\mathrm{X_{b}}$ &  & $-0.127^{***}$ &  &  \\
 &  & $(0.022)$ &  &  \\
lnPatent &  & $0.739^{***}$ &  &  \\
 &  & $(0.147)$ &  &  \\
Fisher's test &  &  & \multicolumn{2}{c}{$0.262^{**}$}  \\
lnPopulation & $-0.448$ & $-0.773^{**}$ & $-0.218$ & $-0.596$ \\
 & $(0.296)$ & $(0.290)$ & $(0.357)$ & $(0.421)$ \\
lnRetail & $0.179^{**}$ & $0.177^{**}$ & $0.362^{***}$ & $0.084$ \\
 & $(0.066)$ & $(0.064)$ & $(0.100)$ & $(0.097)$ \\
Vgdp & $-0.013^{**}$ & $-0.013^{**}$ & $-0.019^{**}$ & $-0.014^{*}$ \\
 & $(0.004)$ & $(0.004)$ & $(0.006)$ & $(0.006)$ \\
lnGovernment & $-0.151$ & $0.066$ & $-0.119$ & $-0.104$ \\
 & $(0.092)$ & $(0.096)$ & $(0.112)$ & $(0.121)$ \\
Law & $0.004$ & $0.004$ & $-0.015$ & $0.008$ \\
 & $(0.006)$ & $(0.006)$ & $(0.009)$ & $(0.008)$ \\
lnFirst & $0.036$ & $0.102$ & $-0.161$ & $0.280^{*}$ \\
 & $(0.078)$ & $(0.078)$ & $(0.087)$ & $(0.134)$ \\
lnPower & $0.118$ & $0.086$ & $0.027$ & $0.147$ \\
 & $(0.063)$ & $(0.062)$ & $(0.089)$ & $(0.079)$ \\
Unemployment & $0.049$ & $0.060^{*}$ & $0.026$ & $0.047$ \\
 & $(0.027)$ & $(0.026)$ & $(0.032)$ & $(0.035)$ \\
Constants & $36.264$ & $35.823$ & $52.385$ & $73.000^{*}$ \\
 & $(20.885)$ & $(21.407)$ & $(27.445)$ & $(29.959)$ \\
Province & $\checkmark$ & $\checkmark$ & $\checkmark$ & $\checkmark$ \\
Time Trend & $\checkmark$ & $\checkmark$ & $\checkmark$ & $\checkmark$ \\
$R^{2}$ & $0.139$ & $0.210$ & $0.323$ & $0.185$ \\
Number & $420$ & $420$ & $140$ & $280$ \\
\bottomrule
\end{tabular}
\end{table}

\noindent\textbf{Regional Analysis and Discussion}: The analysis of this paper reveals that provinces with higher innovation capabilities are predominantly concentrated in the eastern coastal regions of China. This concentration is inextricably linked to the eastern region's development pattern centered on “innovation” as a core objective. The region's economic prosperity and advanced production technologies, driven by the concentration of new manufacturing enterprises and industrial chains, enable more rational allocation of production factors. This facilitates accelerated elimination of outdated production capacity, promotes the desirable development of industrial structures and enhances total factor productivity. Consequently, the impact of exchange rate factors on export trade in this region appears to be slightly muted.

Meanwhile, several provinces with high innovation capabilities are located in the western regions of China. These areas boast vast territorial space, abundant natural resources and unique conditions for industrial development. Decades of accumulation have fostered a cluster of mature industrial bases with comprehensive sectors and strong industrial agglomeration. Traditional industries such as energy and power, metallurgy and chemicals, and equipment manufacturing have established foundational strengths. Moreover, during the “\textit{Third Front}” and “\textit{New Third Front}” construction periods of China, the state strategically deployed universities and research institutions across the western region. Cities like \textit{Chongqing} and \textit{Chengdu} possess research capabilities that exert nationwide influence, hence securing a certain advantage in innovation.

Provinces with relatively low innovation capabilities are predominantly concentrated in the central region of China. The export competitiveness of this area will be significantly influenced by exchange rate fluctuations. This is because the central region lags behind the eastern region in terms of economic development level, utilization of advanced technologies, rational allocation of labor resources and innovation environment or conditions. However, it possesses a solid agricultural foundation, abundant energy and mineral resources  with deep manufacturing roots. This region has long supplied labor and energy resources to the eastern coastal areas. Its industrial structure is dominated by relatively simple midstream and upstream equipment manufacturing, endowing it with unique conditions and capabilities for exporting products with lower added value.

Therefore, the western region possesses significant advantages in both export trade and innovative development. Relevant policies should intensify efforts to foster the growth of innovative industries in the western region, strengthen exchanges and cooperation with enterprises in the eastern region, and promote the sharing of advanced technologies. Furthermore, the advantages of the central region should be gradually leveraged to expand the reach of its manufacturing industries, enabling their diffusion into the west. This will further enhance the export trade conditions and capabilities of the western region, facilitating the formation of a development pattern driven by both innovation output and scale (efficiency) output.

\subsection{Further Analysis}
As an extension of Section 4.1, we plot the curve-fitting results between the export competitiveness ($RCA$) and the measurement of exchange rate volatility ($D.[EXRATE]$) separately for the Group-a and Group-b samples.

Figures 16 and 17 respectively illustrate the nonlinear relationship between provincial export competitiveness and exchange rate volatility (the absolute value of the first-order difference in the bilateral exchange rate) in Group-a and Group-b. In each figure, the left graph presents samples from regions with low innovation capacity, while the right graph shows samples from regions with high innovation capacity. We observe that regardless of regional innovation capacity, $RCA$ and $D.[EXRATE]$ exhibit a typical inverted U-shaped relationship, which is consistent with the results in Figure 10. Moreover, the magnitude of export competitiveness only affects the position of the inverted U-shaped curve, not its shape. Hence, although our empirical analysis demonstrates that export competitiveness in regions with higher innovation capabilities does not directly depend on exchange rates, exchange rate fluctuations still exert some influence on export competitiveness. Significant exchange rate volatility hinders the enhancement of export competitiveness by reducing RCA. Particularly in regions with high innovation capabilities, we observe an overall negative correlation between export competitiveness $RCA$ and exchange rate volatility $D.[EXRATE]$.

\begin{figure}[htbp]
\centering
\includegraphics[width=14cm]{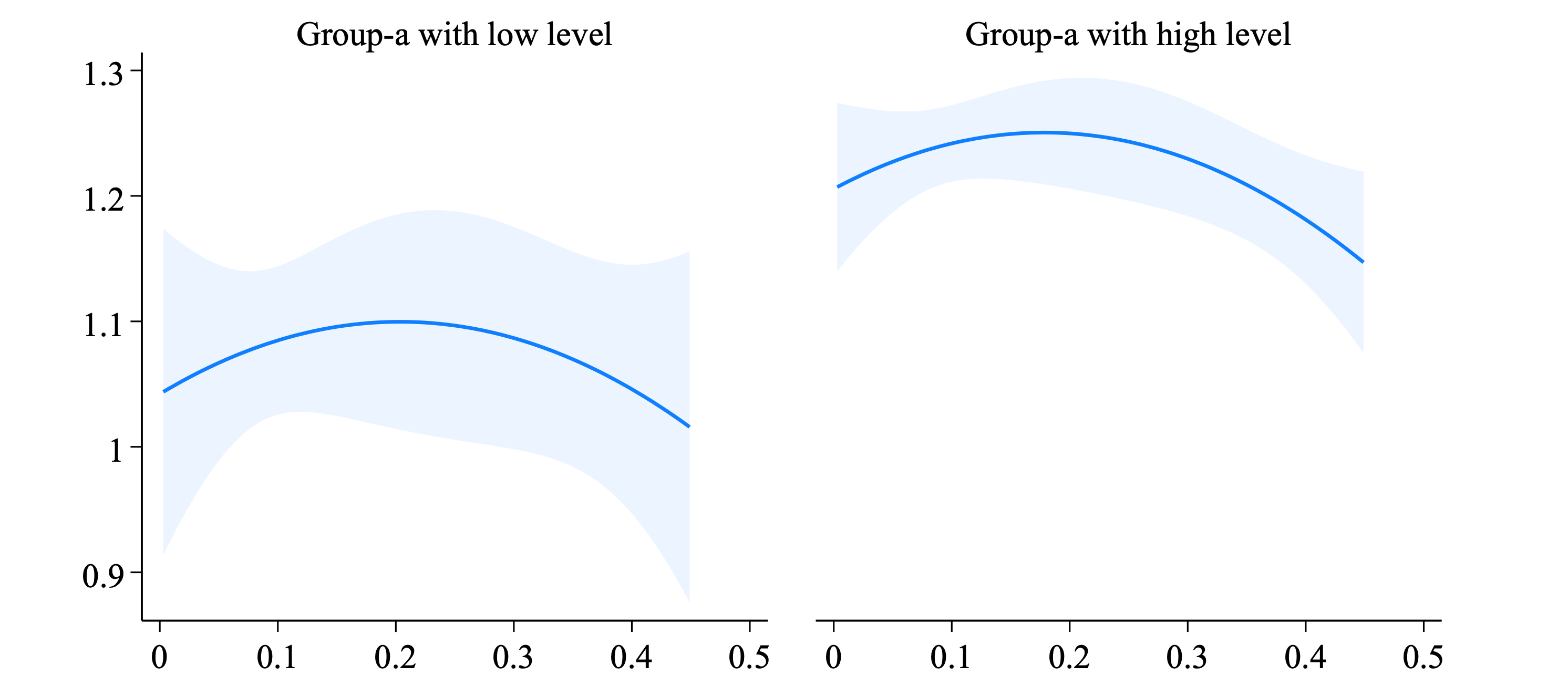}
\caption{\label{fig:F16}RCA and D.[EXRATE] in Group-a}
\end{figure}
\vspace{-0.5cm}
\begin{figure}[htbp]
\centering
\includegraphics[width=14cm]{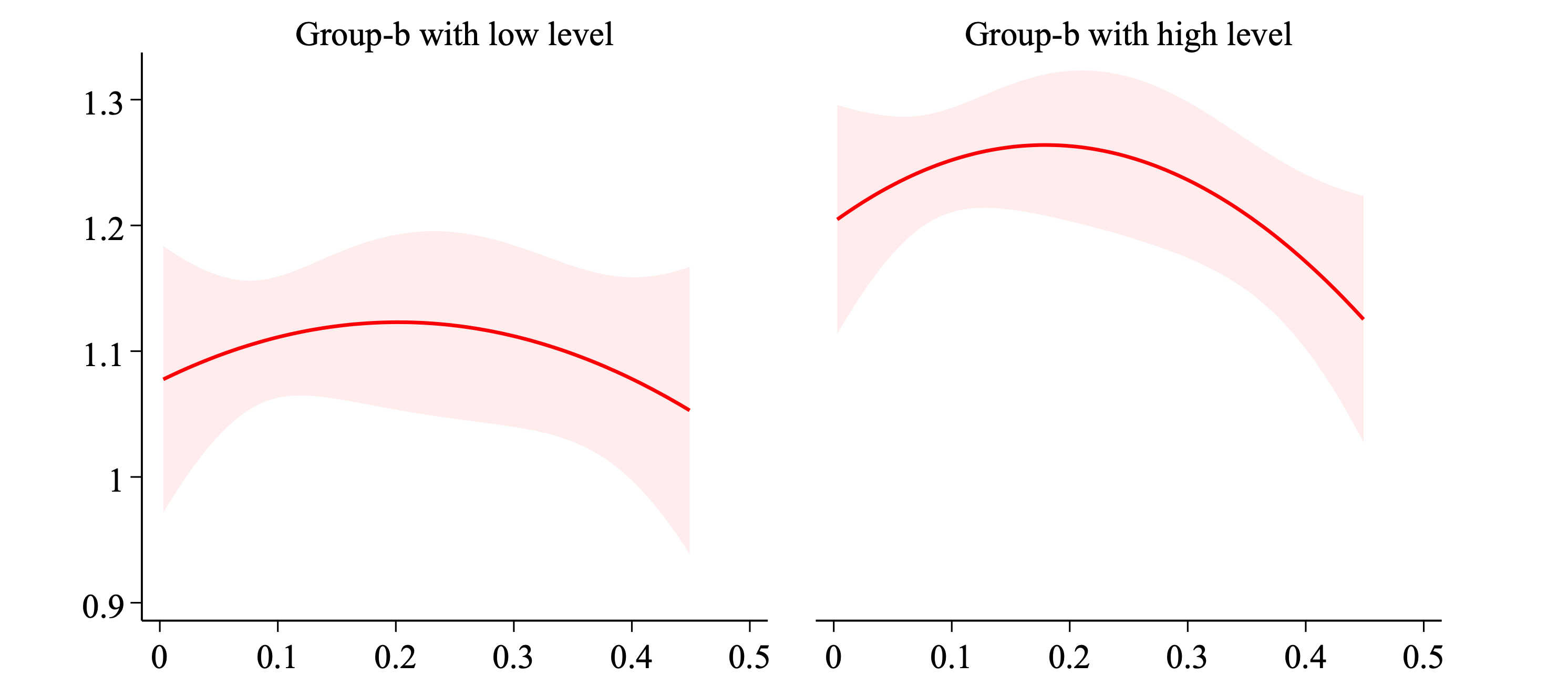}
\caption{\label{fig:F17}RCA and D.[EXRATE] in Group-b}
\end{figure}
\vspace{-0.5cm}

\section{Optimal Policy According to Exchange Rate Fluctuations}\label{Section 5}
\subsection{Linear Trend Aversion}
Certain economic variables exist in three states: growth, decline and neutrality. Each state can have a positive impact on the economy. However, when an economic variable remains persistently in one state, this is termed a linear trend. Since all three states can exert positive effects, in practice we exhibit aversion to linear trends, i. e., \textit{Linear Trend Aversion} (LTA). In this section, we aim to illustrate that linear trends exhibit positive feedback characteristics. That is, an economic variable in a linear trend is more likely to persist in that trend, making it difficult to transition to other states. Therefore, due to this aversion to linear trends, policy intervention becomes imperative.

In the scenario described in this paper, as shown in Figure 5, the USD-CNY bilateral exchange rate exhibited a sustained linear appreciation trend from 2008 to 2015. Consequently, the People's Bank of China implemented exchange rate reform policies at the end of 2015 to attempt altering this state. Through theoretical and empirical analysis, we demonstrate that the policy was effective. However, what changes would occur in the exchange rate's linear trend if no reform policies had been implemented? Specifically, we need to analyze the stability (variance) of the exchange rate's linear trend in the absence of policy. If it is stable (low variance) under natural conditions, the exchange rate would have a high probability of continuing in this state, making policy intervention essential. The opposite holds true.

Referring to \citet{jeanblanc1993impulse}, \citet{mundaca1998optimal}, \citet{cadenillas1999optimal}, \citet{cadenillas2000classical}, \citet{ferrari2020singular} and \citet{gwee2025risk}, we model exchange rate fluctuations from a more general perspective as a Geometric Brownian Motion (GBM). It consists of two components: The first component, with a ratio of $\theta_0\in(0,1)$, represents exchange rate fluctuations influenced by endogenous deterministic factors such as domestic economic policies. The second component, with a ratio of $(1-\theta_0)\in(0,1)$, reflects the impact of exogenous uncertain factors on exchange rate fluctuations, such as international political conditions. This component is modeled by Standard Brownian Motion (SBM). Thus, we get the general dynamics of the exchange rate $S_t$ as follows:
\[dS_t=\theta_0S_tdt+(1-\theta_0)S_tdW_t\]

According to \textit{Ito's Lemma}, let $x=\log(S_t)$, then we get:
\[d\log(S_t)=\left(\theta_0-\frac{1}{2}(1-\theta_0)^2\right)dt+(1-\theta_0)dW_t=\left(-\frac{1}{2}\theta_0^2+2\theta_0-\frac{1}{2}\right)dt+(1-\theta_0)dW_t\]

We denote:
\[\Sigma=\left(-\frac{1}{2}\theta_0^2+2\theta_0-\frac{1}{2}\right),\Omega=(1-\theta_0)\]

Then:
\[dx=\Sigma dt+\Omega dW_t\]

Let $p(x)$ denote the probability density function of $x=\log(S_t)$, satisfying the following \textit{Kolmogorov Forward Equation} (KFE):
\[0=-\Sigma\frac{\partial p(x)}{\partial x}+\frac{1}{2}\Omega^2\left(\frac{\partial^2p(x)}{\partial x^2}\right)-\mathcal{D}_ip(x)+\mathcal{D}_i\delta(x)\]

Solving the KFE, we get the analytical form of $p(x)$:
\[\begin{gathered}
p(x)=A_0exp\left(\frac{\Sigma+\sqrt{\Sigma^2+2\mathcal{D}_i\Omega^2}}{\Omega^2}x\right),x<0\\p(x)=A_0exp\left(\frac{\Sigma-\sqrt{\Sigma^2+2\mathcal{D}_i\Omega^2}}{\Omega^2}x\right),x>0
\end{gathered}\]

We standardize $p(x)$:
\[A_0\left[\int_{-\infty}^0exp\left(\frac{\Sigma+\sqrt{\Sigma^2+2\mathcal{D}_i\Omega^2}}{\Omega^2}x\right)dx+\int_0^\infty exp\left(\frac{\Sigma-\sqrt{\Sigma^2+2\mathcal{D}_i\Omega^2}}{\Omega^2}x\right)dx\right]=1\]

Hence:

\[A_0=\frac{\mathcal{D}_i}{\sqrt{\Sigma^2+2\mathcal{D}_i\Omega^2}}\]

And we get:
\[\begin{gathered}
p(x)=\frac{\mathcal{D}_i}{\sqrt{\Sigma^2+2\mathcal{D}_i\Omega^2}}exp\left(\frac{\Sigma+\sqrt{\Sigma^2+2\mathcal{D}_i\Omega^2}}{\Omega^2}x\right),x<0\\p(x)=\frac{\mathcal{D}_i}{\sqrt{\Sigma^2+2\mathcal{D}_i\Omega^2}}exp\left(\frac{\Sigma-\sqrt{\Sigma^2+2\mathcal{D}_i\Omega^2}}{\Omega^2}x\right),x>0
\end{gathered}\]

Here, $\mathcal{D}_i$ represents exchange rate dissipation. For the 2008–2015 time period shown in Figure 5, we need to simulate the trend of exchange rate appreciation, meaning we set $\mathcal{D}_i > 0$ to indicate exchange rate appreciation. In the scenario we discussed, for a linear appreciation trend, ideally $\mathcal{D}_i$ should be a function of the exchange rate state $x$, i.e., $\mathcal{D}_i(x)$. However, since any existing analytic approach in economics literature, regardless of Laplace transform suggested by \citet{gabaix2016dynamics} or eigenvalue-eigenfunction decomposition taken by \citet{alvarez2022analytic}, is not an ideal technique to handle such a case. A currently feasible approach involves applying “\textit{Path Integral}” methods by using \textit{Feynman-Kac} analysis. Nevertheless, this would significantly deviate from the scope of this paper and increase analytical complexity. Hence, we employ numerical simulation to model the linear appreciation of the exchange rate.

\begin{figure}[htbp]
\centering
\includegraphics[width=8cm]{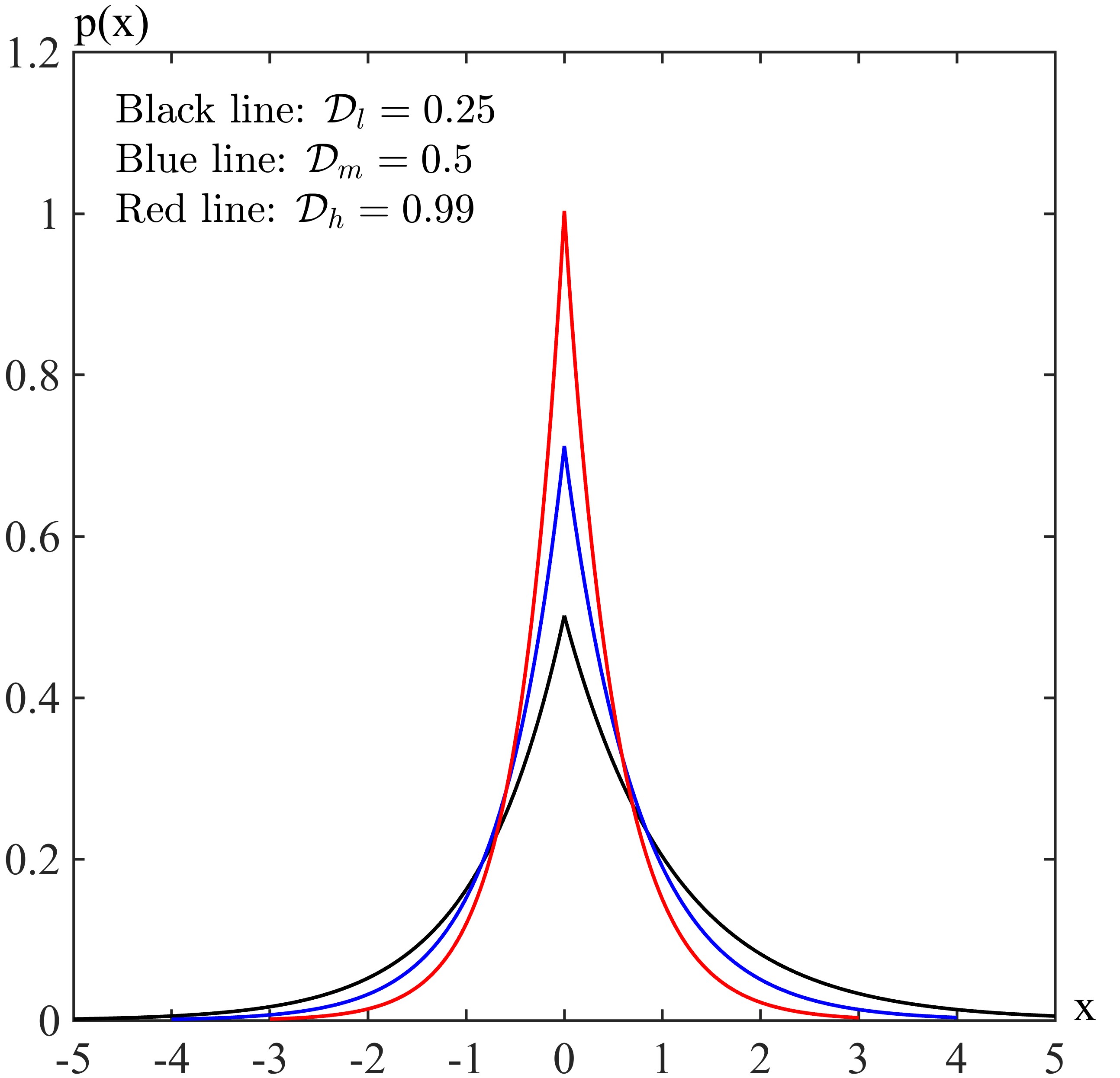}
\caption{\label{fig:F18}Distribution of Exchange Rate Fluctuation with Increasing Dissipation}
\end{figure}

We set $\theta_0=0.3$, $\mathcal{D}_i\in\{\mathcal{D}_l,\mathcal{D}_m,\mathcal{D}_h\}$ and $(\mathcal{D}_l,\mathcal{D}_m,\mathcal{D}_h)=(0.25,0.5,0.99)$, this indicates that the dissipation of the exchange rate is increasing, meaning the exchange rate is in a state of continuous appreciation. Then, we get Figure 18.

According to Figure 18, we observe that the variance in the distribution of exchange rate fluctuation gradually decreases as the dissipation coefficient increases. This implies a positive feedback effect in the linear appreciation trend of the exchange rate: When the exchange rate appreciates linearly, the variance in its state distribution diminishes, indicating that this state tends toward stability. And this suggests that when the exchange rate is in a linear appreciation state, if there is no policy intervention, the exchange rate will have a high probability of continuing to maintain the linear appreciation state and will be difficult to shift to the other two states (depreciation or neutrality). Therefore, due to linear trend aversion, policy is imperative.

\subsection{The Exporters Problem}
In Section 5.1, we attempted to explain that when exchange rate fluctuations exhibit a pronounced linear trend, particularly in the case of extreme appreciation, the variance in the exchange rate distribution gradually diminishes. This implies a higher probability that the exchange rate will continue following its prior linear trend. Under such circumstances, relying solely on market forces to regulate the exchange rate process becomes challenging, making policy intervention imperative. Now, returning to the context of bilateral export trade, we will analyze how to formulate an optimal policy according to exchange rate fluctuations.

We consider a simple binational trade model where domestic exporters supply three types of products to foreign importers: Low-type products ($x_t^L$), Neutral-type products ($x_t^N$) and High-type products ($x_t^H$). Assume that the representative firm produces each of these three types of products with labor input $L_t^0$ and labor compensation $L_t^0$. In each economic cycle, the firm exports these three types of products to foreign markets at their respective prices to generate profits. Therefore, the representative firm problem involves solving the following Bellman equation:
\[Q_t^E(x_t^L,x_t^N,x_t^H)=\max_{c_t,L_t^0,x_{t+1}^L,x_{t+1}^N,x_{t+1}^H}\left[U(X_t)-L_t^0+\rho_x\mathbb{E}\left[V_{t+1}\left(x_{t+1}^L,x_{t+1}^N,x_{t+1}^H\right)\right]\right]\]

s.t.:
\[X_t=p_t^L(x_t^L-x_{t+1}^L)+p_t^N(x_t^N-x_{t+1}^N)+p_t^H(x_t^H-x_{t+1}^H)+L_t^0\]

Then, we obtain:
\[Q_t^E(x_t^L,x_t^N,x_t^H)=\max_{c_t,L_t^0,\\x_{t+1}^L,x_{t+1}^N,x_{t+1}^H}\begin{Bmatrix}U(X_t)+p_t^L(x_t^L-x_{t+1}^L)+p_t^N(x_t^N-x_{t+1}^N)\\+p_t^H(x_t^H-x_{t+1}^H)-c_t+\rho_x\mathbb{E}[V_{t+1}(x_{t+1}^L,x_{t+1}^N,x_{t+1}^H)]\end{Bmatrix}\]

Where:
\[\frac{\partial Q_t^E(x_t^L,x_t^N,x_t^H)}{\partial x_t^L}=p_t^L,\frac{\partial Q_t^E(x_t^L,x_t^N,x_t^H)}{\partial x_t^N}=p_t^N,\frac{\partial Q_t^E(x_t^L,x_t^N,x_t^H)}{\partial x_t^H}=p_t^H\]

Now, we calculate the first-order conditions.

FOC:
\[\begin{gathered}
\rho_x\frac{\partial V_{t+1}(x_{t+1}^L,x_{t+1}^N,x_{t+1}^H)}{\partial x_{t+1}^L}=p_t^L\\\rho_x\frac{\partial V_{t+1}(x_{t+1}^L,x_{t+1}^N,x_{t+1}^H)}{\partial x_{t+1}^N}=p_t^N\\\rho_x\frac{\partial V_{t+1}(x_{t+1}^L,x_{t+1}^N,x_{t+1}^H)}{\partial x_{t+1}^H}=p_t^H
\end{gathered}\]

Then, we assume the importing country exhibits four types of preferences for products: (\textbf{A}) Importing only low-type and neutral-type products, with a proportion $\alpha_{LN}\in(0,1)$. (\textbf{B}) Importing only high-type and neutral-type products, with a proportion $\alpha_{NH}\in(0,1)$. (\textbf{C}) Importing all three types of products, with a proportion $\alpha_{LNH}\in(0,1)$. (\textbf{D}) Importing none of these three types of products, with a proportion of $\alpha_0=(1-\alpha_{LN}-\alpha_{NH}-\alpha_{LNH})\in(0,1)$. Subject to these four constraints, the exporting country's current value function can be expressed as:
\[V_t(x_t^L,x_t^N,x_t^H)=\begin{Bmatrix}\alpha_{LN}[u(x_t^1)+Q_t^E(x_t^L-x_t^{l_1},x_t^N-x_t^{n_1},x_t^H)]\\+\alpha_{NH}[u(x_t^2)+Q_t^E(x_t^L,x_t^N-x_t^{n_2},x_t^H-x_t^{h_2})]\\+\alpha_{LNH}[u(x_t^3)+Q_t^E(x_t^L-x_t^{l_3},x_t^N-x_t^{n_3},x_t^H-x_t^{h_3})]\\+(1-\alpha_{LN}-\alpha_{NH}-\alpha_{LNH})[Q_t^E(x_t^L,x_t^N,x_t^H)]\end{Bmatrix}\]

Then, we can get:
\[V_t(x_t^L,x_t^N,x_t^H)=\begin{Bmatrix}\alpha_{LN}[u(x_t^1)-p_t^Lx_t^{l_1}-p_t^Nx_t^{n_1}]\\+\alpha_{NH}[u(x_t^2)-p_t^Nx_t^{n_2}-p_t^Hx_t^{h_2}]\\+\alpha_{LNH}[u(x_t^3)-p_t^Lx_t^{l_3}-p_t^Nx_t^{n_3}-p_t^Hx_t^{h_3}]\end{Bmatrix}+\alpha_0[Q_t^E(x_t^L,x_t^N,x_t^H)]\]

Now, we denote:
\[V_{t}(x_{t}^{L},x_{t}^{N},x_{t}^{H})=\mathbb{Q}_{t}^{LN}+\mathbb{Q}_{t}^{NH}+\mathbb{Q}_{t}^{LHN}+\alpha_{0}[Q_{t}^{E}(x_{t}^{L},x_{t}^{N},x_{t}^{H})]\]

And:
\[\begin{gathered}
\mathbb{C}_t^1=p_t^Lx_t^L+p_t^Nx_t^N\\\mathbb{C}_t^2=p_t^Hx_t^H+p_t^Nx_t^N\\\mathbb{C}_t^3=p_t^Lx_t^L+p_t^Nx_t^N+p_t^Hx_t^H
\end{gathered}\]

Then we can calculate the first-order conditions.

FOC:
\[\begin{gathered}
\frac{\partial V_t(x_t^L,x_t^N,x_t^H)}{\partial x_t^L}=\frac{\partial\mathbb{Q}_t^{LN}}{\partial x_t^L}+\frac{\partial\mathbb{Q}_t^{LHN}}{\partial x_t^L}+\alpha_0\frac{\partial Q_t^E(x_t^L,x_t^N,x_t^H)}{\partial x_t^L}\\\frac{\partial V_t(x_t^L,x_t^N,x_t^H)}{\partial x_t^L}=\alpha_{LN}\times\frac{\partial u(x_t^1)}{\partial x_t^L}\times\frac{\partial x_t^1}{\partial\mathbb{C}_t^1}\times\frac{\partial\mathbb{C}_t^1}{\partial x_t^L}+\alpha_{LNH}\times\frac{\partial u(x_t^3)}{\partial x_t^L}\times\frac{\partial x_t^3}{\partial\mathbb{C}_t^3}\times\frac{\partial\mathbb{C}_t^3}{\partial x_t^L}+\alpha_0p_t^L\\\frac{\partial V_t(x_t^L,x_t^N,x_t^H)}{\partial x_t^L}=\alpha_{LN}\hat{u}(x_t^1)p_t^L+\alpha_{LNH}\hat{u}(x_t^3)p_t^L+\alpha_0p_t^L
\end{gathered}\]

Similarly:
\[\begin{gathered}
\frac{\partial V_t(x_t^L,x_t^N,x_t^H)}{\partial x_t^H}=\frac{\partial\mathbb{Q}_t^{NH}}{\partial x_t^H}+\frac{\partial\mathbb{Q}_t^{LHN}}{\partial x_t^H}+\alpha_0\frac{\partial Q_t^E(x_t^L,x_t^N,x_t^H)}{\partial x_t^H}\\\frac{\partial V_t(x_t^L,x_t^N,x_t^H)}{\partial x_t^H}=\alpha_{NH}\times\frac{\partial u(x_t^2)}{\partial x_t^H}\times\frac{\partial x_t^2}{\partial\mathbb{C}_t^2}\times\frac{\partial\mathbb{C}_t^2}{\partial x_t^H}+\alpha_{LNH}\times\frac{\partial u(x_t^3)}{\partial x_t^H}\times\frac{\partial x_t^3}{\partial\mathbb{C}_t^3}\times\frac{\partial\mathbb{C}_t^3}{\partial x_t^H}+\alpha_0p_t^H\\\frac{\partial V_t(x_t^L,x_t^N,x_t^H)}{\partial x_t^H}=\alpha_{NH}\hat{u}(x_t^2)p_t^H+\alpha_{LNH}\hat{u}(x_t^3)p_t^H+\alpha_0p_t^H
\end{gathered}\]

And:
\[\begin{gathered}
\frac{\partial V_t(x_t^L,x_t^N,x_t^H)}{\partial x_t^N}=\frac{\partial\mathbb{Q}_t^{LN}}{\partial x_t^N}+\frac{\partial\mathbb{Q}_t^{NH}}{\partial x_t^N}+\frac{\partial\mathbb{Q}_t^{LHN}}{\partial x_t^N}+\alpha_0\frac{\partial Q_t^E(x_t^L,x_t^N,x_t^H)}{\partial x_t^N}\\\frac{\partial V_t(x_t^L,x_t^N,x_t^H)}{\partial x_t^N}=\alpha_{LN}\hat{u}(x_t^1)p_t^N+\alpha_{NH}\hat{u}(x_t^2)p_t^N+\alpha_{LNH}\hat{u}(x_t^3)p_t^N+\alpha_0p_t^N
\end{gathered}\]

Therefore, we can get:
\[\begin{gathered}
p_{t}^{L}=\rho_{x}\alpha_{LN}\hat{u}(x_{t+1}^{1})p_{t+1}^{L}+\rho_{x}\alpha_{LNH}\hat{u}(x_{t+1}^{3})p_{t+1}^{L}+\rho_{x}\alpha_{0}p_{t+1}^{L}\\p_{t}^{H}=\rho_{x}\alpha_{NH}\hat{u}(x_{t+1}^{2})p_{t+1}^{H}+\rho_{x}\alpha_{LNH}\hat{u}(x_{t+1}^{3})p_{t+1}^{H}+\rho_{x}\alpha_{0}p_{t+1}^{H}\\p_{t}^{N}=\rho_{x}\alpha_{LN}\hat{u}(x_{t+1}^{1})p_{t+1}^{N}+\rho_{x}\alpha_{NH}\hat{u}(x_{t+1}^{2})p_{t+1}^{N}+\rho_{x}\alpha_{LNH}\hat{u}(x_{t+1}^{3})p_{t+1}^{N}+\rho_{x}\alpha_{0}p_{t+1}^{N}
\end{gathered}\]

Under the condition of market clearance for export products, the growth rates of the three types of products satisfy:
\[\delta_{t+1}^L=\frac{x_{t+1}^L}{x_t^L}-1,\delta_{t+1}^N=\frac{x_{t+1}^N}{x_t^N}-1,\delta_{t+1}^H=\frac{x_{t+1}^H}{x_t^H}-1\]

Hence:
\[\begin{gathered}
(\delta_{t+1}^L+1)p_t^Lx_t^L=\rho_x\alpha_{LN}\hat{u}(x_{t+1}^1)p_{t+1}^Lx_{t+1}^L+\rho_x\alpha_{LNH}\hat{u}(x_{t+1}^3)p_{t+1}^Lx_{t+1}^L+\rho_x\alpha_0p_{t+1}^Lx_{t+1}^L\\(\delta_{t+1}^H+1)p_t^Hx_t^H=\rho_x\alpha_{NH}\hat{u}(x_{t+1}^2)p_{t+1}^Hx_{t+1}^H+\rho_x\alpha_{LNH}\hat{u}(x_{t+1}^3)p_{t+1}^Hx_{t+1}^H+\rho_x\alpha_0p_{t+1}^Hx_{t+1}^H\\(\delta_{t+1}^N+1)p_t^Nx_t^N=\begin{bmatrix}\rho_x\alpha_{LN}\hat{u}(x_{t+1}^1)p_{t+1}^Nx_{t+1}^N+\rho_x\alpha_{NH}\hat{u}(x_{t+1}^2)p_{t+1}^Nx_{t+1}^N\\+\rho_x\alpha_{LNH}\hat{u}(x_{t+1}^3)p_{t+1}^Nx_{t+1}^N+\rho_x\alpha_0p_{t+1}^Nx_{t+1}^N\end{bmatrix}
\end{gathered}\]

Finally, we can obtain an analytical solution for the revenue dynamics:
\[\begin{gathered}
\frac{p_{t+1}^Lx_{t+1}^L}{p_t^Lx_t^L}=\frac{\delta_{t+1}^L+1}{\rho_x\alpha_{LN}\hat{u}(x_{t+1}^1)+\rho_x\alpha_{LNH}\hat{u}(x_{t+1}^3)+\rho_x\alpha_0}\\\frac{p_{t+1}^Hx_{t+1}^H}{p_t^Hx_t^H}=\frac{\delta_{t+1}^H+1}{\rho_x\alpha_{NH}\hat{u}(x_{t+1}^2)+\rho_x\alpha_{LNH}\hat{u}(x_{t+1}^3)+\rho_x\alpha_0}\\\frac{p_{t+1}^Nx_{t+1}^N}{p_t^Nx_t^N}=\frac{\delta_{t+1}^N+1}{\rho_x\alpha_{LN}\hat{u}(x_{t+1}^1)+\rho_x\alpha_{NH}\hat{u}(x_{t+1}^2)+\rho_x\alpha_{LNH}\hat{u}(x_{t+1}^3)+\rho_x\alpha_0}
\end{gathered}\]

We assume that under market clearance for export products, the growth rates of the three types of products are exogenous. Simultaneously, the discount rate and the proportion of the importing country's preference for the commodities are also exogenous. Then, the change in returns for the three types of products depends solely on $\hat{u}(x_{t+1}^1)$, $\hat{u}(x_{t+1}^2)$ and $\hat{u}(x_{t+1}^3)$. That is, the change in returns for exported products is related to the first-order derivative of the utility function with respect to a certain product under a specific export product combination.

In the scenario discussed in this paper, due to linear trend aversion (LTA), all three exchange rate states exert positive effects on the economy. Therefore, we assume that the utility derived from export activities by the exporting country differs across various exchange rate states. We further set that the exchange rate variable $S_t$ can indirectly influence the utility obtained by the exporting country from exporting the three types of products by affecting export competitiveness, namely $\hat{u}(x_{t+1}^1(S_t))$, $\hat{u}(x_{t+1}^2(S_t))$ and $\hat{u}(x_{t+1}^3(S_t))$, ultimately affecting export revenue. Therefore, here, the revenue from export products is solely a function of the bilateral exchange rate $S_t$.

\subsection{The Central Bank Problem}
In our binational trade model, domestic exporters supply three types of products to foreign importers: Low-type products ($x_t^L$), Neutral-type products ($x_t^N$) and High-type products ($x_t^H$). The proportions of these three types of products, $q_L, q_N, q_H\in(0,1)$, satisfy the normalization condition $q_L+q_N+q_H=1$. The relative magnitude of $q_{j\in\{L,N,H\}}$ reflects the comparative advantage of the exporting country's export capacity. Based on Section 5.1, we define the bilateral exchange rate as a more general Geometric Brownian Motion (GBM):
\[dS_t=\mu_SS_tdt+\sigma_SS_tdW_t\]

Where $\mu_S\in\mathbb{R}$ is the drift rate, reflecting the exchange rate trend. When $\mu_S<0$, the bilateral exchange rate exhibits an appreciation trend, when $\mu_S>0$, it exhibits a depreciation trend. $\sigma_S\in(0,+\infty)$ is the volatility, reflecting exchange rate fluctuations. $\sigma_S\rightarrow0$ indicates increasingly stable exchange rate fluctuations. $W_t$ is a Standard Brownian Motion (SBM).

To simplify the problem, we assume that the revenue for each of the three types of products is a constant function of export volume and export price, which is exogenously given, without considering exchange rate factors. After considering exchange rate fluctuations, the impact of the exchange rate $S_t$ on the three types of products differs:

(\textbf{A}) For low-type products, they have an advantage when the exchange rate depreciates, and the export revenue increases. However, when the exchange rate appreciates, they no longer possess an export advantage.

(\textbf{B}) For neutral-type products, these can be defined as essential products in the importing country. Therefore, exchange rates only affect their export prices but not export volumes. When the exchange rate appreciates, such products gain an advantage, leading to increased export revenues. Conversely, export revenues decrease under the opposite scenario.

(\textbf{C}) For high-type products, the theoretical and empirical analysis sections of the article have already demonstrated that innovation reduces the dependence of exports on exchange rates. Therefore, the returns on high-type products are not directly affected by exchange rate factors. However, since exchange rate volatility could represent uncertainty in exogenous (domestic and international) economic conditions to a certain extent, greater and more intense fluctuations may indicate negative uncertainties disrupting normal economic operations (such as geopolitical conflicts, technological embargoes or changes in trade and tariff treaties). In such cases, exports of innovative products would be affected. Moreover, as shown in Figures 16 and 17, when bilateral exchange rate volatility increases, the export competitiveness of innovative products may face a risk of decline, exhibiting an overall negative trend. Therefore, we define that exchange rate volatility affects the returns of high-type products, establishing a negative correlation between the returns of high-type products and bilateral exchange rate volatility $\sigma_S$.

To simplify the problem and obtain feasible analytical solutions, we define the revenue functions $R_{j\in\{L,N,H\},t}$ for the three types of products as following:
\[\begin{gathered}
R_{Lt}=q_La_SS_t\\R_{Nt}=q_Nb_S\frac{1}{S_t}\\R_{Ht}=q_H(c_S-d_S\sigma_S)
\end{gathered}\]

Where $a_S, b_S, c_S, d_S>0$ represent constant return functions. Then the total return at time $t$ is as follows:
\[\Pi_t=q_La_SS_t+q_Nb_S\frac{1}{S_t}+q_H(c_S-d_S\sigma_S)\]

The central bank solves the following optimization problem by adjusting $\mu_S$ and $\sigma_S$ through monetary policy:
\[V_t(\mu_S,\sigma_S)=\max_{\mu_S,\sigma_S}\mathbb{E}\left[\int_0^\infty e^{-\rho_0t}\Pi_tdt\right]\]

Hence:
\[V_{t}(\mu_{S},\sigma_{S})=\int_{0}^{\infty}e^{-\rho_{0}t}\mathbb{E}[\Pi_{t}]dt=\int_{0}^{\infty}e^{-\rho_{0}t}\mathbb{E}[R_{Lt}]dt+\int_{0}^{\infty}e^{-\rho_{0}t}\mathbb{E}[R_{Nt}]dt+\int_{0}^{\infty}e^{-\rho_{0}t}\mathbb{E}[R_{Ht}]dt\]

We set the initial exchange rate level to $S_0$, then:
\[\mathbb{E}[S_t]=S_0e^{\mu_St},\mathbb{E}\left[\frac{1}{S_t}\right]=\frac{1}{S_0}e^{(-\mu_S+\sigma_S^2)t}\]

Therefore, we get:
\[V_{t}(\mu_{S},\sigma_{S})=\frac{q_{L}a_{S}S_{0}}{\rho_{0}-\mu_{S}}+\frac{q_{N}b_{S}}{S_{0}(\rho_{0}+\mu_{S}-\sigma_{S}^{2})}+\frac{q_{H}(c_{S}-d_{S}\sigma_{S})}{\rho_{0}}\]

s.t.:
\[\begin{gathered}
\rho_{0}>0\\\sigma_{S}\geq0\\\rho_{0}>\mu_{S}\\\rho_{0}+\mu_{S}-\sigma_{S}^{2}>0
\end{gathered}\]

FOC:
\[\begin{gathered}
\frac{\partial V_t(\mu_S,\sigma_S)}{\partial\mu_S}=0\\\frac{q_La_SS_0}{(\rho_0-\mu_S)^2}=\frac{q_Nb_S}{S_0(\rho_0+\mu_S-\sigma_S^2)^2}\\\frac{\partial V_t(\mu_S,\sigma_S)}{\partial\sigma_S}=0\\\frac{2\sigma_Sq_Nb_S}{S_0(\rho_0+\mu_S-\sigma_S^2)^2}=\frac{q_Hd_S}{\rho_0}
\end{gathered}\]

We denote:
\[K=\sqrt{\frac{q_Nb_S}{q_La_SS_0^2}}=\frac{\rho_0+\mu_S-\sigma_S^2}{\rho_0-\mu_S}\]

Then:
\[\mu_S=\frac{\rho_0(K-1)+\sigma_S^2}{1+K}\]

Now, we denote:
\[M=\frac{q_{H}d_{S}S_{0}}{q_{N}b_{S}\rho_{0}}=\frac{2\sigma_{S}}{(\rho_{0}+\mu_{S}-\sigma_{S}^{2})^{2}}\]

Substituting $K$:
\[\begin{gathered}
\rho_0+\mu_S-\sigma_S^2=K(\rho_0-\mu_S)\\(\rho_0-\mu_S)^2=\frac{2\sigma_S}{K^2M}\\\rho_0-\mu_S=\rho_0-\frac{\rho_0(K-1)+\sigma_S^2}{1+K}=\frac{2\rho_0-\sigma_S^2}{1+K}
\end{gathered}\]

Hence:
\[\left(\frac{2\rho_{0}-\sigma_{S}^{2}}{1+K}\right)^{2}=\frac{2\sigma_{S}}{K^{2}M}\]

Where:
\[A=\frac{2(1+K)^2}{K^2M}\]

And we can get:
\[\begin{gathered}
A\sigma_{S}=(2\rho_{0}-\sigma_{S}^{2})^{2}\\\sigma_{S}^{4}-4\rho_{0}\sigma_{S}^{2}-A\sigma_{S}+4\rho_{0}^{2}=0
\end{gathered}\]

We will employ the \textit{Newton-Raphson Iteration} algorithm to solve the quartic equation for $\sigma_S$. After obtaining the optimal solution $\sigma_S^*$, we will substitute it into the equation to calculate the optimal solution $\mu_S^*$.

\subsection{Parameter Calibration and Numerical Simulation}
We categorize export trade into three structures based on the comparative advantages of exporting countries' export capabilities:

The first structure involves the trade dominated by low-type products $(q_L^-,q_N^-,q_H^-)$ and is defined as labor-intensive exports, where $q_L^->q_H^-, q_L^->q_N^-$. It is low-type trade structure. The second structure is the trade dominated by exports of high-type products $(q_L^+,q_N^+,q_H^+)$, defined as technologically advanced exports where $q_H^+>q_L^+, q_H^+>q_N^+$. It is high-type trade structure. The third structure is a balanced export trade $(q_L^0,q_N^0,q_H^0)$, where $q_L^0\approx q_N^0\approx q_H^0$. It is neutral-type trade structure. Furthermore, we define this type of export trade shift as neutral, exhibiting neither a tendency toward labor-intensive exports $max\{q_L,q_N,q_H\}=q_L$ nor a tendency toward technology-intensive exports $max\{q_L,q_N,q_H\}=q_H$. It would mainly focus on exporting neutral-type products $max\{q_L,q_N,q_H\}=q_N$. And the initial (standard) constant revenue function $(a_S^0,b_S^0,d_S^0)$ is set as shown in Table 6.

\begin{table}[!ht]
\centering
\caption{Standard Economic Parameter Setting}
\label{6}
\small
\setlength{\tabcolsep}{18pt}
\renewcommand{\arraystretch}{1.5}
\begin{tabular}{@{}l c c@{}}
\toprule
\textbf{Parameter} & \textbf{Notation} & \textbf{Value} \\
\midrule
Discount Rate & $\rho_{0}$ & 0.05 \\

Initial Exchange Rate & $S_{0}$ & 1 \\

Profit of Low-type Product & $a_{S}^{0}$ & 1 \\

Profit of Neutral-type Product & $b_{S}^{0}$ & 1 \\

Profit of High-type Product & $d_{S}^{0}$ & 1 \\

\multirow{3}{*}{Low-type Trade Structure} & $q_{L}^{-}$ & 0.7 \\
                                          & $q_{N}^{-}$ & 0.2 \\
                                          & $q_{H}^{-}$ & 0.1 \\

\multirow{3}{*}{High-type Trade Structure} & $q_{L}^{+}$ & 0.1 \\
                                           & $q_{N}^{+}$ & 0.2 \\
                                           & $q_{H}^{+}$ & 0.7 \\

\multirow{3}{*}{Neutral-type Trade Structure} & $q_{L}^{0}$ & 0.3 \\
                                              & $q_{N}^{0}$ & 0.4 \\
                                              & $q_{H}^{0}$ & 0.3 \\
\bottomrule
\end{tabular}
\end{table}

According to Table 6, we can use the algorithm to calculate the optimal parameters $\sigma_S^*$ and $\mu_S^*$ for exchange rate fluctuations under three trade structures with the given profit function combination $(a_S^0,b_S^0,d_S^0)$, utility discount rate $\rho_0$ and initial exchange rate level $S_0$, as shown in Table 7:

\begin{table}[htbp]
\centering
\caption{Optimal Parameter $\sigma_S^*$ and $\mu_S^*$}
\label{7}
\small
\renewcommand{\arraystretch}{1.5}
\begin{tabularx}{\textwidth}{l >{\centering\arraybackslash}X c c}
\toprule
\textbf{Trade Structure} & \textbf{Parameter Group} & $\bm{\sigma_{S}^{*}}$ & $\bm{\mu_{S}^{*}}$ \\
\midrule
Low-type Trade Structure & 
$\left\{ S_{0},\rho_{0},a_{S}^{0},b_{S}^{0},d_{S}^{0},q_{L}^{-},q_{N}^{-},q_{H}^{-} \right\}$ & 
0.006062 & -0.015143 \\
\addlinespace[0.3cm]
High-type Trade Structure & 
$\left\{ S_{0},\rho_{0},a_{S}^{0},b_{S}^{0},d_{S}^{0},q_{L}^{+},q_{N}^{+},q_{H}^{+} \right\}$ & 
0.098098 & 0.012565 \\
\addlinespace[0.3cm]
Neutral-type Trade Structure & 
$\left\{ S_{0},\rho_{0},a_{S}^{0},b_{S}^{0},d_{S}^{0},q_{L}^{0},q_{N}^{0},q_{H}^{0} \right\}$ & 
0.021343 & 0.003801 \\
\bottomrule
\end{tabularx}
\end{table}

Based on the results in Table 7, we find that for $\sigma_S^*$, while reduced exchange rate volatility unilaterally benefits exports of high-type products, the optimal scenario exhibits the highest exchange rate volatility in the high-type trade structure. This indicates that such a trade structure possesses greater resilience and flexibility to absorb more exogenous shocks and hence can withstand stronger exchange rate fluctuations. The exchange rate volatility is minimal in the low-type trade structure, indicating that this trade structure is relatively fragile compared to the other two trade structures. It requires more stable exchange rate fluctuations to maintain its optimal profit state.

For $\mu_S^*$, both the high-type and neutral-type trade structures exhibit a slight depreciation trend in bilateral exchange rates, i.e., $\mu_S^*>0$. However, for the low-type trade structure, although exchange rate depreciation unilaterally benefits exports of low-type products, under optimal conditions, the bilateral exchange rate in the low-type structure shows an appreciation trend, i.e., $\mu_S^*<0$. This indicates that reliance on low-type products exports may lead to “impoverished growth”, which means increased export volumes but deteriorating conditions of trade, resulting in declining real income. Therefore, the exchange rate should be gradually appreciated under optimal strategy in order to force enterprises to shift towards the production and $R\&D$ of neutral-type or high-type products, improve the conditions of trade, then promote economic structural transformation and the transition of comparative advantage.

According to the parameter settings in Table 6, we further change the proportions of the three types of products in the three trade structures to analyze the evolutions of the optimal parameters $\sigma_S^*$ and $\mu_S^*$. We set $q_L,q_N,q_H\in(0,1)$ as variables, so the optimal parameters for the bilateral exchange rate become endogenous to them, namely $\sigma_S^*(q_L,q_N,q_H)$ and $\mu_S^*(q_L,q_N,q_H)$. Figures 19, 20, and 21 respectively illustrate the changes in optimal parameters for the low-type, high-type and neutral-type trade structures.

Based on the overall situations of the three figures, we observe that: For the optimal trade conditions $\sigma_S^*(q_L,q_N,q_H)$, increasing the export proportions of low-type and neutral-type products reduces exchange rate volatility, which makes the exchange rate more stable. Conversely, increasing exports of high-type products will increase exchange rate volatility. Meanwhile, for the optimal trade condition $\mu_S^*(q_L,q_N,q_H)$, changes in the export proportion of high-type products have a weak impact on exchange rate trends. Only when the proportion of high-type products is already high and continues to increase its proportion will make the exchange rate show a slight depreciation trend. However, changes in the export proportions of low-type and neutral-type products significantly impact exchange rate trends. Increasing the proportion of low-type products leads to a pronounced appreciation trend, while increasing the proportion of neutral-type products results in a pronounced depreciation trend.

Therefore, from an overall perspective, if we increase the export proportions of the three types of products separately, based on the results presented by bilateral exchange rates and the respective profit functions of the three types of products, this would be detrimental to their unilateral profit growth. We have two possible explanations for this: 

The first is to avoid the emergence of “Dutch disease” (where the prosperity of one sector leads to the decline of others, making the economic structure monolithic and vulnerable). If exports of a specific product type expand excessively, it may cause economic resources (such as labor and capital, etc) to concentrate in that sector, squeezing the survival and development of other sectors or enterprises. Therefore, the optimal exchange rate policy should promote balanced economic structure rather than unilateral exchange rate adjustments based solely on trade advantages.

The second concerns the use of exchange rate mechanisms. Under optimal conditions, policy should leverage exchange rate mechanisms to promote trade structure transformation rather than reinforce existing comparative advantages. For example: Increasing the export proportion of low-type products can reduce exchange rate volatility and create an appreciating trend in the exchange rate. This would unilaterally benefit exports of high-type and neutral-type products, thus promoting the upgrading of low-type products, the production and $R\&D$ of high-type products. Then, enabling a gradual transition from a low-type trade structure to a high-type or neutral-type trade structure.

\begin{figure}[htbp]
\centering
\includegraphics[width=15.5cm]{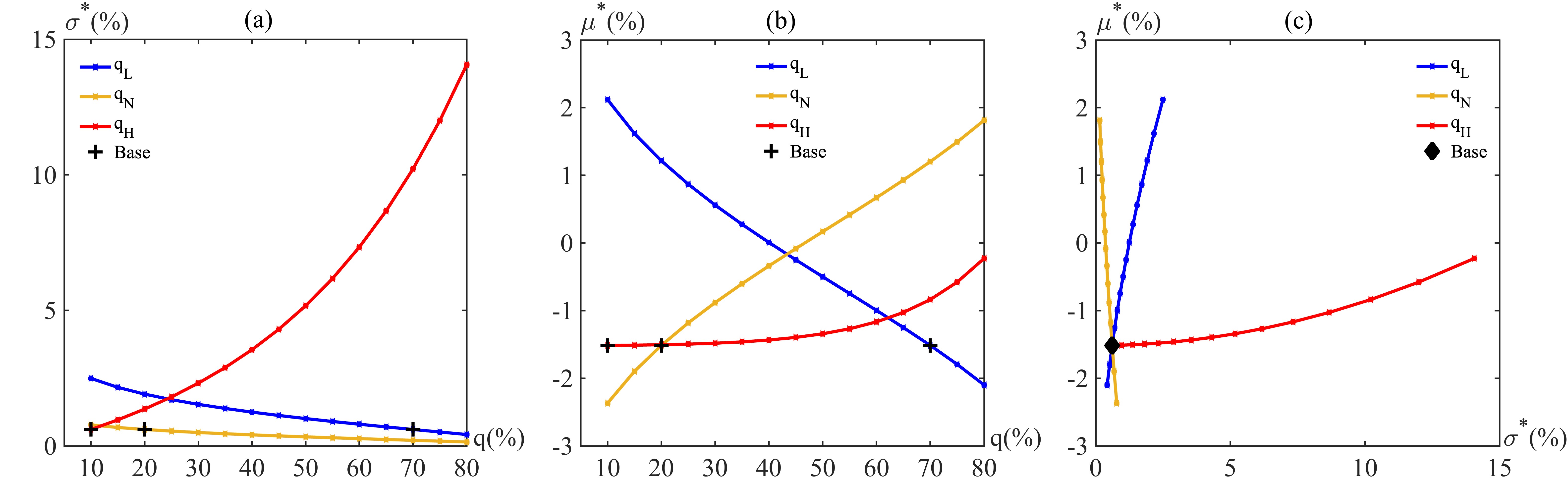}
\caption{\label{fig:F19}The Evolution of Optimal Parameter in Low-type Trade Structure}
\end{figure}
\vspace{-0.3cm}
\begin{figure}[htbp]
\centering
\includegraphics[width=15.5cm]{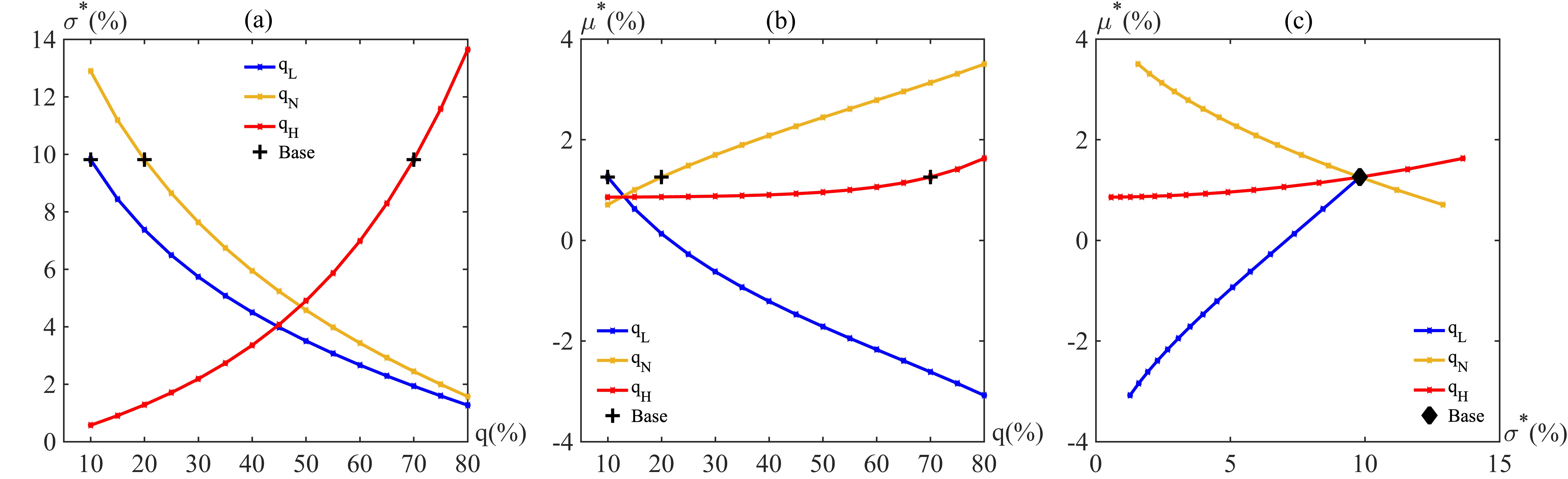}
\caption{\label{fig:F20}The Evolution of Optimal Parameter in High-type Trade Structure}
\end{figure}
\vspace{-0.3cm}
\begin{figure}[htbp]
\centering
\includegraphics[width=15.5cm]{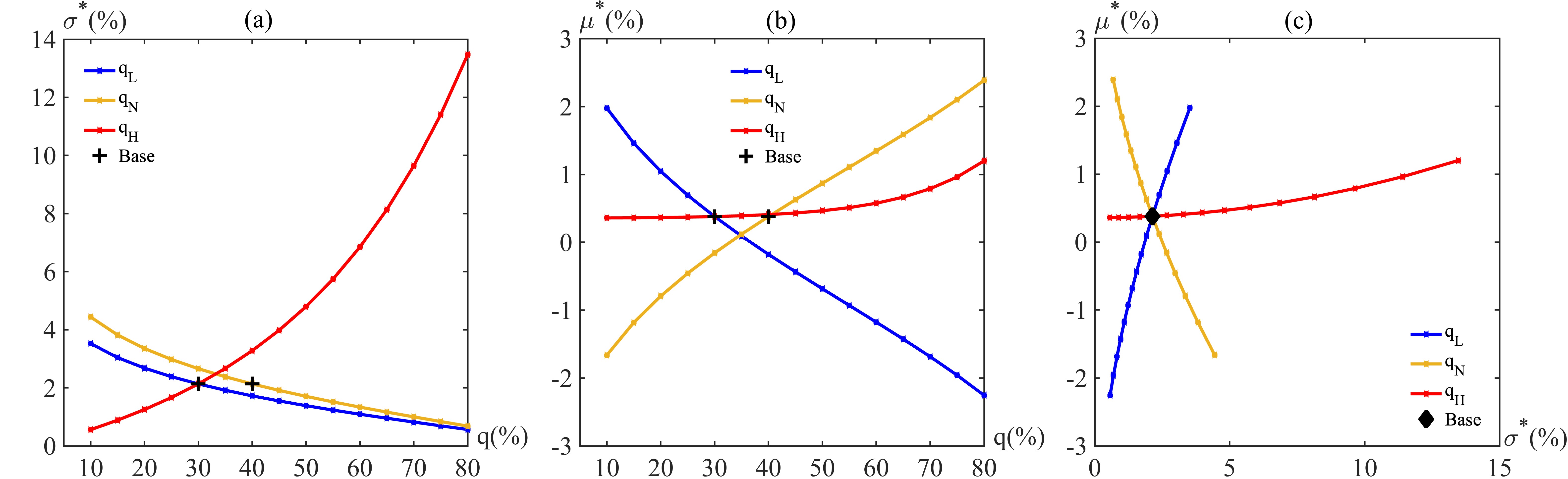}
\caption{\label{fig:F21}The Evolution of Optimal Parameter in Neutral-type Trade Structure}
\end{figure}

Additionally, as shown in Figures 19-a, 20-a, and 21-a, the marginal effect of changing product proportions on exchange rate fluctuations is heterogeneous across different trade structures for the optimal exchange rate volatility $\sigma_S^*$. In the high-type trade structure, increasing the export proportion of low-type and neutral-type products influences exchange rate volatility with the most pronounced marginal effect. In this type of trade structure, raising the export proportion of low-type and neutral-type products can significantly reduce exchange rate volatility. However, for both low-type and neutral-type trade structures, the marginal effect of increasing the export proportions of low-type and neutral-type products on exchange rate volatility is relatively weak. Particularly in the low-type trade structure, increasing the export proportions of low-type and neutral-type products will have the smallest effect on reducing exchange rate volatility. 

This shows that in the trade structure where innovation serves as a comparative advantage, appropriately increasing the export proportions of low-type and neutral-type products can effectively reduce exchange rate volatility and promote stable exchange rate fluctuations. Consequently, this unilaterally benefits the growth of export revenues for high-type products, which can promote the production and $R\&D$ of innovative products to a certain extent and strengthen the competitive advantage in innovation.

\subsection{The Sensitivity Analysis}
Based on Table 6, we further perform a simple linear transformation on the standard economic parameters $(\rho_0,S_0,a_S^0,b_S^0,d_S^0)$ to analyze the sensitivity of the optimal volatility and optimal drift rate $(\sigma_S^*,\mu_S^*)$ in the exchange rate fluctuations to these economic parameters. Where:
\[(\rho_{0}^{\kappa},S_{0}^{\kappa},a_{S}^{\kappa},b_{S}^{\kappa},d_{S}^{\kappa})=\kappa\times(\rho_{0},S_{0},a_{S}^{0},b_{S}^{0},d_{S}^{0}),\kappa\in(0.5,1.5)\]

As shown in Figures 22-a, 23-a, and 24-a: Increasing $\rho_0^{\kappa}$ and $d_S^{\kappa}$ significantly raises $\sigma_S^*$, indicating that higher utility discount rates and high-type product revenue coefficients are detrimental to stable exchange rate fluctuations. Increases in $a_S^{\kappa}$ and $b_S^{\kappa}$ markedly reduce $\sigma_S^*$, indicating that higher revenue coefficients for low-type and neutral-type products promote stable exchange rate fluctuations. Regarding the initial exchange rate level $S_0^{\kappa}$, it does not directly indicate that $\sigma_S^*$ exhibits strong sensitivity to $S_0^{\kappa}$. In high-type and neutral-type trade structures, an increase in $S_0^{\kappa}$ slightly raises $\sigma_S^*$. However, in low-type trade structures, an increase in $S_0^{\kappa}$ reduces $\sigma_S^*$ to a certain extent.

And according to Figures 22-b, 23-b, and 24-b: We find that the sensitivity of $\mu_S^*$ to the high-type product revenue coefficient $d_S^{\kappa}$ is weak. Only in the high-type trade structure does an increase in $d_S^{\kappa}$ lead to an increase in $\mu_S^*$. Regarding the initial exchange rate level $S_0^{\kappa}$ and the low-type product revenue coefficient $a_S^{\kappa}$, they show a clear negative correlation with $\mu_S^*$. This implies that increasing the initial exchange rate level and the low-type product revenue coefficient can lead to an appreciation trend in the exchange rate. For the neutral-type product revenue coefficient $b_S^{\kappa}$, it exhibits a positive correlation with $\mu_S^*$ across all three trade structures, indicating that increasing neutral-type revenue coefficient promotes exchange rate depreciation. However, regarding the utility discount rate $\rho_0^{\kappa}$, the sensitivity of $\sigma_S^*$ to $\rho_0^{\kappa}$ shows heterogeneity depending on the trade structure. In the low-type trade structure, an increase in $\rho_0^{\kappa}$ reduces $\mu_S^*$, promoting exchange rate appreciation. In contrast, in the other two trade structures, an increase in $\rho_0^{\kappa}$ increases $\mu_S^*$, promoting exchange rate depreciation.

\begin{figure}[!ht]
\centering
\includegraphics[width=15.5cm]{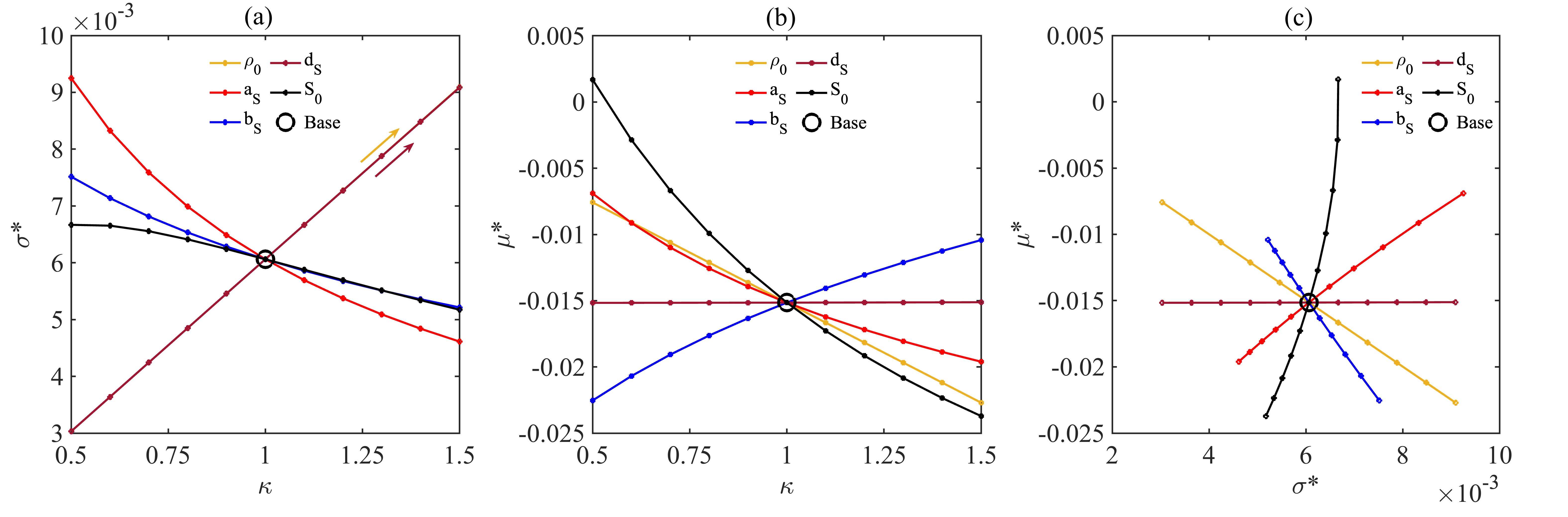}
\vspace{-0.8cm}
\caption{\label{fig:F22}The Sensitivity of Optimal Parameter in Low-type Trade Structure}
\end{figure}
\vspace{-0.3cm}
\begin{figure}[!ht]
\centering
\includegraphics[width=15.5cm]{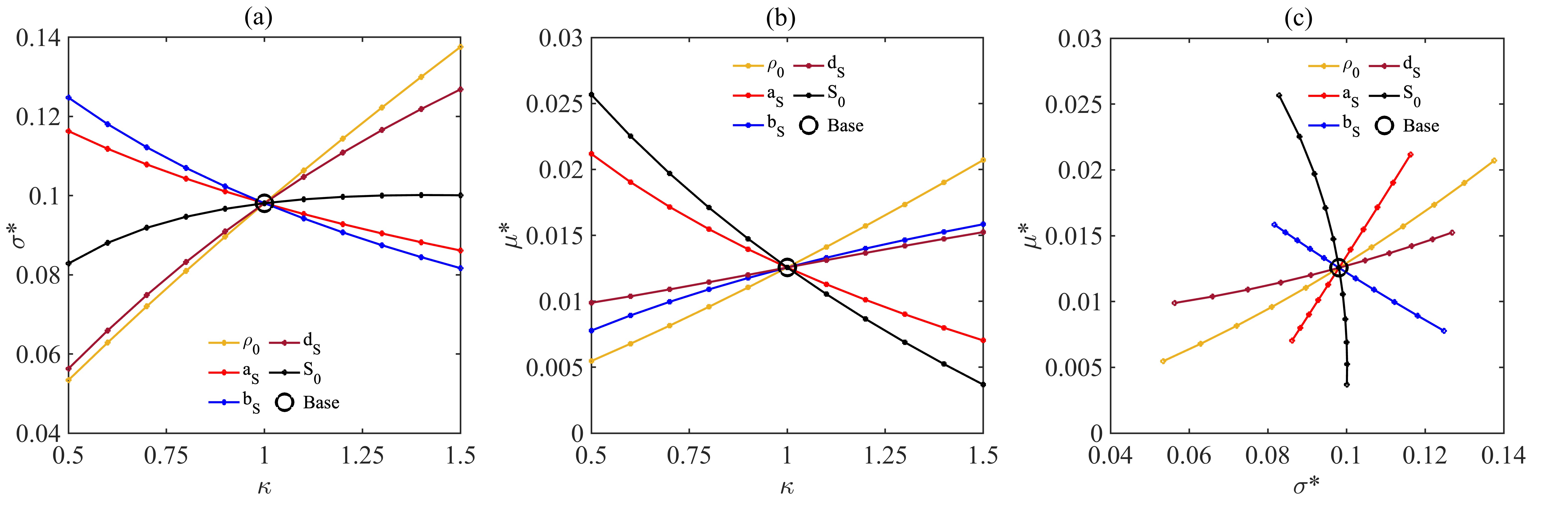}
\vspace{-0.8cm}
\caption{\label{fig:F23}The Sensitivity of Optimal Parameter in High-type Trade Structure}
\end{figure}
\vspace{-0.3cm}
\begin{figure}[!ht]
\centering
\includegraphics[width=15.5cm]{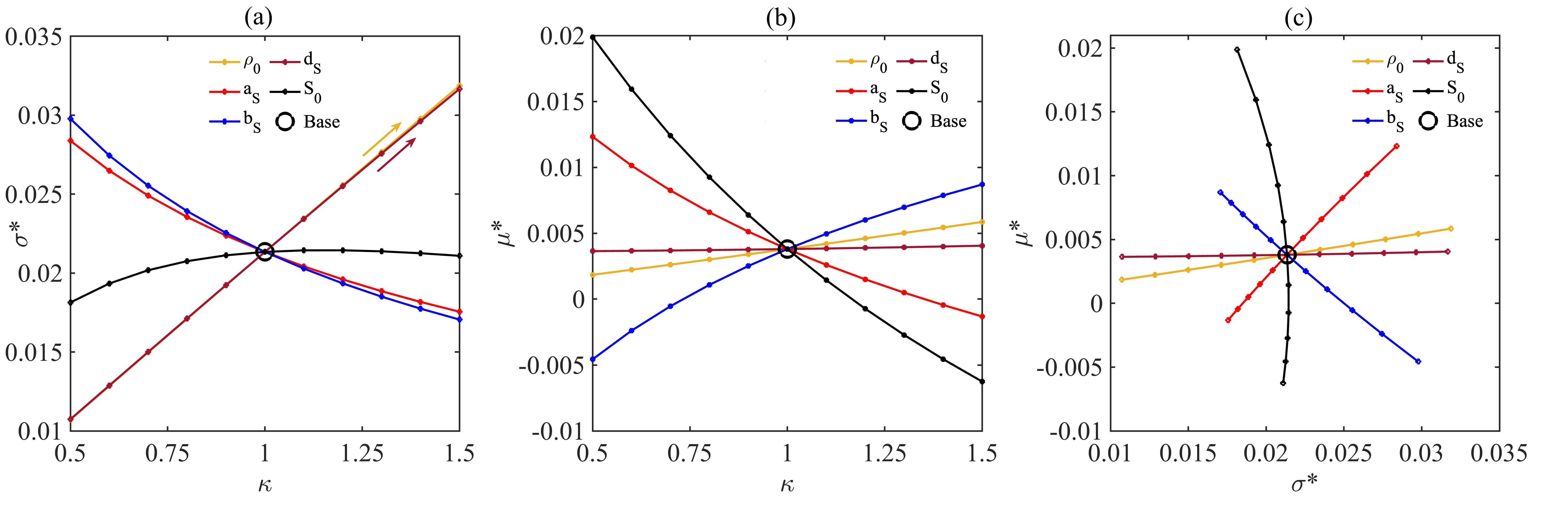}
\vspace{-0.8cm}
\caption{\label{fig:F24}The Sensitivity of Optimal Parameter in Neutral-type Trade Structure}
\end{figure}
\vspace{-0.3cm}

\subsection{Wealth Distribution According to Trade Structure}
Heterogeneous agents model is widely applied in research on topics such as household income, public tax, firm wealth distribution and monetary policy, etc. However, the application of this model to binational trade issues is relatively rare. In particular, there has been little research on the dynamic evolution of firm wealth distribution in different export trade structures. For specific references of this model, see: \cite{huggett1993risk}, \cite{aiyagari1994uninsured}, \cite{krusell1998income}, \cite{adrian2010liquidity}, \cite{brunnermeier2014macroeconomic}, \cite{moll2014productivity}, \cite{achdou2014partial}, \cite{buera2015aggregate}, \cite{gabaix2016dynamics}, \cite{kaplan2018monetary}, \cite{ahn2018inequality}, \cite{achdou2020mean}, \cite{achdou2022income}, \cite{wr2022sentiment}, \cite{bilal2023solving}, \cite{fernandez2023financial}, \cite{moll2025mean}, \cite{maxted2025present}, \cite{le2025optimal}, \cite{gomez2025wealth}, \cite{campbell2025sustainability} and \cite{hu2025big}. In Section 5.6, we will study the firm wealth distribution of exporters in different trade structures by referring to some of the methods described in the literature above.

In the scenario discussed in our paper, firm wealth $N_t$ comes from two types of assets: Risk-free Assets and Risky Assets. We assume firms allocate a proportion $\theta_n\in (0,1)$ of their wealth to risk-free assets, with a return rate $\psi_n>0$. Similarly, they allocate the remaining proportion $(1-\theta_n)$ to risky assets, with a return rate $\omega_n>0$. Here, the accumulation of risk-free assets is related to the interest rate $r_n$, consumption $C_n$ and the expected profit $\mathbb{E}[\Pi_t]$ earned by the firm through export trade. The risk assets follow a Standard Brownian Motion $dW_t$. Thus, the wealth accumulation equation for firms in an exporting country is as following:
\[dN_t=\theta_n\psi_n(r_n+\mathbb{E}[\Pi_t]-C_n)N_tdt+(1-\theta_n)\omega_nN_tdW_t\]

We  denote $x^n=\log(N_t)$. Then, according to the \textit{Ito's Lemma}, we obtain:
\[dx^n=\left(\theta_n\psi_nr_n+\theta_n\psi_n\mathbb{E}[\Pi_t]-\theta_n\psi_nC_n-\frac{1}{2}(\omega_n-\theta_n\omega_n)^2\right)dt+(\omega_n-\theta_n\omega_n)dW_t\]

We denote:
\[\begin{gathered}
\Sigma^n=\theta_n\psi_nr_n+\theta_n\psi_n\mathbb{E}[\Pi_t]-\theta_n\psi_nC_n-\frac{1}{2}(\omega_n-\theta_n\omega_n)^2\\\Omega^n=\omega_n-\theta_n\omega_n
\end{gathered}\]

And:
\[\mathbb{E}[\Pi_t]=q_La_SS_0e^{\mu_St}+q_Nb_S\frac{1}{S_0}e^{(-\mu_S+\sigma_S^2)t}+q_H(c_S-d_S\sigma_S)\]

Then:
\[dx_t^n=\Sigma^ndt+\Omega^ndW_t\]

Let the probability density function of $x^n=\log(N_t)$ be denoted as $p(x^n)$, and it satisfies the following \textit{Kolmogorov Forward Equation} (KFE):
\[0=-\Sigma^n\frac{\partial p(x^n)}{\partial x^n}+\frac{1}{2}(\Omega^n)^2\left(\frac{\partial^2p(x^n)}{\partial(x^n)^2}\right)-\mathcal{D}p(x^n)+\mathcal{D}\delta(x^n)\]

Then, referring to the method in Section 5.1, we can solve the analytical form of the firm wealth distribution:
\[\begin{gathered}
p(x^n)=A_nexp\left(\frac{\Sigma^n+\sqrt{(\Sigma^n)^2+2\mathcal{D}(\Omega^n)^2}}{(\Omega^n)^2}x^n\right),x^n<0\\p(x^n)=A_nexp\left(\frac{\Sigma^n-\sqrt{(\Sigma^n)^2+2\mathcal{D}(\Omega^n)^2}}{(\Omega^n)^2}x^n\right),x^n>0
\end{gathered}\]

Finally, according to $\int_{-\infty}^0p(x^n)dx^n+\int_0^\infty p(x^n)dx^n=1$, we get:
\[\begin{gathered}
p(x^n)=\frac{\mathcal{D}}{\sqrt{(\Sigma^n)^2+2\mathcal{D}(\Omega^n)^2}}exp\left(\frac{\Sigma^n+\sqrt{(\Sigma^n)^2+2\mathcal{D}(\Omega^n)^2}}{(\Omega^n)^2}x^n\right),x^n<0\\p(x^n)=\frac{\mathcal{D}}{\sqrt{(\Sigma^n)^2+2\mathcal{D}(\Omega^n)^2}}exp\left(\frac{\Sigma^n-\sqrt{(\Sigma^n)^2+2\mathcal{D}(\Omega^n)^2}}{(\Omega^n)^2}x^n\right),x^n>0
\end{gathered}\]

We assume that the economic parameters related to asset returns are exogenous, setting $\theta_n=0.6,\psi_n=0.02,\omega_n=0.2,r_n=0.05,C_n=0.1,\mathcal{D}=0.5$. Substituting the standard economic return parameters from Table 6 (supplement: $c_S^0=1$) and the respective optimal parameters $(\sigma_S^*,\mu_S^*)$ under the three types of trade structures, then we can get the expectation of return in each trade structure:
\[\begin{gathered}
\mathbb{E}[\Pi_t^-(\sigma_S^*,\mu_S^*)]=q_L^-a_S^0S_0e^{\mu_S^*t}+q_N^-b_S^0\frac{1}{S_0}e^{(-\mu_S^*+(\sigma_S^*)^2)t}+q_H^-(c_S^0-d_S^0\sigma_S^*)\\\mathbb{E}[\Pi_t^+(\sigma_S^*,\mu_S^*)]=q_L^+\alpha_S^0S_0e^{\mu_S^*t}+q_N^+b_S^0\frac{1}{S_0}e^{(-\mu_S^*+(\sigma_S^*)^2)t}+q_H^+(c_S^0-d_S^0\sigma_S^*)\\\mathbb{E}[\Pi_t^0(\sigma_S^*,\mu_S^*)]=q_L^0a_S^0S_0e^{\mu_S^*t}+q_N^0b_S^0\frac{1}{S_0}e^{(-\mu_S^*+(\sigma_S^*)^2)t}+q_H^0(c_S^0-d_S^0\sigma_S^*)
\end{gathered}\]

Figures 25, 26, and 27 respectively illustrate the wealth distribution among the firms in low-type, high-type and neutral-type trade structures. 

Based on Figures 25-a, 26-a, and 27-a, we observe that in the short term, the wealth distributions of firms across the three trade structures all resemble normal distributions, and the dynamic shift patterns of the wealth distribution curves are stable and homogeneous. However, according to Figures 25-b, 26-b, and 27-b, in the scenario of long term, the dynamic shift patterns of the wealth distribution curves exhibit heterogeneous characteristics:

(\textbf{A}) The wealth distribution curve ultimately converges over time, which is consistent with the general patterns observed in dynamic simulations. However, the convergence rate exhibits a heterogeneity depending on the trade structure, with the low-type trade structure converging the fastest and the neutral-type trade structure converging the slowest.

(\textbf{B}) The wealth distribution curve shifts dynamically over time, transitioning from an approximately normal distribution curve to a right-skewed distribution curve, with both the mean and variance of wealth distribution increasing. This indicates that in each type of trade structure, the mean wealth of firms gradually increases over time, but it also leads to more serious wealth inequality.

(\textbf{C}) The rate of increase in the right-tail thickness of wealth distribution (i.e., the rate in which the mean and variance of wealth distribution grow) exhibits a threshold. Prior to this threshold, the right-tail thickness of wealth distribution will continue to increase over time, signifying a persistent rise in both the mean and variance of wealth distribution. After the threshold, the wealth distribution enters a convergence process, with the economy gradually approaching a steady state. Simultaneously, the magnitude of the threshold exhibits heterogeneity depending on the trade structure. In the low-type trade structure, the right-tail thickness of the wealth distribution reaches its maximum between $t\in(200,400)$. In the high-type trade structure, the right-tail thickness of the wealth distribution peaks between $t\in(400,600)$. And in the neutral-type trade structure, the right-tail thickness of the wealth distribution peaks between $t\in(1000,1500)$.

\vspace{+0.2cm}
\begin{figure}[!ht]
\centering
\includegraphics[width=13cm]{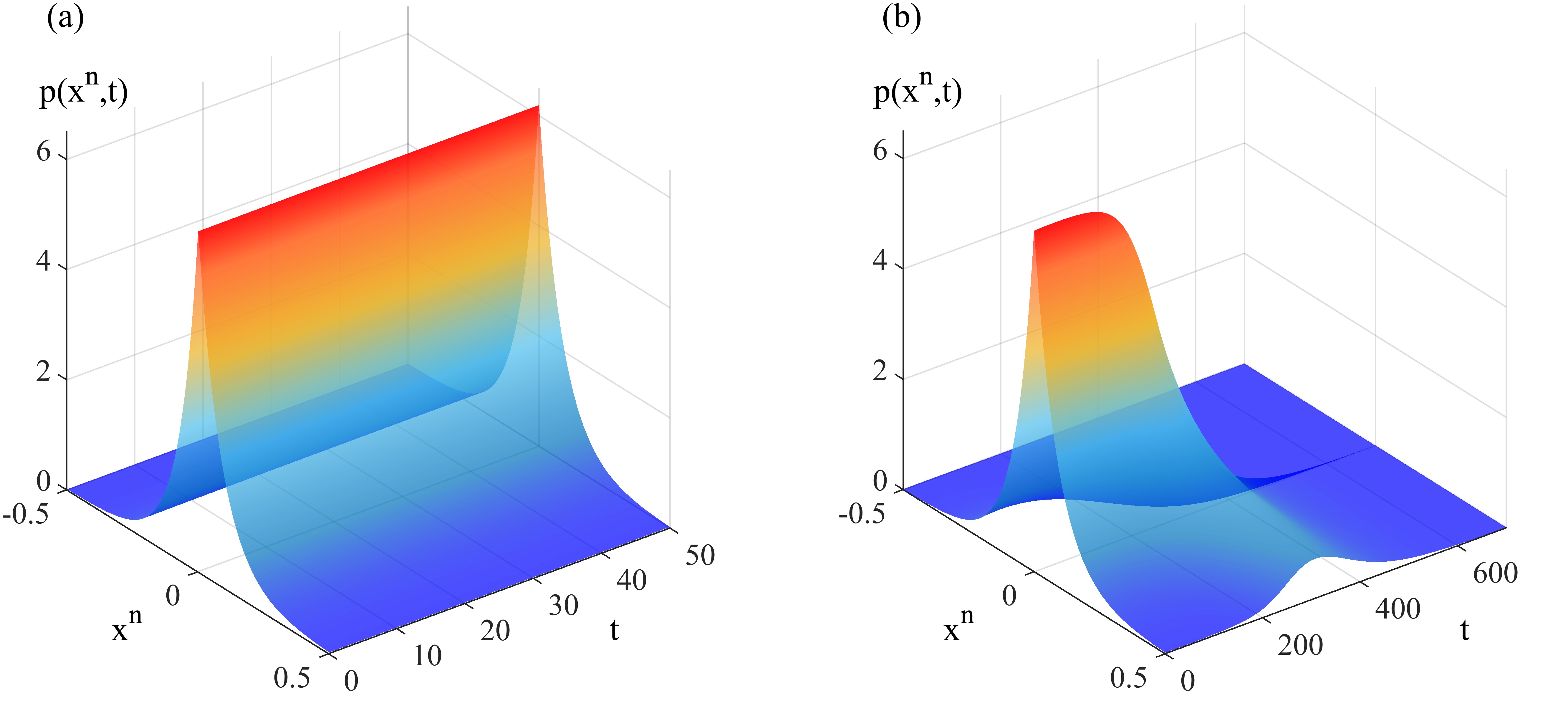}
\caption{\label{fig:F25}The Dynamic of Wealth Distribution in Low-type Trade Structure}
\end{figure}
\vspace{-0.2cm}
\begin{figure}[!ht]
\centering
\includegraphics[width=13cm]{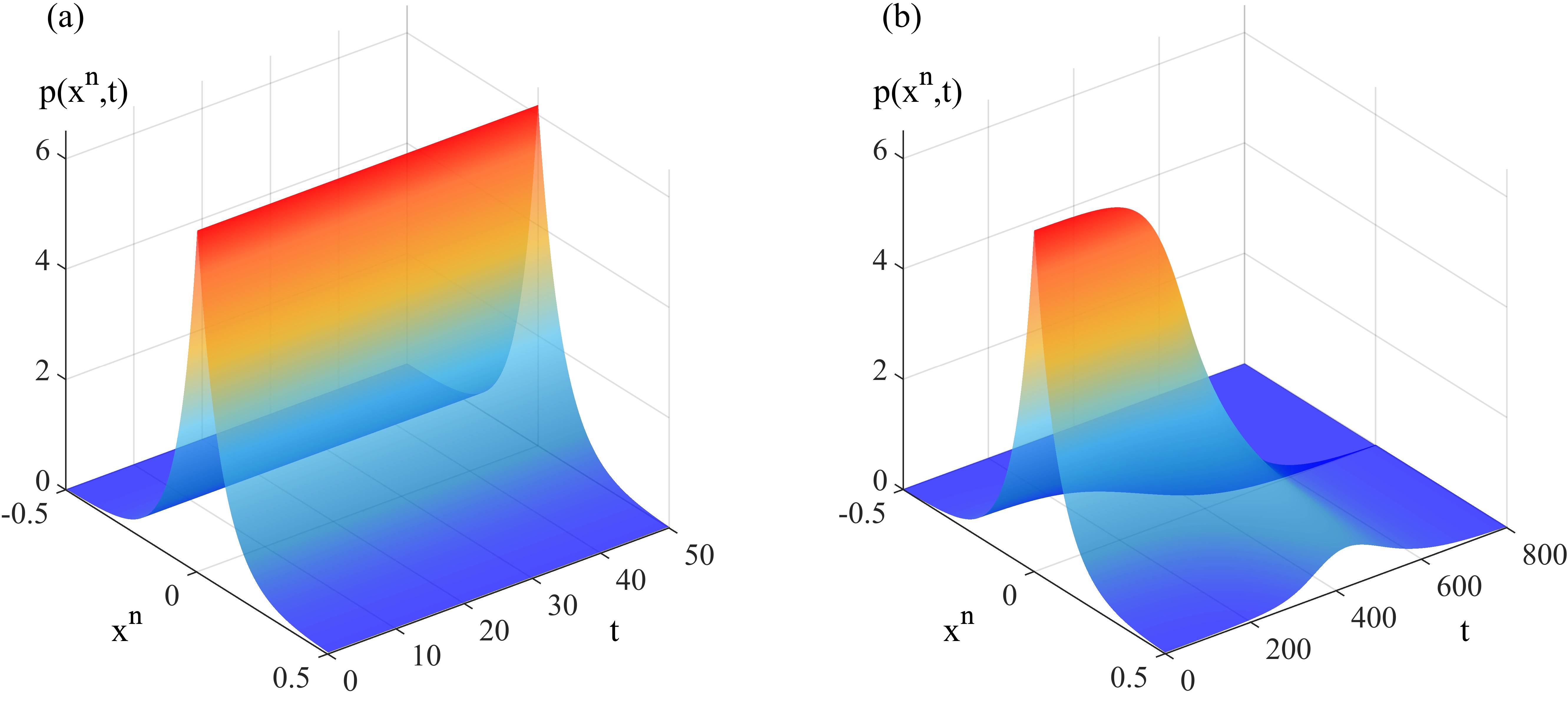}
\caption{\label{fig:F26}The Dynamic of Wealth Distribution in High-type Trade Structure}
\end{figure}
\vspace{-0.2cm}
\begin{figure}[!ht]
\centering
\includegraphics[width=13cm]{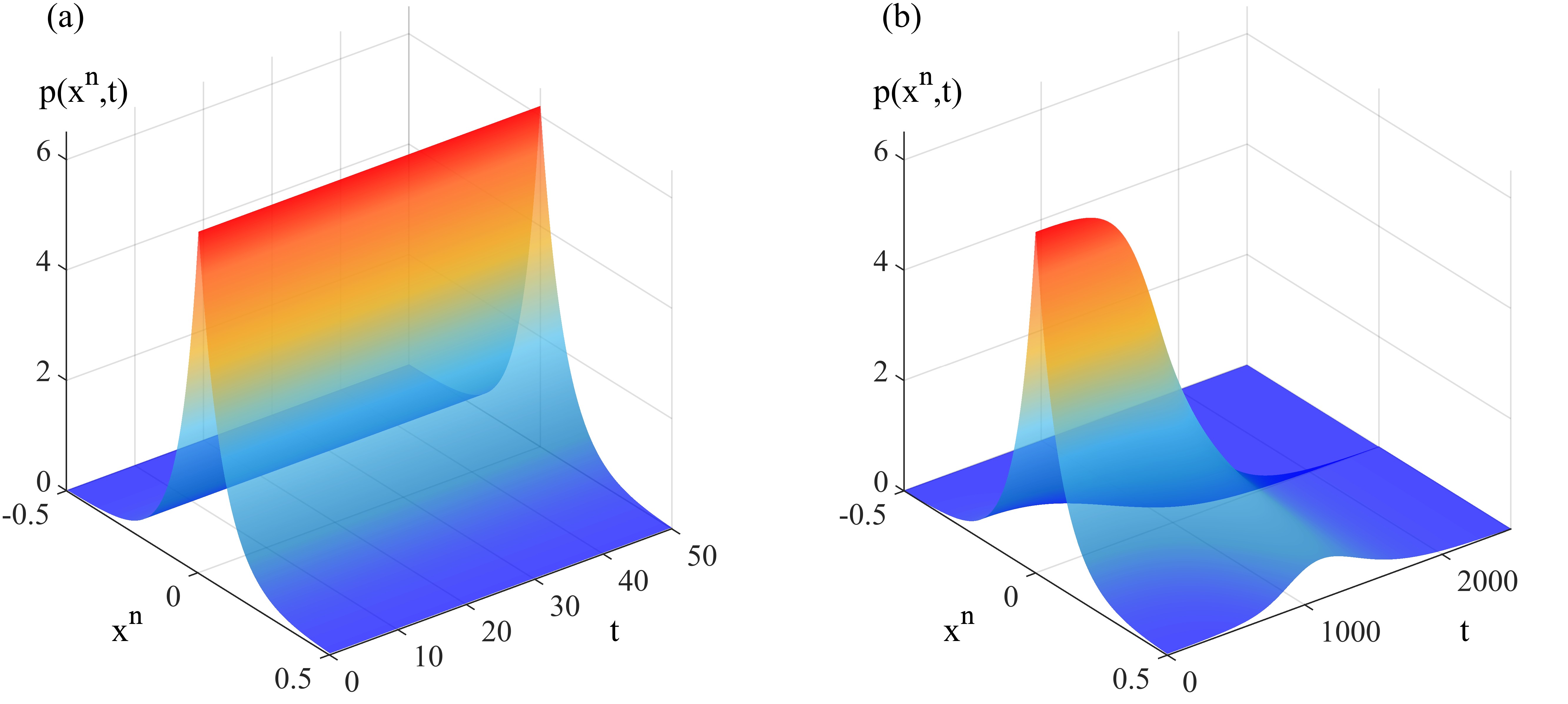}
\caption{\label{fig:F27}The Dynamic of Wealth Distribution in Neutral-type Trade Structure}
\end{figure}

\section{International Trade and Labor Market}\label{Section 6}
\subsection{Review and Further Discussion}
Generally speaking, the exchange rate variable differs from economic variables such as GDP and TFP because it possesses more instrumental attributes than target attributes. From a conventional economic development perspective, the currency exchange rate is often viewed as a tool for adjusting external balance. An excessive trade surplus should be addressed by appreciating the currency to reduce exports and increase imports, while an excessive trade deficit should be countered by depreciating the currency to boost exports and reduce imports, ultimately achieving external trade equilibrium. However, from the perspective of a national development strategy, an obsession with achieving a single objective through exchange rates causes the exchange rate to dominate the strategic framework, transforming it into a critical tool for implementing national strategy. This creates a rigid demand for the tool. Such rigid demand carries substantial risks for the economy, potentially leading to systemic risks in achieving other macroeconomic objectives.

For instance, sustained appreciation may lead to a contraction in exports while simultaneously attracting “Hot Money” flows driven by arbitrage opportunities, which can inflate asset price bubbles. When the arbitrage incentives dissipate, capital flows ultimately reverse, posing a significant risk of abrupt and sharp exchange rate adjustments. Therefore, it is essential to formulate monetary policies that ensure long-term, stable fluctuations in the exchange rate.

However, the instrumental role of exchange rates also faces the possibility of becoming ineffective. In today's world, exogenous factors such as intense competition among economies and geopolitical conflicts have become deeply embedded in the global economic operating process. For export trade within economic globalization, two primary forms of impact currently exist: The first concerns trade costs, primarily manifested as tariff frictions. The second involves core technology restrictions or industrial chain decoupling.

For the first type of impact, currency depreciation could serve as a tool to offset tariff increases, unless universal “reciprocal tariffs” are implemented on a large scale globally. Only then could exchange rate changes effectively counteract tariffs on an equivalent basis. However, this scenario is unrealistic in the current international economic landscape. Recent tariff developments, particularly since April 2025, reveal that bilateral tariffs and tariff escalations between the United States and most global economies are largely asymmetric (\citealp{mehrotra2025tariffs}; \citealp{kalemli2026global}). Consequently, most economies struggle to offset the impact of excessive tariff asymmetry through currency depreciation. Against this backdrop where currency depreciation struggles to offset tariff costs, coupled with the fact that tariffs themselves lead to a contraction in the global economy, particularly in total trade volume (\citealp{auclert2025macroeconomics}; \citealp{adao2026world}), a significant and unfavorable appreciation of the exchange rate could become a critical variable that overwhelms foreign trade enterprises. This risk needs to be carefully considered.

And the second type of shock essentially involves the fragmentation of globalized markets. Geopolitical rivalries have fueled major developed economies' fears of losing core technological advantages and critical resources, leading to the worldwide application of national security. This is the root cause of market segmentation. In a fragmented global market, arbitrage mechanisms cease to exist, and the ability of exchange rates to adjust prices is consequently lost. Therefore, under such circumstances, even with expectations of currency appreciation, foreign capital is unable to flow into industries that are prohibited or required to decouple.

From this perspective, under specific conditions, exchange rates may lose part of their function in adjusting external economic balances through relative price changes in the market. In other words, the exchange rate mechanism in international trade could potentially fail. This is because: In the past, countries deeply engaged in the global division of labor, where competitiveness centered on efficiency and comparative advantage. Today, however, trade policy orientations have undergone a fundamental shift, and the logic of global trade is being profoundly restructured. The principle of prioritizing efficiency is no longer the sole criterion for measuring competitiveness. Nations no longer pursue division of labor and lean production at all costs, instead, considerations of value alignment, geopolitical relations, and the stability of industrial and supply chains have taken precedence. This situation renders traditional theories of international division of labor, the principle of production efficiency and the Marshall-Lerner condition partially ineffective within the current global trade structure.

Based on these issues, Sections 2 to 5 of our paper provide analytical and empirical discussions on the relevant topics. We find that under the prerequisite of stable exchange rate fluctuations, currency depreciation can enhance provincial export competitiveness. Furthermore, innovation development plays a substitutive role in the exchange rate mechanism for boosting export competitiveness. In regions with strong innovation levels, the exchange rate ceases to be the primary factor influencing export competitiveness.

Furthermore, regarding optimal monetary policy (exchange rate policy)\footnote{Theoretically, the international consensus on optimal exchange rate selection (including exchange rate regime arrangements and policy operations) is that: There isn't a single exchange rate choice suitable for all countries and all periods within a single country, as any exchange rate choice carries both advantages and disadvantages. Therefore, this paper focuses solely on analyzing optimal exchange rate policy from the perspective of export commodity revenues and exchange rate fluctuations. We assume that the market always exhibits three distinct functional relationships between export returns and exchange rate fluctuation parameters.}, we attempt to model the economic associations between three categories of export commodities and exchange rates to construct a heterogeneous trade structure. Based on these models, we analyze the optimal parameters for exchange rate fluctuations in each trade structure. We find that the optimal exchange rate policy is not to maintain the comparative advantage of each trade structure, but rather to promote the dynamic transformation of the trade structure. This can not only generate optimal returns but also reduce the pressure of wealth inequality.

However, the above analysis relies on a core assumption, i.e., a stable supply side. This can also be understood as disregarding random fluctuations on the production side, treating production as exogenous (i.e., a constant production function), and assuming a constant supply of sufficient and stable products to meet export demands (i.e., if optimal returns can be achieved under a certain export pattern, there must exist sufficient commodities or combinations of commodities that can be allocated to satisfy this export pattern). While this approach enables concise and efficient analytical solutions, its practical relevance depends on specific external conditions, such as modeling rational export trade during stable and prosperous economic stages.

\begin{figure}[htbp]
    \centering
    \begin{minipage}[b]{1\textwidth}
        \centering
        \includegraphics[width=\textwidth]{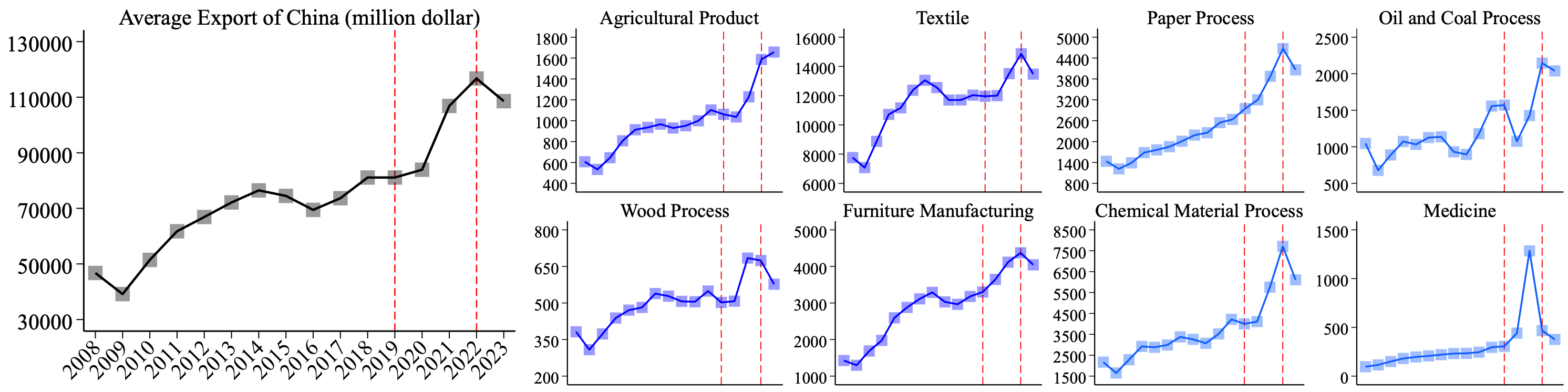}
    \end{minipage}
    \hfill
    \begin{minipage}[b]{1\textwidth}
        \centering
        \includegraphics[width=\textwidth]{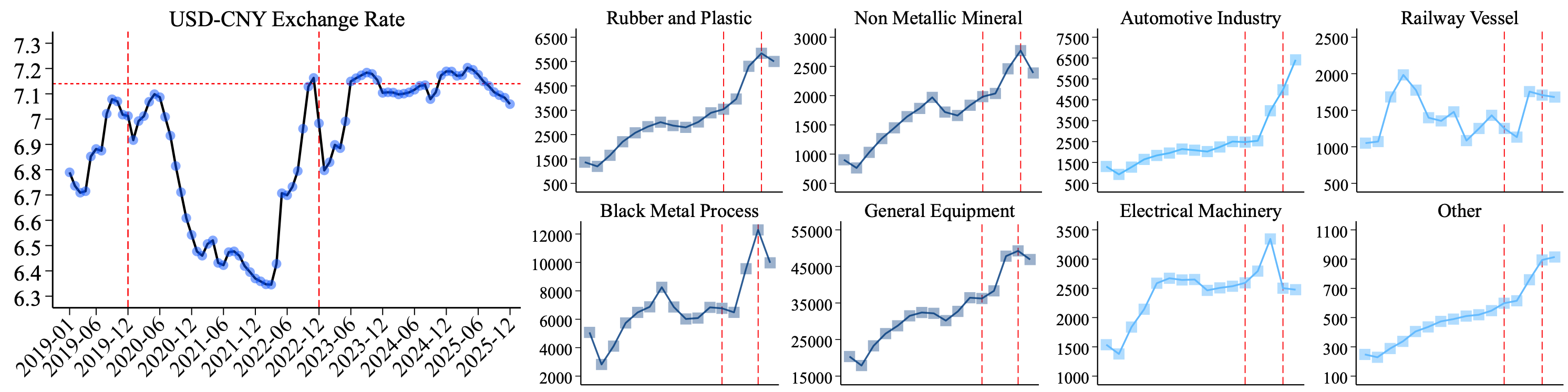}
    \end{minipage}
    \caption{The Situations of Export Trade and Exchange Rate}
    \label{fig:F28}
\end{figure}

Figure 28 presents\footnote{The data in Section 6 are sourced from the official website of the National Bureau of Statistics, the official websites of provincial governments, the China statistical yearbook and provincial statistical yearbooks.} the provincial export trade performance of China's manufacturing sector (across 30 provinces) from 2008 to 2023, along with monthly fluctuations in the USD-CNY exchange rate from 2019 to 2025. We observe that manufacturing export trade remained relatively stable from 2008 to 2019, both at the national average level across all industries and for each individual industry.

However, after 2019, the trend in manufacturing export trade deviated significantly from previous patterns: Manufacturing export value rose sharply from 2019 to 2022, a trend particularly evident in sectors like chemical material processing and medicine, etc. Nevertheless, examining exchange rate data from this period reveals that the USD-CNY exchange rate did not consistently depreciate. From December 2019 to June 2022, the USD-CNY exchange rate maintained an appreciating trend. Then, since 2022, manufacturing exports have experienced a brief overall decline. Meanwhile, the USD-CNY exchange rate has maintained a depreciation trend from December 2022 to December 2023. This indicates that the exchange rate mechanism for export trade has noticeably weakened since 2019. Hence, we must also consider the impact of dynamic random fluctuations on export trade from the production side.

\subsection{The Labor Market Fluctuations}
In this section, we will utilize Chinese data to analyze employment situations and wage levels in society, attempting to define some fundamental and critical variables in labor market fluctuations. Figure 29 shows the total industrial employment and the employment in each of the three industrial sectors across China (31 provinces) from 2008 to 2024.

According to the literature, the production of export commodities is fundamentally derived from labor factors (\citealp{krugman1980scale}; \citealp{melitz2003impact}; \citealp{melitz2008market}; \citealp{amiti2014importers}; \citealp{caliendo2019trade}; \citealp{AERI2021}). Hence, when the labor market structure remains stable, we can reasonably assume that the production of export commodities is exogenous, as shown in Figure 29 for the period 2008–2019. However, two types of exogenous shocks after 2019 significantly impacted the employment structure. The first shock was the global pandemic that occurred in early 2020, known as the COVID-19 shock. This caused a sharp decline in overall national employment over the following years. The second shock was the formal application of artificial intelligence at the end of 2022, known as the AI shock\footnote{Specifically, the launch of ChatGPT (GPT-3.5), a conversational AI model officially released by the American AI research institute OpenAI on November 30, 2022}. This caused long-term fluctuations in national employment over the subsequent years.

Regarding heterogeneity, the COVID-19 shock had a relatively weak impact on employment in the second industry. However, its effects on employment in the first and third industries were pronounced. It disrupted the long-term upward trend in the third industry's employment, causing a sharp short-term decline. Simultaneously, it accelerated the rate of decline in the first industry's employment, worsening its employment situation. Nevertheless, following the AI shock, employment in the third industry returned to its previous upward path. While employment in the first industry showed some improvement compared to its previous state, it still exhibited an overall downward trend. Employment in the second industry remained subject to minor fluctuations.

\begin{figure}[htbp]
\centering
\includegraphics[width=15cm]{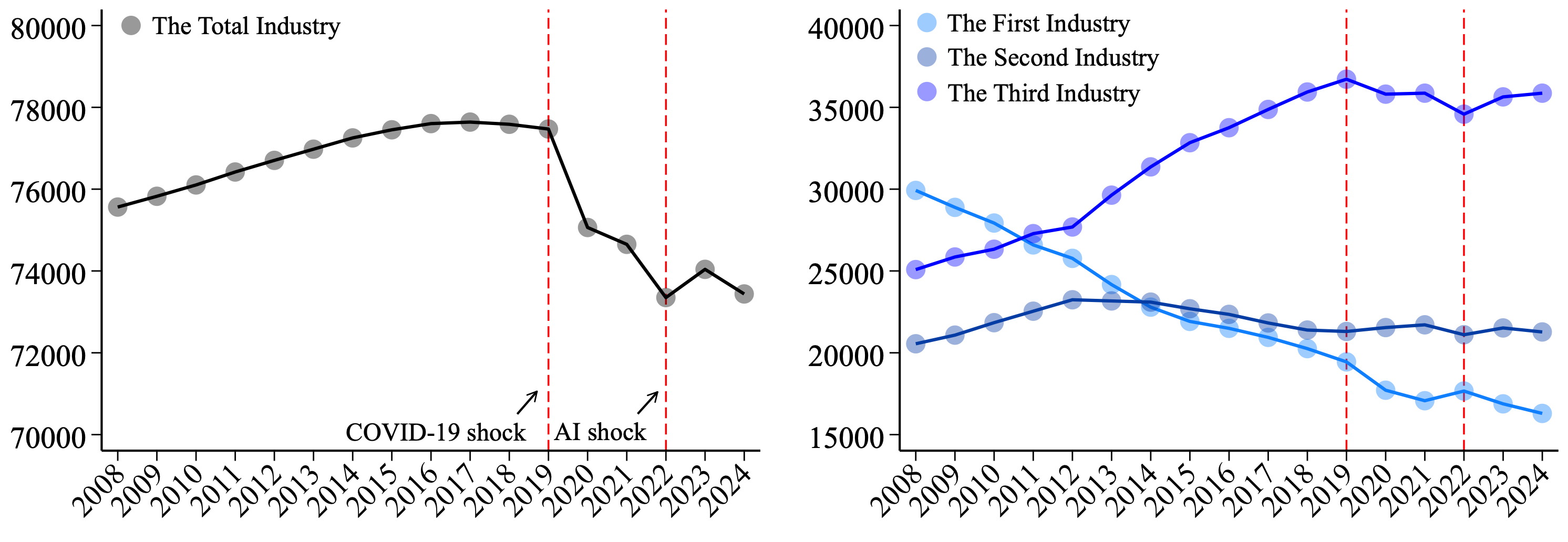}
\caption{\label{fig:F29}The Situations of Employment (10000 people)}
\end{figure}

Therefore, based on Figure 29, as a study of the production side, we should consider two consequences in the labor market triggered by these two types of exogenous shocks: First, not all labor supply can be fully allocated to production. This represents friction between the total labor supply and the effective labor supply, which means the labor factor that can be allocated to production constitutes a proportion of the total labor supply. Second, the labor factor may undergo long-term fluctuations, involving the transition from human resources (traditional labor) to human capital (highly skilled, new-type labor). This transition enhances the marginal productivity of the labor factor, thus impacting employment.

For example: If the demand for production in a particular field remains unchanged, allocating human capital to unsaturated employment can achieve the same effect as allocating human resources to saturated employment. Furthermore, we can identify three key variables in labor market fluctuations: The first is the total labor supply. The second is the labor factor, i.e., the effective labor supply available for production. The third is the marginal productivity of labor, i.e., the value of labor factor.

\begin{figure}[htbp]
    \centering
    \begin{minipage}[b]{1\textwidth}
        \centering
        \includegraphics[width=\textwidth]{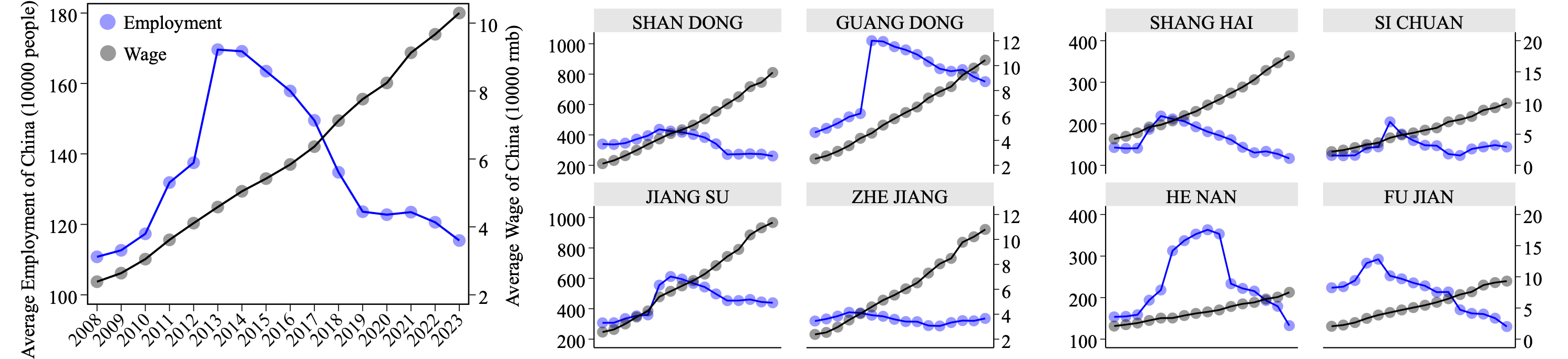}
    \end{minipage}
    \hfill
    \begin{minipage}[b]{1\textwidth}
        \centering
        \includegraphics[width=\textwidth]{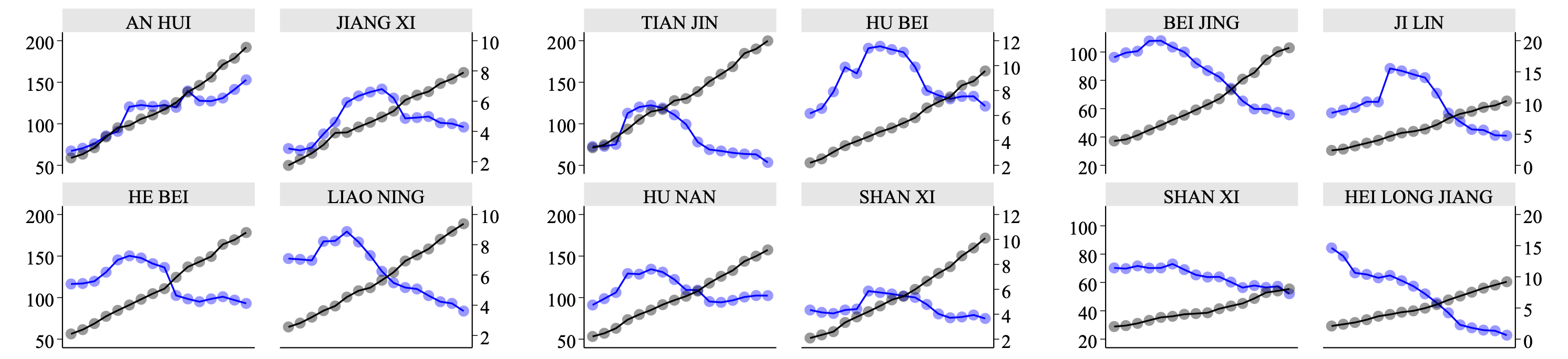}
    \end{minipage}
    \hfill
    \begin{minipage}[b]{1\textwidth}
        \centering
        \includegraphics[width=\textwidth]{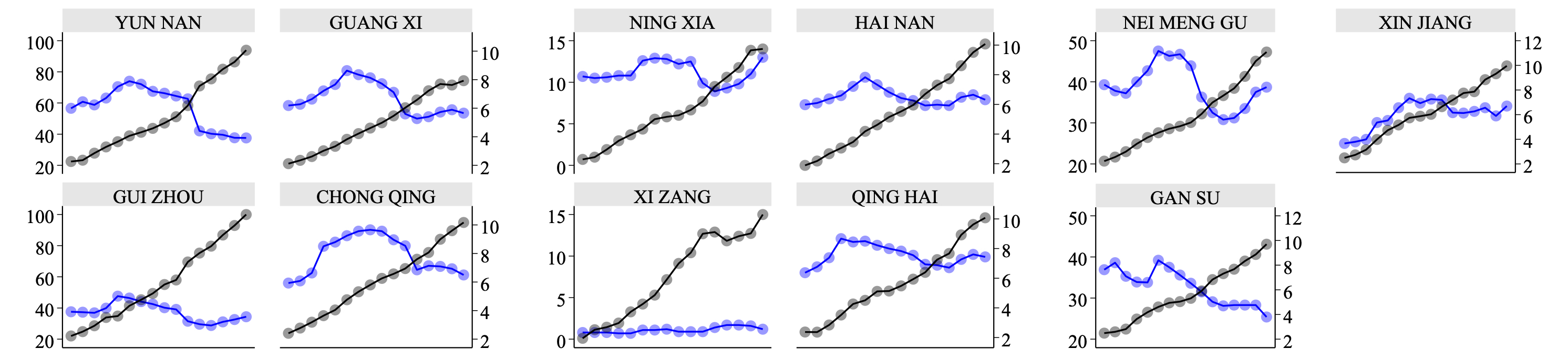}
    \end{minipage}
    \caption{The Situations of Employment and Wage in Manufacturing Labor Market}
    \label{fig:F30}
\end{figure}

Taking the manufacturing labor market as an example, Figure 30 illustrates the average employment and average wages of urban manufacturing workers nationwide (across 31 provinces) from 2008 to 2023, along with provincial heterogeneity. Regarding wage levels (black curve), both the heterogeneous sample and the overall average sample exhibit similar patterns, with average wages maintaining a steady annual increase. Regarding employment levels (blue curve), most heterogeneous samples and the overall average sample exhibit an inverted U-shaped trend. Some provinces show stable fluctuations (e.g., ZHE JIANG), while a minority maintain an upward trend (e.g., AN HUI).

Therefore, based on historical data from China's manufacturing labor market, we observe the following overall characteristics: First, the number of employed individuals exhibits a declining trend. In other words, the labor factor available for production represents a proportion of the total labor supply, and this proportion diminishes with time. Second, the evolution of employment levels and average wages starts to follow an “X” pattern from a specific year, indicating an enhancement in the marginal productivity of labor, that is, the growth in the value per unit of labor factor input. In the following sections, based on these two realities, we will model the labor market fluctuations from the perspective of stochastic process and develop an economic model for export trade that incorporates labor market shocks.

\section{The Labor Market Shock and Aggregate Production}\label{Section 7}
\subsection{The Labor Market}
\noindent\textbf{Heterogeneous Dynamics}: In this section, we will analyze the heterogeneity of labor factors and construct their dynamics, then attempt to model potential economic shocks. Based on the theory in Section 6, we model the aggregate labor factor dynamics $L_t$ in the labor market as a composite of three types of economic variables related to labor factors. These are the total labor supply $l_t$, the labor factor proportion $s_{it}$, and the price level of labor (wages) $p_t$. We model them using standard \textit{Geometric Brownian Motion} (GBM). That is:
\[L_t=l_t s_{it} p_t\]

Similar methods could refer to \cite{gomez2025wealth} and \cite{QJE2026}. Specifically:

First, the total labor supply $l_t$ represents the aggregate labor supply available to society during period $t$, i.e., the total number of employed individuals in the labor market. Its random component follows standard Brownian motion $W_t^l$, which simultaneously represents the general fluctuations in the labor market.

Second, the labor factor proportion $s_{it}$ represents the actual labor factor allocated to industry $i$ during period $t$, i.e., the number of employed individuals within a specific industry. We assume that all fluctuations in economic variables occurring in the labor market are homogeneous, stemming from the general labor market fluctuations. In other words, they are correlated rather than independent. Therefore, the random component of $s_{it}$ follows standard Brownian motion $W_t^s$. We assume $W_t^s$ originates from the aggregate labor market fluctuation $W_t^l$, with their correlation coefficient being $\rho_a\in [-1,0)\cup(0,1]$. Meanwhile, we need to set $s_{it}$ to exhibit a decreasing trend, representing the first type of shock in the labor market. Specifically, if such a shock occurs in period $t$, the labor factor (effective labor supply) in industry $i$ will constitute a proportion of the total labor supply. The stronger the shock, the smaller this proportion becomes, indicating reduced labor factor allocation and a smaller employment scale.

Third, the price of labor $p_t$ represents the wage that can be paid for the labor factor in period $t$, i.e., the value of labor. Similarly, its random component is the standard Brownian motion $W_t^p$. We assume $W_t^p$ originates from the aggregate labor market fluctuation $W_t^l$, with their correlation coefficient being $\rho_b\in [-1,0)\cup(0,1]$. We will treat changes in the drift parameter of $p_t$ as the second type of shock in the labor market. That is, if such a shock occurs in period $t$, the wage level of labor will be altered. This implies that both the value of labor and the marginal productivity of labor are changed following this type of shock. Ultimately, we assume that the two types of shocks mentioned above are independent, that is, $W_t^s$ and $W_t^p$ are mutually independent. They can occur independently or simultaneously, but the occurrence of one type of shock does not directly cause the occurrence of the other.

Hence, for $l_t$, we have:
\[\frac{dl_t}{l_t}=\mu_ldt+\sigma_ldW_t^l\]

s.t.:
\[\begin{gathered}
\mu_{l}>0,\sigma_{l}>0
\end{gathered}\]

And, for $s_{it}$, we get:
\[\frac{ds_{it}}{s_{it}}=-\mu_sdt+\sigma_sdW_t^s\]

s.t.:
\[\begin{gathered}
\mu_s>0,\sigma_s>0\\d[W_t^l,W_t^s]=\rho_adt
\end{gathered}\]

Similarly, we get $p_t$:
\[\frac{dp_t}{p_t}=\mu_pdt+\sigma_pdW_t^p\]

s.t.:
\[\begin{gathered}
\mu_{p}>0,\sigma_{p}>0\\d[W_t^l,W_t^p]=\rho_bdt\\d[W_t^s,W_t^p]=0
\end{gathered}\]

Then, we calculate $L_{it}=l_t s_{it}$:
\[\begin{gathered}dL_{it}=s_{it}dl_t+l_tds_{it}+d[l_t,s_{it}]\\s_{it}dl_t=s_{it}l_t(\mu_ldt+\sigma_ldW_t^l)\\l_{t}ds_{it}=l_{t}s_{it}(-\mu_{s}dt+\sigma_{s}dW_{t}^{s})\\d[l_t,s_{it}]=\left(l_{t}\sigma_{l}dW_{t}^{l}\right)(s_{it}\sigma_{s}dW_{t}^{s})=l_{t}s_{it}\sigma_{l}\sigma_{s}d\left[W_{t}^{l},W_{t}^{s}\right]=l_{t}s_{it}\sigma_{l}\sigma_{s}\rho_{a}dt\end{gathered}\]

Hence:
\[\frac{dL_{it}}{L_{it}}=(-\mu_s+\mu_l+\sigma_l\sigma_s\rho_a)dt+\sigma_ldW_t^l+\sigma_sdW_t^s\]

Finally, we can get $L_t=L_{it} p_t$:
\[\begin{gathered}dL_t=p_tdL_{it}+L_{it}dp_{it}+d[L_{it},p_t]\\p_tdL_{it}=p_tL_{it}\left((-\mu_s+\mu_l+\sigma_l\sigma_s\rho_a)dt+\sigma_ldW_t^l+\sigma_sdW_t^s\right)\\L_{it}dp_{it}=L_{it}p_t(\mu_pdt+\sigma_pdW_t^p)\\d[L_{it},p_t]=(L_{it}\sigma_ldW_t^l)(p_t\sigma_pdW_t^p)+(L_{it}\sigma_sdW_t^s)(p_t\sigma_pdW_t^p)\\d[L_t,p_{t}]=L_{it}p_{t}\sigma_{l}\sigma_{p}d[W_{t}^{l},W_{t}^{p}]+L_{it}p_{t}\sigma_{s}\sigma_{p}d[W_{t}^{s},W_{t}^{p}]=L_{it}p_{t}\sigma_{l}\sigma_{p}\rho_{b}dt\end{gathered}\]

That is:
\[\frac{dL_t}{L_t}=(-\mu_s+\mu_l+\mu_p+\sigma_l\sigma_s\rho_a+\sigma_l\sigma_p\rho_b)dt+\sigma_ldW_t^l+\sigma_sdW_t^s+\sigma_pdW_t^p\]

Now, based on \cite{LNAER2023} and \cite{LNAERinsight2023}, we assume that $dW_t^l$, $dW_t^s$ and $dW_t^p$ are complete positively correlated, that is $\rho_a = \rho_b=1$, and:

\[\begin{gathered}
\frac{dW_t^s}{dW_t^l}=\theta_s\in(0,1)\\\frac{dW_t^p}{dW_t^l}=\theta_p\in(0,1)
\end{gathered}\]

Then, we rewrite the equations:
\[\begin{gathered}\frac{dl_t}{l_t}=\mu_ldt+\sigma_ldW_t^l\\\frac{ds_{it}}{s_{it}}=-\mu_sdt+\sigma_s\theta_sdW_t^l\\\frac{dp_t}{p_t}=\mu_pdt+\sigma_p\theta_pdW_t^l\end{gathered}\]

That is:
\[\frac{dL_t}{L_t}=(-\mu_s+\mu_l+\mu_p+\sigma_l\sigma_s+\sigma_l\sigma_p)dt+(\sigma_l+\sigma_s\theta_s+\sigma_p\theta_p)dW_t^l\]

Therefore, the aggregate dynamics of effective labor market could be written as:
\[dL_{t}=L_{t}(-\mu_{s}+\mu_{l}+\mu_{p}+\sigma_{l}\sigma_{s}+\sigma_{l}\sigma_{p})dt+L_{t}(\sigma_{l}+\sigma_{s}\theta_{s}+\sigma_{p}\theta_{p})dW_{t}^{l}\]

s.t.:
\[\begin{gathered}\mu_{s},\mu_{l},\mu_{p}>0\\\sigma_{s},\sigma_{l},\sigma_{p}>0\\\theta_{s},\theta_{p}\in(0,1)\end{gathered}\]

$L_t$ represents the aggregate labor dynamics allocated to final goods production, which concentrates the value and structure of the labor market\footnote{The structure of the labor market determines whether labor can be allocated rationally and effectively according to social production needs. Generally speaking, a balanced labor market structure aligns with social production demands, as normal development involves the simultaneous satisfaction and advancement of all economic sectors, rather than focusing solely on a specific field. For instance, within a typical industrial structure, we require a certain scale of human resources to be allocated to labor-intensive traditional industries, while also needing a certain scale of human capital to be allocated to technology-driven new industries. If labor factors are allocated within specific industries incorrectly, that is, if there is a mismatch of labor factors, it will cause labor remuneration and economic outcomes to deviate significantly from expectations, ultimately hindering industrial development. This would be unstable and unsustainable.}.

If the mean value of $L_t$ is large, it indicates that the scale of labor factor allocation in this labor market is substantial, with high marginal productivity, making it a high-value labor market. Conversely, if the mean value of $L_t$ is small, it indicates that the scale of labor factor allocation in this labor market is limited, with low marginal productivity, making it a low-value labor market. 

Meanwhile, if the variance of $L_t$ is large, it indicates that the labor market structure is highly differentiated (extreme), facing high inequality pressures. If the variance of $L_t$ is small, it indicates that the labor market structure is concentrated (balanced), facing low inequality pressures.

Hence, according to \textit{Ito’s Lemma}, let $z=\mathrm{log}(L_t)$, then:
\[dz=\left(\left(-\mu_{s}+\mu_{l}+\mu_{p}+\sigma_{l}\sigma_{s}+\sigma_{l}\sigma_{p}\right)-\frac{1}{2}\left(\sigma_{l}+\sigma_{s}\theta_{s}+\sigma_{p}\theta_{p}\right)^{2}\right)dt+(\sigma_{l}+\sigma_{s}\theta_{s}+\sigma_{p}\theta_{p})dW_{t}^{l}\]

We denote:
\[\begin{gathered}
\Sigma_L=\left(-\mu_s+\mu_l+\mu_p+\sigma_l\sigma_s+\sigma_l\sigma_p\right)-\frac{1}{2}\left(\sigma_l+\sigma_s\theta_s+\sigma_p\theta_p\right)^2\\\Omega_L=\sigma_l+\sigma_s\theta_s+\sigma_p\theta_p
\end{gathered}\]

Hence:
\[dz=\Sigma_Ldt+\Omega_LdW_t^l\]

We set $g(z)$ is the probability density function of $z=\mathrm{log}(L_t)$, satisfying the KFE:
\[0=-\Sigma_{L}\frac{\partial g(z)}{\partial z}+\frac{1}{2}(\Omega_{L})^{2}\left(\frac{\partial^{2}g(z)}{\partial z^{2}}\right)-\mathcal{D}_{0}g(z)+\mathcal{D}_{0}\delta(z)\]

Solving the KFE, we get the analytical form of $g(z)$:
\[\begin{gathered}
g(z)=C_0exp\left(\frac{\Sigma_L+\sqrt{(\Sigma_L)^2+2\mathcal{D}_0(\Omega_L)^2}}{(\Omega_L)^2}z\right),z<0\\g(z)=C_0exp\left(\frac{\Sigma_L-\sqrt{(\Sigma_L)^2+2\mathcal{D}_0(\Omega_L)^2}}{(\Omega_L)^2}z\right),z>0
\end{gathered}\]

Then:
\[\int_{-\infty}^{0}g\left(z\right)dz+\int_{0}^{\infty}g\left(z\right)dz=1\]

Therefore:
\[C_{0}=\frac{\mathcal{D}_{0}}{\sqrt{(\Sigma_{L})^{2}+2\mathcal{D}_{0}(\Omega_{L})^{2}}}\]

Now, for $z=\mathrm{log}(L_t)$, we have:
\[\begin{gathered}g(z)=\frac{\mathcal{D}_{0}}{\sqrt{(\Sigma_{L})^{2}+2\mathcal{D}_{0}(\Omega_{L})^{2}}}exp\left(\frac{\Sigma_{L}+\sqrt{(\Sigma_{L})^{2}+2\mathcal{D}_{0}(\Omega_{L})^{2}}}{(\Omega_{L})^{2}}z\right),z<0\\g(z)=\frac{\mathcal{D}_{0}}{\sqrt{(\Sigma_{L})^{2}+2\mathcal{D}_{0}(\Omega_{L})^{2}}}exp\left(\frac{\Sigma_{L}-\sqrt{(\Sigma_{L})^{2}+2\mathcal{D}_{0}(\Omega_{L})^{2}}}{(\Omega_{L})^{2}}z\right),z>0\end{gathered}\]

s.t.:
\[\begin{gathered}\Sigma_{L}=\left(-\mu_{s}+\mu_{l}+\mu_{p}+\sigma_{l}\sigma_{s}+\sigma_{l}\sigma_{p}\right)-\frac{1}{2}\left(\sigma_{l}+\sigma_{s}\theta_{s}+\sigma_{p}\theta_{p}\right)^{2}\\\Omega_L=\sigma_l+\sigma_s\theta_s+\sigma_p\theta_p\\\mu_s,\mu_l,\mu_p>0\\\sigma_s,\sigma_l,\sigma_p>0\\\theta_s,\theta_p\in(0,1)\end{gathered}\]

Regarding shocks $\mathbb{K}_j$, our discussion of the labor market in this paper identifies three types of shocks $\mathbb{K}_j \in \{\mathbb{K}_1, \mathbb{K}_2, \mathbb{K}_3\}$. The first and second types of shocks have already been discussed in the preceding analysis, while the third type of shock is constituted by both of them.

First, the first type of shock impacts the allocation of labor factors, altering the scale of labor factors allocated or invested in final goods production without changing their value. That is, $\mathbb{K}_1=\{\mu_s^i,\mu_p^0\},\mu_s^i \in(\mu_s^-,\mu_s^+), \mu_s^+>\mu_s^->0$. Here, $\mu_p^0>0$ is a constant representing a fixed coefficient for the drift of wage levels dynamics. $\mu_s^i>0$ indicates that the coefficient for the drift of labor factor allocation proportions dynamics is a variable. $\mu_s^i \rightarrow \mu_s^-$ denotes decreasing dissipation in labor factor allocation, while $\mu_s^i \rightarrow \mu_s^+$ indicates increasing dissipation in labor factor allocation.

Secondly, the second type of shock impacts the value of labor factors, altering the wage levels of allocated labor factors without changing their allocation scale, i.e.,  $\mathbb{K}_2=\{\mu_s^0,\mu_p^i\},\mu_p^i \in(\mu_p^-,\mu_p^+), \mu_p^+>\mu_p^->0$. Here, $\mu_s^0>0$ is a constant representing a fixed coefficient for the drift of labor factor allocation proportions dynamics, while $\mu_p^i>0$ denotes a variable coefficient for the drift of wage levels dynamics. $\mu_p^i \rightarrow \mu_p^-$ indicates lower labor factor value, whereas $\mu_p^i \rightarrow \mu_p^+$ denotes higher labor factor value.

Finally, the third type of shock involves the transition from human resources to human capital. This transformation alters both the allocation of labor factors and wage levels, thus changing the marginal productivity of labor factors. Specifically, $\mathbb{K}_3=\{\mu_s^i,\mu_p^i\}$, where $\mu_s^i \in (\mu_s^-, \mu_s^+)$ and $\mu_p^i \in (\mu_p^-, \mu_p^+)$. We denote $\mathbb{K}_3^i \in (\mathbb{K}_3^-, \mathbb{K}_3^+)$, with $\mathbb{K}_3^-=\{\mu_s^-, \mu_p^-\}$ and $\mathbb{K}_3^+=\{\mu_s^+, \mu_p^+\}$. Therefore, when $\mathbb{K}_3^i \rightarrow \mathbb{K}_3^-$, it indicates that the majority of labor allocation in final goods production is human resources, with lower marginal productivity of labor. When $\mathbb{K}_3^i \rightarrow \mathbb{K}_3^+$, it indicates that the majority of labor allocation in final goods production is human capital, with higher marginal productivity of labor.

We define the third type of shock as a “human capital shock”, and we identify three types of human capital shocks in the paper: The first is a “stable human capital shock”, where $\mu_s^i=\mu_p^i$. This represents the most ideal state of human capital shocks, signifying equilibrium between the dissipation of labor factors and wage growth. The second type is a “positive human capital shock”, where $\mu_p^i>\mu_s^i$. This implies that wage growth will be stronger than the dissipation of labor factors. The third type is a “negative human capital shock”, where $\mu_p^i<\mu_s^i$. This implies that wage growth will be weaker than the dissipation of labor factors.

\noindent\textbf{Calibration and Simulation}: Table 8 presents the values of parameters in the aggregate labor factor dynamics $L_t$. For specific data values, please refer to \hyperref[Appendix C]{Appendix C}.

Figures 31 to 35 respectively illustrate the transitions in the distribution of the aggregate labor factor dynamics $L_t$ following the shocks:

Figure 31 shows the transition in the distribution of $L_t$ following shock $\mathbb{K}_1$. We observe that the labor factor allocation shock causes the aggregate labor factor distribution to gradually shift from a right-skewed distribution toward a normal distribution, ultimately becoming left-skewed. This implies that shock $\mathbb{K}_1$ reduces $\mathbb{E}[L_t]$, leading to a gradual decline in the value of labor market. Simultaneously, when shock $\mathbb{K}_1$ reaches a certain state, the distribution of $L_t$ transitions to a normal distribution. This implies that the trend of $\mathrm{Var}[L_t]$ follows a U-shaped pattern, indicating the possibility of the labor market structure shifting toward a balanced state following shock $\mathbb{K}_1$.
\vspace{+0.3cm}
\begin{figure}[!ht]
\centering
\includegraphics[width=8cm]{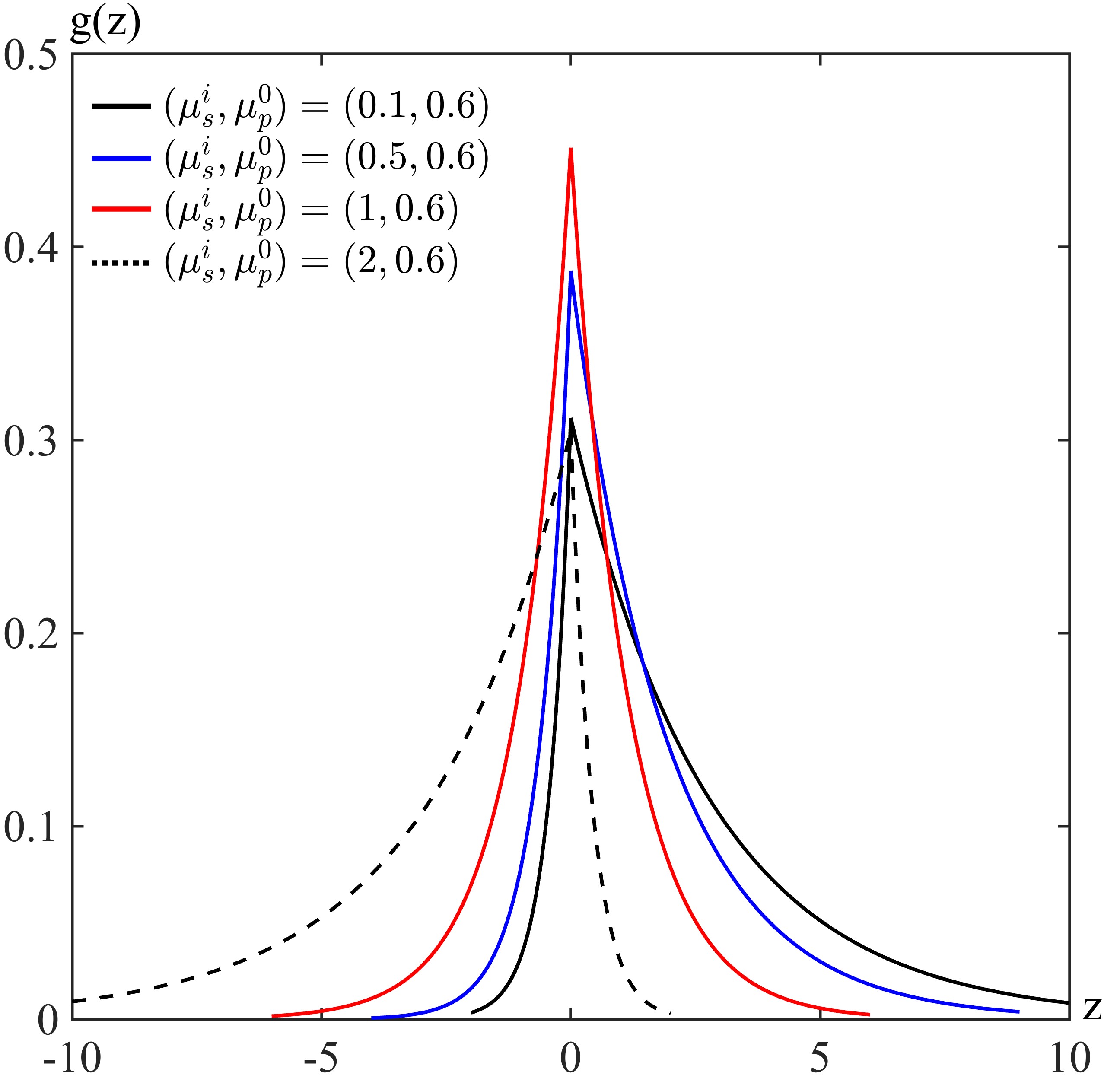}
\caption{\label{fig:F31}The Dynamic Transitions of $z$ after $\mathbb{K}_1$}
\end{figure}

Figure 32 illustrates the transition in the distribution of $L_t$ following the shock $\mathbb{K}_2$. We observe that the impact of the labor factor value shock on the overall distribution of labor factor dynamics is linear. It gradually shifts the distribution of $L_t$ from an approximately normal distribution to a typical right-skewed distribution, increasing both $\mathbb{E}[L_t]$ and $\mathrm{Var}[L_t]$. This indicates that while the value of the labor market continues to rise after shock $\mathbb{K}_2$, inequality in the labor market structure also progressively intensifies.
\vspace{+0.3cm}
\begin{figure}[!ht]
\centering
\includegraphics[width=8cm]{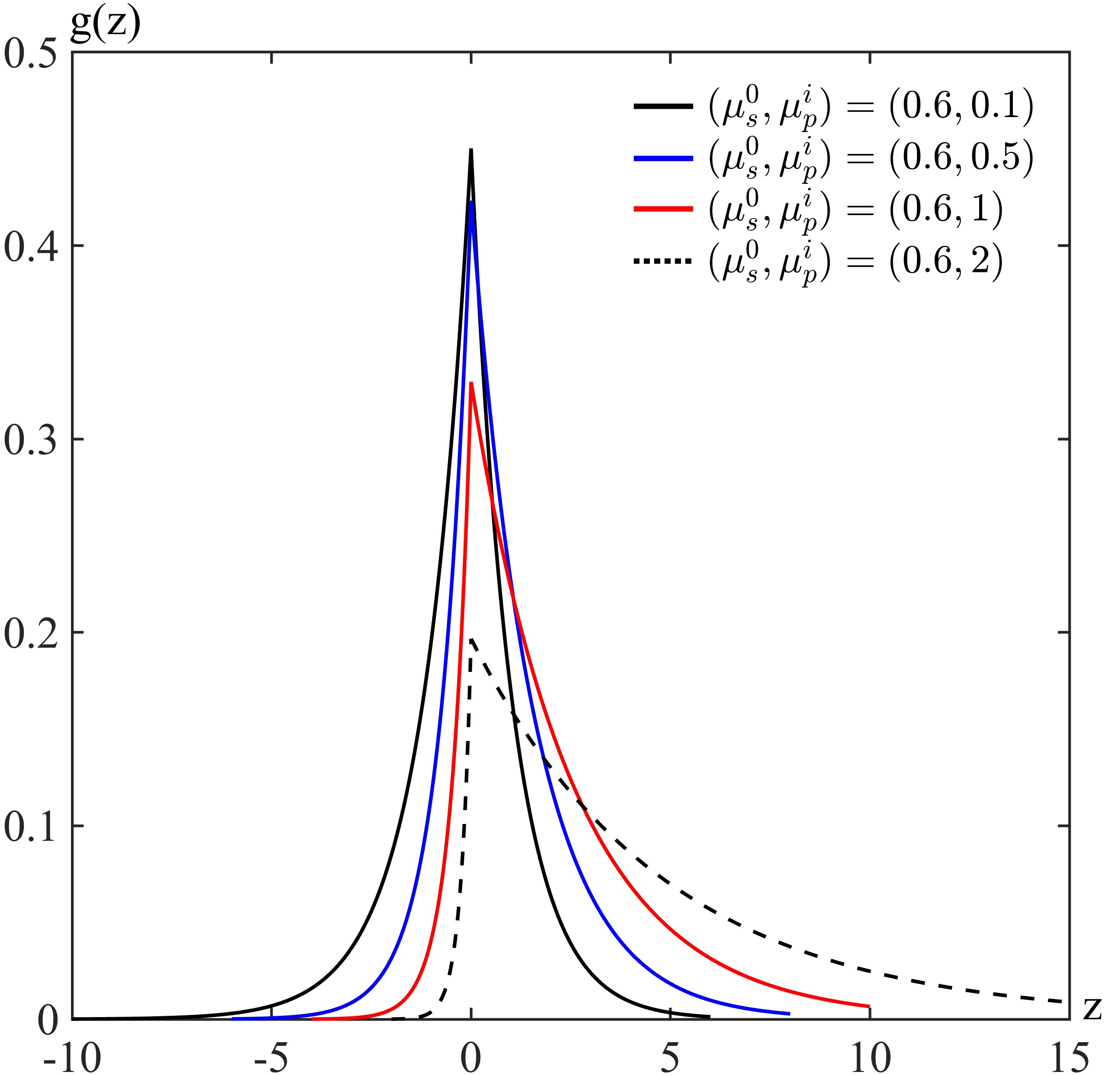}
\caption{\label{fig:F32}The Dynamic Transitions of $z$ after $\mathbb{K}_2$}
\end{figure}

Figures 33 to 35 illustrate the transition in the distribution of $L_t$ following a shock $\mathbb{K}_3$. Figure 33 shows the distribution of $L_t$ after a stable human capital shock. We observe that under this scenario, the overall fluctuations in labor factor dynamics remain stable, and the shock does not cause a dynamic transition in the distribution of $L_t$. Figure 34 presents the distribution of $L_t$ following a positive human capital shock. Here, the transition pattern resembles that after shock $\mathbb{K}_2$, where both $\mathbb{E}[L_t]$ increases and $\mathrm{Var}[L_t]$ continuously expands. Figure 35 illustrates the distribution of $L_t$ following a negative human capital shock. In this scenario, the transition pattern in the distribution of $L_t$ resembles that observed after shock $\mathbb{K}_1$, where both $\mathbb{E}[L_t]$ gradually decreases and $\mathrm{Var}[L_t]$ has a minimum value.
\vspace{+0.3cm}
\begin{figure}[!ht]
\centering
\includegraphics[width=8cm]{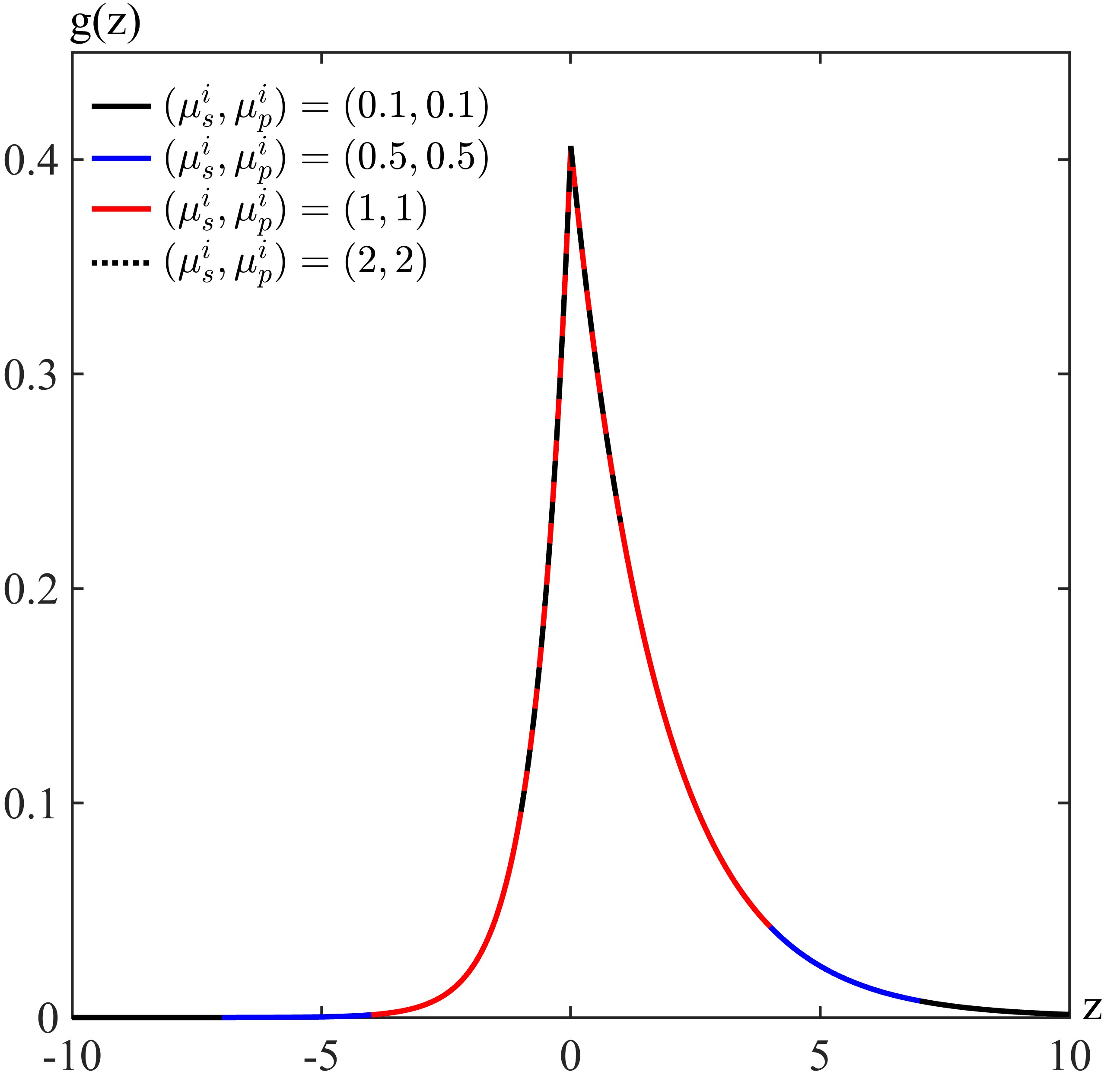}
\caption{\label{fig:F33}The Dynamic Transitions of $z$ after $\mathbb{K}_3^a$}
\end{figure}
\vspace{-0.1cm}
\begin{figure}[htbp]
\centering
\includegraphics[width=8cm]{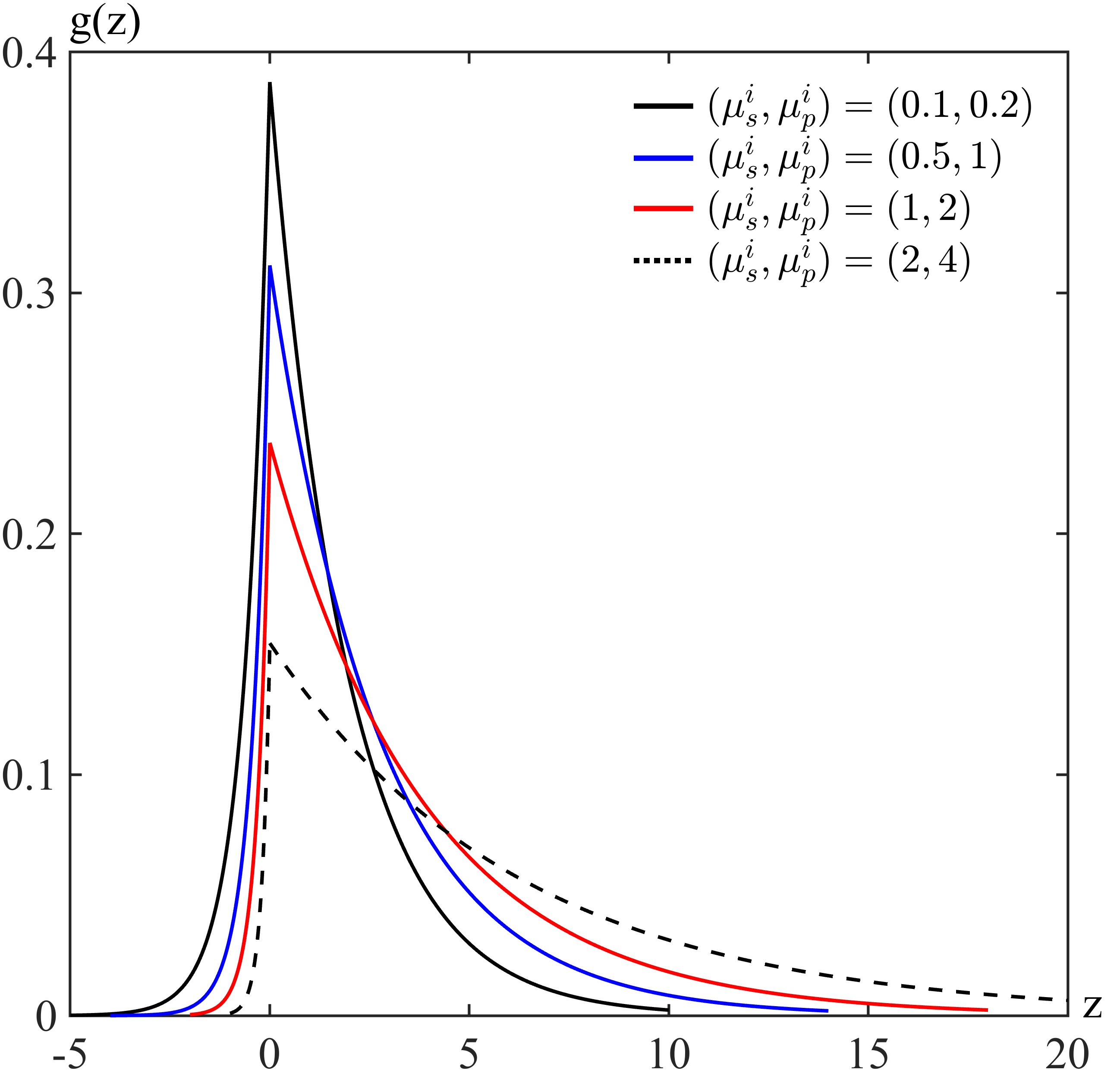}
\caption{\label{fig:F34}The Dynamic Transitions of $z$ after $\mathbb{K}_3^b$}
\end{figure}
\begin{figure}[!ht]
\centering
\includegraphics[width=8cm]{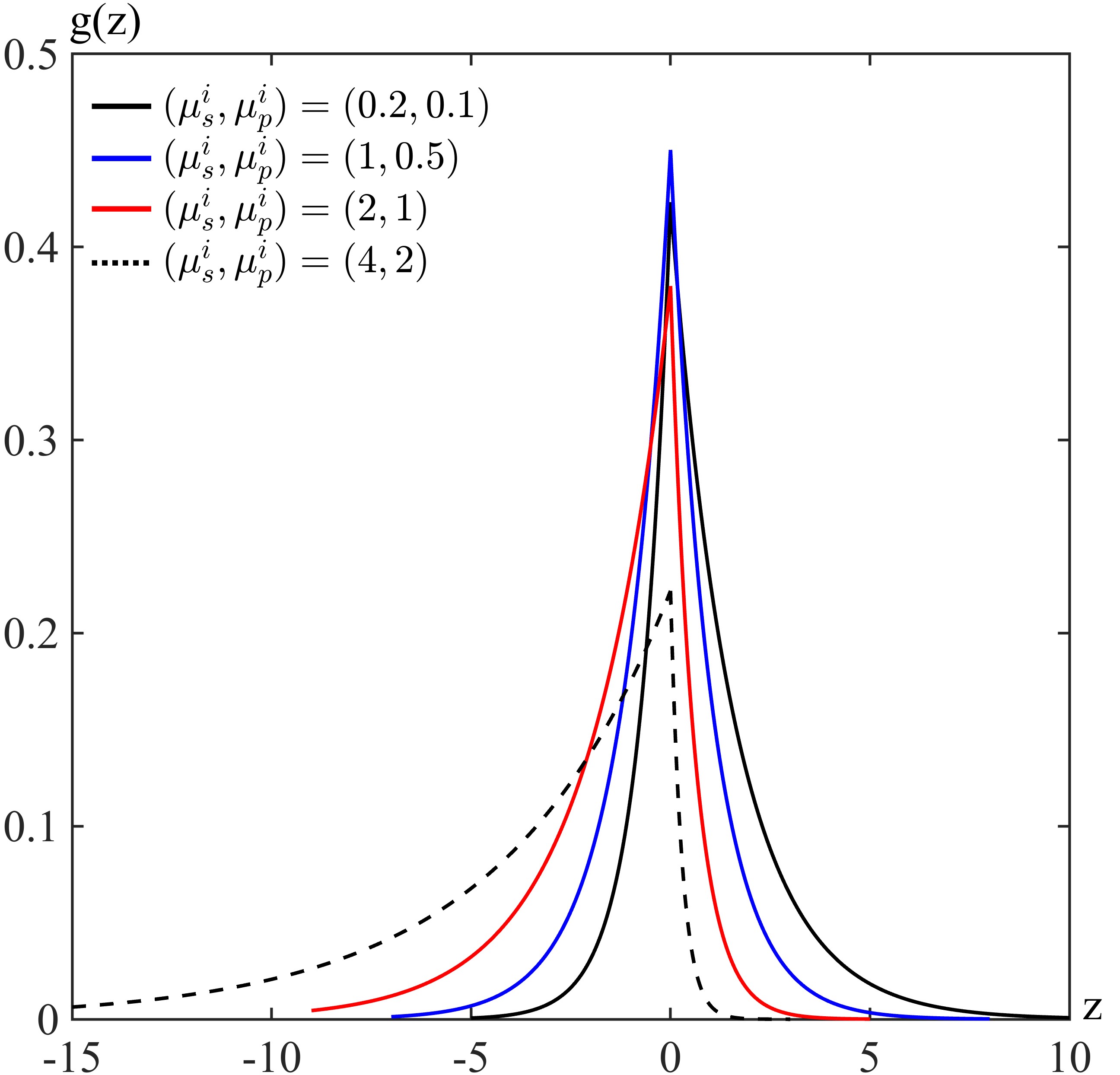}
\caption{\label{fig:F35}The Dynamic Transitions of $z$ after $\mathbb{K}_3^c$}
\end{figure}

\subsection{The Aggregate Production}
\noindent\textbf{Heterogeneous Endowments}: For the production of exports, we model it to be determined by and only by the aggregate labor factor $L_t$ and the ability $\lambda_i$ to utilize production factors or factor combinations $i$ beyond labor. For instance, the ability to utilize capital and technology, etc. Regarding this, we further set that trade “sanction” or “cooperation” will directly impact $\lambda_i$. For example, a country subject to trade sanctions will face export restrictions, supply chain disruptions, industrial chain disengagement, and critical technology blockades. These will directly affect the ability to utilize production factors $i$ other than the labor factor, and vice versa.

Referring to \cite{pastor2020political}, the production function $Y_{i,t+1}$ can be written as:
\[Y_{i,t+1}=e^{\lambda_i+\varepsilon_{i,t+1}+\varepsilon_{t+1}}L_t\]

Where:
\[\begin{gathered}
\lambda_{i}{\sim}\mathcal{N}\left(\lambda_{0},\sigma_{\lambda}^{2}\right)\\\varepsilon_{i,t+1}{\sim}\mathcal{N}\left(-\frac{1}{2}\sigma_{1}^{2},\sigma_{1}^{2}\right)\\\varepsilon_{t+1}{\sim}\mathcal{N}\left(-\frac{1}{2}\sigma_{2}^{2},\sigma_{2}^{2}\right)
\end{gathered}\]

Specifically, $\varepsilon_{i,t+1}$ represents a heterogeneous shock, while $\varepsilon_{t+1}$ denotes the aggregate shock. Hence, we obtain:
\[\mathbb{E}[e^{\varepsilon_{i,t+1}}]=\mathbb{E}[e^{\varepsilon_{t+1}}]=1\]

Now, we assume that production factor $i$ is generated from the set $\mathcal{F}_t$. The scale of production factor $i$ is $m_t=\int_{i\in\mathcal{F}_t}di\in(0,1)$. And the scale of all other production factors in $\mathcal{F}_t$, excluding factor $i$, can be expressed as $1-m_t$.

Next, we can calculate the aggregate production function $Y_{t+1}$:
\[Y_{t+1}=\int_{i\in\mathcal{F}_t}Y_{i,t+1}di=L_te^{\varepsilon_{t+1}}\int_{i\in\mathcal{F}_t}e^{\lambda_i+\varepsilon_{i,t+1}}di\]

And:
\[\int_{i\in\mathcal{F}_t}e^{\lambda_i+\varepsilon_{i,t+1}}di=m_t\mathbb{E}[e^{\lambda_i+\varepsilon_{i,t+1}}|i\in\mathcal{F}_t]=m_t\mathbb{E}[e^{\lambda_i}|i\in\mathcal{F}_t]\mathbb{E}[e^{\varepsilon_{i,t+1}}|i\in\mathcal{F}_t]=m_t\mathbb{E}[e^{\lambda_i}|i\in\mathcal{F}_t]\]

Hence:
\[Y_{t+1}=L_te^{\varepsilon_{t+1}}m_t\mathbb{E}[e^{\lambda_i}|i\in\mathcal{F}_t]\]

Now, we define the probability density function of $\lambda_i$ as $\phi(\lambda_i;\lambda_0,\sigma_\lambda^2)$. Additionally, for the ability to utilize all production factors $j$ in $\mathcal{F}_t$ except factors $i$, we assume $\lambda_j{\sim}\mathcal{N}(\lambda_0,\sigma_\lambda^2)$, thus $m_t^k = 1-m_t$. Then:
\[\begin{gathered}
m_t^k=\int_{k_t}^\infty\phi\left(\lambda_j;\lambda_0,\sigma_\lambda^2\right)d\lambda_j=1-\Phi(k_t;\lambda_0,\sigma_\lambda^2)\\\mathbb{E}[e^{\lambda_i}|i\in\mathcal{F}_t]=\frac{1}{m_t^k}\int_{k_t}^\infty e^{\lambda_i}\phi(\lambda_i;\lambda_0,\sigma_\lambda^2)d\lambda_i=\frac{e^{\lambda_0+\frac{1}{2}\sigma_\lambda^2}\left(1-\Phi(k_t;\lambda_0+\sigma_\lambda^2,\sigma_\lambda^2)\right)}{m_t^k}
\end{gathered}\]

That is:
\[Y_{t+1}=L_te^{\varepsilon_{t+1}}m_te^{\lambda_0+\frac{1}{2}\sigma_\lambda^2}\frac{\left(1-\Phi(k_t;\lambda_0+\sigma_\lambda^2,\sigma_\lambda^2)\right)}{1-\Phi(k_t;\lambda_0,\sigma_\lambda^2)}\]

We denote:
\[\lambda_t^k=k_t-\lambda_0\]

Hence:
\[Y_{t+1}(L_t,\lambda_t^k)=L_te^{\varepsilon_{t+1}}m_te^{\lambda_0+\frac{1}{2}\sigma_\lambda^2}\frac{\left(1-\Phi(\lambda_t^k;\sigma_\lambda^2,\sigma_\lambda^2)\right)}{1-\Phi(\lambda_t^k;0,\sigma_\lambda^2)}\]

Here, we denote $f\left(\lambda_{t}^{k}\right)=\frac{\left(1-\Phi(\lambda_{t}^{k};\sigma_{\lambda}^{2},\sigma_{\lambda}^{2})\right)}{1-\Phi(\lambda_{t}^{k};0,\sigma_{\lambda}^{2})}$, then:
\[\frac{\partial f\left(\lambda_t^k\right)}{\partial\lambda_t^k}=\frac{-\phi(\lambda_t^k;\sigma_\lambda^2,\sigma_\lambda^2)[1-\Phi(\lambda_t^k;0,\sigma_\lambda^2)]+[1-\Phi(\lambda_t^k;\sigma_\lambda^2,\sigma_\lambda^2)]\phi(\lambda_t^k;0,\sigma_\lambda^2)}{\left[1-\Phi(\lambda_t^k;0,\sigma_\lambda^2)\right]^2}\]

And:
\[\frac{\phi(\lambda_t^k;0,\sigma_\lambda^2)}{1-\Phi(\lambda_t^k;0,\sigma_\lambda^2)}>\frac{\phi(\lambda_t^k-\sigma_\lambda^2;0,\sigma_\lambda^2)}{1-\Phi(\lambda_t^k-\sigma_\lambda^2;0,\sigma_\lambda^2)}=\frac{\phi(\lambda_t^k;\sigma_\lambda^2,\sigma_\lambda^2)}{1-\Phi(\lambda_t^k;\sigma_\lambda^2,\sigma_\lambda^2)}\]

Since the above inequality always holds, we can conclude that:
\[\frac{\partial f(\lambda_t^k)}{\partial\lambda_t^k}>0\]

That is: For $\lambda_t^H>\lambda_t^k>\lambda_t^L$, we have $Y_{t+1}^H(L_t,\lambda_t^H)$ and $Y_{t+1}^L(L_t,\lambda_t^L)$, where $Y_{t+1}^H(L_t,\lambda_t^H)>Y_{t+1}^L(L_t,\lambda_t^L)$. Accordingly, we will simplify the aggregate production function $Y_{t+1}(L_t,\lambda_t^k)$ and define the impacts of “sanction” and “cooperation” in trade.

We set $f(\lambda_t^k)=\alpha_k\lambda_t^k$, where $\alpha_k>0$ and $|\beta_k|<1$, that is:
\[\begin{gathered}
\lambda_{t+1}^k=\beta_k\lambda_t^k+\delta_t^k\\\delta_t^k{\sim}\mathcal{N}(0,\omega_0^2)
\end{gathered}\]

Now, we formally define trade sanction and trade cooperation within aggregate production. We establish three states of aggregate production $Y_{t+1}^i$, i.e., $Y_{t+1}^i\in\{Y_{t+1}^L,Y_{t+1}^0,Y_{t+1}^H\}$, which originate from three types of shocks $\mathbb{H}_j\in\{\mathbb{H}_1,\mathbb{H}_2,\mathbb{H}_3\}$.

Aggregate production $Y_{t+1}^0$ represents the state following a neutral shock $\mathbb{H}_1=\{\delta_t^0,\lambda_t^0\}$. It indicates that traders have not faced trade sanctions or accepted trade cooperation\footnote{For another possible scenario, even if traders have faced trade sanctions or accepted trade cooperation, they possess an absolute leading position in their ability to utilize production factors based on their comparative advantage. which fully offsets the negative effects of trade sanctions or renders the positive effects of trade cooperation insignificant. The same remains true below.}. The trader's ability to utilize production factors will fluctuate stably around the mean. Aggregate production $Y_{t+1}^L$ represents the state following the trade sanction shock $\mathbb{H}_2=\{\delta_t^-,\lambda_t^-\}$. This indicates that traders face trade sanctions, leading to a decline in their ability to utilize production factors (reduced mean) and increased uncertainty (increased variance). Aggregate production $Y_{t+1}^H$ represents the state following the trade cooperation shock $\mathbb{H}_3=\{\delta_t^+,\lambda_t^+\}$. It indicates that traders have chosen to engage in trade cooperation, leading to enhanced ability to utilize production factors (increased mean) and improved stability (reduced variance).

Specifically, we assume that shock $\mathbb{H}_1$ occurs at $T_0$, shock $\mathbb{H}_2$ occurs at $T_A$, and shock $\mathbb{H}_3$ occurs at $T_B$, with $T_0\neq T_A\neq T_B$. Then, we have the following two signal functions:
\[\begin{gathered}
\mathbb{I}_A(t)=\begin{cases}1,\quad T_A\leq t\\0,T_0\leq t<T_A\end{cases}\\\mathbb{I}_B(t)=\begin{cases}1,\quad T_B\leq t\\0,T_0\leq t<T_B\end{cases}
\end{gathered}\]

For the same trader at time $t$, shocks $\mathbb{H}_1$ and $\mathbb{H}_2$ cannot occur simultaneously, meaning that $\mathbb{I}_A(t)\times \mathbb{I}_B(t)=0$ holds for all time $t$. Furthermore, traders experience only one-off shocks rather than continuous shocks, implying that the effects of shocks are convergent.

Then, when shock $\mathbb{H}_1$ occurs, $\mathbb{I}_A(t)=\mathbb{I}_B(t)=0$:
\[\lambda_{t+1}^0=\beta_k\lambda_t^0+\delta_t^0\]

s.t.:
\[\begin{gathered}
\delta_t^0{\sim}\mathcal{N}(0,\omega_0^2)\\\mathbb{E}[\lambda_t^0]=0\\\mathrm{Var}[\lambda_t^0]=\frac{\omega_0^2}{1-\beta_k}
\end{gathered}\]

Given $\lambda_{T_0}^0$, the analytical solution is:

\[\lambda_t^0(t>T_0)=\beta_k^{t-T_0}\lambda_{T_0}^0+\sum_{s_t=T_0}^{t-1}\beta_k^{t-1-s_t}\delta_s^0\]

When shock $\mathbb{H}_2$ occurs, $\mathbb{I}_A(t)=1$:
\[\lambda_{t+1}^-=\beta_k\lambda_t^-+\delta_t^--c_Ae^{-\alpha_A(t-T_A)}\mathbb{I}_A(t)\]

s.t.:
\[\begin{gathered}\delta_t^-\sim\mathcal{N}\left(-c_Ae^{-\alpha_A(t-T_A)}\mathbb{I}_A(t),\omega_0^2[1+e^{-\gamma_A(t-T_A)}]^2\mathbb{I}_A(t)+\omega_0^2(1-\mathbb{I}_A(t))\right)\\c_{A}>0\\\alpha_{A}>0\\\gamma_{A}>0\\\mathbb{E}[\lambda_t^-]<\mathbb{E}[\lambda_t^0]\\\mathrm{Var}[\lambda_t^-]>\mathrm{Var}[\lambda_t^0]\end{gathered}\]

Given $\lambda_{T_A}^-$, we can obtain the analytical solution:
\[\lambda_t^-(t>T_A)=\beta_k^{t-T_A}\lambda_{T_A}^-\sum_{s_t=T_A}^{t-1}\beta_k^{t-1-s_t}\left[c_Ae^{-\alpha_A(s_t-T_A)}-\delta_{s_t}^-\right]\]

Similarly, when shock $\mathbb{H}_3$ occurs, $\mathbb{I}_B(t)=1$:
\[\lambda_{t+1}^+=\beta_k\lambda_t^++\delta_t^++c_Be^{-\alpha_B(t-T_B)}\mathbb{I}_B(t)\]

s.t.:
\[\begin{gathered}\delta_{t}^{+}{\sim}\mathcal{N}\left(c_{B}e^{-\alpha_{B}(t-T_{B})}\mathbb{I}_{B}(t),\omega_{0}^{2}\big[1-e^{-\gamma_{B}(t-T_{B})}\big]^2\mathbb{I}_{B}(t)+\omega_0^2\big(1-\mathbb{I}_{B}(t)\big)\right)\\c_{B}>0\\\alpha_{B}>0\\\gamma_{A}>0\\\mathbb{E}[\lambda_t^+]>\mathbb{E}[\lambda_t^0]\\\mathrm{Var}[\lambda_t^+]<\mathrm{Var}[\lambda_t^0]\end{gathered}\]

Therefore, given $\lambda_{T_B}^+$, we can obtain the analytical solution:
\[\lambda_t^+(t>T_B)=\beta_k^{t-T_B}\lambda_{T_B}^++\sum_{s_t=T_B}^{t-1}\beta_k^{t-1-s_t}\left[c_Be^{-\alpha_B(s_t-T_B)}+\delta_{s_t}^+\right]\]

Figure 36 is a descriptive illustration explaining the shock settings, presenting the partial dynamics of the ability $\lambda_t^k$ to utilize production factors for three types of traders after being subjected to shocks $\mathbb{H}_j\in\{\mathbb{H}_1,\mathbb{H}_2,\mathbb{H}_3\}$ respectively. The first type of trader experiences shock $\mathbb{H}_1$, with its dynamics represented by the black curve. The second type of trader faces shock $\mathbb{H}_2$, depicted by the blue curve. The third type of trader encounters shock $\mathbb{H}_3$, shown by the red curve.
\begin{figure}[htbp]
\centering
\includegraphics[width=10cm]{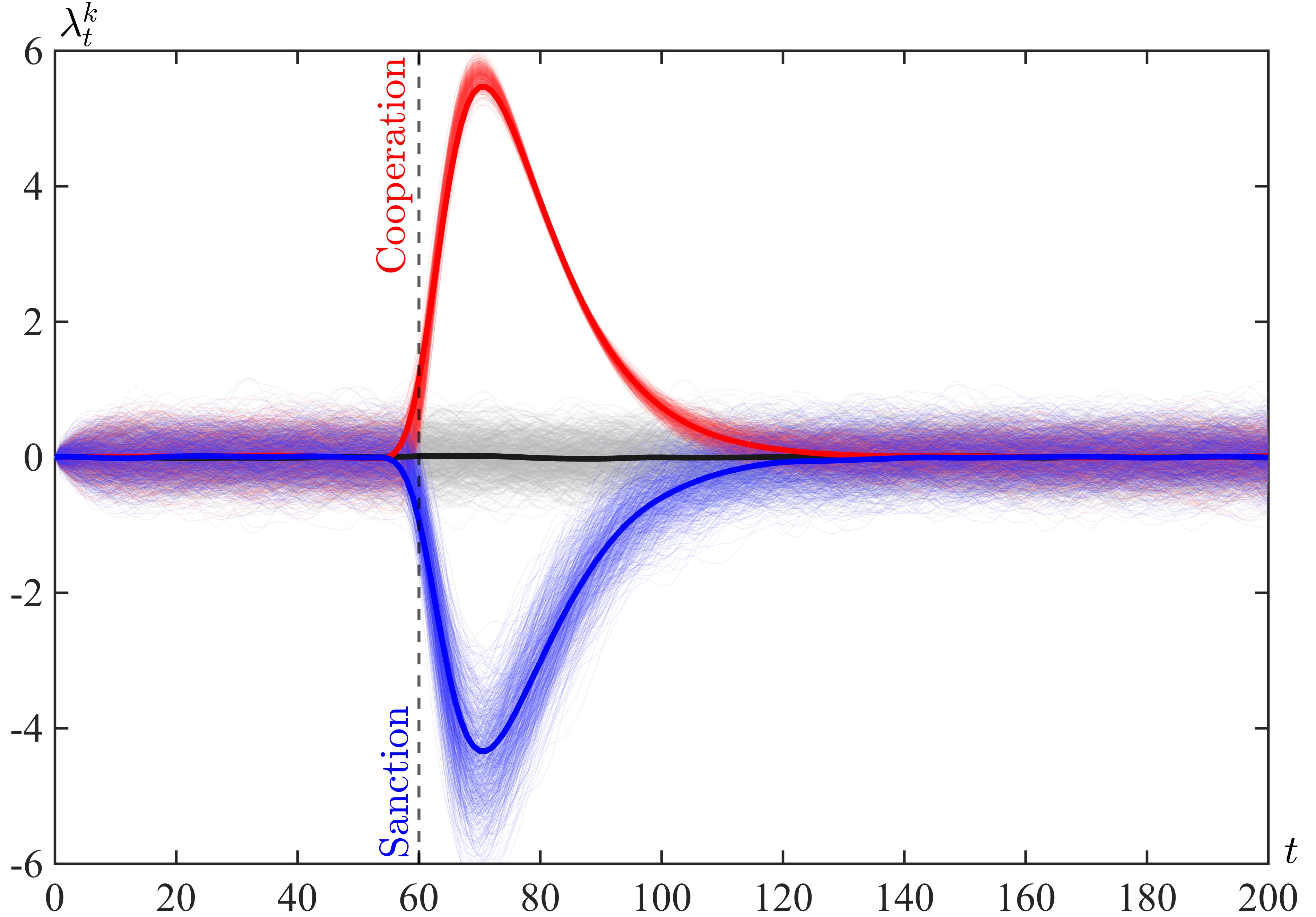}
\caption{\label{fig:F36}The Descriptive Illustration of $\mathbb{H}_j\in\{\mathbb{H}_1,\mathbb{H}_2,\mathbb{H}_3\}$}
\end{figure}

In summary:
\[\lambda_{t+1}^k=\beta_k\lambda_t^k+\delta_t^k-c_Ae^{-\alpha_A(t-T_A)}\mathbb{I}_A(t)+c_Be^{-\alpha_B(t-T_B)}\mathbb{I}_B(t)\]

s.t.:
\[\begin{gathered}\delta_{t}^{k}{\sim}\mathcal{N}\left({\binom{c_{B}e^{-\alpha_{B}(t-T_{B})}\mathbb{I}_{B}(t)}{-c_{A}e^{-\alpha_{A}(t-T_{A})}\mathbb{I}_{A}(t)}},{\binom{\omega_{0}^{2}(1-\mathbb{I}_{B}(t)-\mathbb{I}_{A}(t))+\omega_{0}^{2}{\left[1-e^{-\gamma_{B}(t-T_{B})}\right]}^{2}\mathbb{I}_{B}(t)}{+\omega_{0}^{2}{\left[1+e^{-\gamma_{A}(t-T_{A})}\right]}^{2}\mathbb{I}_{A}(t)}}\right)\\c_{A},c_{B}>0\\\alpha_{A},\alpha_{B}>0\\\gamma_{A},\gamma_{B}>0\end{gathered}\]

Finally, for the aggregate production function, we have:
\[Y_{t+1}=\alpha_km_te^{\lambda_0+\frac{1}{2}\sigma_\lambda^2+\varepsilon_{t+1}}\lambda_t^kL_t\]

s.t.:
\[\begin{gathered}dL_t=L_t(-\mu_s+\mu_l+\mu_p+\sigma_l\sigma_s+\sigma_l\sigma_p)dt+L_t(\sigma_l+\sigma_s\theta_s+\sigma_p\theta_p)dW_t^l\\\lambda_{t+1}^k=\beta_k\lambda_t^k+\delta_t^k-c_Ae^{-\alpha_A(t-T_A)}\mathbb{I}_A(t)+c_Be^{-\alpha_B(t-T_B)}\mathbb{I}_B(t)\\\varepsilon_{t+1}\thicksim\mathcal{N}\left(-\frac{1}{2}\sigma_2^2,\sigma_2^2\right)\\m_t\in(0,1)\\\alpha_{k}>0\end{gathered}\]

Meanwhile, two different types of shocks will respectively impact the labor market and the ability to utilize production factors, ultimately leading to the transitions in aggregate production:
\[\begin{gathered}
\mathbb{K}_j\in\left\{\mathbb{K}_1=\{\mu_s^i,\mu_p^0\},\mathbb{K}_2=\{\mu_s^0,\mu_p^i\},\mathbb{K}_3=\{\mu_s^i,\mu_p^i\}\right\}\\\mathbb{H}_j\in\left\{\mathbb{H}_1=\{\delta_t^0,\lambda_t^0\},\mathbb{H}_2=\{\delta_t^-,\lambda_t^-\},\mathbb{H}_3=\{\delta_t^+,\lambda_t^+\}\right\}
\end{gathered}\]

\section{Model Economy: Trade Dynamics and Equilibrium}\label{Section 8}
This section will construct a continuous-time dynamic stochastic general equilibrium (DSGE) model to examine the dynamic effects of labor market fluctuations on domestic consumption and imported consumption under two policy scenarios, i.e., trade sanction and trade cooperation. The model incorporates four core stochastic processes: Labor factor dynamics, the ability to utilize production factors, iceberg costs (or a set of factors affecting prices), and exogenous productivity shocks (aggregate shocks).

We set households to have CES preferences, and the production function for final goods is influenced by the aforementioned factors simultaneously. In this section, our model assumes households do not engage in intertemporal optimization, thus eliminating the need for solving Euler equations\footnote{We do not consider interest rates or savings, consumption is entirely exogenous to current income, or the consumption and income ratio is exogenous, it depends solely on the trade environment}. Similarly, production does not involve dynamic optimization, it is calculated solely based on the current production function. Thus, the model essentially constitutes a recursive system of stochastic differential equations, eliminating the need to solve expected equations. 

Finally, due to the model's nonlinearity and inclusion of multidimensional random shocks, analytical solutions are theoretically unattainable. Therefore, we employ Monte Carlo methods to numerically simulate the discrete economic system (setting the number of fitting paths to $10000$) and compute the expected paths of consumption variables across periods.

\subsection{The Households}
The consumption of a representative household is modeled by the CES function, incorporating domestic goods $C_{Ht}$ and imported goods (foreign goods) $C_{Ft}$:
\[C_t=\left[\gamma^\frac{1}{\theta}C_{Ht}^\frac{\theta-1}{\theta}+(1-\gamma)^\frac{1}{\theta}C_{Ft}^\frac{\theta-1}{\theta}\right]^\frac{\theta}{\theta-1}\]

s.t.:
\[\begin{gathered}\theta>1\\\gamma\in(0,1)\end{gathered}\]

Here, $\theta$ represents the elasticity of substitution between domestic and imported goods, while $\gamma$ denotes the consumer's preference weight for domestic goods. Now, given the domestic goods price $P_{Ht}$ and the imported goods price $P_{Ft}$, households choose consumption $C_{Ht}$ for domestic goods and $C_{Ft}$ for imported goods to minimize total expenditure while satisfying total consumption $C_t$:

\[\min_{C_{Ht},C_{Ft}}[P_{Ht}C_{Ht}+P_{Ft}C_{Ft}]\]

The Lagrangian function:
\[\mathcal{L}=P_{Ht}C_{Ht}+P_{Ft}C_{Ft}-\lambda_t\left[\left(\gamma^{\frac{1}{\theta}}C_{Ht}^{\frac{\theta-1}{\theta}}+(1-\gamma)^{\frac{1}{\theta}}C_{Ft}^{\frac{\theta-1}{\theta}}\right)^{\frac{\theta}{\theta-1}}-C_t\right]\]

FOC:
\[\begin{gathered}
\frac{\partial\mathcal{L}}{\partial C_{Ht}}=P_{Ht}-\lambda_t\frac{\theta}{\theta-1}\left(\gamma^{\frac{1}{\theta}}C_{Ht}^{\frac{\theta-1}{\theta}}+(1-\gamma)^{\frac{1}{\theta}}C_{Ft}^{\frac{\theta-1}{\theta}}\right)^{\frac{1}{\theta-1}}\gamma^{\frac{1}{\theta}}\frac{\theta-1}{\theta}C_{Ht}^{\frac{1}{\theta}}=0\\\frac{\partial\mathcal{L}}{\partial C_{Ft}}=P_{Ft}-\lambda_t\frac{\theta}{\theta-1}\left(\gamma^{\frac{1}{\theta}}C_{Ht}^{\frac{\theta-1}{\theta}}+(1-\gamma)^{\frac{1}{\theta}}C_{Ft}^{\frac{\theta-1}{\theta}}\right)^{\frac{1}{\theta-1}}(1-\gamma)^{\frac{1}{\theta}}\frac{\theta-1}{\theta}C_{Ft}^{-\frac{1}{\theta}}=0
\end{gathered}\]

Since:
\[\gamma^\frac{1}{\theta}C_{Ht}^\frac{\theta-1}{\theta}+(1-\gamma)^\frac{1}{\theta}C_{Ft}^\frac{\theta-1}{\theta}=C_t^\frac{\theta-1}{\theta}\]

Then, we get:
\[\begin{gathered}
P_{Ht}=\lambda_t\gamma^{\frac{1}{\theta}}\left(\frac{C_t}{C_{Ht}}\right)^{\frac{1}{\theta}}\\P_{Ft}=\lambda_t(1-\gamma)^{\frac{1}{\theta}}\left(\frac{C_t}{C_{Ft}}\right)^{\frac{1}{\theta}}
\end{gathered}\]

That is:
\[\begin{gathered}\frac{C_{Ht}}{C_t}=\gamma\left(\frac{\lambda_t}{P_{Ht}}\right)^\theta\\\frac{C_{Ft}}{C_t}=(1-\gamma)\left(\frac{\lambda_t}{P_{Ft}}\right)^\theta\\\left(\frac{C_{Ht}}{C_t}\right)^{\frac{\theta-1}{\theta}}=\gamma^{\frac{\theta-1}{\theta}}\left(\frac{\lambda_t}{P_{Ht}}\right)^{\theta-1}\\\left(\frac{C_{Ft}}{C_t}\right)^{\frac{\theta-1}{\theta}}=(1-\gamma)^{\frac{\theta-1}{\theta}}\left(\frac{\lambda_t}{P_{Ft}}\right)^{\theta-1}\\\gamma^{\frac{1}{\theta}}\left(\frac{C_{Ht}}{C_t}\right)^{\frac{\theta-1}{\theta}}=\gamma\left(\frac{\lambda_t}{P_{Ht}}\right)^{\theta-1}\\(1-\gamma)^{\frac{1}{\theta}}\left(\frac{C_{Ft}}{C_t}\right)^{\frac{\theta-1}{\theta}}=(1-\gamma)\left(\frac{\lambda_t}{P_{Ft}}\right)^{\theta-1}\end{gathered}\]

Substituting into the CES function, we obtain:
\[\begin{gathered}\left[\gamma\left(\frac{\lambda_t}{P_{Ht}}\right)^{\theta-1}+(1-\gamma)\left(\frac{\lambda_t}{P_{Ft}}\right)^{\theta-1}\right]^{\frac{\theta}{\theta-1}}=1\\\gamma\left(\frac{\lambda_t}{P_{Ht}}\right)^{\theta-1}+(1-\gamma)\left(\frac{\lambda_t}{P_{Ft}}\right)^{\theta-1}=1\\\lambda_t^{\theta-1}\left[\gamma P_{Ht}^{-(\theta-1)}+(1-\gamma)P_{Ft}^{-(\theta-1)}\right]=1\end{gathered}\]

Hence:
\[\lambda_t=\left[\gamma P_{Ht}^{1-\theta}+(1-\gamma)P_{Ft}^{1-\theta}\right]^{\frac{1}{1-\theta}}\]

For the minimum total expenditure $E_t=P_{Ht}C_{Ht}+P_{Ft}C_{Ft}$, we have:
\[E_t=P_{Ht}\gamma C_t\left(\frac{\lambda_t}{P_{Ht}}\right)^\theta+P_{Ft}(1-\gamma)C_t\left(\frac{\lambda_t}{P_{Ft}}\right)^\theta=C_t\lambda_t^\theta\lambda_t^{1-\theta}=C_t\lambda_t\]

So, the consumer price index (CPI) for household consumption is $\lambda_{t}$, denoted as $P_t$:
\[P_t=\lambda_t=\left[\gamma P_{Ht}^{1-\theta}+(1-\gamma)P_{Ft}^{1-\theta}\right]^{\frac{1}{1-\theta}}\]

We standardize domestic goods prices to 1, i.e., $P_{Ht}=1$. The domestic price of imported goods is jointly determined by the exchange rate $\mathcal{E}_t$ and the iceberg costs $\mathcal{B}_t$. Setting the foreign currency price of international imports as an exogenous constant equal to 1, the domestic price of imported goods is:
\[P_{Ft}=\mathcal{E}_t\mathcal{B}_t=\tau_{it}\]

Here, we define the exchange rate $\mathcal{E}_t$ and the iceberg costs $\mathcal{B}_t$ as a set of factors $\tau_{it}$ that influence prices, which are solely related to the international trade environment.

We define $\tau_{it}$ as standard Geometric Brownian Motion, then:
\[d\tau_{it}=\tau_{it}\mu_\tau^idt+\tau_{it}\sigma_\tau dW_t^\tau\]

Where: $\mu_{\tau}^i\in(\mu_{\tau}^-, \mu_{\tau}^+), \mu_{\tau}^+>0>\mu_{\tau}^-$ is the drift term, representing the instantaneous rate of change in the price factor dynamics. It is influenced solely by exogenous trade policies (i.e., the trade environment). $\sigma_{\tau}>0$ denotes the instantaneous volatility of the price factor dynamics, which is a constant. $W_t^{\tau}$ represents standard Brownian motion. Therefore:
\[P_t=[\gamma+(1-\gamma)\tau_{it}^{1-\theta}]^{\frac{1}{1-\theta}}\]

Now, we can obtain the Hicks demand function for households:
\[\begin{gathered}
P_{Ht}=\lambda_t\gamma^{\frac{1}{\theta}}\left(\frac{C_t}{C_{Ht}}\right)^{\frac{1}{\theta}}\\P_{Ft}=\lambda_t(1-\gamma)^{\frac{1}{\theta}}\left(\frac{C_t}{C_{Ft}}\right)^{\frac{1}{\theta}}\\P_{t}=\lambda_{t}=[\gamma+(1-\gamma)\tau_{it}^{1-\theta}]^{\frac{1}{1-\theta}}
\end{gathered}\]

That is:
\[\begin{gathered}
C_{Ht}=\gamma\left(\frac{P_t}{P_{Ht}}\right)^\theta C_t=\gamma P_t^\theta C_t\\C_{Ft}=(1-\gamma)\left(\frac{P_t}{P_{Ft}}\right)^\theta C_t=(1-\gamma)\tau_{it}^{-\theta}P_t^\theta C_t
\end{gathered}\]

We assume that total household consumption $E_t$ is a proportion $\psi_i \in(0,1)$ of aggregate production $Y_t$ . Similarly, $\psi_i=\{\psi_L,\psi_H\},\psi_H>\psi_L$, depends solely on the international trade environment. Thus, $E_t = P_t C_t = \psi_i Y_t$, and:
\[\begin{gathered}
C_{Ht}=\gamma P_t^\theta C_t=\gamma P_t^\theta\frac{\psi_iY_t}{P_t}=\gamma P_t^{\theta-1}\psi_iY_t\\C_{Ft}=(1-\gamma)\tau_{it}^{-\theta}P_t^\theta C_t=(1-\gamma)\tau_{it}^{-\theta}P_t^\theta\frac{\psi_iY_t}{P_t}=(1-\gamma)\tau_{it}^{-\theta}P_t^{\theta-1}\psi_iY_t
\end{gathered}\]

\subsection{The Production Department}
We have already discussed the details of the aggregate production function in Section 7. We will now proceed with further specifications and necessary simplifications. Assuming no intertemporal optimization behavior, final goods production and all random shocks occur in period $t$, the production function is given by $Y_t$:
\[Y_t=\alpha_km_te^{\lambda_0+\frac{1}{2}\sigma_\lambda^2+\varepsilon_t}\lambda_t^kL_t\]

Where:

$\alpha_k>0$ is a constant representing the positive proportional relationship between the ability to utilize production factors and the production of final goods. $m_t\in(0,1)$ denotes the scale coefficient of production factors. We further assume that $m_t$ is jointly influenced by cost factors and the trade environment, i.e., $m_{it}=v_i \tau_{it}^{-\kappa}$, where $v_i \in (0,1)$ measures the impact of trade sanction and trade cooperation on the scale of production factors, and $\kappa \in (0,1)$ measures the sensitivity of production factor scale to trade costs. $\lambda_0$ is the long-term mean of production factor utilization ability, and $\sigma_{\lambda}$ is its variance, both of which are constants.
$\varepsilon_{t}$ represents aggregate shocks. Here, we assume aggregate shocks are exogenous and arise solely from fluctuations in factor utilization ability. Since $\mathbb{E}[e^{\varepsilon_t}]=1$, therefore we can get $\varepsilon_{t}\thicksim\mathcal{N}\left(-\frac{1}{2}\sigma_{\lambda}^2,\sigma_{\lambda}^2\right)$.

And $L_t$ is the labor factor dynamics:
\[dL_t=L_t\mu_Ldt+L_t\sigma_LdW_t^L\]

Where:

$\mu_L$ is the drift term, representing the instantaneous rate of change in the labor factor. $\sigma_L>0$ denotes the instantaneous volatility of the labor factor, and $W_t^L$ is a standard Brownian motion. Here, we set $\sigma_L>0$ as a constant. Based on Section 7, we define two primary types of shocks occurring in the labor factor dynamics, leading to $\mu_L\in(\mu_L^-,\mu_L^+)$ and $\mu_L^+>\mu_L^0>\mu_L^-$. First scenario: $\mu_L=\mu_L^0$. This indicates no shock occurred, or that the value shock and allocation shock of the labor factor have neutralized each other, resulting in stable fluctuations of the labor factor dynamics. Second scenario: $\mu_L>\mu_L^0,\mu_L\rightarrow\mu_L^+$. This indicates a positive shock occurred, where the effect of the value shock is stronger than that of the allocation shock. The third scenario, $\mu_L^0 > \mu_L, \mu_L \rightarrow \mu_L^-$, indicates a negative shock where the impact of the value shock is weaker than that of the allocation shock.

Then, $\lambda_t^k$ denotes the ability dynamics to utilize production factors :
\[\lambda_{t+1}^k=\beta_k\lambda_t^k+\delta_t^k-c_Ae^{-\alpha_A(t-T_A)}\mathbb{I}_A(t)+c_Be^{-\alpha_B(t-T_B)}\mathbb{I}_B(t)\]

s.t.:
\[\delta_t^k{\sim}\mathcal{N}\left(\binom{c_Be^{-\alpha_B(t-T_B)}\mathbb{I}_B(t)}{-c_Ae^{-\alpha_A(t-T_A)}\mathbb{I}_A(t)},\binom{\omega_0^2(1-\mathbb{I}_B(t)-\mathbb{I}_A(t))+\omega_0^2[1-e^{-\gamma_B(t-T_B)}]^2\mathbb{I}_B(t)}{+\omega_0^2[1+e^{-\gamma_A(t-T_A)}]^2\mathbb{I}_A(t)}\right)\]

Where:

$\beta_k\in(0,1)$ is the persistence parameter, representing the serial correlation in the ability to utilize production factors. $\delta_t^k\in\{\delta_t^-,\delta_t^+\}$ denotes the distribution of the ability to utilize production factors under trade sanction and trade cooperation environments. $\mathbb{I}_A,\mathbb{I}_B=\{0,1\}$ are signal functions indicating trade sanction or trade cooperation. $c_A,c_B>0$ denote the intensity of trade shocks. $\alpha_A,\alpha_B>0$ represent the decline rate of trade shocks. $\gamma_A,\gamma_B>0$ indicate the volatility intensity of trade shocks.

\subsection{The Equilibrium}
We will analyze international trade in small open economies under two distinct trade environments, and then respectively compute dynamic stochastic general equilibrium models. These two scenarios fundamentally stem from the effects of trade sanction and trade cooperation on households behavior and production sector behavior. We define the parameter set influencing households and production sector behavior as the collection $\mathbb{X}_i=\{\tau_{it},\psi_i,m_{it},\delta_t^k\}$.

The first scenario is trade sanction, denoted as $\mathbb{X}_{Sanction}=\{\tau_{Ht},\psi_L,m_{Lt},\delta_t^-\}$. In this case, households face increased consumption costs and reduced willingness to consume, resulting in a smaller share of consumption from aggregate production. For the production sector, the ability to utilize production factors is constrained. Consequently, the scale of production factors allocated to final goods production is reduced, the mean of the distribution for factor utilization ability decreases, and the variance increases. This indicates weaker and more unstable factor utilization ability. Specifically:
\[\begin{gathered}d\tau_{Ht}=\tau_{Ht}\mu_\tau^+dt+\tau_{Ht}\sigma_\tau dW_t^\tau\\\psi_i=\psi_L,\psi_L\in(0,1)\\m_{Lt}=v_L\tau_{Ht}^{-\kappa},v_L\in(0,1)\\\lambda_{t+1}^-=\beta_k\lambda_t^-+\delta_t^--c_Ae^{-\alpha_A(t-T_A)}\\\delta_t^-\sim\mathcal{N}\left(\left(-c_Ae^{-\alpha_A(t-T_A)}\right),\left(\omega_0^2[1+e^{-\gamma_A(t-T_A)}]^2\right)\right)\end{gathered}\]

The second scenario is trade cooperation, denoted as $\mathbb{X}_{Cooperation}=\{\tau_{Lt},\psi_H,m_{Ht},\delta_t^+\}$. In this case, households experience reduced consumption costs and increased willingness to consume, resulting in a higher proportion of consumption from aggregate production. For the production sector, the utilization ability of production factors improves due to cooperation. Consequently, a larger scale of production factors can be allocated to final goods production. The mean of the distribution for production factor utilization ability increases while its variance decreases, indicating stronger and more stable utilization ability. Specifically:
\[\begin{gathered}d\tau_{Lt}=\tau_{Lt}\mu_\tau^-dt+\tau_{Lt}\sigma_\tau dW_t^\tau\\\psi_i=\psi_H,\psi_H\in(0,1)\\m_{Ht}=v_{H}\tau_{Lt}^{-\kappa},v_{H}\in(0,1)\\\lambda_{t+1}^+=\beta_k\lambda_t^++\delta_t^++c_Be^{-\alpha_B(t-T_B)}\\\delta_t^+{\sim}\mathcal{N}\left(\left(c_Be^{-\alpha_B(t-T_B)}\right),\left(\omega_0^2{[1-e^{-\gamma_B(t-T_B)}]^2}\right)\right)\end{gathered}\]

Therefore, the equilibrium of this economic system is the aggregate set of the following variables:
\[\left\{L_t,\lambda_t^k,\tau_{it},\varepsilon_t,\delta_t^k,W_t^L,W_t^\tau,Y_t,P_t,C_{Ht},C_{Ft}\right\}_{t\in\mathbb{N}}\]

Define the state variable $\mathbb{V}_t=(L_t, \lambda_t^k, \tau_{it})$, the shock variable $\mathbb{W}_t=(\varepsilon_t, \delta_t^k, W_t^L, W_t^{\tau})$, and the endogenous variable $\mathbb{U}_t=(Y_t, P_t, C_{Ht}, C_{Ft})$. Here, we employ the Euler-Maruyama method to make the discrete state variable with a time step of $\Delta t$. Given initial state $\mathbb{V}_0=(L_0, \lambda_0^k, \tau_{i0})$ and exogenous shock sequence $\{\mathbb{W}_t\}_{t=0}^T$, there exists and only exists one recursive equilibrium\footnote{First, $P_t$ is uniquely determined by $\tau_{it}$. Second, $Y_t$ is uniquely determined by $L_t, \lambda_t^k, \tau_{it}, \varepsilon_t$. Third, $P_t$, $Y_t$, and $\tau_{it}$ uniquely determine $C_{Ht}$ and $C_{Ft}$, respectively.} is the state sequence $\{\mathbb{V}_t\}_{t=0}^{T+1}$ and the endogenous variable sequence $\{\mathbb{U}_t\}_{t=0}^T$, satisfying for any $t\geq0$:
\[\begin{gathered}P_t=\left[\gamma+(1-\gamma)\tau_{it}^{1-\theta}\right]^{\frac{1}{1-\theta}}\\Y_t=\alpha_kv_i\tau_{it}^{-\kappa}e^{\lambda_0+\frac{1}{2}\sigma_\lambda^2+\varepsilon_t}\lambda_t^kL_t\\C_{Ht}=\gamma P_t^{\theta-1}\psi_iY_t\\C_{Ft}=(1-\gamma)\tau_{it}^{-\theta}P_t^{\theta-1}\psi_iY_t\end{gathered}\]

Where:
\[\begin{gathered}
L_{t+1}=L_t[1+\mu_L\Delta t+\sigma_L\Delta W_t^L]\\\lambda_{t+1}^k=\beta_k\lambda_t^k+\delta_t^k\Delta t\\\tau_{i,t+1}=\tau_{i,t}[1+\mu_{\tau}\Delta t+\sigma_{\tau}\Delta W_{t}^{\tau}]
\end{gathered}\]

\subsection{The Analysis of Dynamic Stochastic General Equilibrium}
In this section, we will employ Monte Carlo simulation methods to analyze the DSGE model, and then further discuss and study the dynamics of exports. We set parameter values within reasonable ranges based on the conditions of each economic variable, as shown in Table 9. Specific details regarding parameter values are provided in \hyperref[Appendix D]{Appendix D}. 

Figure 37 and Figure 38 illustrate the DSGE models for domestic goods consumption $C_{Ht}$ and imported goods consumption $C_{Ft}$ under trade sanction and trade cooperation policy scenarios respectively, following a labor factor shock.

We find that under trade sanction policy, following a labor factor shock, $C_{Ht}$ and $C_{Ft}$ decline rapidly and remain below zero ($C_{Ht}<0,C_{Ft}<0$) after a certain time $t_0$ before gradually converging. This indicates that labor factor shocks simultaneously depress domestic goods consumption and imported goods consumption. Furthermore, if a positive shock occurs during labor factor fluctuations, i.e., $\mu_L > \mu_L^0$ and $\mu_L \rightarrow \mu_L^+$, the convergence rates of $C_{Ht}$ and $C_{Ft}$ slow down. This implies that at a certain period after the shock and before reaching convergence, the positive shock exerts a stronger inhibition on consumption. Similarly, if a negative shock occurs in labor factor fluctuations, that is, $\mu_L^0 > \mu_L$ and $\mu_L \rightarrow \mu_L^-$, then the convergence rates of $C_{Ht}$ and $C_{Ft}$ accelerate. This implies that during a specific period after the shock and before reaching the convergence state, the negative shock exerts a weaker inhibition on consumption.
\begin{figure}[htbp]
\centering
\includegraphics[width=15cm]{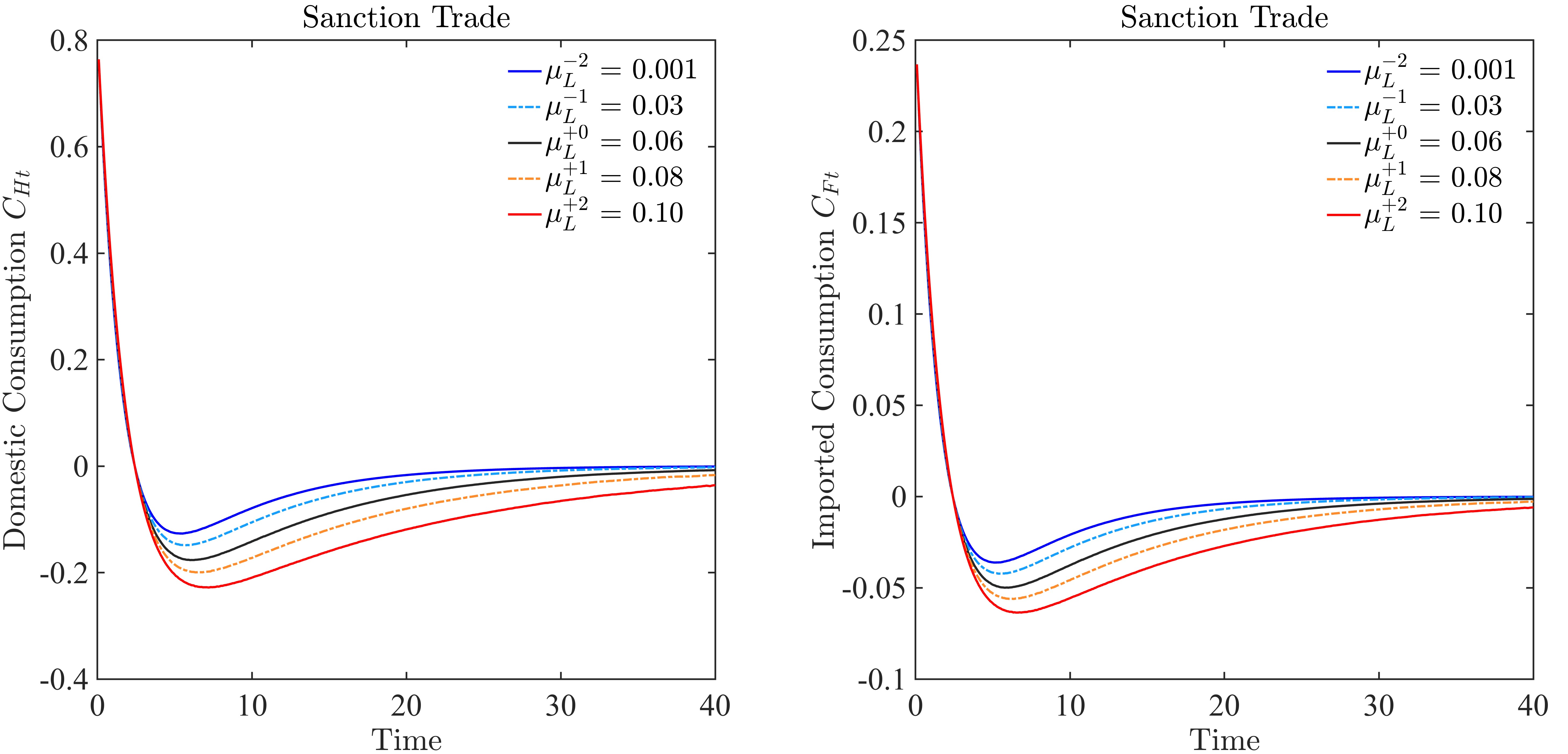}
\caption{\label{fig:F37}The Dynamic Transitions of Consumption in Sanction Trade}
\end{figure}
\vspace{-0.1cm}
\begin{figure}[htbp]
\centering
\includegraphics[width=15cm]{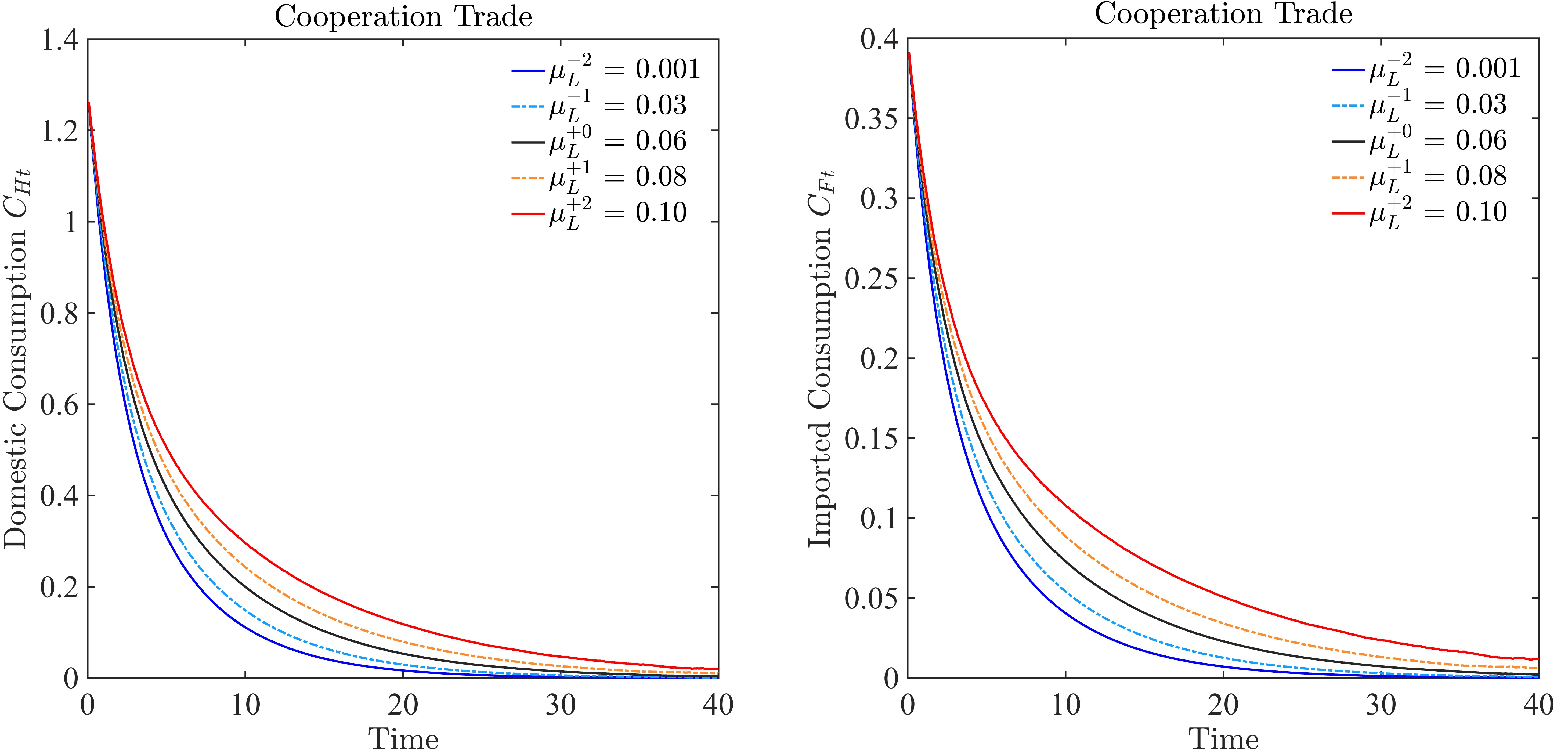}
\caption{\label{fig:F38}The Dynamic Transitions of Consumption in Cooperation Trade}
\end{figure}

Conversely, under trade cooperation policy, labor factor shocks cause $C_{Ht}$ and $C_{Ft}$ to decline. However, during convergence, $C_{Ht}>0$ and $C_{Ft}>0$ are consistently maintained, indicating that labor factor shocks simultaneously increase both domestic and imported goods consumption. Moreover, if a positive shock occurs during labor factor fluctuations, i.e., $\mu_L > \mu_L^0$ and $\mu_L \rightarrow \mu_L^+$, the convergence rates of $C_{Ht}$ and $C_{Ft}$ will slow down. This implies that at a specific stage after the shock but before reaching convergence, the positive shock generates a greater boost to consumption. Similarly, if a negative shock occurs in labor factor fluctuations, that is, $\mu_L^0 > \mu_L$ and $\mu_L \rightarrow \mu_L^-$, then the convergence rates of $C_{Ht}$ and $C_{Ft}$ will accelerate. This implies that at a certain period after the shock and before convergence is reached, the negative shock will have a smaller boost on consumption.

To conclude, we compare the dynamic transitions in equilibrium consumption under the two trade policies, as shown in the graphs of Figure 39. We find that under trade sanction, labor shocks inhibit household consumption. Under trade cooperation, labor shocks enhance household consumption. Furthermore, regardless of whether trade cooperation or sanction are implemented, an increase in $\mu_L$ intensifies the impact of labor shocks, making their effect on consumption both stronger and more persistent.
\begin{figure}[htbp]
    \centering
    \begin{minipage}[b]{1\textwidth}
        \centering
        \includegraphics[width=\textwidth]{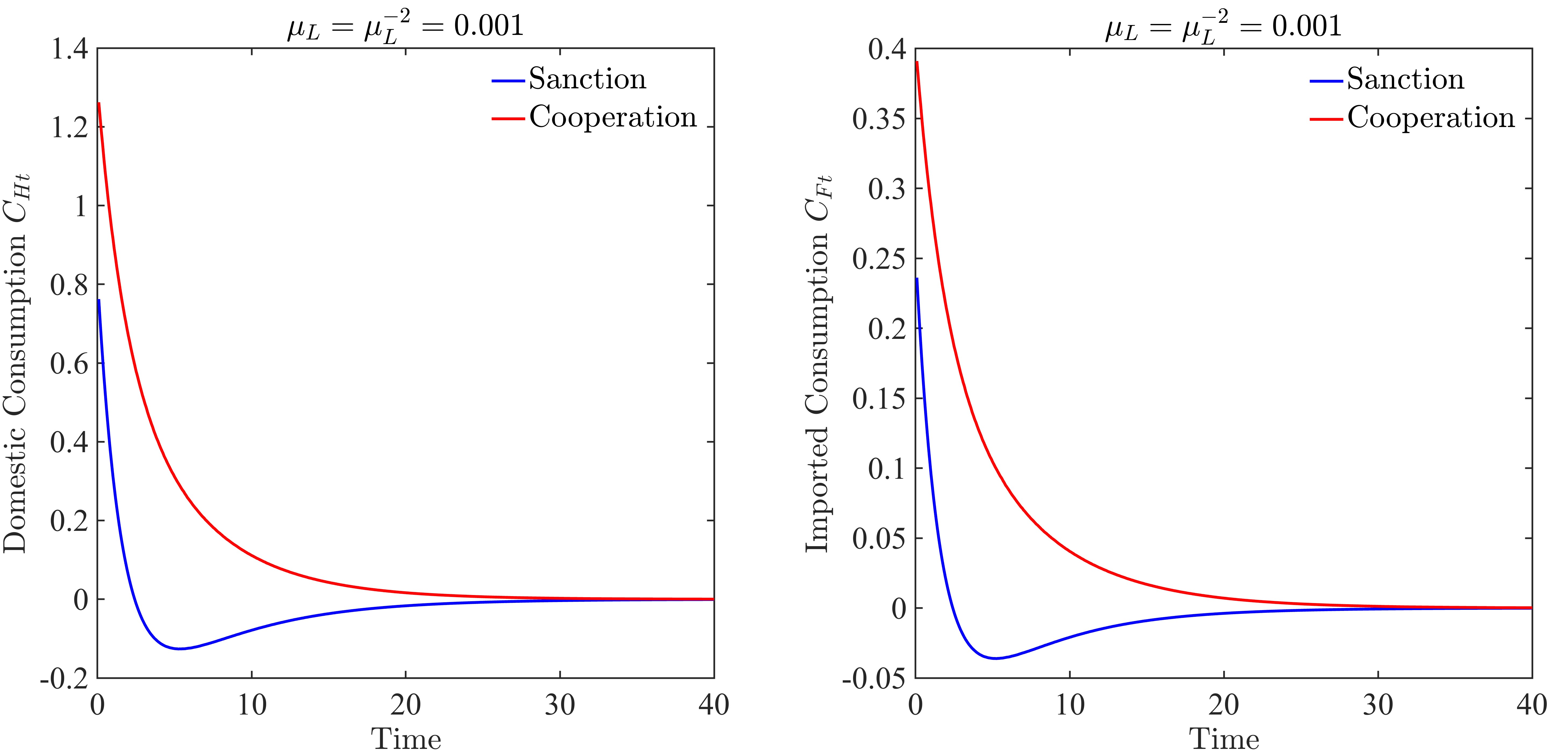}
    \end{minipage}
    \hfill
    \begin{minipage}[b]{1\textwidth}
        \centering
        \includegraphics[width=\textwidth]{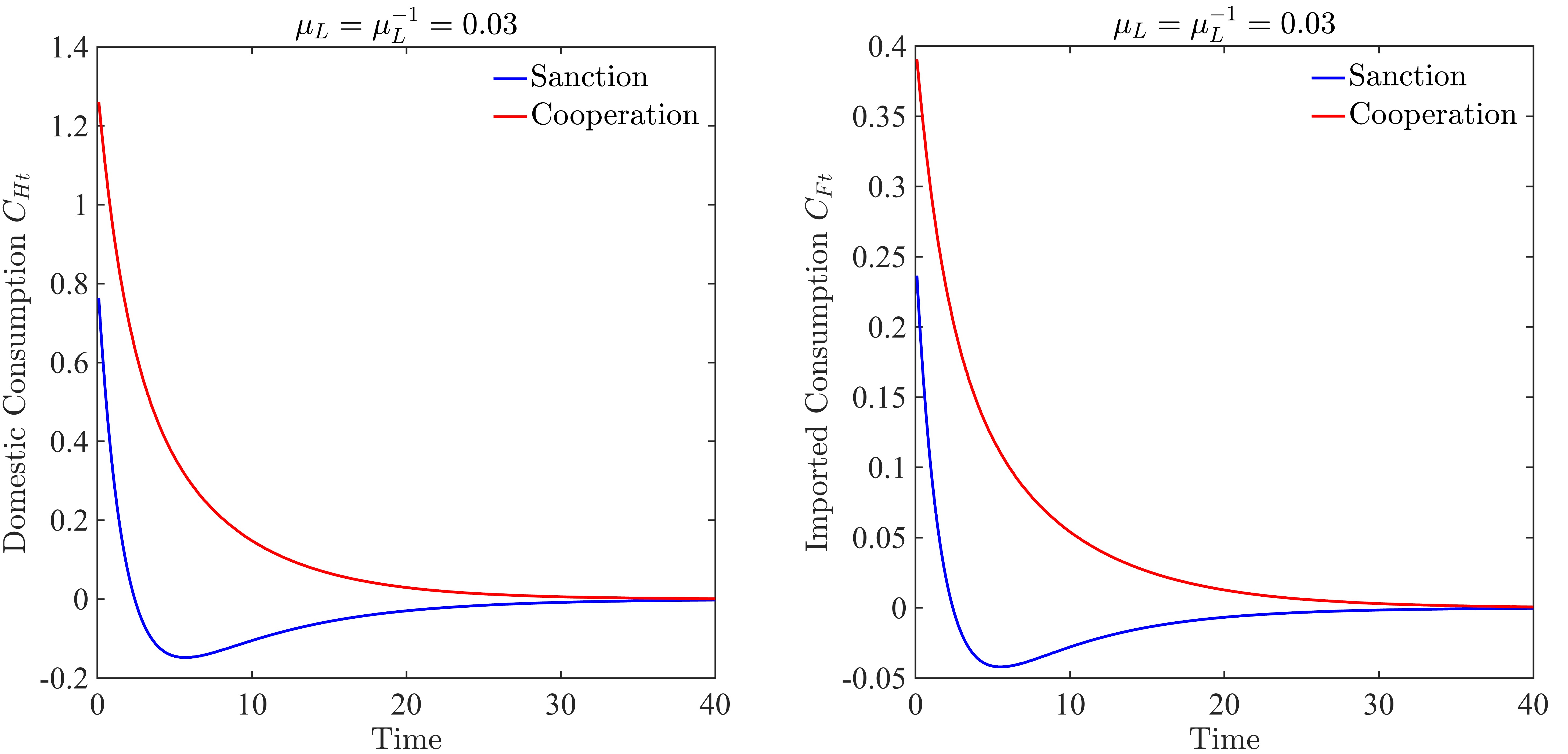}
    \end{minipage}
\end{figure}

\begin{figure}[htbp]
    \centering
    \begin{minipage}[b]{1\textwidth}
        \centering
        \includegraphics[width=\textwidth]{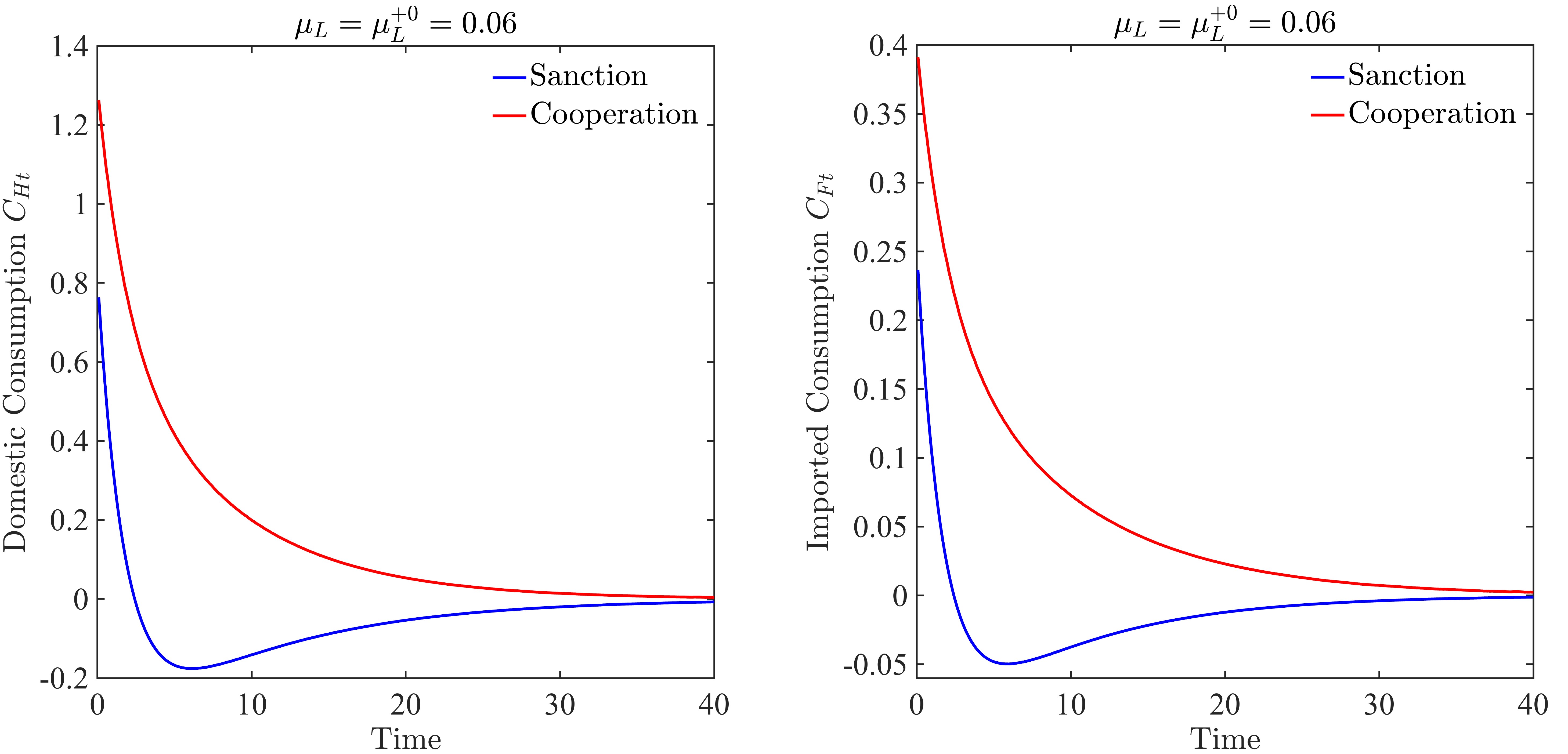}
    \end{minipage}
    \hfill
    \begin{minipage}[b]{1\textwidth}
        \centering
        \includegraphics[width=\textwidth]{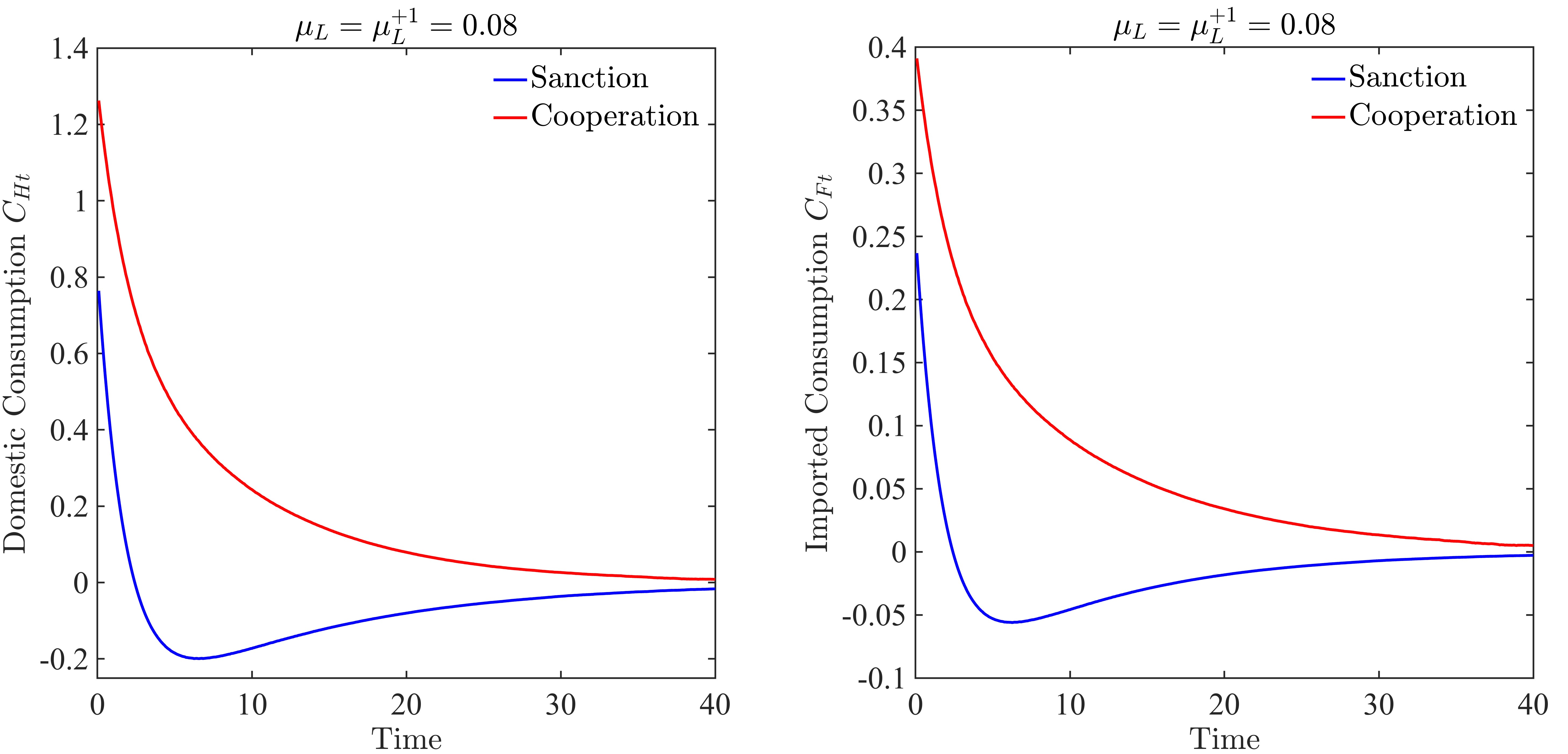}
    \end{minipage}
    \hfill
    \begin{minipage}[b]{1\textwidth}
        \centering
        \includegraphics[width=\textwidth]{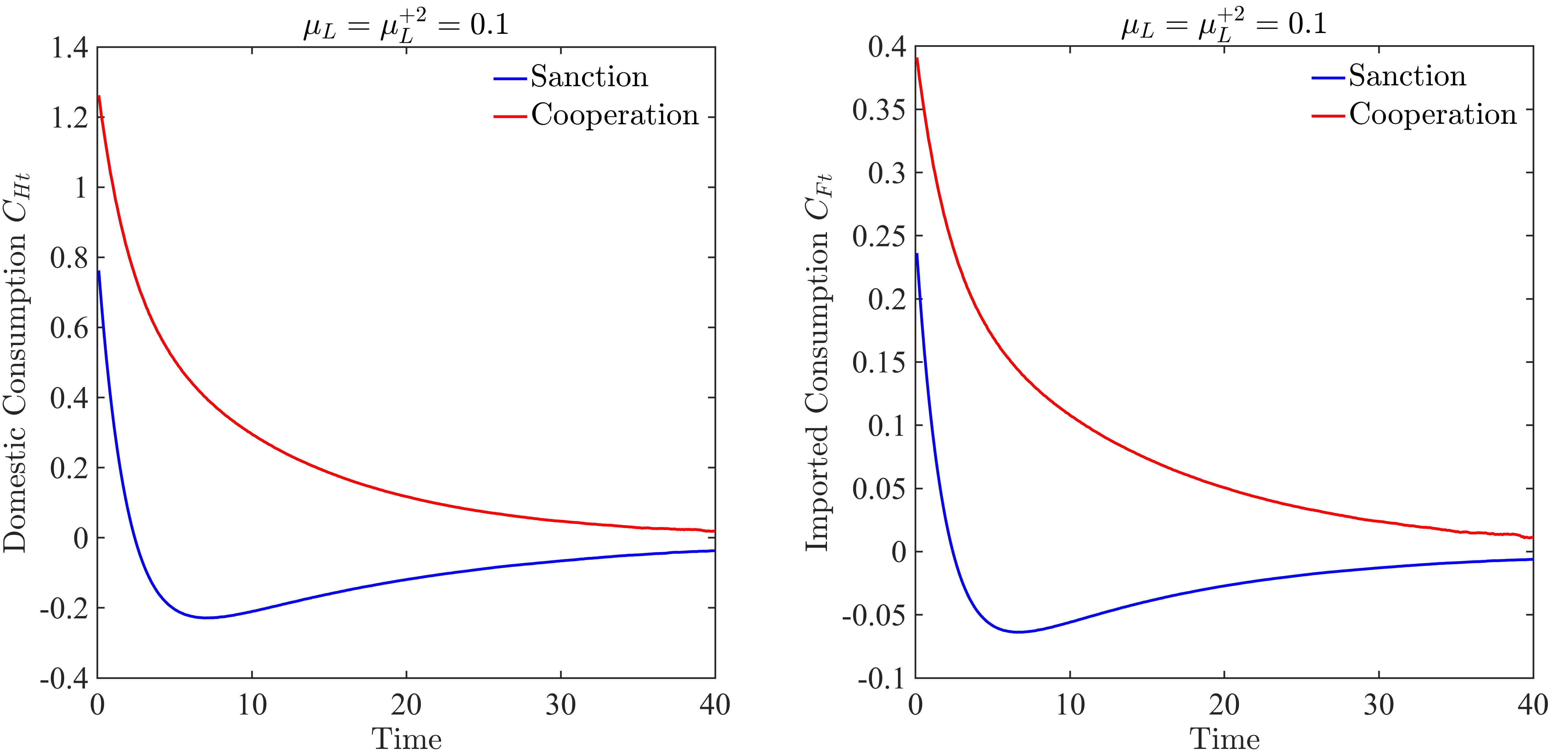}
    \end{minipage}
    \caption{The Dynamic Transitions of Consumption with $\mu_L^{k}\in(\mu_L^-,\mu_L^+)$}
    \label{fig:F39}
\end{figure}

Based on the simulation results from Figures 37 to 39, we can simply summarize the general impact of labor factor fluctuations on household consumption. Specifically, we will examine the effect of labor factor shocks on household imported consumption $C_{Ft}$, and we propose Hypothesis 4:

\begin{Hypothesis}
Under a certain trade policy, if the labor factor shock is positive, its impact on imported consumption will be strengthened. If the labor factor shock is negative, its impact on imported consumption will be weakened.
\end{Hypothesis}

\subsection{The Export Dynamics}
In this section, we will attempt to analyze what potential effects the consumption of domestic goods $C_{Ht}$ and imported goods $C_{Ft}$ may have on export dynamics following a labor factor shock. In Section 7.1, we assumed household consumption constitutes a proportion of aggregate production. Here, we define the aggregate production will be allocated to domestic consumption $C_{Ht}$ and export $X_t$, that is:
\[Y_t=C_{Ht}+X_t\]

Hence:
\[X_t=Y_t-C_{Ht}=Y_t[1-\gamma P_t^{\theta-1}\psi_0]\]

Where: $\gamma\in(0,1)$ represents domestic consumption preferences. $\psi_0\in(0,1)$ denotes the propensity to consume, i.e., the proportion of consumption relative to aggregate production. Now, we express the variables as logarithmic deviations from a deterministic steady state, and denote them as: 
$y_{t}=\ln(Y_{t}/Y_{ss}),l_{t}=\ln(L_t/L_{ss}),c_t=\ln(C_t/C_{ss}),x_t=\ln(X_t/X_{ss}),p_t=\ln(P_t/P_{ss}),\tau_{st}=\ln(\tau_{it}/\tau_{iss})$

Taking the logarithm of the aggregate production function $Y_{t}=\alpha_{k}v_{i}\tau_{it}^{-\kappa}e^{\lambda_{0}+\frac{1}{2}\sigma_{\lambda}^{2}+\varepsilon_{t}}\lambda_{t}^{k}L_{t}$, we obtain:
\[\ln Y_t=\ln\alpha_k+\ln v_i-\kappa\ln\tau_{it}+\lambda_0+\frac{1}{2}\sigma_\lambda^2+\varepsilon_t+\ln\lambda_t^k+\ln L_t\]

Assuming $\varepsilon_t=0$ at steady state, then $y_t$ can be written as:
\[y_t=-\kappa\tau_{st}+\lambda_t^k+l_t+\varepsilon_t\]

Similarly, we linearize the price index $P_t=[\gamma+(1-\gamma)\tau_{it}^{1-\theta}]^{\frac{1}{1-\theta}}$ to obtain:
\[\ln P_t=\frac{1}{1-\theta}\ln\left[\gamma+(1-\gamma)\tau_{it}^{1-\theta}\right]\]

Let $\xi=\ln\tau_{it}$, then $e^{\xi}=\tau_{it}$, therefore:
\[\ln P_t=\frac{1}{1-\theta}\ln\left[\gamma+(1-\gamma)e^{\xi(1-\theta)}\right]\]

Taking the first-order Taylor expansion at the steady state $\xi_{ss}=\ln\tau_{iss}$ and performing the calculation, we obtain:

\[\begin{gathered}\ln P_t\approx\ln P_{ss}+\frac{d}{d\xi}\left(\frac{1}{1-\theta}\ln\left[\gamma+(1-\gamma)e^{\xi(1-\theta)}\right]\right)_{\xi=\xi_{ss}}(\xi-\xi_{ss})\\\frac{d}{d\xi}\left(\frac{1}{1-\theta}\ln\left[\gamma+(1-\gamma)e^{\xi(1-\theta)}\right]\right)=\frac{1}{1-\theta}\frac{(1-\gamma)(1-\theta)e^{\xi(1-\theta)}}{\gamma+(1-\gamma)e^{\xi(1-\theta)}}=\frac{(1-\gamma)e^{\xi(1-\theta)}}{\gamma+(1-\gamma)e^{\xi(1-\theta)}}\\\frac{d}{d\xi}\left(\frac{1}{1-\theta}\ln\left[\gamma+(1-\gamma)e^{\xi(1-\theta)}\right]\right)_{\xi=\xi_{ss}}=\frac{(1-\gamma)e^{\xi_{ss}(1-\theta)}}{\gamma+(1-\gamma)e^{\xi_{ss}(1-\theta)}}\\\ln\frac{P_t}{P_{ss}}\approx\frac{(1-\gamma)\tau_{iss}^{1-\theta}}{\gamma+(1-\gamma)\tau_{iss}^{1-\theta}}\ln\frac{\tau_{it}}{\tau_{iss}}\end{gathered}\]

That is:
\[\begin{gathered}
p_t=\phi_p\tau_{st}\\\phi_p\equiv\frac{(1-\gamma)\tau_{iss}^{1-\theta}}{\gamma+(1-\gamma)\tau_{iss}^{1-\theta}}\in(0,1)
\end{gathered}\]

Now, we linearize the export $X_t$ using a similar approach:
\[\ln X_t=\ln Y_t+\ln\left[1-\gamma P_t^{\theta-1}\psi_0\right]\]

We let:
\[\begin{gathered}
P_t=P_{ss}e^{p_t}\\f(P_t)=1-\gamma\psi_0P_t^{\theta-1}
\end{gathered}\]

Here, we assume that at steady state, $p_t \rightarrow p=0$. Then, performing a first-order Taylor approximation of $\ln f(P_t)$ at the steady-state $P_{ss}$:
\[\begin{gathered}
\ln f(P_t)\approx\ln f(P_{ss}e^p)+\frac{f^{\prime}(P_{ss}e^p)P_{ss}e^p}{f(P_{ss}e^p)}(p_t-p)=\ln f(P_{ss})+\frac{f^{\prime}(P_{ss})P_{ss}}{f(P_{ss})}p_t\\\frac{f^{\prime}(P_{ss})P_{ss}}{f(P_{ss})}=\frac{-(\theta-1)\gamma\psi_0P_{ss}^{\theta-2}P_{ss}}{1-\gamma\psi_0P_{ss}^{\theta-1}}=-\frac{(\theta-1)\gamma\psi_0P_{ss}^{\theta-1}}{1-\gamma\psi_0P_{ss}^{\theta-1}}
\end{gathered}\]

We denote:
\[\phi_x\equiv\frac{(\theta-1)\gamma\psi_0P_{ss}^{\theta-1}}{1-\gamma\psi_0P_{ss}^{\theta-1}}\]

Hence:
\[x_t=y_t-\phi_xp_t=l_t+\lambda_t^k+\varepsilon_t-\kappa\tau_{st}-\phi_x\phi_p\tau_{st}\]

Accordingly, we find that fluctuations in labor factors will directly impact export dynamics. Now, we need to consider a household's intertemporal optimization behavior\footnote{This analysis of household intertemporal optimization is not directly related to the DSGE analysis above. We consider household intertemporal optimization here solely to decompose the function $\mathbb{F}$, i.e., $\mathbb{F}:c_t \rightarrow x_t$.} and attempt to establish a link between export dynamics and consumption dynamics.

The intertemporal utility function for a representative household is:
\[U_0=\mathbb{E}_0\sum_{t=0}^\infty\beta^t\left[\frac{C_t^{1-\sigma}}{1-\sigma}-\chi\frac{L_t^{1+\varphi}}{1+\varphi}\right]\]

s.t.:
\[\begin{gathered}C_t=\left[\gamma^{\frac{1}{\theta}}C_{Ht}^{\frac{\theta-1}{\theta}}+(1-\gamma)^{\frac{1}{\theta}}C_{Ft}^{\frac{\theta-1}{\theta}}\right]^{\frac{\theta}{\theta-1}}\\P_tC_t\leq W_tL_t\\\beta\in(0,1)\\\sigma>0\\\varphi>0\\\chi\geq0\end{gathered}\]

Where $C_t$ is the composite consumption of household, and $P_t$ denotes the CPI. Then:
\[\mathcal{L}=\mathbb{E}_0\sum_{t=0}^\infty\beta^t\left[\frac{C_t^{1-\sigma}}{1-\sigma}-\chi\frac{L_t^{1+\varphi}}{1+\varphi}-\lambda_t(P_tC_t-W_tL_t)\right]\]

FOC:
\[\begin{gathered}
\frac{\partial\mathcal{L}}{\partial C_t}=\beta^t\left[C_t^{-\sigma}-\lambda_tP_t\right]=0\\\frac{\partial\mathcal{L}}{\partial L_t}=\beta^t\left[-\chi L_t^\varphi+\lambda_tW_t\right]=0
\end{gathered}\]

Hence, we get:
\[\begin{gathered}\lambda_t=\frac{C_t^{-\sigma}}{P_t}\\\chi L_t^\varphi=\lambda_tW_t\\\chi L_t^\varphi=\frac{C_t^{-\sigma}}{P_t}W_t\end{gathered}\]

Taking the logarithm of the above equation and subtracting the steady state, we obtain:
\[\begin{gathered}
\ln\chi+\varphi\ln L_t=-\sigma\ln C_t-\ln P_t+\ln W_t\\\varphi(L_t-L_{ss})=-\sigma(C_t-C_{ss})-(P_t-P_{ss})+(W_t-W_{ss})\\\varphi l_t=-\sigma c_t-p_t+w_t
\end{gathered}\]

Under perfect competition, the first-order condition for profit maximization in the production sector is that the marginal product of labor equals the actual wage rate. Therefore:
\[\frac{\partial Y_t}{\partial L_t}=\alpha_kv_i\tau_{it}^{-\kappa}e^{\lambda_0+\frac{1}{2}\sigma_\lambda^2+\varepsilon_t}\lambda_t^k=\frac{W_t}{P_t}\]

Taking the logarithm of the above equation and subtracting the steady state:
\[w_t-p_t=-\kappa\tau_{st}+\lambda_t^k+\varepsilon_t\]

Combine the two equations:
\[\begin{cases}\varphi l_t=-\sigma c_t-p_t+w_t\\w_t-p_t=-\kappa\tau_{st}+\lambda_t^k+\varepsilon_t\end{cases}\]

We can solve:
\[l_t^*=\frac{-\sigma c_t-\kappa\tau_{st}+\lambda_t^k+\varepsilon_t}{\varphi}\]

Substitute $x_t$:
\[\begin{gathered}
x_t=\frac{-\sigma c_t-\kappa\tau_{st}+\lambda_t^k+\varepsilon_t}{\varphi}+\lambda_t^k+\varepsilon_t-\kappa\tau_{st}-\phi_x\phi_p\tau_{st}\\x_t=-\left(\frac{\sigma}{\varphi}\right)c_t-\left(\frac{\kappa+\varphi\kappa+\varphi\phi_x\phi_p}{\varphi}\right)\tau_{st}+\left(\frac{1+\varphi}{\varphi}\right)\lambda_t^k+\left(\frac{1+\varphi}{\varphi}\right)\varepsilon_t
\end{gathered}\]

And:
\[\frac{\partial x_t}{\partial c_t}=-\left(\frac{\sigma}{\varphi}\right)<0\]

Therefore, we observe that if both consumption and exports change after a stochastic shock and ultimately converge to a steady state, then when consumption deviates positively from the current state, exports will deviate negatively from the current state. For example, taking the labor factor shock discussed in this paper, under a policy of trade cooperation, if fluctuations in the labor factor positively impact imported consumption, then after the shock and before convergence to the steady state, this shock will strengthen its enhancement of imported consumption while weakening its enhancement of exports. Now, we can conclude these and propose Hypothesis 5:

\begin{Hypothesis}
Under a certain trade policy, if the labor factor shock is positive, it will strengthen the impact on imported consumption and weaken the impact on exports. If the labor factor shock is negative, it will weaken the impact on imported consumption and strengthen the impact on exports.
\end{Hypothesis}

\section{Empirical Evidence}\label{Section 9}
\subsection{Reality}
Here, we investigate national and provincial international trade data of China from 2001 to 2024. Figure 40 presents the trends in Chinese export value ($10^6$ million dollars), import value ($10^6$ million dollars), and the absolute value of the trade balance (difference). We observe a general pattern in the trends of these three variables: Short-term fluctuations accompanied by long-term growth. This also implies that while both export and import trade have prospered, they simultaneously face greater pressure to maintain a balanced international trade.

Figure 41 illustrates the international trade situation across 30 Chinese provinces during four distinct periods\footnote{These four periods correspond to three global economic shocks: The Financial Crisis (2008), COVID-19 (2019), and Artificial Intelligence (2023).}. The horizontal axis displays the logarithm of the absolute value of the trade balance, representing the degree of international trade equality. The vertical axis shows the logarithm of either export or import value, with blue samples indicating export trade and gray samples representing import trade. We observe that provincial heterogeneity aligns with the national pattern. Specifically, during each period, both export and import trade exhibit a clear positive correlation with the degree of international trade balance. This indicates that simultaneous improvements in exports and imports exacerbate trade imbalances. Higher export and import values tend to correspond with greater levels of international trade inequality.

\begin{figure}[htbp]
\centering
\includegraphics[width=12cm]{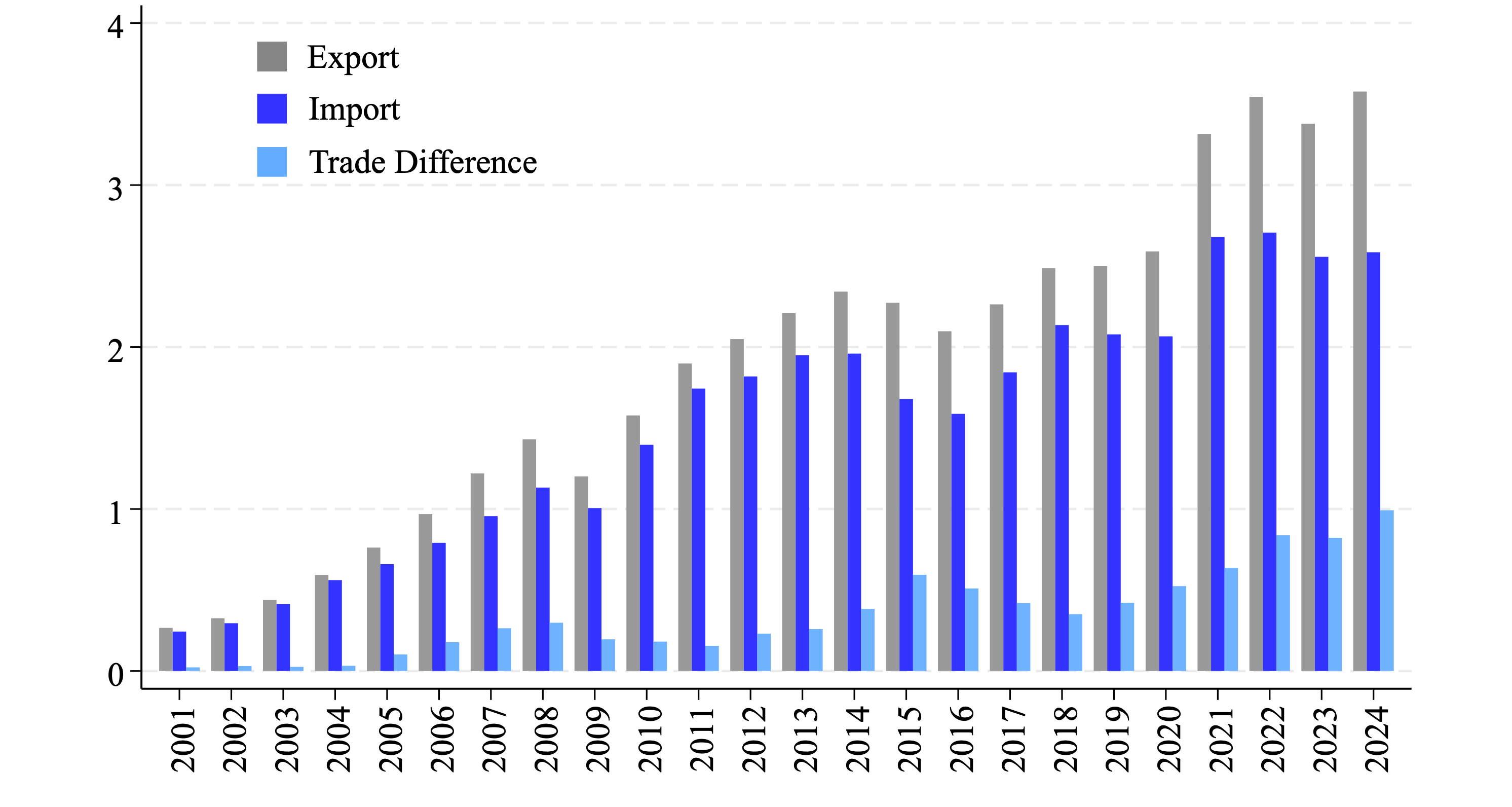}
\caption{\label{fig:F40}The International Trade of China ($10^6$ million dollars)}
\end{figure}

\begin{figure}[ht!]
    \centering
    \begin{minipage}[b]{0.9\textwidth}
        \centering
        \includegraphics[width=\textwidth]{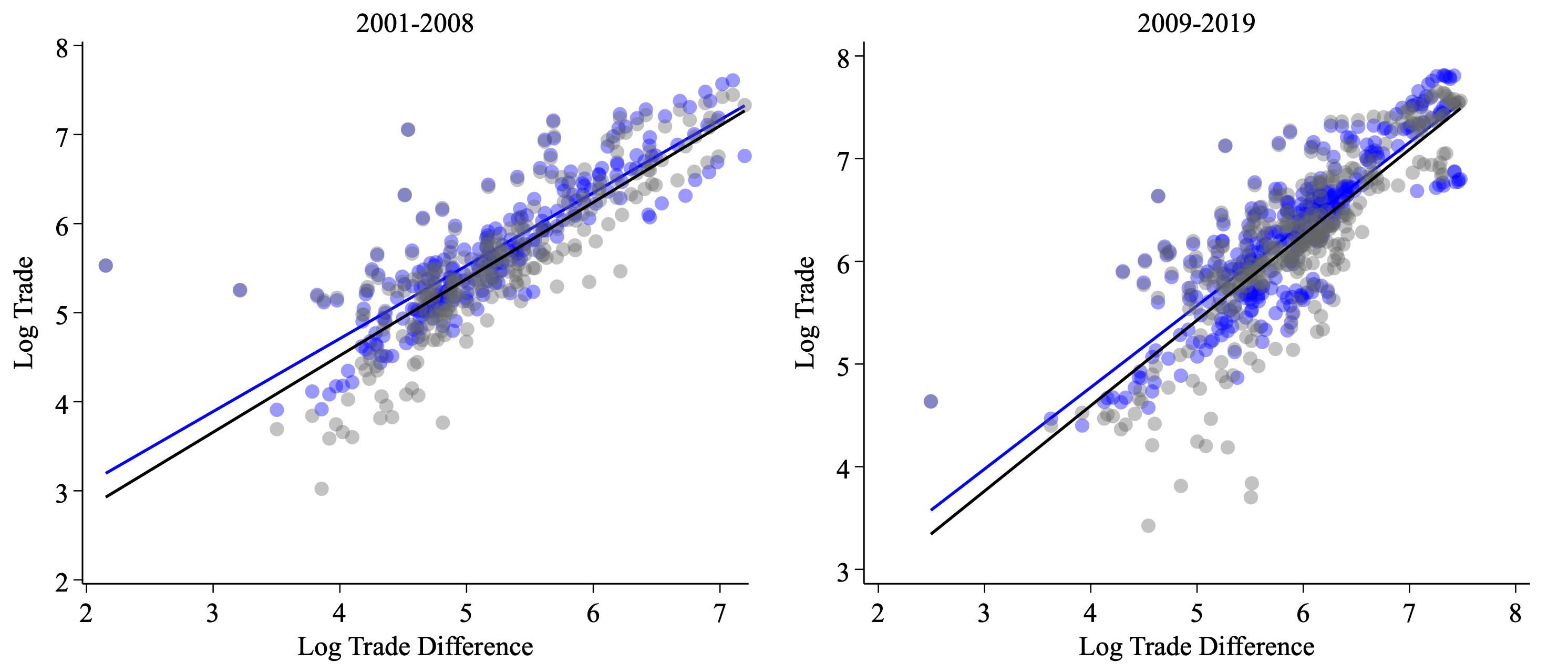}
    \end{minipage}
    \hfill
    \begin{minipage}[b]{0.9\textwidth}
        \centering
        \includegraphics[width=\textwidth]{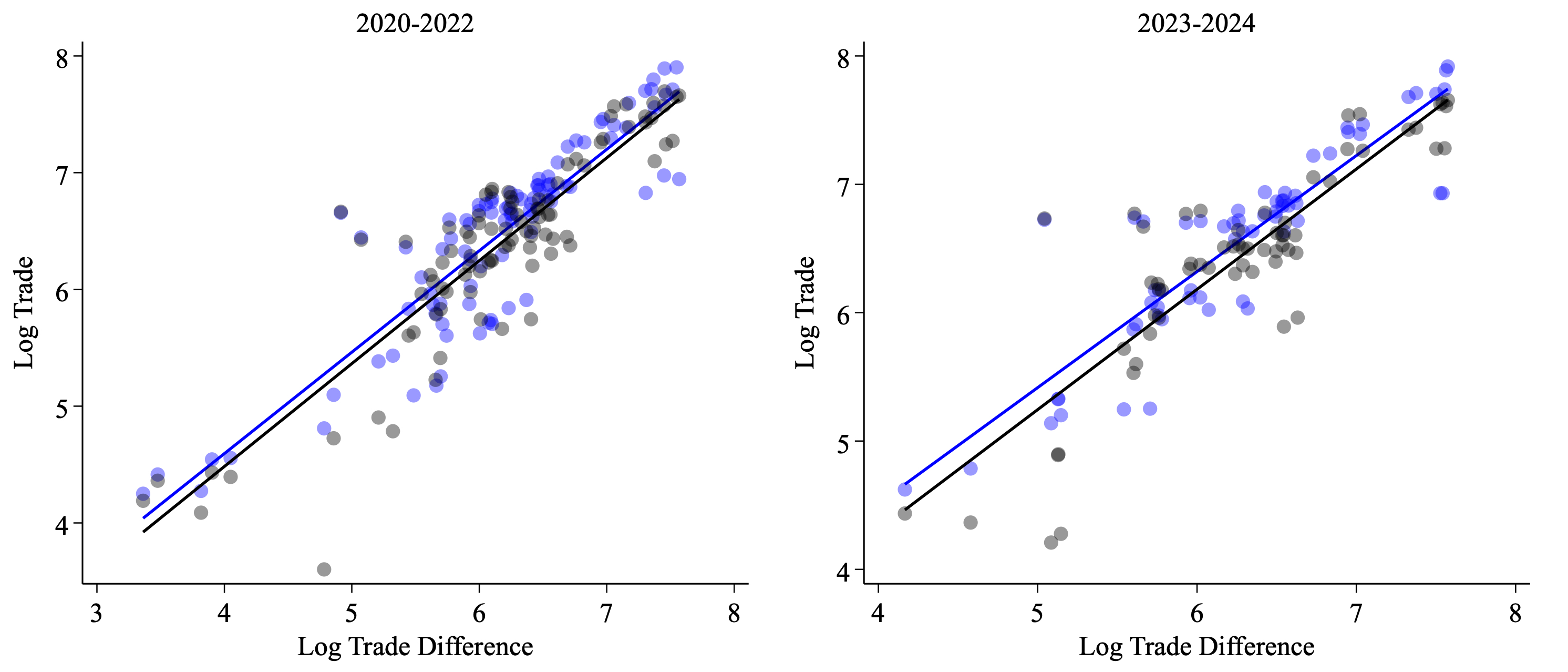}
    \end{minipage}
\caption{The International Trade of Chinese Provinces}
    \label{fig:F41}
\end{figure}

\subsection{Research Framework}
Our empirical work primarily aims to demonstrate Hypothesis 4 and Hypothesis 5, analyzing the impact of fluctuations in Chinese labor markets on international trade. We selected macroeconomic data from 30 Chinese provinces during $2008-2023$ as our research sample\footnote{This data is sourced from the Statistical Yearbook of China, Provincial Statistical Yearbooks, The National Bureau of Statistics, Provincial Government Websites, and Provincial Statistics Bureaus.}. The dependent variables are the logarithms of provincial exports ($10^4$ dollars) and imports ($10^4$ dollars). The independent variable is provincial manufacturing urban labor market fluctuations $L_t$, which is composed of the number of employed persons in provincial manufacturing urban areas ($10^4$ persons) and the average wage of employed persons in provincial manufacturing urban areas (yuan RMB). Specifically, our econometric equation is as follows:
\[\begin{gathered}
L_t=l_ts_{jt}p_t\\Trade_{it,i\in\{Export,Import\}}=\mu_0+\mu_{it,i\in\{\mu^-,\mu^+\}}L_t+\sum_iProvince_i+\sum_tYear_i+\varepsilon_{it}
\end{gathered}\]

Where: $l_t$ represents the aggregate employment in society, $s_{jt}$ denotes labor factor allocation shocks in the labor market, and thus $l_t s_{jt}$ represents the fluctuation of employment within industry $j$ following a labor factor allocation shock. In this paper, industry $j$ refers to manufacturing. $p_t$ represents labor factor value shocks in the labor market, i.e., fluctuations in average wages. 
\begin{figure}[htbp]
\centering
\includegraphics[width=13cm]{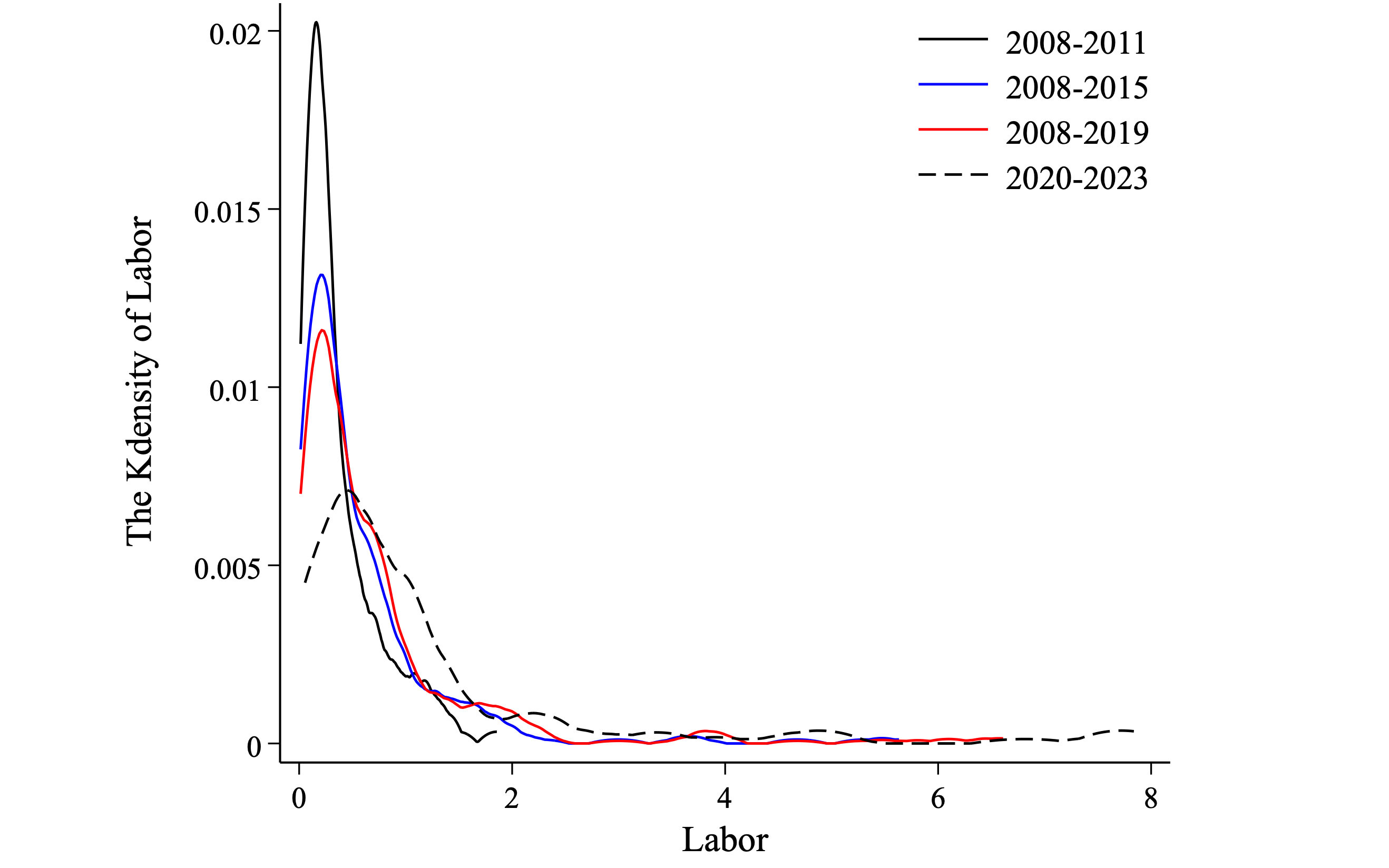}
\caption{\label{fig:F42}The Probability Density Distribution of Provincial Labor Market}
\end{figure}

Figure 42 presents the probability density distribution function ($10^{-5}$) of provincial manufacturing urban labor market fluctuations $L_t$ ($10^7$) over time. We observe that this dynamic pattern aligns with Figures 32 and 34, This indicates that from 2008 to 2023, the manufacturing labor markets across 30 Chinese provinces experienced positive shocks, where the impact on labor factor value was stronger than that on labor factor allocation. Specifically, taking 2020 as the shock year, we observe a significant increase in both the mean and variance of labor market fluctuations $L_t$ after 2020.

Therefore, we will employ a two-way fixed effects model to measure the impact of labor market fluctuations on international trade. Based on Figures 28, 29, and 42, we define the labor market shock occurring in 2020. Subsequently, we conduct separate regression models for international trade conditions before and after 2020. Finally, we test Hypothesis 4 and Hypothesis 5 by comparing the magnitude of $\mu_{it,i\in\{\mu^-,\mu^+\}},\mu^+>\mu^-$.

\subsection{Research Result}
Figure 43, Table 10 and Table 11 present the experimental results. In Figure 43, the blue samples represent international trade data from 2008 to 2019, while the red samples represent international trade data from 2020 to 2023. We observe that both provincial export trade and provincial import trade exhibit a strong positive correlation with fluctuations in the provincial manufacturing labor market of China. Now, we calculate the regression coefficients to test the two Hypotheses.
\begin{figure}[htbp]
\centering
\includegraphics[width=14cm]{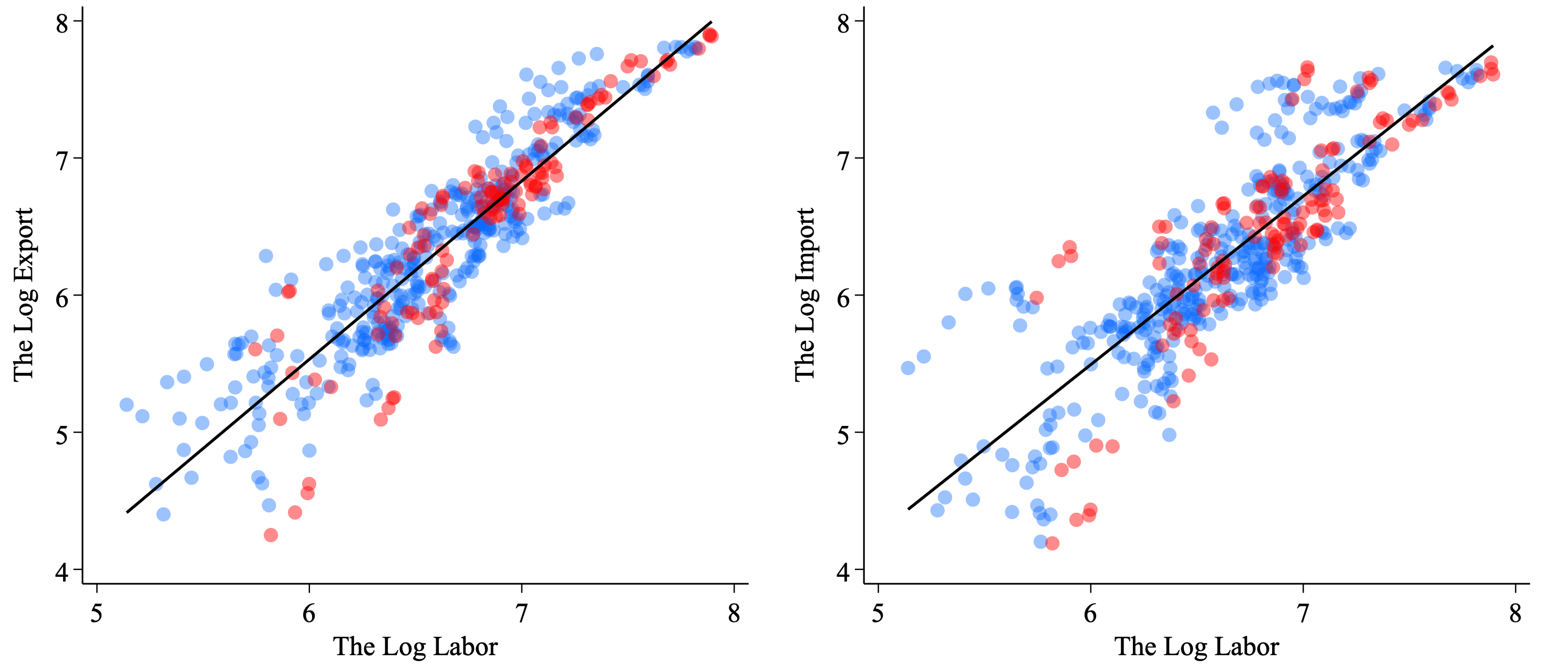}
\caption{\label{fig:F43}The International Trade of Chinese Provinces with Labor Market Fluctuations}
\end{figure}

Based on the aggregate sample from Tables 10 and Table 11 covering the period 2008–2023, we find that $L_t$ has a positive effect on both exports and imports, i.e., $\mu_{it,i\in\{\mu^-,\mu^+\}}>0$. Therefore, we can confirm that the international trade of China overall operated under a policy of trade cooperation between 2008 and 2023. This environment enabled fluctuations in the manufacturing labor factor market to contribute beneficially to improvements in both exports and imports.

According to the dynamic transition of $L_t$ in Figure 42, we know that $L_t$ is experiencing a positive shock originating from the labor factor market. In Table 11, since $\mu^{+}_{t\geq 2020}=0.624>\mu^{-}_{t\leq 2019}=0.322$, we can confirm that under trade cooperation policy, positive shock can strengthen import growth and increase import consumption. Hypothesis 4 is proven.

Furthermore, since $\mu^{+}_{t\geq 2020}=0.624>\mu^{-}_{t\leq 2019} =0.322$ in Table 11, and $\mu^{+}_{t\geq 2020}=0.485<\mu^{-}_{t\leq 2019}=0.668$ in Table 10, we can confirm that within trade cooperation policy, positive shock strengthens import growth while weakening export growth, which proves Hypothesis 5.

\setcounter{table}{9}
\begin{table}[ht!]
\centering
\caption{The Regression of Export}
\label{10:export}
\small                
\setlength{\tabcolsep}{8pt} 
\renewcommand{\arraystretch}{1.2} 
\begin{tabular}{lcccc}
\toprule
 & \multicolumn{4}{c}{Export} \\
\cmidrule(lr){2-5}
 & 2008--2023 & 2008--2023 & 2008--2019 & 2020--2023 \\
\midrule
\multirow{1}{*}{$L_t$} 
 & $1.301^{***}$ & $0.662^{***}$ & $0.668^{***}$ & $0.485^{*}$ \\
 & (0.028)       & (0.100)        & (0.115)        & (0.210)        \\
\multirow{1}{*}{Constant} 
 & $-2.271^{***}$ & $1.985^{**}$   & $1.951^{**}$   & $3.142^{*}$    \\
 & (0.184)        & (0.627)        & (0.721)        & (1.422)        \\
Province & $\times$   & $\checkmark$ & $\checkmark$ & $\checkmark$ \\
Year     & $\times$   & $\checkmark$ & $\checkmark$ & $\checkmark$ \\
$R^2$      & 0.822      & 0.587         & 0.591        & 0.605        \\
Number   & 480        & 480           & 360          & 120          \\
\bottomrule
\end{tabular}
\end{table}

\setcounter{table}{10}
\begin{table}[ht!]
\centering
\caption{The Regression of Import}
\label{11:import}
\small                
\setlength{\tabcolsep}{8pt}  
\renewcommand{\arraystretch}{1.2}  
\begin{tabular}{lcccc}
\toprule
 & \multicolumn{4}{c}{Import} \\
\cmidrule(lr){2-5}
 & 2008--2023 & 2008--2023 & 2008--2019 & 2020--2023 \\
\midrule
\multirow{1}{*}{$L_t$}
 & $1.228^{***}$ & $0.269^{**}$ & $0.322^{**}$ & $0.624^{***}$ \\
 & (0.036)       & (0.084)      & (0.100)      & (0.171)      \\
\multirow{1}{*}{Constant}
 & $-1.875^{***}$ & $4.324^{***}$ & $3.991^{***}$ & $2.152$      \\
 & (0.238)        & (0.523)       & (0.627)       & (1.160)      \\
Province & $\times$   & $\checkmark$ & $\checkmark$ & $\checkmark$ \\
Year     & $\times$   & $\checkmark$ & $\checkmark$ & $\checkmark$ \\
$R^2$      & 0.712      & 0.632        & 0.585        & 0.586        \\
Number   & 480        & 480          & 360          & 120          \\
\bottomrule
\end{tabular}
\end{table}

\section{Conclusion}
In this paper, we design two chapters to discuss trade dynamics with heterogeneous fluctuations, attempting to contribute new insights to macroeconomic theories concerning international trade.

The first chapter (Section 2–Section 5) analyzes export trade dynamics and monetary (exchange rate) policy. Using provincial economic data from China between 2008 and 2021, we demonstrate the effectiveness of monetary policy in regulating exchange rates and its role in enhancing export competitiveness. Simultaneously, we validate the substitution effect of innovation mechanisms for exchange rate mechanisms in boosting export competitiveness. Based on this, we construct a heterogeneous export trade structure model to explore optimal monetary policy.

The second chapter (Section 6–Section 9) analyzes general trade dynamics and heterogeneous fluctuations in small open economies. We employ stochastic processes to model labor allocation shocks, labor value shocks, and factor utilization capability shocks to construct production functions. Subsequently, we define economies under trade cooperation policy and trade sanction policy, and study how heterogeneous fluctuations in production affect international trade in these two types of economies. Using provincial economic data from China during 2008–2023, we demonstrate the impact of economic shocks from the labor market on both export and import trade.

Specifically, we find that:

(\textbf{A}) The depreciation of USD-CNY exchange rate\footnote{This chapter selects the USD-CNY real exchange rate as the core independent variable based on the following four considerations: (1) Using a key currency as a representative variable helps avoid the subjectivity inherent in setting weights for multilateral exchange rate indices, clearly illustrating the cyclical relationship where micro-level variables (exchange rates) influence macroeconomic outcomes (export trade). (2) The U.S. dollar serves as the primary currency for pricing and settlement in international trade, lending representativeness to the microfoundations of the exchange rate mechanism affecting export competitiveness examined in this study. (3) Fluctuations in the USD-CNY bilateral exchange rate serve as a core observational window for analyzing monetary policy transmission. Monetary policy adjustments to exchange rates are most concentrated and clearly reflected in the USD-CNY exchange rate. (4) As a key driver of the global financial cycle, the Federal Reserve's monetary policy has undergone numerous unconventional adjustments and decisions in recent years. This uncertainty constitutes an exogenous shock to both the RMB exchange rate and the autonomy of Chinese monetary policy.} within the 6-7 range could significantly enhance provincial export competitiveness. Moreover, export competitiveness is markedly influenced by CNY exchange rate volatility. In the short term, stabilizing CNY exchange rate fluctuations within a range of less than 0.2 RMB can steadily promote the improvement of export competitiveness. 

(\textbf{B}) The “8.11 Exchange Rate Reform” policy effectively regulated the flexibility of CNY exchange rate, expanded its market space, and reversed the single linear trend before the policy shock. After the implementation of monetary policy, CNY exchange rate maintains long term stability within a narrow fluctuation range, which indirectly enhances provincial export competitiveness.

(\textbf{C}) Innovative mechanisms for enhancing provincial export competitiveness can significantly replace exchange rate mechanisms. Innovative development can reduce export competitiveness's dependence on exchange rates. From a regional heterogeneity perspective, exchange rates cease to be the primary factor influencing export competitiveness in the areas with high innovation capabilities. These regions significantly improve their export competitiveness through the manufacturing of high value added products with robust innovation capabilities and new economic patterns. Conversely, for regions with low innovation capabilities, exchange rates remain a key factor affecting export competitiveness due to their reliance on low value added products and traditional economic patterns.

(\textbf{D}) We obtain the analytical probability density distribution of the exchange rate through using KFE and find that when exchange rates exhibit a linear appreciation trend, this state gradually stabilizes, making transition to other states increasingly difficult. Consequently, relying solely on market forces to control exchange rates becomes challenging under such conditions, which is another dimension demonstrating the essentiality of monetary policy. We further construct a model of heterogeneous export trade structures to explore optimal policy according to exchange rate fluctuations. We find that optimal monetary policy should leverage exchange rate mechanisms to promote dynamic transitions in trade structures rather than unilaterally strengthen existing comparative advantages. Furthermore, we observe that in heterogeneous trade structures, both the mean and variance of wealth distribution among exporting firms gradually increase over time, indicating that as exporting firms' wealth rises, they also face higher wealth inequality. However, wealth inequality grows slowest in a balanced trade structure (i.e., neutral-type trade structure). Therefore, policy should encourage balanced trade development by comprehensively enhancing the comparative advantage of all types of products in the domestic export structure. This approach mitigates inequality in wealth accumulation among firms in the export structure and extends the time interval for wealth growth.

(\textbf{E}) Positive and negative shocks exist in the labor factor market. If the labor factor allocation shock is weaker than the labor factor value shock, this constitutes a positive shock, where average wage growth exceeds unemployment. If the labor factor allocation shock is stronger than the labor factor value shock, this constitutes a negative shock, where unemployment exceeds average wage growth. Our application of the theoretical model reveals that, whether the international trade is under sanction policy or cooperation policy, a positive shock in the labor factor market can strengthen its impact on import trade while weakening its impact on export trade, and vice versa. Empirical analysis examines Chinese data and validates the theoretical hypotheses. We find that Chinese international trade overall operated under a trade cooperation policy from 2008 to 2023. Furthermore, taking 2020 as the shock year, a positive shock occurred in the labor market, which strengthened improvements in provincial import trade while weakening improvements in provincial export trade.

\singlespacing
\setlength{\bibsep}{2pt}
\bibliographystyle{plainnat}
\bibliography{main}

\begin{thebibliography}{67}
\providecommand{\natexlab}[1]{#1}
\providecommand{\url}[1]{\texttt{#1}}
\expandafter\ifx\csname urlstyle\endcsname\relax
  \providecommand{\doi}[1]{doi: #1}\else
  \providecommand{\doi}{doi: \begingroup \urlstyle{rm}\Url}\fi

\bibitem[Acharya et~al.(2023)Acharya, Challe, and Dogra]{Acharya}
Sushant Acharya, Edouard Challe, and Keshav Dogra.
\newblock Optimal monetary policy according to hank.
\newblock \emph{American Economic Review}, 113\penalty0 (7), 2023.

\bibitem[Achdou and Lauri{\`e}re(2020)]{achdou2020mean}
Yves Achdou and Mathieu Lauri{\`e}re.
\newblock Mean field games and applications: Numerical aspects.
\newblock \emph{arXiv preprint arXiv:2003.04444}, 2020.

\bibitem[Achdou et~al.(2014)Achdou, Buera, Lasry, Lions, and Moll]{achdou2014partial}
Yves Achdou, Francisco~J Buera, Jean-Michel Lasry, Pierre-Louis Lions, and Benjamin Moll.
\newblock Partial differential equation models in macroeconomics.
\newblock \emph{Philosophical Transactions of the Royal Society A: Mathematical, Physical and Engineering Sciences}, 372\penalty0 (2028), 2014.

\bibitem[Achdou et~al.(2022)Achdou, Han, Lasry, Lions, and Moll]{achdou2022income}
Yves Achdou, Jiequn Han, Jean-Michel Lasry, Pierre-Louis Lions, and Benjamin Moll.
\newblock Income and wealth distribution in macroeconomics: A continuous-time approach.
\newblock \emph{The Review of Economic Studies}, 89\penalty0 (1):\penalty0 45--86, 2022.

\bibitem[Ad{\~a}o et~al.(2026)Ad{\~a}o, Becko, Costinot, and Donaldson]{adao2026world}
Rodrigo Ad{\~a}o, John~Sturm Becko, Arnaud Costinot, and Dave Donaldson.
\newblock A world trading system for whom? evidence from global tariffs.
\newblock Working Paper 34658, NBER, 2026.

\bibitem[Adrian and Shin(2010)]{adrian2010liquidity}
Tobias Adrian and Hyun~Song Shin.
\newblock Liquidity and leverage.
\newblock \emph{Journal of Financial Intermediation}, 19\penalty0 (3):\penalty0 418--437, 2010.

\bibitem[Afrouzi et~al.(2026)Afrouzi, Blanco, Drenik, and Hurst]{QJE2026}
Hassan Afrouzi, Andres Blanco, Andres Drenik, and Erik Hurst.
\newblock A theory of how workers keep up with inflation.
\newblock \emph{The Quarterly Journal of Economics}, 2026.

\bibitem[Ahn et~al.(2018)Ahn, Kaplan, Moll, Winberry, and Wolf]{ahn2018inequality}
SeHyoun Ahn, Greg Kaplan, Benjamin Moll, Thomas Winberry, and Christian Wolf.
\newblock When inequality matters for macro and macro matters for inequality.
\newblock \emph{NBER Macroeconomics Annual}, 32\penalty0 (1):\penalty0 1--75, 2018.

\bibitem[Aiyagari(1994)]{aiyagari1994uninsured}
S~Rao Aiyagari.
\newblock Uninsured idiosyncratic risk and aggregate saving.
\newblock \emph{The Quarterly Journal of Economics}, 109\penalty0 (3):\penalty0 659--684, 1994.

\bibitem[Alfaro et~al.(2018)Alfaro, Cunat, Fadinger, and Liu]{alfaro2018real}
Laura Alfaro, Alejandro Cunat, Harald Fadinger, and Yanping Liu.
\newblock The real exchange rate, innovation and productivity: heterogeneity, asymmetries and hysteresis.
\newblock Working Paper 24633, NBER, 2018.

\bibitem[Alvarez and Lippi(2022)]{alvarez2022analytic}
Fernando Alvarez and Francesco Lippi.
\newblock The analytic theory of a monetary shock.
\newblock \emph{Econometrica}, 90\penalty0 (4):\penalty0 1655--1680, 2022.

\bibitem[{\'A}lvarez and L{\'o}pez(2009)]{alvarez2009skill}
Roberto {\'A}lvarez and Ricardo~A L{\'o}pez.
\newblock Skill upgrading and the real exchange rate.
\newblock \emph{World Economy}, 32\penalty0 (8):\penalty0 1165--1179, 2009.

\bibitem[Amiti et~al.(2014)Amiti, Itskhoki, and Konings]{amiti2014importers}
Mary Amiti, Oleg Itskhoki, and Jozef Konings.
\newblock Importers, exporters, and exchange rate disconnect.
\newblock \emph{American Economic Review}, 104\penalty0 (7):\penalty0 1942--1978, 2014.

\bibitem[Atkeson and Burstein(2008)]{atkeson2008trade}
Andrew Atkeson and Ariel Burstein.
\newblock Trade costs, pricing to market, and international relative prices.
\newblock \emph{American Economic Review}, 98\penalty0 (5):\penalty0 1998--2031, 2008.

\bibitem[Auclert et~al.(2025)Auclert, Rognlie, and Straub]{auclert2025macroeconomics}
Adrien Auclert, Matthew Rognlie, and Ludwig Straub.
\newblock The macroeconomics of tariff shocks.
\newblock Working Paper 33726, NBER, 2025.

\bibitem[Auer et~al.(2021)Auer, Burstein, and Lein]{Auer}
Raphael Auer, Ariel Burstein, and Sarah~M. Lein.
\newblock Exchange rates and prices: Evidence from the 2015 swiss franc appreciation.
\newblock \emph{American Economic Review}, 111\penalty0 (2):\penalty0 652–86, 2021.

\bibitem[Bilal(2023)]{bilal2023solving}
Adrien Bilal.
\newblock Solving heterogeneous agent models with the master equation.
\newblock Working Paper 31103, NBER, 2023.

\bibitem[Blaum(2017)]{blaum2017importing}
Joaquin Blaum.
\newblock Importing, exporting and aggregate productivity in large devaluations.
\newblock In \emph{2017 Meeting Papers}. Society for Economic Dynamics, 2017.

\bibitem[Brunnermeier and Sannikov(2014)]{brunnermeier2014macroeconomic}
Markus~K Brunnermeier and Yuliy Sannikov.
\newblock A macroeconomic model with a financial sector.
\newblock \emph{American Economic Review}, 104\penalty0 (2):\penalty0 379--421, 2014.

\bibitem[Buera and Moll(2015)]{buera2015aggregate}
Francisco~J Buera and Benjamin Moll.
\newblock Aggregate implications of a credit crunch: The importance of heterogeneity.
\newblock \emph{American Economic Journal: Macroeconomics}, 7\penalty0 (3):\penalty0 1--42, 2015.

\bibitem[Cadenillas and Zapatero(1999)]{cadenillas1999optimal}
Abel Cadenillas and Fernando Zapatero.
\newblock Optimal central bank intervention in the foreign exchange market.
\newblock \emph{Journal of Economic Theory}, 87\penalty0 (1):\penalty0 218--242, 1999.

\bibitem[Cadenillas and Zapatero(2000)]{cadenillas2000classical}
Abel Cadenillas and Fernando Zapatero.
\newblock Classical and impulse stochastic control of the exchange rate using interest rates and reserves.
\newblock \emph{Mathematical Finance}, 10\penalty0 (2):\penalty0 141--156, 2000.

\bibitem[Caliendo et~al.(2019)Caliendo, Dvorkin, and Parro]{caliendo2019trade}
Lorenzo Caliendo, Maximiliano Dvorkin, and Fernando Parro.
\newblock Trade and labor market dynamics: General equilibrium analysis of the china trade shock.
\newblock \emph{Econometrica}, 87\penalty0 (3):\penalty0 741--835, 2019.

\bibitem[Campbell and Martin(2025)]{campbell2025sustainability}
John~Y Campbell and Ian~WR Martin.
\newblock Sustainability in a risky world.
\newblock \emph{American Economic Review: Insights}, 7\penalty0 (2):\penalty0 196--212, 2025.

\bibitem[Cimoli et~al.(2013)Cimoli, Fleitas, and Porcile]{cimoli2013technological}
Mario Cimoli, Sebastian Fleitas, and Gabriel Porcile.
\newblock Technological intensity of the export structure and the real exchange rate.
\newblock \emph{Economics of Innovation and New Technology}, 22\penalty0 (4):\penalty0 353--372, 2013.

\bibitem[Ekholm et~al.(2012)Ekholm, Moxnes, and Ulltveit-Moe]{ekholm2012manufacturing}
Karolina Ekholm, Andreas Moxnes, and Karen~Helene Ulltveit-Moe.
\newblock Manufacturing restructuring and the role of real exchange rate shocks.
\newblock \emph{Journal of International Economics}, 86\penalty0 (1):\penalty0 101--117, 2012.

\bibitem[Engel and West(2005)]{engel2005exchange}
Charles Engel and Kenneth~D West.
\newblock Exchange rates and fundamentals.
\newblock \emph{Journal of Political Economy}, 113\penalty0 (3):\penalty0 485--517, 2005.

\bibitem[Feenstra and Romalis(2014)]{feenstra2014international}
Robert~C Feenstra and John Romalis.
\newblock International prices and endogenous quality.
\newblock \emph{The Quarterly Journal of Economics}, 129\penalty0 (2):\penalty0 477--527, 2014.

\bibitem[Fern{\'a}ndez-Villaverde et~al.(2023)Fern{\'a}ndez-Villaverde, Hurtado, and Nuno]{fernandez2023financial}
Jes{\'u}s Fern{\'a}ndez-Villaverde, Samuel Hurtado, and Galo Nuno.
\newblock Financial frictions and the wealth distribution.
\newblock \emph{Econometrica}, 91\penalty0 (3):\penalty0 869--901, 2023.

\bibitem[Ferrari and Vargiolu(2020)]{ferrari2020singular}
Giorgio Ferrari and Tiziano Vargiolu.
\newblock On the singular control of exchange rates.
\newblock \emph{Annals of Operations Research}, 292\penalty0 (2):\penalty0 795--832, 2020.

\bibitem[Foster et~al.(2008)Foster, Haltiwanger, and Syverson]{foster2008reallocation}
Lucia Foster, John Haltiwanger, and Chad Syverson.
\newblock Reallocation, firm turnover, and efficiency: Selection on productivity or profitability?
\newblock \emph{American Economic Review}, 98\penalty0 (1):\penalty0 394--425, 2008.

\bibitem[Fukui et~al.(2025)Fukui, Nakamura, and Steinsson]{fukui2025macroeconomic}
Masao Fukui, Emi Nakamura, and J{\'o}n Steinsson.
\newblock The macroeconomic consequences of exchange rate depreciations.
\newblock \emph{The Quarterly Journal of Economics}, 140\penalty0 (4):\penalty0 3015--3065, 2025.

\bibitem[Fung(2008)]{fung2008large}
Loretta Fung.
\newblock Large real exchange rate movements, firm dynamics, and productivity growth.
\newblock \emph{Canadian Journal of Economics}, 41\penalty0 (2):\penalty0 391--424, 2008.

\bibitem[Gabaix et~al.(2016)Gabaix, Lasry, Lions, and Moll]{gabaix2016dynamics}
Xavier Gabaix, Jean-Michel Lasry, Pierre-Louis Lions, and Benjamin Moll.
\newblock The dynamics of inequality.
\newblock \emph{Econometrica}, 84\penalty0 (6):\penalty0 2071--2111, 2016.

\bibitem[Gomez(2025)]{gomez2025wealth}
Matthieu Gomez.
\newblock Wealth inequality and asset prices.
\newblock \emph{The Review of Economic Studies}, 92\penalty0 (6):\penalty0 3924--3967, 2025.

\bibitem[Gwee and Zervos(2025)]{gwee2025risk}
Justin Gwee and Mihail Zervos.
\newblock A risk-sensitive ergodic singular stochastic control problem.
\newblock \emph{arXiv preprint arXiv:2509.09835}, 2025.

\bibitem[Hu et~al.(2021)Hu, Parsley, and Tan]{hu2021exchange}
Cui Hu, David Parsley, and Yong Tan.
\newblock Exchange rate induced export quality upgrading: A firm-level perspective.
\newblock \emph{Economic Modelling}, 98:\penalty0 336--348, 2021.

\bibitem[Hu(2025)]{hu2025big}
Yongheng Hu.
\newblock How big data dilutes cognitive resources, interferes with rational decision-making and affects wealth distribution?
\newblock \emph{arXiv preprint arXiv:2508.20435}, 2025.

\bibitem[Huggett(1993)]{huggett1993risk}
Mark Huggett.
\newblock The risk-free rate in heterogeneous-agent incomplete-insurance economies.
\newblock \emph{Journal of economic Dynamics and Control}, 17\penalty0 (5-6):\penalty0 953--969, 1993.

\bibitem[Itskhoki and Mukhin(2021)]{itskhoki2021exchange}
Oleg Itskhoki and Dmitry Mukhin.
\newblock Exchange rate disconnect in general equilibrium.
\newblock \emph{Journal of Political Economy}, 129\penalty0 (8):\penalty0 2183--2232, 2021.

\bibitem[Jeanblanc~Picqu{\'e}(1993)]{jeanblanc1993impulse}
Monique Jeanblanc~Picqu{\'e}.
\newblock Impulse control method and exchange rate.
\newblock \emph{Mathematical Finance}, 3\penalty0 (2):\penalty0 161--177, 1993.

\bibitem[Jeanneney and Hua(2011)]{jeanneney2011does}
Sylviane~Guillaumont Jeanneney and Ping Hua.
\newblock How does real exchange rate influence labour productivity in china?
\newblock \emph{China Economic Review}, 22\penalty0 (4):\penalty0 628--645, 2011.

\bibitem[Kalemli-Ozcan et~al.(2026)Kalemli-Ozcan, Soylu, and Yildirim]{kalemli2026global}
Sebnem Kalemli-Ozcan, Can Soylu, and Muhammed~A Yildirim.
\newblock Global trade, tariff uncertainty and the us dollar.
\newblock Working Paper 34728, NBER, 2026.

\bibitem[Kaplan et~al.(2018)Kaplan, Moll, and Violante]{kaplan2018monetary}
Greg Kaplan, Benjamin Moll, and Giovanni~L Violante.
\newblock Monetary policy according to hank.
\newblock \emph{American Economic Review}, 108\penalty0 (3):\penalty0 697--743, 2018.

\bibitem[Kim and Vogel(2021)]{AERI2021}
Ryan Kim and Jonathan Vogel.
\newblock Trade shocks and labor market adjustment.
\newblock \emph{American Economic Review: Insights}, 3\penalty0 (1):\penalty0 115–30, 2021.

\bibitem[Krugman(1979)]{krugman1979model}
Paul Krugman.
\newblock A model of balance-of-payments crises.
\newblock \emph{Journal of Money, Credit and Banking}, 11\penalty0 (3):\penalty0 311--325, 1979.

\bibitem[Krugman(1980)]{krugman1980scale}
Paul Krugman.
\newblock Scale economies, product differentiation, and the pattern of trade.
\newblock \emph{American Economic Review}, 70\penalty0 (5):\penalty0 950--959, 1980.

\bibitem[Krusell and Smith(1998)]{krusell1998income}
Per Krusell and Anthony~A Smith, Jr.
\newblock Income and wealth heterogeneity in the macroeconomy.
\newblock \emph{Journal of Political Economy}, 106\penalty0 (5):\penalty0 867--896, 1998.

\bibitem[Le~Grand and Ragot(2025)]{le2025optimal}
Fran{\c{c}}ois Le~Grand and Xavier Ragot.
\newblock Optimal fiscal policy with heterogeneous agents and capital: Should we increase or decrease public debt and capital taxes?
\newblock \emph{Journal of Political Economy}, 133\penalty0 (7), 2025.

\bibitem[Lind and Ramondo(2023{\natexlab{a}})]{LNAER2023}
Nelson Lind and Natalia Ramondo.
\newblock Trade with correlation.
\newblock \emph{American Economic Review}, 113\penalty0 (2):\penalty0 317–53, 2023{\natexlab{a}}.

\bibitem[Lind and Ramondo(2023{\natexlab{b}})]{LNAERinsight2023}
Nelson Lind and Natalia Ramondo.
\newblock Global innovation and knowledge diffusion.
\newblock \emph{American Economic Review: Insights}, 5\penalty0 (4):\penalty0 494–510, 2023{\natexlab{b}}.

\bibitem[Martin and Mejean(2014)]{martin2014low}
Julien Martin and Isabelle Mejean.
\newblock Low-wage country competition and the quality content of high-wage country exports.
\newblock \emph{Journal of International Economics}, 93\penalty0 (1):\penalty0 140--152, 2014.

\bibitem[Maxted et~al.(2025)Maxted, Laibson, and Moll]{maxted2025present}
Peter Maxted, David Laibson, and Benjamin Moll.
\newblock Present bias amplifies the household balance-sheet channels of macroeconomic policy.
\newblock \emph{The Quarterly Journal of Economics}, 140\penalty0 (1):\penalty0 691--743, 2025.

\bibitem[Meese and Rogoff(1983)]{meese1983empirical}
Richard~A Meese and Kenneth Rogoff.
\newblock Empirical exchange rate models of the seventies: Do they fit out of sample?
\newblock \emph{Journal of International Economics}, 14\penalty0 (1-2):\penalty0 3--24, 1983.

\bibitem[Mehrotra and Waugh(2025)]{mehrotra2025tariffs}
Neil Mehrotra and Michael~E Waugh.
\newblock Tariffs, trade wars, and the natural rate of interest.
\newblock Working Paper 34206, NBER, 2025.

\bibitem[Melitz(2003)]{melitz2003impact}
Marc~J Melitz.
\newblock The impact of trade on intra-industry reallocations and aggregate industry productivity.
\newblock \emph{Econometrica}, 71\penalty0 (6):\penalty0 1695--1725, 2003.

\bibitem[Melitz and Ottaviano(2008)]{melitz2008market}
Marc~J Melitz and Gianmarco~IP Ottaviano.
\newblock Market size, trade, and productivity.
\newblock \emph{The Review of Economic Studies}, 75\penalty0 (1):\penalty0 295--316, 2008.

\bibitem[Missio and Gabriel(2016)]{missio2016real}
Fabricio~J Missio and Luciano~F Gabriel.
\newblock Real exchange rate, technological catching up and spillovers in a balance-of-payments constrained growth model.
\newblock \emph{Economia}, 17\penalty0 (3):\penalty0 291--309, 2016.

\bibitem[Moll(2014)]{moll2014productivity}
Benjamin Moll.
\newblock Productivity losses from financial frictions: Can self-financing undo capital misallocation?
\newblock \emph{American Economic Review}, 104\penalty0 (10):\penalty0 3186--3221, 2014.

\bibitem[Moll and Ryzhik(2025)]{moll2025mean}
Benjamin Moll and Lenya Ryzhik.
\newblock Mean field games without rational expectations.
\newblock \emph{arXiv preprint arXiv:2506.11838}, 2025.

\bibitem[Mouradian(2013)]{mouradian2013real}
Florence Mouradian.
\newblock Real exchange rate and quality: product-level evidence from the eurozone.
\newblock In \emph{Afse Meeting}, pages 1--44. Citeseer, 2013.

\bibitem[Mundaca(1998)]{mundaca1998optimal}
Gabriela Mundaca.
\newblock Optimal stochastic intervention control with application to the exchange rate.
\newblock \emph{Journal of Mathematical Economics}, 29, 1998.

\bibitem[Obstfeld(1996)]{obstfeld1996models}
Maurice Obstfeld.
\newblock Models of currency crises with self-fulfilling features.
\newblock \emph{European Economic Review}, 40\penalty0 (3-5):\penalty0 1037--1047, 1996.

\bibitem[P{\'a}stor and Veronesi(2020)]{pastor2020political}
Lubo{\v{s}} P{\'a}stor and Pietro Veronesi.
\newblock Political cycles and stock returns.
\newblock \emph{Journal of Political Economy}, 128\penalty0 (11):\penalty0 4011--4045, 2020.

\bibitem[Tomlin(2014)]{tomlin2014exchange}
Ben Tomlin.
\newblock Exchange rate fluctuations, plant turnover and productivity.
\newblock \emph{International Journal of Industrial Organization}, 35:\penalty0 12--28, 2014.

\bibitem[Verhoogen(2008)]{verhoogen2008trade}
Eric~A Verhoogen.
\newblock Trade, quality upgrading, and wage inequality in the mexican manufacturing sector.
\newblock \emph{The Quarterly Journal of Economics}, 123\penalty0 (2):\penalty0 489--530, 2008.

\bibitem[WR~Martin and Papadimitriou(2022)]{wr2022sentiment}
Ian WR~Martin and Dimitris Papadimitriou.
\newblock Sentiment and speculation in a market with heterogeneous beliefs.
\newblock \emph{American Economic Review}, 112\penalty0 (8):\penalty0 2465--2517, 2022.

\end{thebibliography}

\begin{appendices}
\setstretch{1.2}
\section{Proof of Theorem}\label{Appendix A}

\begin{Theorem}
In the analysis of long panel data, for the benchmark regression model in standard econometrics, if the independent variable is a time-correlated stochastic process, the measured regression coefficient of the independent variable remains consistent with the real regression coefficient even if there isn't a time-fixed effect in the regression model.    
\end{Theorem}

\begin{proof}
Consider a long panel data where the number of individuals $N$ is fixed and the time dimension $T \to \infty$. The regression model of true data is:
\[
y_{it}=\beta x_{it}+u_i+\lambda_t+\epsilon_{it},\quad i=1,\ldots,N,t=1,\ldots,T
\]

Where: $y_{it}$ is the dependent variable. $x_{it}$ is the independent variable, following a stochastic process, here we set it as a unit root process. $u_i$ is the individual fixed effect. $\lambda_t$ is the time fixed effect. $\epsilon_{it}$ is the disturbance term.

Assume $x_{it}$ follows the stochastic process (i.e., unit root process):
\[
x_{it}=x_{i,t-1}+\eta_{it},\quad \eta_{it}\sim\text{i.i.d.}(0, \sigma_\eta^2)
\]

And the initial condition $x_{i0} = O_p(1)$. Then:
\[
\Delta x_{it}=\eta_{it},\quad \eta_{it}\sim\text{i.i.d.}(0,\sigma_\eta^2)
\]

We assume: Time fixed effects $\lambda_t$ are covariance stationary, $\mathbb{E}[\lambda_t]=\mu_\lambda$, $\text{Cov}(\lambda_t,\lambda_{t-k})=\gamma_\lambda(k)$, and $\sum_{k=-\infty}^{\infty}|\gamma_\lambda(k)|<\infty$. Disturbance terms $\epsilon_{it}\sim\text{i.i.d.}(0,\sigma_\epsilon^2)$, independent of $x_{it}$ and $\lambda_t$. Individual effects $u_i$ can be fixed or random effects. $\{\eta_{it}\}$, $\{\epsilon_{it}\}$, $\{\lambda_t\}$ are mutually independent. Cross-sectional dimension $N$ is fixed or $N \to \infty$ with $N/T \to 0$

We consider omitting time fixed effects and estimating:
\[
y_{it} = \beta x_{it} + u_i + \xi_{it},\quad \xi_{it}=\lambda_t + \epsilon_{it}
\]

Using the fixed effects:
\[
\hat{\beta} = \frac{\sum_{i=1}^N \sum_{t=1}^T (x_{it} - \bar{x}_i)(y_{it} - \bar{y}_i)}{\sum_{i=1}^N \sum_{t=1}^T (x_{it} - \bar{x}_i)^2}
\]

Where:
\[\bar{x}_i = \frac{1}{T} \sum_{t=1}^T x_{it},\bar{y}_i = \frac{1}{T} \sum_{t=1}^T y_{it}\]

Substituting the true model into the estimator:
\[
\hat{\beta} = \beta + \frac{\sum_{i=1}^N \sum_{t=1}^T (x_{it} - \bar{x}_i)(\lambda_t - \bar{\lambda})}{\sum_{i=1}^N \sum_{t=1}^T (x_{it} - \bar{x}_i)^2} + \frac{\sum_{i=1}^N \sum_{t=1}^T (x_{it} - \bar{x}_i)(\epsilon_{it} - \bar{\epsilon}_i)}{\sum_{i=1}^N \sum_{t=1}^T (x_{it} - \bar{x}_i)^2} = \beta + A + B
\]

Where:
\[
A=\frac{\sum_{i=1}^N \sum_{t=1}^T (x_{it} - \bar{x}_i)(\lambda_t - \bar{\lambda})}{\sum_{i=1}^N \sum_{t=1}^T (x_{it} - \bar{x}_i)^2},B=\frac{\sum_{i=1}^N \sum_{t=1}^T (x_{it} - \bar{x}_i)(\epsilon_{it} - \bar{\epsilon}_i)}{\sum_{i=1}^N \sum_{t=1}^T (x_{it} - \bar{x}_i)^2}
\]

By the functional central limit theorem:
\[
\frac{1}{\sqrt{T}} x_{i,\lfloor Tr \rfloor} \xrightarrow{d} \sigma_\eta W_i(r)
\]

Where $r\in[0,1]$. And the continuous mapping theorem gives:
\[
\frac{1}{T^2} \sum_{t=1}^T (x_{it} - \bar{x}_i)^2 = \sum_{t=1}^T \left( \frac{x_{it} - \bar{x}_i}{\sqrt{T}} \right)^2 \frac{1}{T} \xrightarrow{d} \sigma_\eta^2 \int_0^1 [W_i(r) - \bar{W}_i]^2 dr
\]

Hence, for the unit root process $x_{it} = x_{i,t-1} + \eta_{it}$ with $\eta_{it} \sim \text{i.i.d.}(0, \sigma_\eta^2)$:

\[
\frac{1}{T^2} \sum_{t=1}^T (x_{it} - \bar{x}_i)^2 \xrightarrow{d} \sigma_\eta^2 \int_0^1 [W_i(r) - \bar{W}_i]^2 dr
\]

Where $W_i(r)$ is a standard Brownian motion, $\bar{W}_i = \int_0^1 W_i(r) dr$.

Define the cross-sectional average process:
\[
\bar{x}_t = \frac{1}{N} \sum_{i=1}^N x_{it}
\]

Since $x_{it}$ are independent unit root processes, by the functional central limit theorem and properties of independent Brownian motions:
\[
\frac{1}{\sqrt{T}} \bar{x}_{\lfloor Tr \rfloor} = \frac{1}{N} \sum_{i=1}^N \frac{1}{\sqrt{T}} x_{i,\lfloor Tr \rfloor} \xrightarrow{d} \frac{1}{N} \sum_{i=1}^N \sigma_\eta W_i(r) = \frac{\sigma_\eta}{\sqrt{N}} W(r)
\]

Where $W_i(r)$ are independent standard Brownian motions, $W(r)$ is a standard Brownian motion.

Hence, the cross-sectional average process $\bar{x}_t$ satisfies:
\[
\frac{1}{\sqrt{T}} \bar{x}_{\lfloor Tr \rfloor} \xrightarrow{d} \frac{\sigma_\eta}{\sqrt{N}} W(r)
\]

Where $W(r)$ is a standard Brownian motion.

Therefore, for each individual $i$:
\[
\sum_{t=1}^T (x_{it} - \bar{x}_i)^2 = O_p(T^2)
\]

Then:
\[
\sum_{i=1}^N \sum_{t=1}^T (x_{it} - \bar{x}_i)^2 = N \times O_p(T^2) = O_p(NT^2)
\]

More precisely:
\[
\frac{1}{NT^2} \sum_{i=1}^N \sum_{t=1}^T (x_{it} - \bar{x}_i)^2 \xrightarrow{d} \frac{\sigma_\eta^2}{N} \sum_{i=1}^N \int_0^1 [W_i(r) - \bar{W}_i]^2 dr
\]

Now, we rewrite the numerator:
\[
\sum_{i=1}^N \sum_{t=1}^T (x_{it} - \bar{x}_i)(\lambda_t - \bar{\lambda}) = \sum_{t=1}^T (\lambda_t - \bar{\lambda}) \sum_{i=1}^N (x_{it} - \bar{x}_i) = N \sum_{t=1}^T (\lambda_t - \bar{\lambda})(\bar{x}_t - \bar{x})
\]

Where:
\[\bar{x} = \frac{1}{T} \sum_{t=1}^T \bar{x}_t\]

Then, we analyze the asymptotic properties of $\sum_{t=1}^T (\lambda_t - \bar{\lambda})(\bar{x}_t - \bar{x})$. By the functional central limit theorem:

\[\begin{gathered}
\frac{1}{\sqrt{T}} \bar{x}_{\lfloor Tr \rfloor}\xrightarrow{d} \frac{\sigma_\eta}{\sqrt{N}} W(r) \\
\frac{1}{\sqrt{T}} \sum_{t=1}^{\lfloor Tr \rfloor} (\lambda_t - \bar{\lambda})\xrightarrow{d} \sigma_\lambda B(r)
\end{gathered}\]

Where $B(r)$ is a standard Brownian motion, independent of $W(r)$.

Since:
\[\begin{gathered}\sum_{t=1}^T (\lambda_t - \bar{\lambda})(\bar{x}_t - \bar{x})=\sum_{t=1}^T (\lambda_t - \bar{\lambda})\bar{x}_t\\O_p((\lambda_t - \bar{\lambda})\bar{x}_t)=O_p(\lambda_t - \bar{\lambda})\times O_p(\bar{x}_t)=O_p(1)\times O_p(\sqrt{t})\\\sum_{t=1}^T\sqrt{t}\approx\int_0^T\sqrt{t}dt=\frac{2}{3}(T)^\frac{3}{2}\end{gathered}\]

Also, by the continuous mapping theorem:
\[
\frac{1}{T^{3/2}} \sum_{t=1}^T (\lambda_t - \bar{\lambda})(\bar{x}_t - \bar{x}) \xrightarrow{d} \frac{\sigma_\eta \sigma_\lambda}{\sqrt{N}} \int_0^1 [W(r) - \bar{W}] dB(r)
\]

Hence, for the stationary process $\lambda_t$ and unit root process $\bar{x}_t$:
\[
\frac{1}{T^{3/2}} \sum_{t=1}^T (\lambda_t - \bar{\lambda})(\bar{x}_t - \bar{x}) \xrightarrow{d} \frac{\sigma_\eta \sigma_\lambda}{\sqrt{N}} \int_0^1 [W(r) - \bar{W}] dB(r)
\]

Where $B(r)$ is a Brownian motion independent of $W(r)$, and: 
\[\sigma_\lambda^2 = \lim_{T\to\infty} \frac{1}{T} \text{Var}\left(\sum_{t=1}^T \lambda_t\right)\]

Therefore:
\[
\sum_{i=1}^N \sum_{t=1}^T (x_{it} - \bar{x}_i)(\lambda_t - \bar{\lambda}) = N \times O_p(T^{3/2}) = O_p(NT^{3/2})
\]

Combining the results:
\[
A = \frac{O_p(NT^{3/2})}{O_p(NT^2)} = O_p(T^{-1/2}) \xrightarrow{p} 0
\]

More precisely:
\[
\sqrt{T} A \xrightarrow{d} \frac{\sigma_\lambda \int_0^1 [W(r) - \bar{W}] dB(r)}{\sigma_\eta \times \frac{1}{N} \sum_{i=1}^N \int_0^1 [W_i(r) - \bar{W}_i]^2 dr}
\]

And for term $B$:
\[
B = \frac{\sum_{i=1}^N \sum_{t=1}^T (x_{it} - \bar{x}_i)(\epsilon_{it} - \bar{\epsilon}_i)}{\sum_{i=1}^N \sum_{t=1}^T (x_{it} - \bar{x}_i)^2}
\]

Since $\epsilon_{it}$ is independent of $x_{it}$ and $\epsilon_{it} \sim \text{i.i.d.}(0, \sigma_\epsilon^2)$, we compute the variance:
\[
\text{Var}\left(\sum_{i=1}^N \sum_{t=1}^T (x_{it} - \bar{x}_i)(\epsilon_{it} - \bar{\epsilon}_i)\right)
= \sum_{i=1}^N \sum_{t=1}^T \text{Var}\left((x_{it} - \bar{x}_i)(\epsilon_{it} - \bar{\epsilon}_i)\right)
\]

Hence:
\[\text{Var}\left(\sum_{i=1}^N \sum_{t=1}^T (x_{it} - \bar{x}_i)(\epsilon_{it} - \bar{\epsilon}_i)\right) = \sum_{i=1}^N \sum_{t=1}^T \mathbb{E}[(x_{it} - \bar{x}_i)^2] \text{Var}(\epsilon_{it} - \bar{\epsilon}_i)\]

For unit root processes: $\mathbb{E}[(x_{it} - \bar{x}_i)^2] = O_p(T)$ and $\text{Var}(\epsilon_{it} - \bar{\epsilon}_i) = O_p(1)$.

Therefore:
\[
\text{Var}\left(\sum_{i=1}^N \sum_{t=1}^T (x_{it} - \bar{x}_i)(\epsilon_{it} - \bar{\epsilon}_i)\right) = O_p(NT^2)
\]

And:
\[
\sum_{i=1}^N \sum_{t=1}^T (x_{it} - \bar{x}_i)(\epsilon_{it} - \bar{\epsilon}_i) = O_p(\sqrt{N}T)
\]

Finally, the denominator is $O_p(NT^2)$, then:
\[
B = \frac{O_p(\sqrt{N}T)}{O_p(NT^2)} = O_p\left(\frac{1}{\sqrt{N}T}\right) \xrightarrow{p} 0
\]

Combining the analysis:
\[
\hat{\beta} = \beta + A + B = \beta + O_p(T^{-1/2}) + O_p\left(\frac{1}{\sqrt{N}T}\right) \xrightarrow{p} \beta
\]

Therefore, when $T \to \infty$, even with omitted time fixed effects, the fixed effects estimator $\hat{\beta}$ remains consistent.

\end{proof}

\section{Policy and Export Inequality}\label{Appendix B}
Based on Figure 6, we conclude that exchange rate policy impacts provincial export competitiveness in two ways: First, it enhances overall export competitiveness, reflected by the increase in the mean of RCA index. Second, it reduces the degree of inequality in provincial export competitiveness development, reflected by the decrease in the variance of RCA index. Section 4.4 of this paper has already demonstrated the first proposition. This section will now attempt to provide relevant analysis and interpretation for the second proposition.

Figure 44 shows the changes in the standard deviation of the RCA index for China's manufacturing sector (across 30 provinces) from 2008 to 2021. We observe that the standard deviation of the RCA index began to decline after reaching its peak in 2014. Following the implementation of the exchange rate policy in 2016, the intensity of this decline accelerated significantly. This preliminary finding suggests that the policy has helped mitigate the inequality in the development of export competitiveness across provincial regions.
\begin{figure}[htbp]
\centering
\includegraphics[width=11cm]{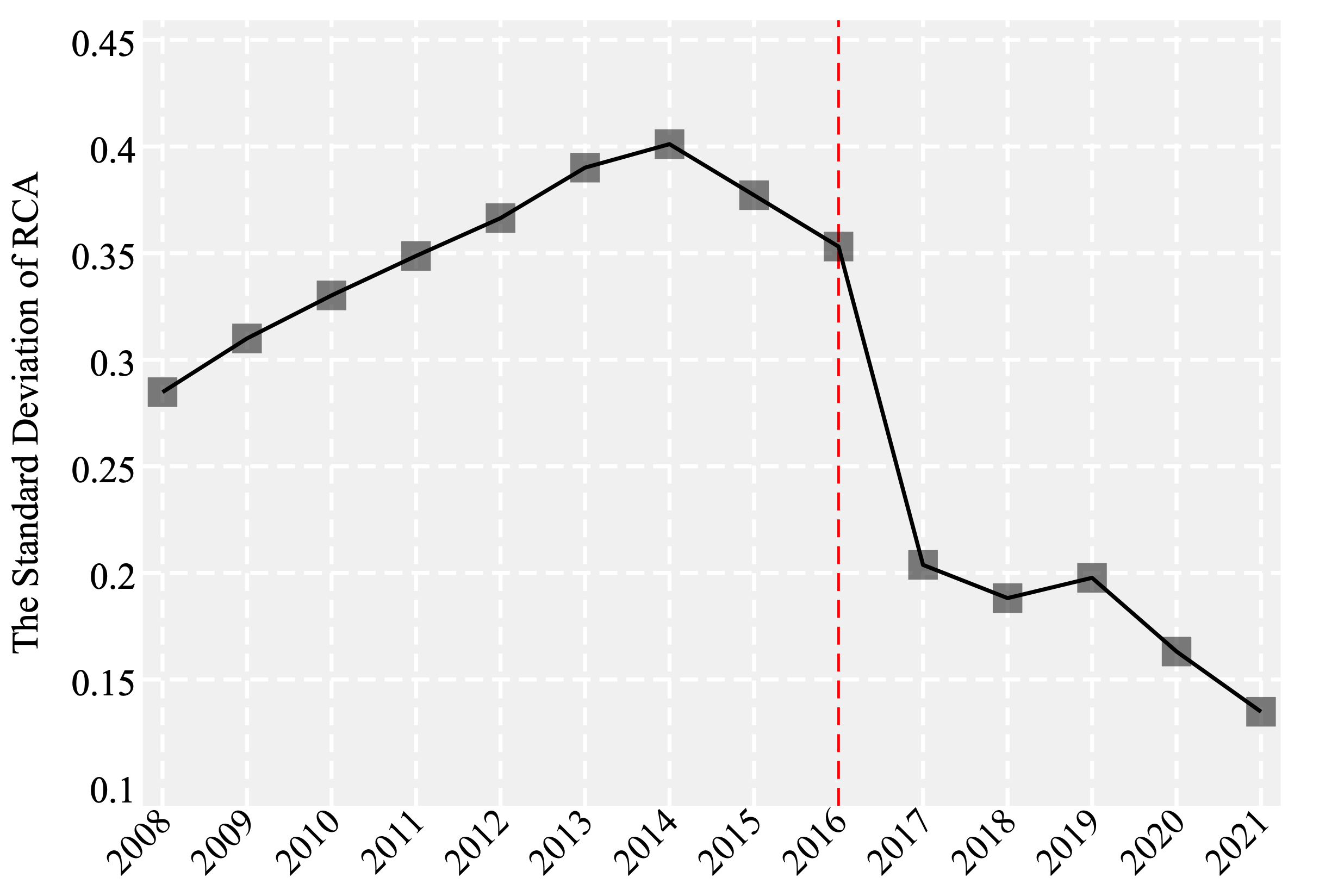}
\caption{\label{fig:F44}The Standard Deviation of RCA}
\end{figure}

Based on this foundation, we conducted empirical research. Given that the article utilizes long panel data, we constructed the provincial export competitiveness development inequality variable as follows to account for sample matching:
\[Y_{it}^{\Delta}=\left|RCA_{it}-\frac{\sum_{j=1}^nRCA_{jt}}{n}\right|\]

According to $Y_{it}^{\Delta}$, we define the variable measuring the inequality in provincial export competitiveness development at year $t$ as the absolute difference between the export competitiveness index $RCA_{it}$ for province $i$ at year $t$ and the average of all provincial export competitiveness indices nationwide for that year $\frac{\sum_{j=1}^{n}RCA_{jt}}{n}$. The remaining variables and grouping methods are referenced from Section 4.4. The experimental results are shown in the figures below.
\begin{figure}[htbp]
\centering
\includegraphics[width=8cm]{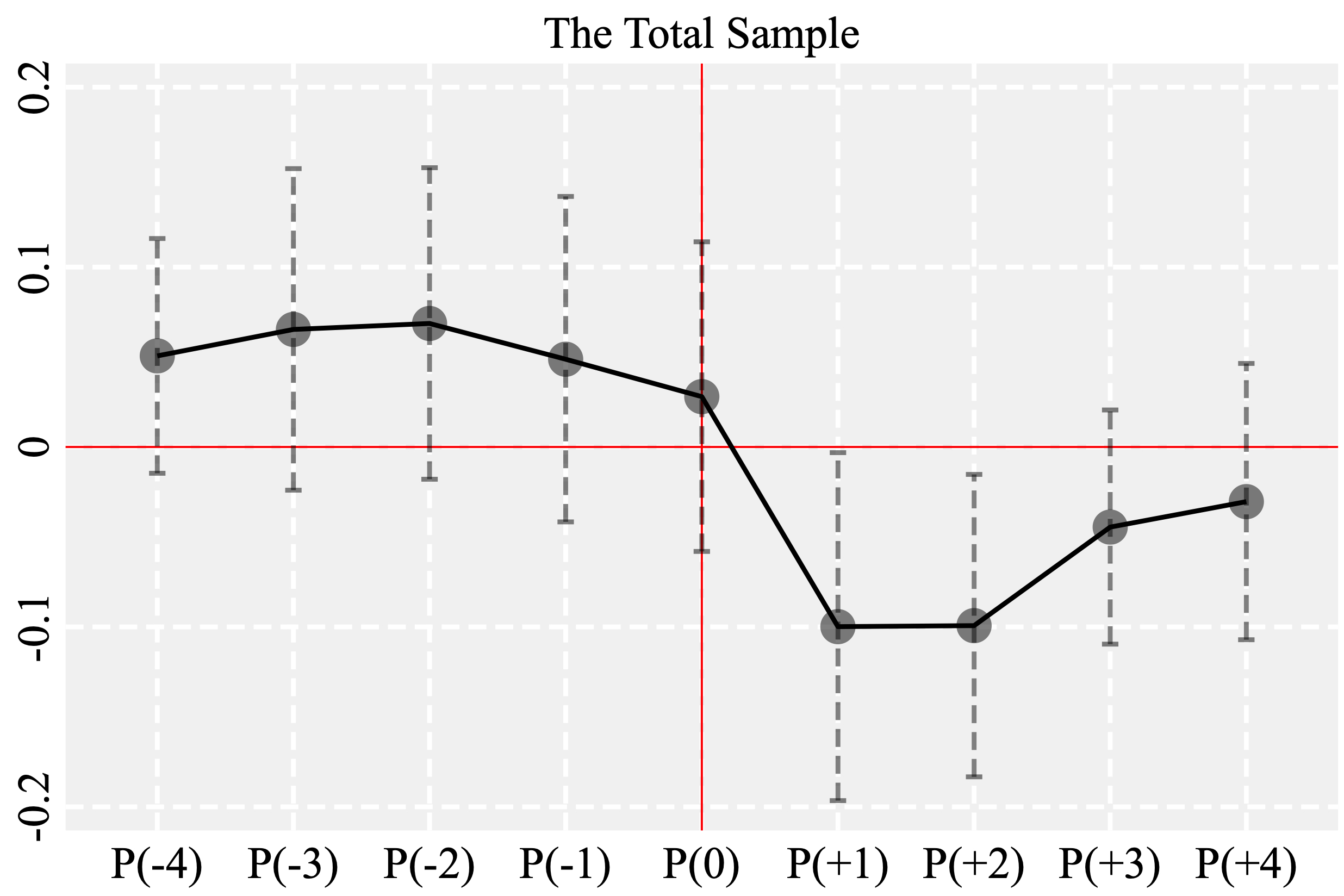}
\caption{\label{fig:F45}The Dynamic Effect of Exchange Rate Policy in Whole Sample}
\end{figure}

Figure 45 illustrates the impact of exchange rate policy on provincial export competitiveness development inequality across the entire sample. We find that the policy is effective, significantly reducing the degree of provincial export competitiveness development inequality (the coefficients for P($+1$) and P(+2) post-policy are both significant, with p-value($+1$)=0.043 and p-value($+2$)=0.022, both less than 0.05). However, the policy effect exhibits limited persistence (the coefficients for P($+3$) and P($+4$) post-policy are not significant).
\begin{figure}[htbp]
\centering
\includegraphics[width=15cm]{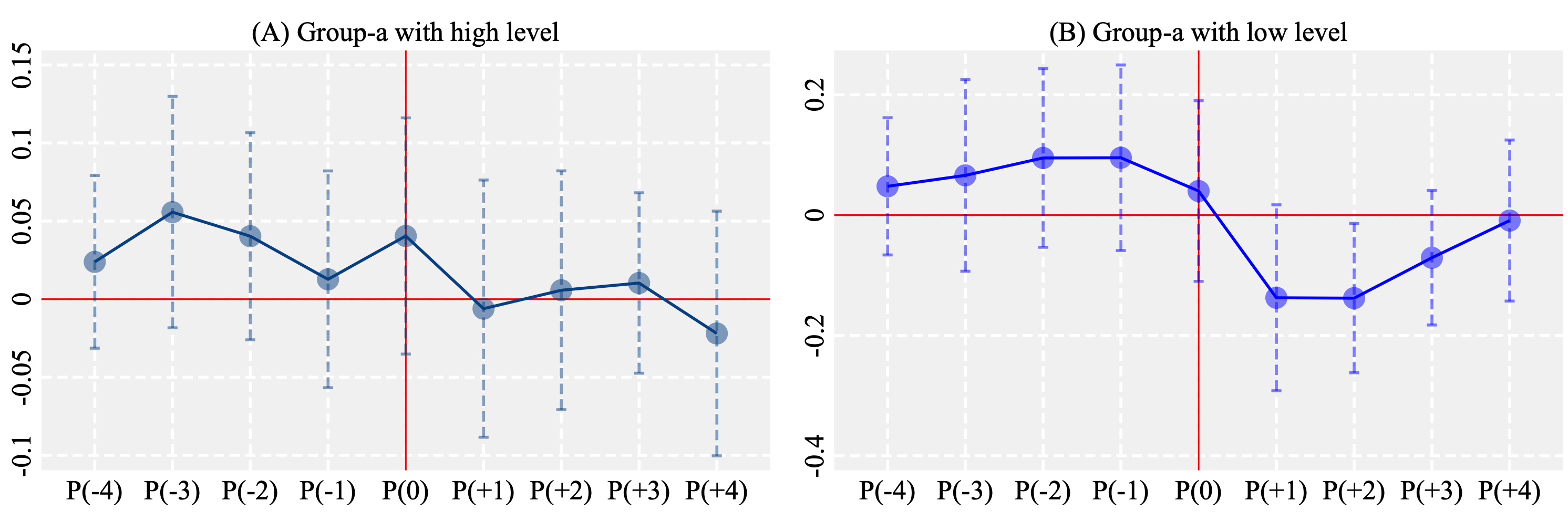}
\caption{\label{fig:F46}The Dynamic Effect of Exchange Rate Policy in Group-a}
\end{figure}
\vspace{-0.5cm}
\begin{figure}[htbp]
\centering
\includegraphics[width=15cm]{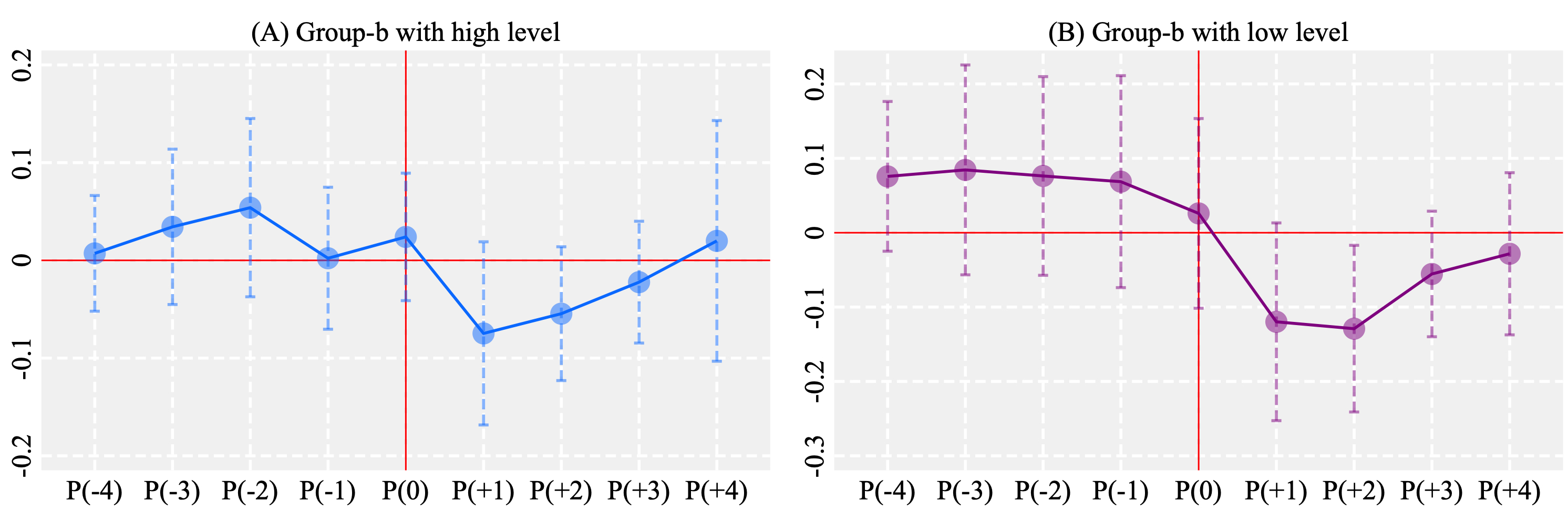}
\caption{\label{fig:F47}The Dynamic Effect of Exchange Rate Policy in Group-b}
\end{figure}

Figure 46 and Figure 47 present results using samples from Group-a and Group-b, respectively. Based on these two figures, we observe that in provinces with high innovation levels, the effectiveness of the policy is extremely terrible. It fails to significantly influence the inequality in the development of provincial export competitiveness. However, in provinces with low innovation levels, the policy remains effective. Although its effects exhibit a lagged characteristic compared to the full sample and persist for a relatively short duration, it can indeed significantly reduce the degree of inequality in provincial export competitiveness development. This aligns with the analytical logic presented in Section 4.4, indicating that in highly innovative regions, the exchange rate mechanism for export competitiveness is weak, and the exchange rate itself, along with exchange rate policy, no longer serves as a primary factor influencing export trade. Conversely, in regions with low innovation levels, the exchange rate mechanism for export competitiveness remains effective. For this section, we can conclude that exchange rate policy mitigates overall inequality in export competitiveness development by reducing the inequality in export competitiveness development in provinces with low innovation levels.

\section{The Parameters of Aggregate Labor Factor}\label{Appendix C}
\setcounter{table}{7}
\begin{table}[ht!]
\centering
\caption{The Parameters of Aggregate Labor Factor}
\label{8:parameters}
\small               
\setlength{\tabcolsep}{20pt} 
\renewcommand{\arraystretch}{1.7} 
\begin{tabular}{lll}
\toprule
Parameter & Notation & Value \\
\midrule
Dissipation Rate & $\mathcal{D}_{0}$ & 0.4 \\
Volatility of $l_t$ & $\sigma_{l}$ & 0.4 \\
Volatility of $s_{it}$ & $\sigma_{s}$ & 0.7 \\
Volatility of $p_t$ & $\sigma_{p}$ & 0.6 \\
Volatility Correlation of $s_{it}$ & $\theta_{s}$ & 0.5 \\
Volatility Correlation of $p_t$ & $\theta_{p}$ & 0.4 \\
Drift of $l_t$ & $\mu_{l}$ & 0.4 \\
\multirow{1}{*}{Labor Allocation Shock} & \multirow{1}{*}{$\mathbb{K}_{1} = \left\{ \mu_{s}^{i},\mu_{p}^{0} \right\}$} & (0.1,0.6) \\
 & & (0.5,0.6) \\
 & & (1,0.6) \\
 & & (2,0.6) \\
\multirow{1}{*}{Labor Value Shock} & \multirow{1}{*}{$\mathbb{K}_{2} = \left\{ \mu_{s}^{0},\mu_{p}^{i} \right\}$} & (0.6,0.1) \\
 & & (0.6,0.5) \\
 & & (0.6,1) \\
 & & (0.6,2) \\
\multirow{1}{*}{Stable Human Capital Shock} & \multirow{1}{*}{$\mathbb{K}_{3} = \left\{ \mu_{s}^{i},\mu_{p}^{i} \mid \mu_{s}^{i} = \mu_{p}^{i} \right\}$} & (0.1,0.1) \\
 & & (0.5,0.5) \\
 & & (1,1) \\
 & & (2,2) \\
\multirow{1}{*}{Positive Human Capital Shock} & \multirow{1}{*}{$\mathbb{K}_{3} = \left\{ \mu_{s}^{i},\mu_{p}^{i} \mid \mu_{p}^{i} > \mu_{s}^{i} \right\}$} & (0.1,0.2) \\
 & & (0.5,1) \\
 & & (1,2) \\
 & & (2,4) \\
\multirow{1}{*}{Negative Human Capital Shock} & \multirow{1}{*}{$\mathbb{K}_{3} = \left\{ \mu_{s}^{i},\mu_{p}^{i} \mid \mu_{s}^{i} > \mu_{p}^{i} \right\}$} & (0.2,0.1) \\
 & & (1,0.5) \\
 & & (2,1) \\
 & & (4,2) \\
\bottomrule
\end{tabular}
\end{table}

\newpage
\section{The Parameters of DSGE}\label{Appendix D}
\setcounter{table}{8}
\begin{table}[ht!]
\centering
\caption{The Parameters of DSGE}
\label{9:parameters}
\small               
\setlength{\tabcolsep}{20pt}
\renewcommand{\arraystretch}{1.7}
\begin{tabular}{lll}
\toprule
Parameter & Notation & Value \\
\midrule
Ability-Production Ratio & $\alpha_{k}$ & $0.8$ \\
Average Ability & $\lambda_{0}$ & $0.8$ \\
Aggregate Volatility & $\sigma_{\lambda}$ & $0.1$ \\
Consumption Share & $\gamma$ & $0.7$ \\
Substitute Elasticity of Product & $\theta$ & $0.8$ \\
Continuity of Ability & $\beta_{k}$ & $0.95$ \\
Cost Elasticity & $\kappa$ & $0.4$ \\
Initial State & $(L_{0},\lambda_{0}^{k},\tau_{i0})$ & $(1,1.5,1.5)$ \\
\multirow{1}{*}{Consumption Ratio Shock} & $\psi_{L}$ & $0.7$ \\
 & $\psi_{H}$ & $0.9$ \\
\multirow{1}{*}{Factor Scale Shock} & $v_{L}$ & $0.7$ \\
 & $v_{H}$ & $0.9$ \\
\multirow{1}{*}{Cost Shock} & $(\mu_{\tau}^{+},\sigma_{\tau})$ & $(0.02,0.02)$ \\
 & $(\mu_{\tau}^{-},\sigma_{\tau})$ & $(-0.02,0.02)$ \\
\multirow{1}{*}{Labor Factor Shock} & $\mu_{L}^{-2}$ & $0.001$ \\
 & $\mu_{L}^{-1}$ & $0.03$ \\
 & $\mu_{L}^{+0}$ & $0.06$ \\
 & $\mu_{L}^{+1}$ & $0.08$ \\
 & $\mu_{L}^{+2}$ & $0.1$ \\
\multirow{1}{*}{Sanction Policy Shock} & $(c_{A},\alpha_{A},\gamma_{A})$ & $(0.35,0.15,0.05)$ \\
 & $\sigma_{L}^{-}$ & $0.04$ \\
\multirow{1}{*}{Cooperation Policy Shock} & $(c_{B},\alpha_{B},\gamma_{B})$ & $(0.35,0.2,0.03)$ \\
 & $\sigma_{L}^{+}$ & $0.05$ \\
\bottomrule
\end{tabular}
\end{table}

\end{appendices}

\end{document}